 \documentclass[final, 3p, hyphens, sort&compress]{elsarticle}
\usepackage{mathrsfs}
\usepackage{amsmath}
\usepackage{amssymb}
\usepackage{multirow}
\usepackage{array}
\newcolumntype{P}[1]{>{\centering\arraybackslash}p{#1}}
\usepackage{longtable} 
\usepackage{pdfpages}
\usepackage{tikz}
\usepackage{color}
\usepackage{graphicx}
\usepackage{epstopdf}
\usepackage{epstopdf}
\usepackage{leftidx}
\usepackage{float}
\usepackage{subcaption}
\usepackage[utf8]{inputenc}
\usepackage{silence} 
\usepackage{babel}
\usepackage{lineno}
\usepackage{comment}
\usepackage{upgreek}
\usepackage{dsfont}
\usepackage{textcomp}
\usepackage{gensymb}
\usepackage{mleftright} 
\usepackage{placeins} 

\DeclareUnicodeCharacter{0302}{é}

\usepackage{ltablex}
\usepackage{booktabs}
\usepackage{pdflscape}
\keepXColumns

\usepackage[utf8]{inputenc}   
\UseRawInputEncoding    

\usepackage{setspace}  
\AtBeginEnvironment{thebibliography}{%
\small
\setstretch{1}
}
\usepackage{xcolor}
\newcommand*{\colorboxed}{}
\def\colorboxed#1#{%
	\colorboxedAux{#1}%
}
\newcommand*{\colorboxedAux}[3]{
	\begingroup
	\colorlet{cb@saved}{.}%
	\color#1{#2}%
	\boxed{%
		\color{cb@saved}%
		#3%
	}%
	\endgroup}

\definecolor{princetonorange}{rgb}{1.0, 0.56, 0.0}

\usepackage{empheq}

\journal{Physics Reports}
\usepackage{units}
\usepackage{enumerate}
\usepackage{url}
\usepackage[colorlinks]{hyperref}
\hypersetup{plainpages=true, 
    breaklinks=true,
	hypertexnames=false, 
	pageanchor=true,
	colorlinks=true,
	linkcolor={blue},
	citecolor={red},
	urlcolor={blue},
	anchorcolor={black}}
\hypersetup{pdfauthor={Name}}

\newcommand{\bra}[1]{\left\langle{#1}\right|}
\newcommand{\ket}[1]{\left|{#1}\right\rangle}

\renewcommand{\eqref}[1]{\mbox{Eq.~(\ref{#1})}}

\newcommand{\figpanel}[2]{Fig.~\hyperref[#1]{\ref*{#1}(#2)}}
\newcommand{\figpanels}[3]{Fig.~\hyperref[#1]{\ref*{#1}(#2)-(#3)}}
\newcommand{\figpanelNoPrefix}[2]{\hyperref[#1]{\ref*{#1}(#2)}}

\newcommand{\be}{\begin{equation}}
\newcommand{\ee}{\end{equation}}
\newcommand{\bea}{\begin{eqnarray}}
\newcommand{\eea}{\end{eqnarray}}
\def\87rb{$^{87}$Rb}

\begin{document}
\begin{frontmatter}

\title{Photonic quantum information with time-bins: Principles and applications}

\author[FirstAddress]{Ashutosh Singh\,\href{https://orcid.org/0000-0001-9628-6127}{\includegraphics[height=2ex]{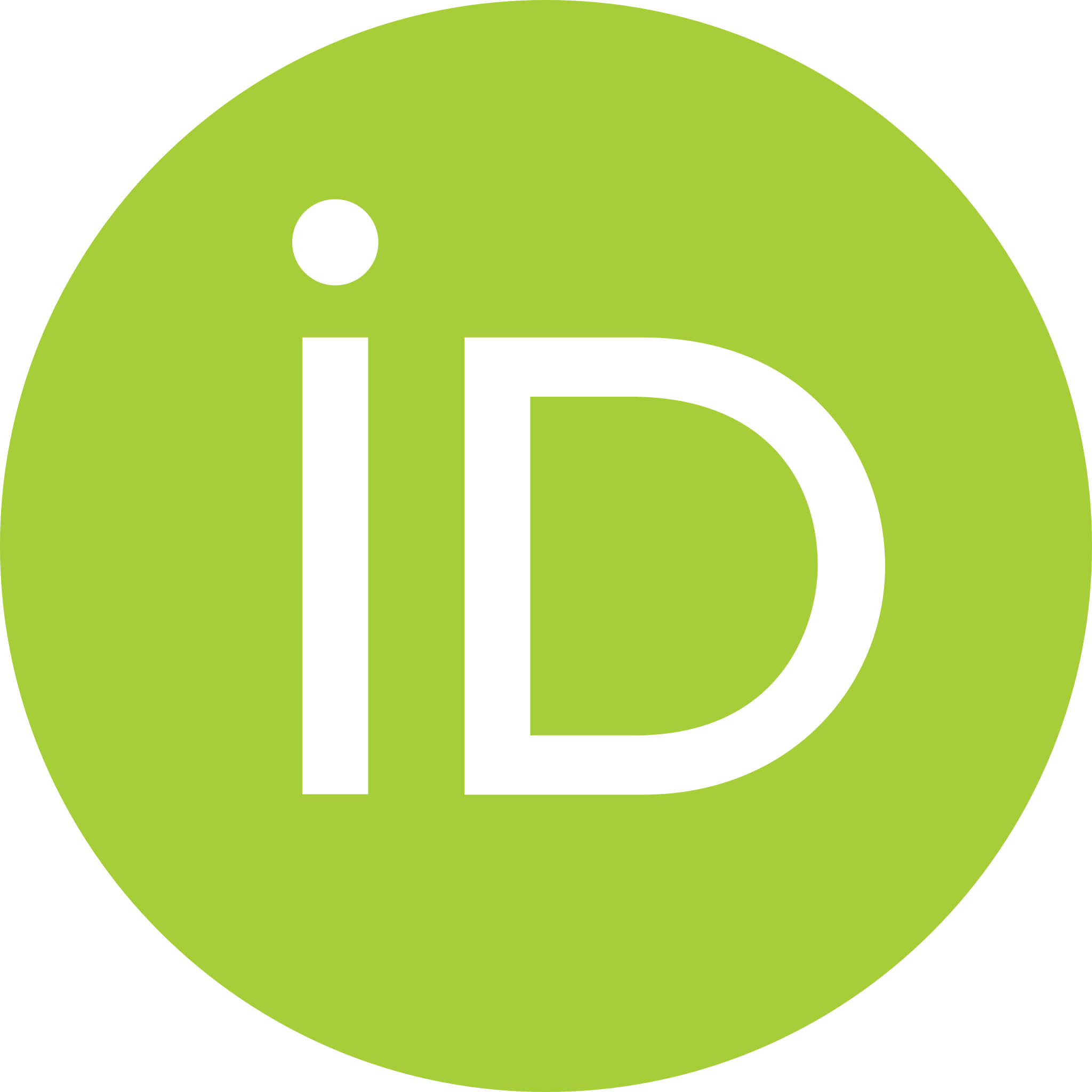}}}
\author[FirstAddress]{Anuj Sethia\,\href{https://orcid.org/0000-0002-7140-3662}{\includegraphics[height=2ex]{orcid.png}}}
\author[FirstAddress]{Leili Esmaeilifar\,\href{https://orcid.org/0000-0001-6557-2550}{\includegraphics[height=2ex]{orcid.png}}}
\author[SecondAddress]{Raju Valivarthi\,\href{https://orcid.org/0000-0002-5422-9340}{\includegraphics[height=2ex]{orcid.png}}}
\author[]{Neil Sinclair$\,\href{https://orcid.org/0000-0003-2348-9497}{\includegraphics[height=2ex]{orcid.png}}^\text{b,c}$}
\author[SecondAddress]{\hspace{25pt}Maria Spiropulu\,\href{https://orcid.org/0000-0001-8172-7081}{\includegraphics[height=2ex]{orcid.png}}}		

 \author[FirstAddress]{Daniel Oblak\,\href{https://orcid.org/0000-0002-0277-3360}{\includegraphics[height=1.5ex]{orcid.png}}\corref{cor}}
 \cortext[cor]{Corresponding authors}
 \ead{doblak@ucalgary.ca}

\address[FirstAddress]{Department of Physics and Astronomy \& Institute for Quantum Science and Technology, University of Calgary, AB T2N 1N4, Canada.}
\address[SecondAddress]{Division of Physics, Mathematics and Astronomy, California Institute of Technology, 1200 E California Blvd., Pasadena, California 91125, USA.}
\address[ThirdAddress]{John A. Paulson School of Engineering and Applied Sciences, Harvard University, 29 Oxford St., Cambridge, Massachusetts 02138, USA.}

\date{\today}

		
\begin{abstract}
Long-range quantum communication, distributed quantum computing and sensing applications require robust and reliable ways to encode the transmitted quantum information. In this context, time-bin encoding has emerged as one of the most promising candidates due to its resilience to mechanical and thermal perturbations, depolarization induced by refractive index changes, and birefringence in fiber optic media. Time-bin quantum bits (qubits) can be produced in a variety of ways, and each experimental implementation calls for different considerations for design parameters, compatibility of optical, electrical, and electro-optical components, as well as the measurement and characterization procedure. 
Here, we provide a comprehensive overview of different experimental methods for preparing and characterizing time-bin qubits (TBQs) in the context of quantum communication protocols, along with an assessment of their advantages and limitations. We discuss challenges in transmitting TBQs over optical fibers and free-space quantum channels, and methods to overcome them, analyze the selection of key parameters of the time-bins, and requirements of components in various experiments. This leads us to explore the preparation and characterization of time-bin entanglement and examine the requirements for interference of time-bins from separate sources. Further, we discuss the preparation and characterization techniques for high-dimensional time-bin states, namely, qudits, and the generation of time-bin entangled qudit pairs. Next, we review the concept of time-energy entanglement, highlighting key experimental realizations. Finally, we present an overview of notable applications of time-bin encoded quantum states, from important quantum communication protocols to photonic quantum computation, before providing our concluding summary and outlook. 
This work provides an accessible introduction for students and researchers planning to employ time-bin encoding in experiments or modelling. It also offers a comprehensive overview of the field, including recent advancements, to help readers stay informed.
\end{abstract}

\end{frontmatter}



\newpage
\tableofcontents

\newpage
\listoffigures

\newpage

\section{Aim and Scope} 
\label{Sec_Scope}
\FloatBarrier

Widespread development of quantum networks will enable a suite of information-theoretically secure communication protocols and the pooling of valuable quantum computing resources across large distances. It is on this backdrop that numerous partnerships between government, academic institutions and industry have started to realize quantum network testbeds across the globe. Along with these, the research and development of technologies and methods for quantum communication are increasing significantly and advancing to the stage at which more and more researchers unfamiliar with the principles of quantum mechanics, e.g., communication and computer engineers, are entering the field. Concise and accurate reviews on key methods and technologies are of value to both students and professionals who want to begin their journey into the rapidly growing and seemingly counterintuitive field of quantum communication.

A key aspect of quantum communication is the encoding of information into quantum bits, a.k.a. qubits. Several approaches to qubit encoding exist, but so-called time-bin qubits (TBQs) are ideally suited for communication through optical fiber, because, unlike polarization or spatially encoded qubits, they are robust against depolarization and spatial mode-mixing effects prevalent in the fiber. As a consequence, TBQs have been employed in many of the recent experimental breakthroughs in the field of quantum communication. With the growth of research and development in quantum information, the number of researchers, engineers, and students actively working with tools and equipment for time-bin generation and detection will increase over the coming years. This paper intends to provide an accessible and detailed overview of the different practical considerations for the preparation, characterization, and manipulation of TBQs in the context of some of the most common quantum information applications. 

The remainder of the paper is structured as follows: In Section (\ref{Sec_Introduction}), we explore various encoding schemes using different degrees of freedom of a single photon, highlighting their advantages and disadvantages. 
We define the TBQ state and outline an experimental method for generating them from a single-photon source. Section (\ref{Sec_Preparation}) reviews techniques for generating TBQs using both single-photon and weak coherent pulse sources. 
We also briefly discuss weak-coherent pulse sources as they are prerequisites for most practical TBQ sources and their associated photon statistics. 
In Section (\ref{Transmission}), we address challenges related to qubit transmissions, such as losses and dispersion, and discuss potential solutions. Section (\ref{Sec_Measurement}) presents methods for characterizing and measuring TBQs. We cover various aspects related to the parameter selection of time-bins in Section (\ref{Sec_ParameterSelection}). 
In Section (\ref{Sec_TBE}), we outline methods for generating and measuring bipartite and multipartite time-bin entanglement. We discuss the Hong-Ou-Mandel (HOM) effect and Bell-state measurement (BSM), which rely on two-photon interference of time-bin encoded states generated from independent sources in Section (\ref{Sec_Interference}). 
We discuss the preparation and characterization techniques for high-dimensional time-bin states -- namely, qudits -- and the generation of time-bin entangled qudit pairs in Section (\ref{TBQudits}), while Section (\ref{Time-Energy}) provides an overview of time-energy entanglement. 
We review applications of time-bins in quantum information applications, such as a few time-bin quantum key distribution (QKD) protocols, and entanglement swapping and teleportation in Section (\ref{Applications}). 
We also discuss and compare their advantages and limitations in relation to polarization encoding schemes and briefly shed light on quantum computing and quantum networks. Finally, in Section (\ref{Conclusion}), we summarize the key takeaways and present concluding remarks.


\section{Introduction} 
\label{Sec_Introduction}
\FloatBarrier

\begin{figure}[htbp!]
    \centering
    \begin{subfigure}{0.47\textwidth}
        \centering
      \includegraphics[width=\textwidth]{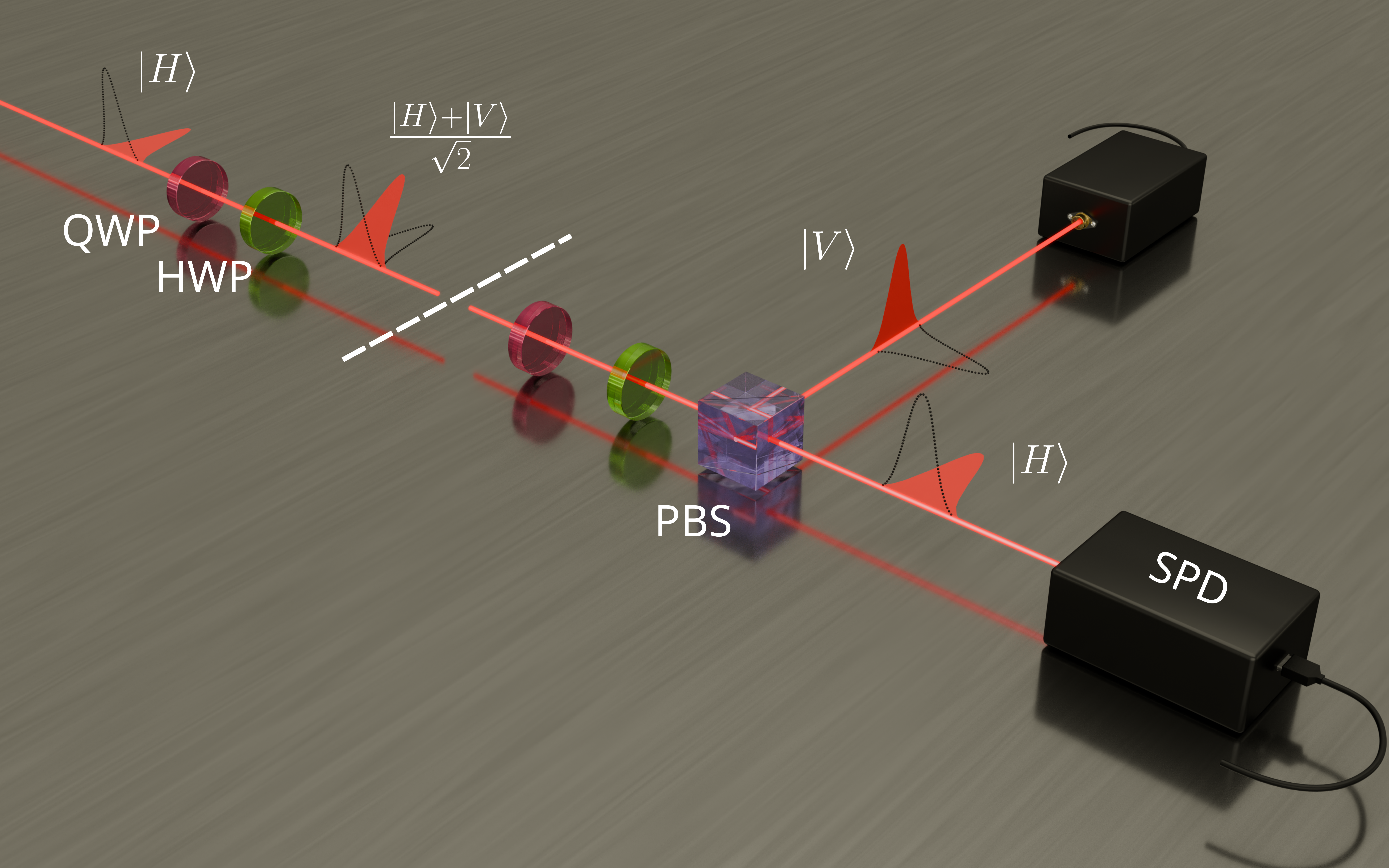} 
        \caption{Polarization encoding: $|0\rangle \equiv |H\rangle,|1\rangle \equiv |V\rangle$.}
        \label{fig:figure1}
    \end{subfigure}
    \begin{subfigure}{0.47\textwidth}
        \centering
        \includegraphics[width=\textwidth]{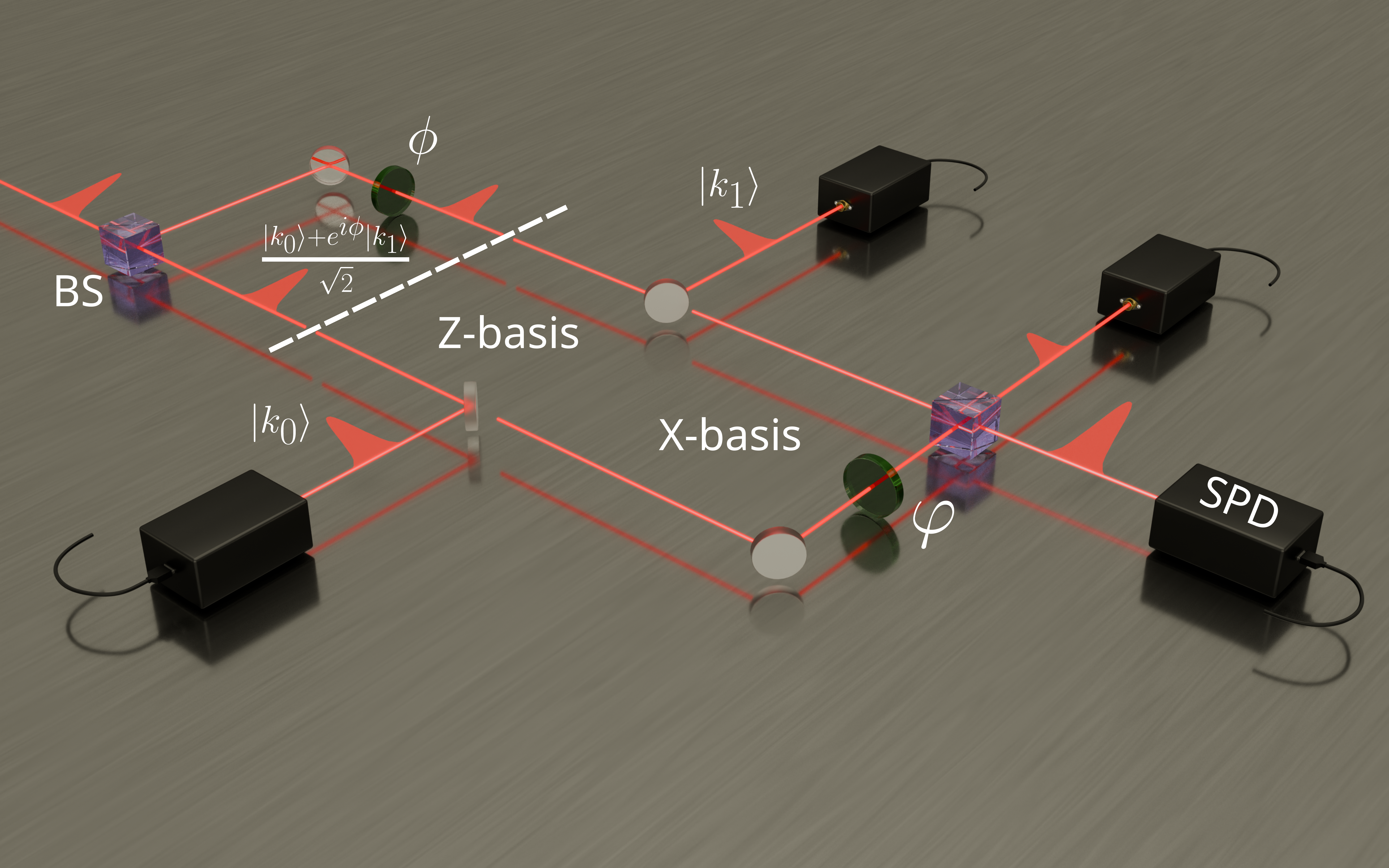}
        \caption{Path encoding: $|0\rangle \equiv |
        \mathbf{k_0}\rangle, |1\rangle \equiv |  \mathbf{k_1}\rangle$,...}
        \label{fig:figure2}
    \end{subfigure}
    \vspace{0.5cm}
    \begin{subfigure}{0.47\textwidth}
        \centering
        \includegraphics[width=\textwidth]{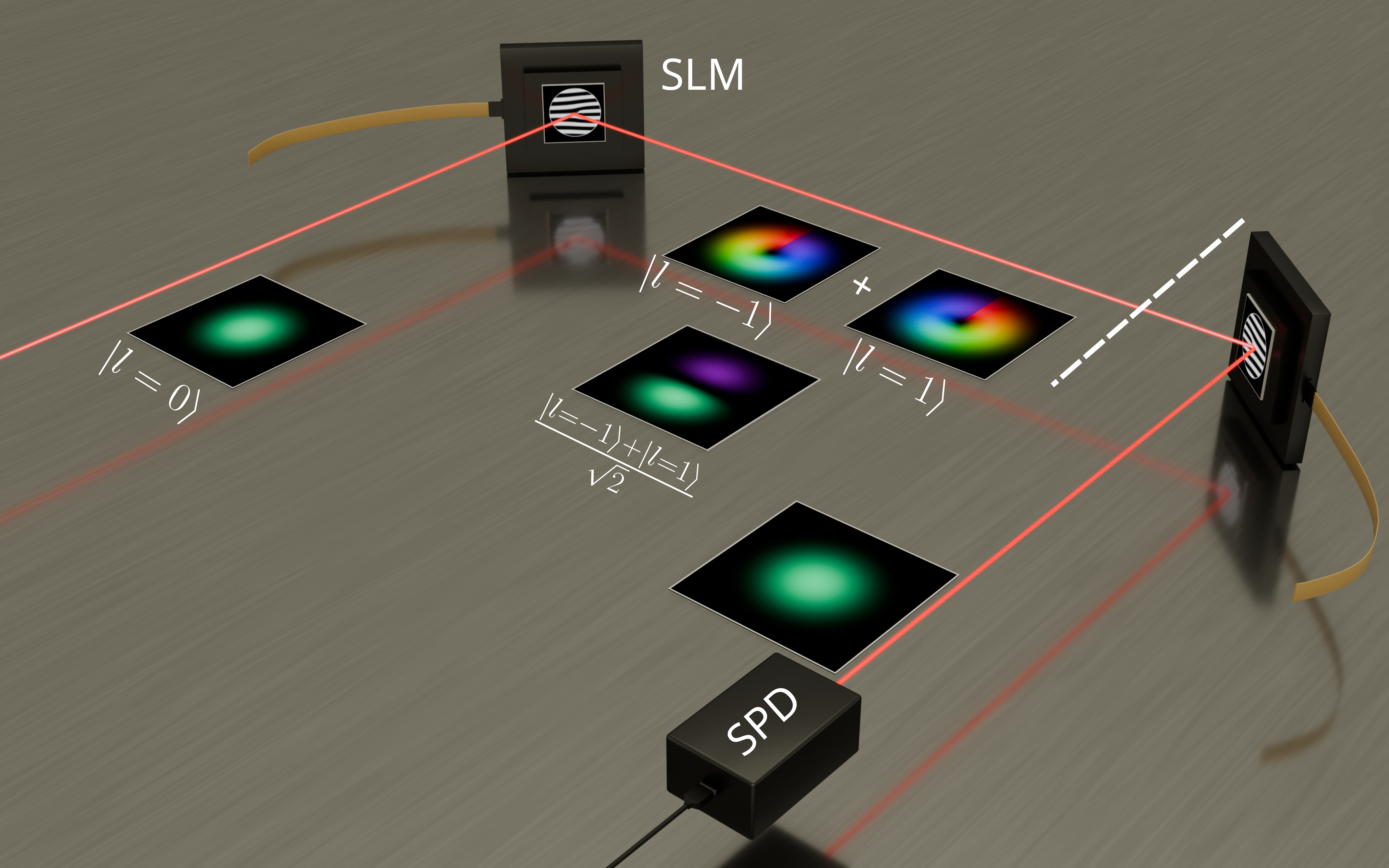} 
      \caption{OAM state encoding: $|0\rangle \equiv |l=0\rangle, |1\rangle \equiv |l=1\rangle$,...}
        \label{fig:figure3}
    \end{subfigure}
    \vspace{-1.5 em}
        \begin{subfigure}{0.47\textwidth}
        \centering 
         \includegraphics[width=\textwidth]{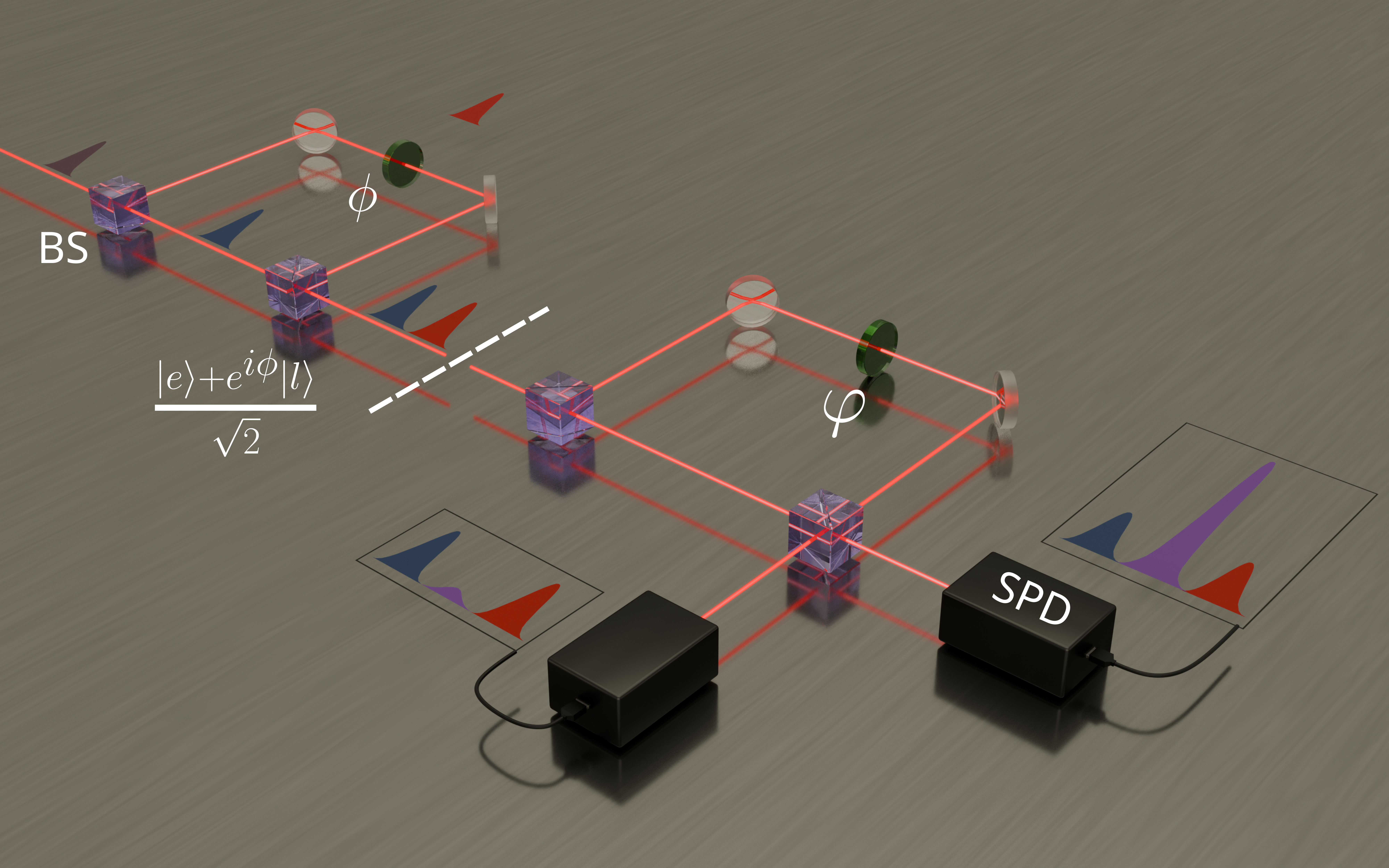}
       \caption{Time-bin encoding: $|0\rangle \equiv |t_0\rangle, |1\rangle \equiv |t_1\rangle$,...}
        \label{fig:figure4}
    \end{subfigure}
     \begin{subfigure}{0.95\textwidth}
        \centering
   \includegraphics[width=\textwidth]{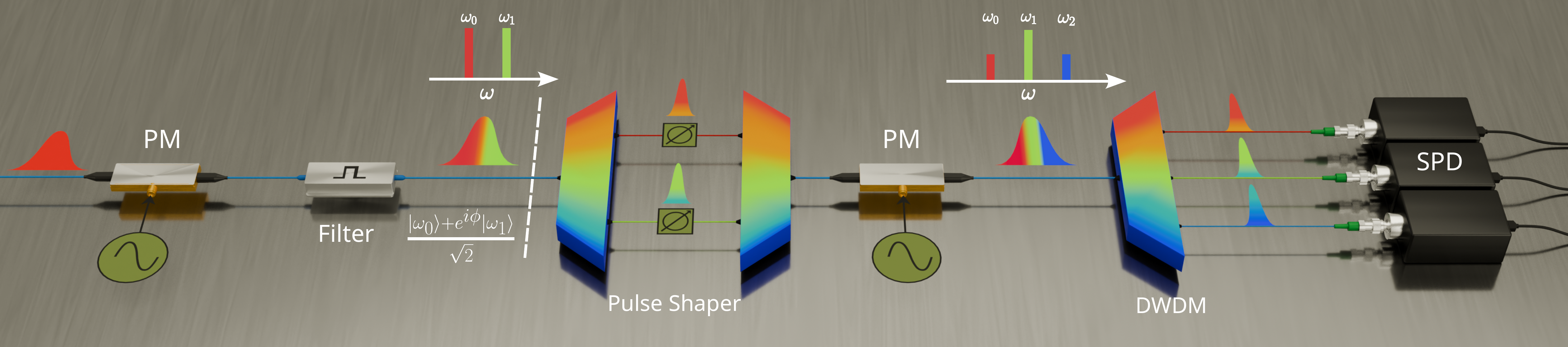}
        \caption{Frequency-bin encoding: $|0\rangle=|\omega_0\rangle, |1\rangle = |\omega_1\rangle$,...}
        \label{fig:figure5}
    \end{subfigure}
    \caption[Schematic of different photonic quantum information encoding schemes.]{Schematic illustration of preparation and measurement systems for different photonic quantum information encoding schemes \cite{QIP_Flamini2018, FB-QI_Lu2023}. A dashed white line separates preparation and measurement setups. BS: Beam splitter, DWDM: Dense wavelength division multiplexer, HWP: Half-wave plate, QWP: Quarter-wave plate, PBS: Polarizing beam splitter, SLM: Spatial light modulator, SPD: Single-photon detector.}
    \label{fig_TBencod}
\end{figure}

Photons, being chargeless and massless particles \cite{Photon_RoyChoudhuri2008}, interact minimally with their environment and can travel long distances \cite{QCom_Juan2017} at the speed of light with little attenuation \cite{Attn_Giggenbach2022}. This makes photons ideal quantum information carriers for long-distance quantum communication networks \cite{QNetwork_Liao2018, QNetwork_Wei2022, QCom_Gisin2007, QCom_Gisin2009, QNetwork_Bassoli2021, QNetwork-BBM92_Fitzke2022, QInt_Kumar2025}, as they transmit information reliably at maximum speed \cite{QIP_Yamamoto2005, QIP_Obrien2009, QIP_Slussarenko2019, QIP_Benyoucef2020} with minimal decoherence from thermal noise \cite{QDecoherence_Zurek2003}. 
Photons also offer several degrees of freedom associated with their spatial distribution, spin or orbital angular momentum, time, and frequency. 
These photonic degrees of freedom are employed to encode information \cite{QIP_Flamini2018, QIP_AbuGhanem2024} in the quantum state of a single photon, resulting in encoding schemes such as the path or spatial mode \cite{Spatial_Solis2013, Spatial_Lima2009}, polarization \cite{PolEncod_Stenholm1996}, orbital angular momentum (OAM) \cite{OAM_Bliokh2015}, time-bin \cite{Ent_Brendel1999}, and frequency-bin \cite{Frequency_Reimer2016, QIP-Freq_Lukens2017, QIP-Freq_Lukens2019}, which are useful in different quantum information processing applications \cite{Book-QIP_chuang, Book-QIP_wilde, Book-QIP_Kok, QCom_Krenn2017}. 
Examples of a few encoding schemes are depicted in Fig. (\ref{fig_TBencod}).  As such, they provide a promising platform for the implementation of quantum information processing (QIP) protocols \cite{QIP_Bennett2000, QIP_Tittel2001, SecQRevolution_MacFarlane2003, QIP_Walmsley2005, QIP_Rudolph2017, QComp_Madsen2022} such as quantum computation \cite{QComp_Knill2001, QComp_Kok2007, QComp_Jeremy2007, QComp_Ladd2010, QComp-TBQ_Humphreys2013,  QComp-Photonic_Zhong2020, QComp_Caleffi2022, QComp_Omkar2022, QComp_Rudolph2023}, QKD \cite{QKD_Gisin2002}, teleportation \cite{Teleportation-Review_Pirandola2015}, superdense coding \cite{SupDens_Bennett1992, SupDens_Mattle1996}, secret sharing \cite{QSS_Hillery1999, QSS_Gaertner2007}, quantum internet \cite{Quinternet_Wehner2018, Quinternet_Rohde2021}, metrology \cite{QMetrology_Marco2022, QMetrology_Emanuele2020}, sensing \cite{QSensing_Degen2017, QSensing_Pirandola2018}, and imaging \cite{QImaging_Berchera2019, QMSI_Dowling2015}.

All the aforementioned encoding schemes exhibit specific advantages and intrinsic limitations \cite{HD-QC_Daniele2019}, particularly in terms of the stability of their basis eigenstates, the feasibility of state preparation, manipulation, and measurement, and their scalability to higher-dimensional Hilbert spaces, as discussed in the following section.

\subsection{Overview of photonic encoding schemes}

Polarization encoding (Fig.~\ref{fig:figure1}), in which quantum information is encoded in orthogonal polarization states of a single photon (e.g., horizontal/vertical or left/right circular polarization), has been used in many free-space experiments due to the ease of operation and abundance of high-quality and cost-effective components for precise state preparation, manipulation, and measurement \cite{Book-QIP_chuang, Book-QIP_wilde} of the polarization state of light \cite{POlQubit_Nicholas2003}. These tasks are typically realized using standard bulk or integrated optical elements such as wave plates, polarizers, beam splitters, and polarization-maintaining components. State preparation, manipulation, and measurement are thus straightforward with these optical tools. However, the photonic polarization degree of freedom offers only two orthogonal basis states; hence, limiting it to realizing qubits \cite{QIP_Tittel2001, Ent-Qdot_Weihs2017}. In contrast, all other encoding schemes mentioned above allow the formation of higher-dimensional quantum systems, so-called qudits (i.e. $d>2$ dimensional states). Moreover, polarization qubits suffer from depolarization or transformation when transmitted through optical fibers or non-stationary free-space channels, respectively, requiring active polarization compensation to recover the state faithfully \cite{PolComp_Peranic2023, PolComp_Alexander2024, PolComp_Niklas2025}. Additionally, the interaction between light and matter is sensitive to photon polarization, posing challenges for polarization qubits in interfacing with atom/matter qubits \cite{LightAtom_Hammerer2010}.

To overcome the dimensionality limitation of polarization encoding, path encoding provides a straightforward approach to realizing high-dimensional quantum states.
Path encoding (Fig.~\ref{fig:figure2}), in which quantum information is encoded in distinct spatial modes or optical paths of a single photon, with each path representing a basis state, is one of the first schemes used for the generation and manipulation of high-dimensional quantum systems \cite{Path_Zukowski1997, Path_Wang2018}. 
In 1997, \.{Z}ukowski et. al. \cite{Path_Zukowski1997} demonstrated how combinations of multiport beam-splitters (BS) could be engineered to exhibit non-classical correlations in higher-dimensional path degree of freedom. State preparation and measurement are typically performed using networks of beam splitters, phase shifters, interferometers, and single-photon detectors (SPDs) to coherently address and resolve individual spatial modes. A significant benefit is that linear optics allows for the implementation of universal operations on path-encoded multidimensional quantum systems, regardless of the dimension \cite{UniversalLO_Carolan2015}. It is also compatible with photonic integrated circuits, making it a popular choice for generating qudits. As path encoding requires a distinct physical waveguide or fiber for each encoded state, it becomes resource-intensive as the number of states increases.

While path encoding allows high-dimensional encoding in spatial modes, orbital angular momentum (OAM) provides an alternative by exploiting the helical phase structure of photons.
The OAM states of light (Fig.~\ref{fig:figure3}) are often utilized to generate high-dimensional quantum states \cite{OAM_Allen1992, OAM_Yao2011, OAM_Molina2007, OAM_Erhard2018}. Photons carrying a non-zero OAM are characterized by a helical phase factor $e^{il\phi}$, where $l$ is the quantum number indicating the amount of OAM $l\hbar$ carried by the photon and $\phi$ is the angle in the plane transverse to the propagation direction. Since $l$ is an unbounded integer, effectively controlling it allows OAM to form a discrete basis for generating high-dimensional states. A range of wave-shaping devices can be employed to modify the wavefront of an initial Gaussian photon, leading to the creation of qudits encoded in orbital angular momentum. Some of these tools include cylindrical lenses, optical cavities, spiral phase plates, holograms or q-plates, and spatial light modulators (SLMs) \cite{QIP_Flamini2018, HD-QC_Daniele2019}. State preparation, manipulation, and measurement are typically realized using wavefront-shaping elements in combination with mode-sorting and projective measurement techniques, such as SLM-based analyzers and interferometric OAM mode sorters. Qubits encoded in OAM states are insensitive to the rotation of the reference frame, making them alignment-free in free-space quantum communication experiments \cite{QCom_Dambrosio2012, QCloning_Bouchard17}. However, integrating the generation and manipulation devices on-chip is challenging, limiting scalability. Also, higher-order OAM states are not well-suited for communication over low-loss single-mode fiber-optic channels; however, transmission of OAM-encoded photons has recently been demonstrated in air-core optical fibers \cite{OAM-ACF_Gregg2015, OAM-HD_Cozzolino2019}.

For long-distance fiber-based quantum communication, temporal degrees of freedom such as time-bin encoding offer both robustness and compatibility with single-mode fibers.
Time-bin encoding (Fig.~\ref{fig:figure4}) is a widely employed technique for generating and distributing qubits and high-dimensional quantum states \cite{Ent-dist_Marcikic2004, HD-QKD_Vagniluca2020}. Its advantages include resilience to environmental effects such as mechanical and thermal perturbations, depolarization induced by refractive index variations, and birefringence in optical fibers. State preparation and measurement are typically performed using unbalanced interferometers, optical modulators, and single-photon detectors. However, one limitation of time-bin encoding is that increasing the dimensionality reduces the maximum repetition rate of the generated qudit states. Notably, some attempts have been made to demonstrate the dimensionality scaling without sacrificing the repetition rate of time-bin entangled qudits \cite{HD-QKD_Yu2025}. Among these encoding schemes, only frequency and time-bin encoding are both high-dimensional and robust while remaining compatible with single-mode fiber transmission. In contrast, OAM and path encodings, though high-dimensional, require spatially multimodal channels.

Complementary to the time-bin, frequency-bin encoding provides a complementary approach, leveraging the spectral domain to achieve similar goals of high-dimensional encoding in a manner that is fully compatible with telecom infrastructure. Frequency-bin encoding (Fig.~\ref{fig:figure5}) makes use of the discrete spectral modes accessible to a single photon (photonic wavepacket) -- restricted to a single spatial mode and polarization mode -- to encode quantum information across the frequency degree of freedom \cite{FB-Ent_Ramelow2009, FB-Ent_Olislager2010, FB-Ent_Morrison2022, FB-Ent-Qudit_Borghi2023, FB-Ent_Tagliavacche2025, FB-Ent_Vinet2026}. A qubit/qudit is encoded in a coherent superposition of frequency bins of fixed spacing, determined by, e.g., an electro-optic modulator or an optical cavity. State preparation, manipulation, and measurement can typically be accomplished using electro-optic modulators, pulse shapers \cite{PulseShaping_Weiner2000, PulseShaping_Weiner2011} and programmable spectral filters, allowing dynamic reconfiguration of the accessible spectral modes. Telecom-wavelength frequency bins are intrinsically compatible with standard single-mode fiber and dense wavelength-division multiplexing (DWDM), allowing for low-loss transmission with high stability \cite{FB-QI_Lu2023}. Frequency bins are also inherently immune to drifts in polarization and relatively robust against mechanical perturbations, and so exhibit excellent phase stability for long-distance propagation. High-dimensional frequency-bin entangled states can readily be produced through spontaneous parametric down-conversion (SPDC) or spontaneous four-wave mixing (SFWM), where conservation of energy ensures natural frequency correlations \cite{FB-HD_Kues2017}. However, despite these favourable properties, practical and efficient methods to perform arbitrary manipulation and analysis of frequency-bin states have proven more elusive. Initial demonstrations relied on nonlinear frequency conversion processes (sum- and difference-frequency generation in $\chi^{(2)}$ media) to directly map between frequency bins. Such schemes require strict phase-matching conditions and can suffer from poor flexibility and scalability due to the need for high pump powers. Techniques based on electro-optic modulation and pulse-shaping can alleviate many of these drawbacks but suffer from insertion loss and added structure from unwanted sidebands, especially when high-order modulation is required to simultaneously address widely spaced bins. Such schemes also tend to rely on cascaded, well-optimized components, resulting in loss buildup. For this reason, methods to directly and efficiently manipulate frequency-bin encoded information are still limited, and coherent manipulation and projective measurement of frequency-bin entangled states remain significantly more challenging than generation. Available techniques are ultimately limited by modulator bandwidth, spectral resolution, detector timing jitter, and total loss, which together constrain the achievable dimensionality and gate fidelity in practice \cite{FB-CNOT_Lu2019}.

\subsection{Time-bin encoding}
\label{Sec_TBEncoding}

The use of the temporal degree of freedom for encoding quantum information was first proposed by Franson in 1989, in the context of testing quantum nonlocality via time-energy entanglement \cite{TE_Franson1989}. The discrete version of this encoding, i.e., time-bin encoding, was later introduced by Brendel et al. in 1999 \cite{Ent_Brendel1999} through the demonstration of pulsed time-energy entangled twin-photon generation, now commonly referred to as a time-bin entangled photon-pair source. Building on this foundation, Tittel et al. demonstrated quantum cryptography over installed optical fiber using time-energy entanglement \cite{BBM92_Tittel2000}, and Marcikic et al. extended the framework to QKD over long distances using time bins \cite{Ent-dist_Marcikic2004}, establishing the practical viability of TBQs for quantum communication.
The subsequent decade saw TBQs become a workhorse of photonic quantum information science, owing to their robustness against fiber-induced birefringence and compatibility with standard telecom infrastructure \cite{RobustDLI_Islam2017}. Early milestones included demonstrations of Bell-state analysis, quantum teleportation \cite{Teleportation_Marcikic2003}, and entanglement swapping \cite{Swap_Riedmatten2005}, each of which exploited the natural resilience of time-bin encoding in deployed fiber networks. A landmark achievement came in 2015, when Hensen et al. leveraged a TBQ-based BSM to realize the first loophole-free Bell-inequality violation \cite{Ent-Bell_Hensen2015}, a result subsequently extended and refined by other groups \cite{TBE-PSL_Francesco2018, Nonlocality_Santagiustina2024}.

Parallel to these foundational experiments, time-bin encoding was adapted to an expanding array of protocols. Humphreys et al. proposed a linear optical quantum computing scheme utilizing TBQs \cite{QComp-TBQ_Humphreys2013}, demonstrating that photonic quantum computing could be realized through a solitary optical path rather than a complex network of spatial modes -- advancing the scalability of photonic quantum processors. More recent work has demonstrated quantum secure direct communication \cite{QSDC-TB_Zhang2022, QSDC-TB_Ahn2024}, long-distance entanglement distribution and QKD over optical fiber \cite{MDI_Liu2013, TBEnt_Dynes2009, QCom_Daniel2022}, and multipartite entanglement preparation \cite{GHZ_Lo2023,GHZ_Davis2023Optica, GHZ_Davis2023CLEO,GHZ_Leili2024}.
Contemporary research continues to expand the frontier of time-bin encoding. Bouchard et al. demonstrated ultrafast measurement of TBQs \cite{QCom_Bouchard2022} and high-dimensional time-bin qudits \cite{TBM_qudit_Bouchard2023}, enabling higher data rates and lower error rates in fiber-based quantum communication. Further recent advances encompass quantum walks \cite{QWalk_Fenwick2024}, novel entanglement generation schemes \cite{TBEnt_Kim2024}, photonic quantum computing \cite{QComp_Bouchard2024}, quantum state processing \cite{QStateProcess_Monika2024}, and robust time-bin manipulation techniques \cite{TB-Robust_Simon2025, TBM_Danese2026}. Collectively, these developments highlight the versatility of TBQs and their pivotal role in advancing practical quantum information processing. 
While broader aspects of time-bin and time-energy entanglement --- including generation, characterization, and interferometric measurement considerations --- are extensively reviewed elsewhere \cite{TB-Review_Yu2024, TBEnt-Review_Xavier2025, TB-Review_Nicola2025, QInterferometer_Jin2024}, the present work provides a conceptual introduction and a systematic review of current developments in time-bin-based photonic quantum information processing, with emphasis on its principles and applications.
As a concise overview of the development and evolution of TBQs, we provide a chronological tally of important milestones and their significance to the relevant applications in Table~\ref{tab:timebin_timeline}.

{\footnotesize
\setlength{\tabcolsep}{4pt}
\renewcommand{\arraystretch}{1.15}
\newcolumntype{L}[1]{>{\raggedright\arraybackslash}p{#1}}
\begin{longtable}{L{0.8cm} L{2.8cm} L{2.8cm} L{4.4cm} L{4.4cm}}
\caption{\color{black} Chronological development of time-energy and  time-bin entanglement, highlighting key demonstrations including  generation, teleportation, entanglement swapping, QKD, and large-scale 
distribution. This overview is not exhaustive. DLI: Delay-Line Interferometer, HD: High-dimensional, QST: Quantum State Tomography, TB: Time-bin, 
TE: Time-energy.}
\label{tab:timebin_timeline} \\
\toprule
\textbf{Year} & \textbf{Milestone} & \textbf{Platform} &
\textbf{Key Demonstration} & \textbf{Significance} \\
\midrule
\endfirsthead
\toprule
\textbf{Year} & \textbf{Milestone} & \textbf{Platform} &
\textbf{Key Demonstration} & \textbf{Significance} \\
\midrule
\endhead
\endfoot
\bottomrule
\endlastfoot
1989 &
Franson's proposal for a Bell test using TE correlations &
Theoretical proposal; intended platform: CW SPDC with bulk crystal &
Proposed Bell test via TE correlations, using one Franson Interferometer, i.e, a DLI, per photon \cite{TE_Franson1989} &
Foundational nonlocal two-photon interference scheme; 
basis for all subsequent TE and TB entanglement experiments \\
\hline

1992-- &
TE Bell tests &
Bulk SPDC, CW pump &
Two-photon interference visibility exceeding the classical 
limit in a Franson scheme (post-selection applied) 
\cite{TE-Bell_Brendel1992} &
First experimental violation of classical correlations 
using TE-entanglement; post-selection loophole present \\
\hline

1998-- &
 TE-entanglement distribution & Telecom fiber (1310\,nm), bulk SPDC &
Bell violation (post-selection) with measurement stations 10.9\,km apart \cite{TE-Dist_Tittel1998} &
Entanglement preserved over km-scale fiber; established telecom-wavelength compatibility \\
\hline

1999 &
Pulsed-pump TE-entanglement ~~~~~~ (TB encoding) &
Pulsed laser pumping bulk SPDC crystal; one DLI per photon &
Franson-type interference with a pulsed pump,  producing photon pairs in discrete early/late temporal modes \cite{Ent_Brendel1999} &
First use of pulsed pump to generate TB entangled photon pairs; groundwork for discrete TB encoding \\ \hline

2000 &
TB-entanglement-based QKD &
Telecom wavelength, bulk SPDC &
BBM92 protocol demonstrated over installed-fiber using TB-entangled photons \cite{BBM92_Tittel2000} &
Established TB-entanglement as a viable platform for quantum-secure communication \\
\hline

2002-- &
HD TB- and TE-entanglement &
Pulsed or CW laser pumping SPDC crystal &
 HD TB-entanglement \cite{HDTB-Ent_Riedmatten2002, Ent-HD_Riedmatten2004}; TE-entangled qutrits \cite{EntQutrit_Thew2004}, and  CGLMP-Bell inequality violation \cite{BellTestQutrit_Thew2004} &
First experimental realizations of HD 
TB- and TE-entanglement characterization, and CGLMP-Bell test  \\
\hline

2002--  &
TB-entanglement distribution & Telecom fiber (1300\,nm, 1550\,nm), bulk SPDC &
Distribution of TB-entangled qubits over 11\,km \cite{Ent-dist_Thew2002}; 50\,km of optical fiber with Bell inequality violation and QKD \cite{Ent-dist_Marcikic2004} &
First long-distance distribution of TB-entangled qubits, establishing viability for fiber-based quantum communication \\
\hline

2003-- &
TB teleportation & Bulk SPDC; linear-optics BSM &
Quantum teleportation of TBQs at 1310\,nm over 2\,km of optical fiber \cite{Teleportation_Marcikic2003, Teleportation-relay_Riedmatten2004} &
First telecom-wavelength quantum relay; compatible with deployed 
fiber infrastructure \\
\hline

2005 &
TB entanglement swapping &
Two independent SPDC sources; linear-optics BSM &
Entanglement swapping between photons from two independent SPDC sources transmitted over 2.2\,km of fiber \cite{Swap_Riedmatten2005} & First TB entanglement swapping with independent sources; demonstrated  feasibility of quantum repeater node operation \\
\hline

2007 &
TB-entangled pairs generated by waveguide sources &
Periodically poled LiNbO$_3$ (PPLN) waveguides; Si waveguides &
Bright, compact, fiber-coupled telecom TB-entangled photon pairs from PPLN waveguides \cite{Ent-dist_Honjo2007} and Si waveguides \cite{Ent-FWM_Takesue2007} &
Improved performance over bulk crystals; key step toward deployable and scalable quantum communication hardware \\
\hline

2007-- &
TB-entanglement distribution over long-distance  &
Up-conversion or self-differencing SPDs &
TB-entanglement distribution over 100\,km \cite{Ent-dist_Honjo2007}, 200\,km  \cite{TBEnt_Dynes2009} of fiber &
Feasibility of TB entanglement distribution over long-distance fiber links \\
\hline

2008 &
TBQ storage in a solid-state quantum memory &
Nd:YVO$_4$ crystal memory (AFC protocol) &
Storage and retrieval of a weak-coherent TBQ preserved after retrieval \cite{LightMatter_Riedmatten2008} &
First solid-state storage of a TBQ; established light-matter interface for quantum repeaters \\
\hline

2011 &
TB/TE entangled photon storage in solid-state quantum memories &
Ti:Tm:LiNbO$_3$ waveguide and Nd:Y$_2$SiO$_5$ crystal 
 (AFC protocol) & 
Storage and retrieval of one photon from a TB/TE-entangled pair, followed by Bell-violation \cite{QMemory_Saglamyurek2011, QMemory_Clausen2011} &
Demonstrations of broadband solid-state TB/TE-entanglement storage; critical step toward multiplexed quantum repeaters \\
\hline

2012 &
Genuine tripartite TE-entanglement &
Cascaded SPDC in bulk nonlinear crystals &
Three-photon TE-entangled state generated via cascaded SPDC and verified by violation of the TE uncertainty relation 
\cite{TE-Tripartite_Shalm2012} &
First genuine tripartite TE entanglement; opens pathways to multi-party quantum communication protocols \\
\hline

2014 &
TBQ entangling gate &
Fast electro-optic $2\times2$ switch, Bulk SPDC &
Bell state generated from two independent TBQ inputs by routing early/late bins via a switch and post-selection 
\cite{Ent-Switch_Takesue2014} &
First two-qubit entangling gate for independent TBQs; enabling scalable multi-qubit TB-entangled state generation \\
\hline

2014-- &
TB-entanglement from quantum dots &
Resonantly excited self-assembled III/V quantum dot &
TB-entangled photon pairs via biexciton-exciton cascade \cite{Ent-Qdot_Jayakumar2014}; entanglement verified by QST &
On-demand TB-entangled photons from a solid-state emitter \\
\hline


2014--15 &
TB-entanglement sources based on integrated-photonic chips &
Silicon photonic crystal CROW, Si micro-ring resonator, and Si$_3$N$_4$ waveguide chip (via SFWM) &
 TB-entanglement using a silicon photonic crystal CROW
\cite{Ent-FWM_Takesue2014}; via SFWM in a Si micro-ring resonator \cite{Ent-FWM_Wakabayashi2015}; fully integrated on-chip source in Si$_3$N$_4$ \cite{Ent-FWM_Xiong2015} &
Established integrated photonic chips as compact, scalable TB-entangled photon sources for quantum communication \\
\hline

2016-- &
HD TB- and TE-entanglement quantification under realistic conditions &
SPDC sources with narrowband spectral filtering; DLI measurement networks &  Information capacity analysis \cite{HD-TB_Thomas2016}; entanglement of formation measurement in HD TB states  \cite{TB-HD-EntChar_Martin2017}; HD TE entanglement stored in crystal \cite{TE-HD-EntChar_Alexey2017} &
Established practical limits and advantages of HD TB- and TE-entanglement under realistic experimental conditions \\
\hline  

2018 &
TB Bell tests: Closure of post-selection loophole & 
SPDC-based TB-entangled source; active electro-optic switching 
in measurement DLI & 
Deterministic early/late path assignment by replacing the first BS with a synchronized optical switching \cite{TBE-PSL_Francesco2018} & Strengthens the validity of TB-entanglement certification with Franson-type Bell-test (general detection and locality loopholes remain open) \\
\hline

2019 &
TB-encoded QKD  over a free-space channel &
Turbulent 1.2\,km atmospheric channel; 
weak-coherent TBQ &
BB84 QKD using a multi-mode TB analyzer to maintain high interferometric visibility despite spatial mode distortion \cite{BB84-FreeSpace_Jin2019} &
Demonstrates the feasibility of free-space TBQ transmission; spatial mode distortion is overcome with a multimode DLI \\ \hline

2019-- &
TB/TE entanglement-based multi-user networks &
Wavelength-multiplexed SPDC and SFWM sources (bulk and integrated) &
Multi-user TE-entanglement swapping \cite{TE-EntSwap_Li2019};  TB-entanglement across  multiplexed users \cite{BBM92_Wen2022}; 
16-user network \cite{QNetwork-TB16user_Huang2025} &
Extended quantum entanglement networks beyond point-to-point links toward scalable multi-user architectures \\
\hline

2019--  &
TBQ interfaces with stationary qubit platforms &
NV centers, SiV centers, and trapped-ion qubits & 
Entanglement of TBQ and NV center spin \cite{NV-center_Tchebotareva2019}; TBQ-mediated entanglement of SiV centers \cite{SCTLS-QNet-SiV_Lukin2024}, and trapped-ions \cite{QNetwork-BaIons_Saha2025} &
Light-matter interfaces with TBQs and stationary qubits; architectures for multi-node quantum repeaters and networks \\ \hline

2023-- &
Genuine multipartite TB-entanglement generation &
Weak-coherent and/or SPDC sources; $2\times2$ optical switch and post-selection &
Three-partite TB-GHZ states from a weak-coherent TBQ and a TB-entangled pair \cite{GHZ_Lo2023, GHZ_Davis2023Optica, GHZ_Davis2023CLEO}, and four-partite TB-GHZ states from two TB-entangled pairs \cite{GHZ_Leili2024} &
Extends TB-entanglement to three and four parties using independent sources; enables multi-node quantum network protocols \\ \hline

2024-25 & 
TBQ-based photonic quantum information processing & 
Kerr effect, temporal photonic lattice,  HOM interference & 
Programmable quantum circuits \cite{QComp_Bouchard2024}, quantum state processing \cite{QStateProcess_Monika2024}, quantum information protocols \cite{TB-Robust_Simon2025}  &
Enables reconfigurable TB quantum state processing for scalable photonic quantum information
\\ \hline
\end{longtable}}

\subsubsection{Time-bin encoding definitions}
\label{Sec_TBEncodingDefs}

Ideally, the time bins are defined as completely orthogonal temporal modes. Quantum information can then be encoded in a photon whose temporal state is in either of those modes or some superposition thereof \cite{Ent_Brendel1999}. For qubits specifically, two temporal modes are involved. As the binary information is encoded in the computational basis of the arrival time of the photons or in the superposition basis as the phase between the time bins, it serves as a reliable quantum information carrier. Consider a single-photon wave function that is in a superposition state of two separated temporal modes \cite{QST_Sedziak2020} given as
\begin{equation}
    \psi(t) = \tilde{\alpha}\, u(t) +  \tilde{\beta}\, u (t-\tau_{el}),
\end{equation}
where $u(t)$ and $u(t-\tau_{el})$ are the envelopes of the temporal modes centred at $t=0$ and $t=\tau_{el}$, respectively, described by the common temporal mode envelope, i.e., wavepacket, function $u$. The factors $\tilde{\alpha}$ and $\tilde{\beta}$ are complex scalars satisfying the normalization condition $|\tilde{\alpha}|^2+|\tilde{\beta}|^2=1$.

The time-bin basis states, labeled $|e\rangle$ for \textit{early} and $|l\rangle$ for \textit{late}, are defined as
\begin{equation}
\begin{split}
    |e\rangle & = \int_{-\infty}^{\infty} dt\, u(t) |t\rangle,\\
    |l\rangle & = \int_{-\infty}^{\infty} dt\, u(t-\tau_{el}) |t\rangle,
    \end{split}
    \label{eq_earlylatedef}
\end{equation}
where $|t\rangle$ denotes the state of a single photon localized at time $t$. 

\begin{figure} [ht!]
    \centering
    \subfloat[\centering \label{fig_TBplot}]{{\includegraphics[trim={0cm 0cm 0cm 0cm},clip, width=0.5\columnwidth]{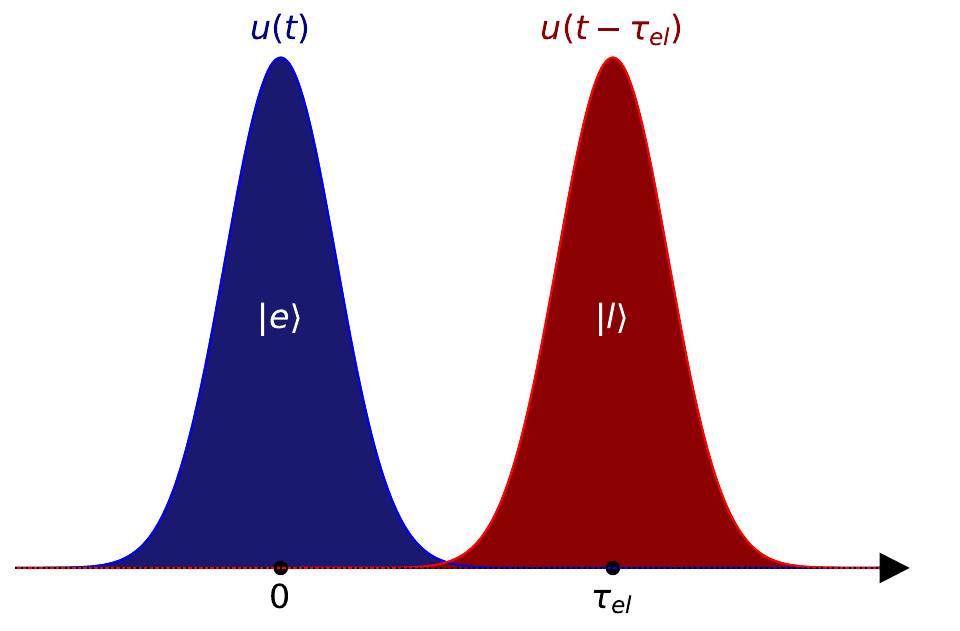} }}
    \qquad
    \subfloat[\centering \label{fig_TB-Bloch}]{{\includegraphics[width=0.42\columnwidth]{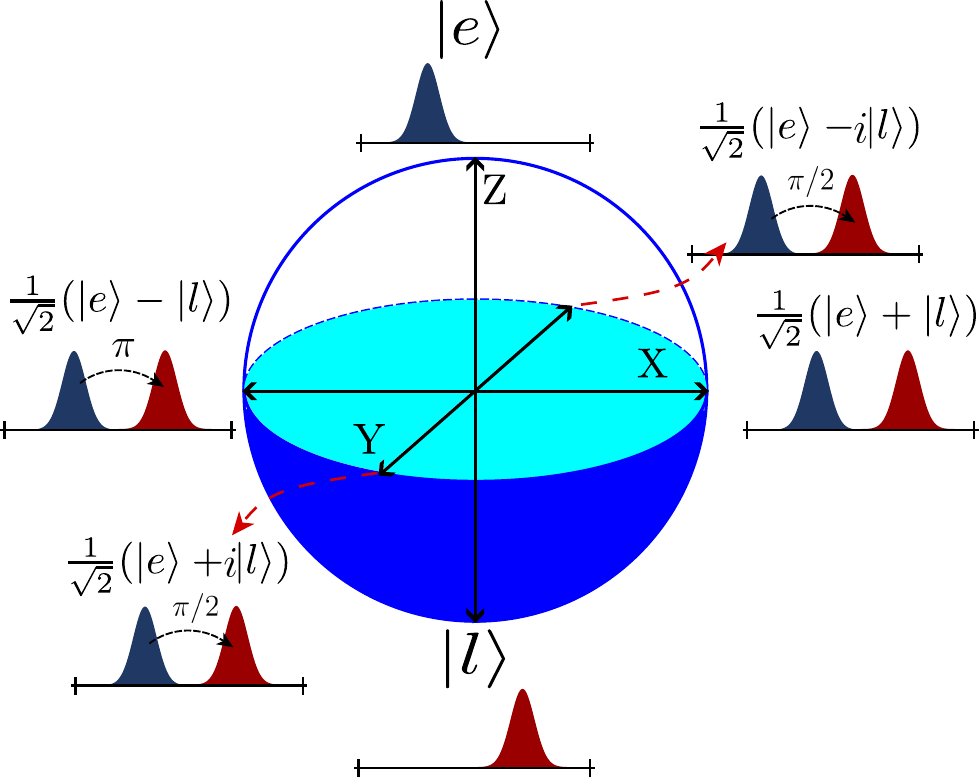} }}
    \caption [Time-bin qubit definition and its Bloch sphere representation.] {(a) A TBQ is defined as a single photon delocalized in two wave packets located in the time interval $[0,\tau_{el}]$. (b) TBQ basis states on a Bloch sphere. TBQ: Time-bin qubit.}
\end{figure}

The overlap of the time-bin basis states is given by $\langle e|l \rangle = \int_{-\infty}^{\infty} dt \, u^*(t)\, u(t-\tau_{el})$, which should ideally be zero for orthogonal basis states. However, for a Gaussian-shaped time-bin envelope function
$u(t)=\frac{1}{\sqrt{\sigma\sqrt{\pi}}} \exp{\left(\frac{-t^2}{2\sigma^2}\right)}$, the overlap becomes $
 \langle e|l\rangle = \exp{\left(-\frac{\tau_{el}^2}{4\sigma^2}\right)}$. This indicates that the basis states are not perfectly orthogonal, but their overlap can be made infinitesimally small by properly choosing the parameters $\tau_{el}$ and $\sigma$ \cite{QST_Sedziak2020}. For example, the time-bin state shown in Fig. (\ref{fig_TBplot}) has $\tau_{el}/\sigma = 6$, which yields $\langle e|l\rangle=1.2 \times 10^{-4}$. Therefore, for all practical purposes, $|e\rangle$ and $|l\rangle$ in Fig. (\ref{fig_TBplot}) can be considered orthogonal basis states, and one can formally write the TBQ state as: $|\psi\rangle = \tilde{\alpha}\, |e\rangle + \tilde{\beta}\, |l\rangle$.

Different temporal mode envelope functions $u$ will rely on different characteristic temporal widths, e.g., $\sigma$ for the Gaussian function. For an unambiguous definition of the temporal width of the time bins, designated by $\Delta \tau$, we employ the full width at half maximum (FWHM) of the envelope function. For Gaussian-shaped time-bins, for example $\Delta \tau$ and $\sigma$ are related as $\Delta \tau = 2.355~\sigma$. 

\begin{figure}[ht!]
\begin{center}
\includegraphics[clip, trim= 10cm  5cm  0cm  6cm, width=0.8\columnwidth]{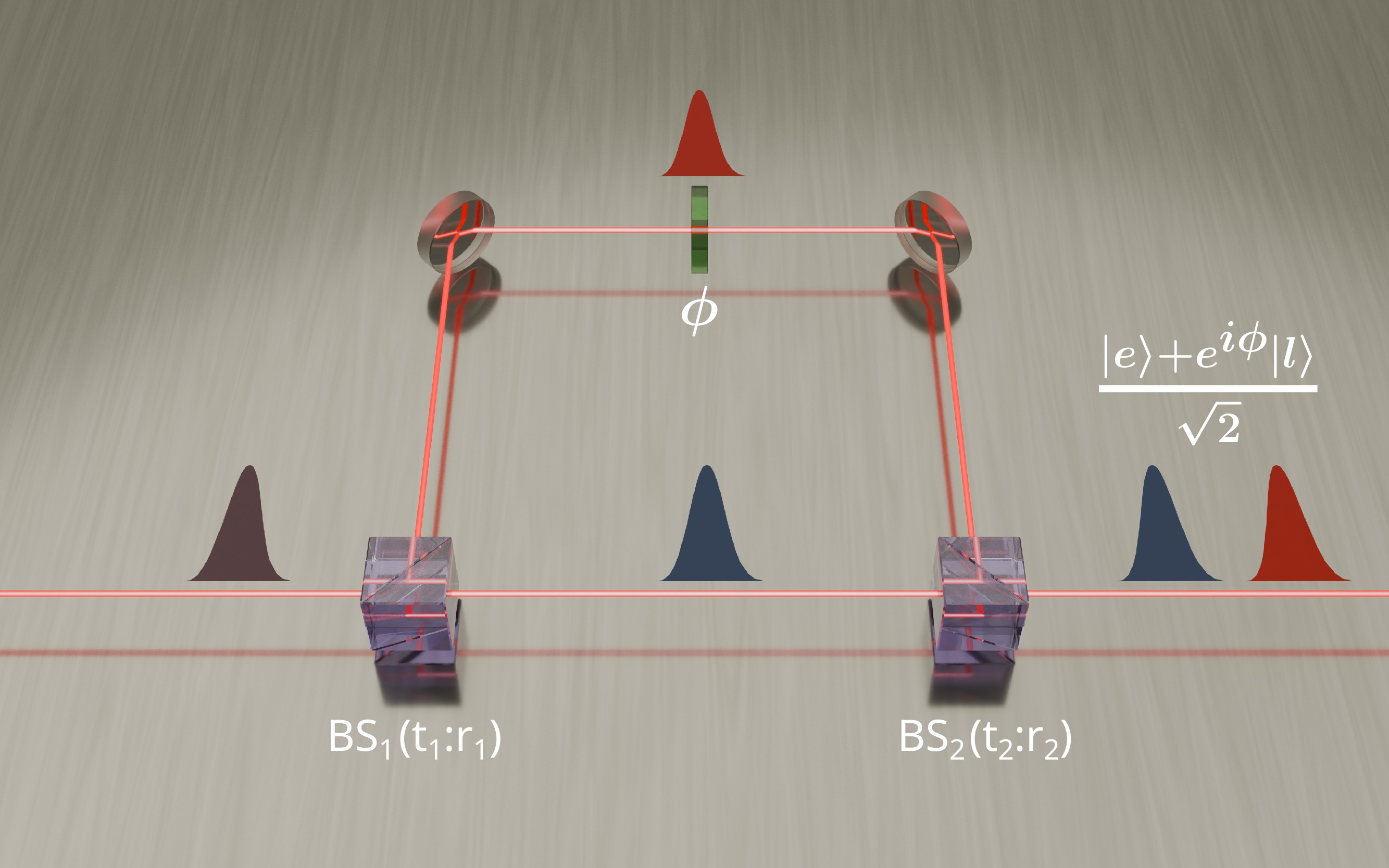}
\caption[Time-bin qubit preparation using a delay line interferometer.]{TBQ preparation using Delay Line Interferometer (DLI) in the Unbalanced Mach-Zehnder Interferometer (UMZI) configuration. A pulse containing a single photon incident on the UMZI can take short and long paths, which results in a superposition state of early and late time-bin states. Note that the output time-bin state is illustrated according to the two modes' temporal configuration as in Fig.~(\ref{fig_TBplot}).}
\label{fig_TBP}
\end{center}
\end{figure}

For an intuitive understanding of the preparation of a TBQ from an experimental point of view, let us consider an optical mode containing a single photon with temporal mode $u(t)$ incident on a Delay Line Interferometer (DLI), also referred to as ``Franson Interferometer'' \cite{TE_Franson1989}, as shown in Fig.~(\ref{fig_TBP}). Since the two interferometric paths are unequal, the photon will exit the interferometer in a superposition state of two modes separated by time $\tau_{el}$, defining the `early' and `late' time-bin basis states, as introduced in Eq.~(\ref{eq_earlylatedef}). The created single-photon superposition state is described by
\begin{equation}
\ket{\psi} = \alpha \ket{e}+\beta e^{i\phi}\ket{l}.
\end{equation}
The real-valued probability amplitudes; $\alpha$ and $\beta$, depend on the splitting ratios of the two beam-splitters of the UMZI as $\alpha= t_1 t_2$ and $\beta= r_1 r_2$, as shown in Fig.~(\ref{fig_TBP}), where $t_1(t_2)$ is the transmittance and $r_1 (r_2)$ is the reflectance of the electromagnetic field amplitude by first (second) beam splitter, namely $\text{BS}_1~(\text{BS}_2)$. Note that this setup inevitably allows the photon to ``leak'' to the unused output of the UMZI, unless the second BS is replaced by a fast optical switch, as discussed in Section~(\ref{Sec_Ent-switch}). The phase factor $\phi$ depends on the path length difference of the two arms of the interferometer, including the phase shifter setting. As an example, most commonly used time-bin states with their mapping on the Bloch sphere are depicted in Fig.~(\ref{fig_TB-Bloch}).


The temporal separation between the early and late time bins ($\tau_{el}$) is determined by the path length difference in DLI. The path difference should be larger than the temporal duration of individual pulses (or time bins; $\Delta \tau$) to ensure the orthogonality of the base states. In other words, the coherence length of the input temporal mode must be shorter than the path length difference between the unequal arms of the DLI. In addition, maintaining the constant relative phase of the qubit ($\phi$) is important as discussed in Section~(\ref{Sec_Measurement}).

Practically, the path fluctuations of the DLI must be much smaller than the wavelength of the photon to generate a phase-stable TBQ, leading to practical design considerations that we discuss in Section (\ref{Sec_MeasInt}). Note that a weak-coherent pulse with a mean photon number less than one ($\mu<1$) can also be used at the input of the interferometer to create a time-bin state, but this can only approximately mimic the qubit since the photon number is statistically distributed as discussed in Section (\ref{Sec_WCPS}). Note, finally, that although the qubit state is encoded in the temporal mode, all other photonic degrees of freedom --- polarization, spatial, spectral --- must be consistent for the two time bins, e.g., the polarization of the early time bin must be the same as that of the late. If this is not the case, the time-bin state may be distorted during transmission or vulnerable to side-channel attacks in privacy protocols, such as QKD.

In the following sections, we discuss in detail the options and implications for preparing (Section~\ref{Sec_Preparation}), transmitting (Section~\ref{Transmission}), and detecting (Section~\ref{Sec_Measurement}) TBQs, leading to a summary of all considerations for choosing the time-bin parameters in Section~(\ref{Sec_ParameterSelection}).



\section{Time-bin qubit preparation}
\label{Sec_Preparation}

\begin{figure} [ht!]
    \centering
    \includegraphics[width=0.95\linewidth]{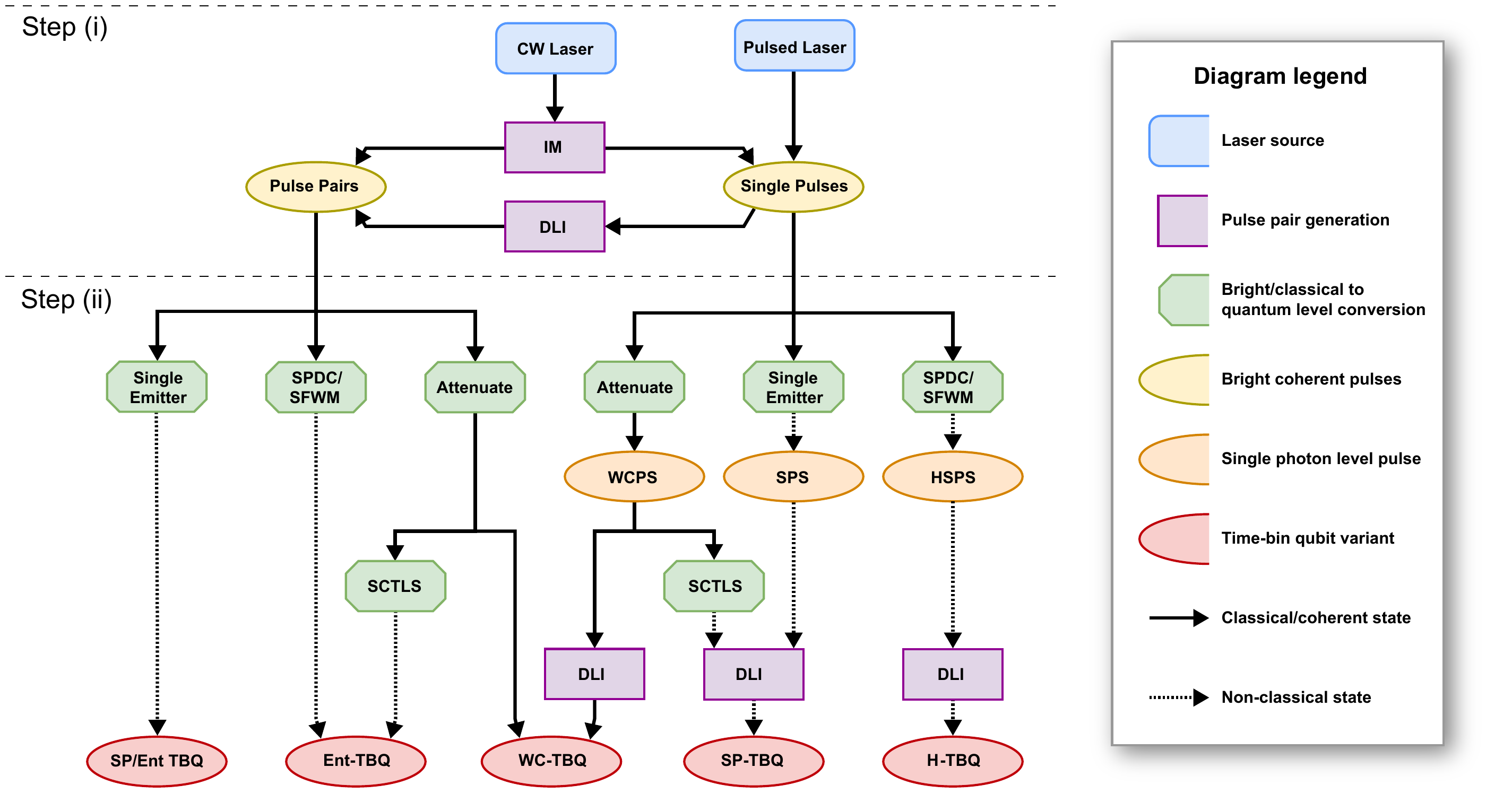}
    \caption[Flowchart of different methods for TBQ generation.]{The flow chart illustrates different methods for preparing TBQs using continuous-wave (CW) and pulsed laser sources (blue boxes) to create a train of single pulses or pulse pairs (yellow boxes) as a first stage [step (i)]. This defines the two main paths of the subsequent stage (step(ii)) based on either single pulses or pulse pairs, with intermediate components and atomic systems to create the TBQs. Green boxes indicate specific interfaces (attenuation, single-emitters, SPDC = Spontaneous Parametric Down Conversion, SFWM = Spontaneous Four-Wave Mixing, SCTLS = Strongly-Coupled Two-Level Systems) applied to a classical input optical signals to convert them to either single-photon level or non-classical optical pulses. Devices that generate the time-bin superpositions (i.e., IM = intensity modulation or DLI = Delay Line Interferometers), in either Step (i) or Step (ii), are illustrated as purple boxes. Orange ovals indicate different single-photon states (WCPS = Weak-Coherent-Pulse Source, SPS = Single-Photon Source, and HSPS = Heralded-Single-Photon Source) and red boxes denote different final types/variants of TBQs (WC-TBQ = Weak-Coherent TBQ, SP-TBQ = Single-Photon TBQ, H-TBQ = Heralded TBQ, and Ent-TBQ = Entangled TBQ). The arrows indicate the statistics of the photonic state, i.e., solid line for classical/coherent states and dashed line for non-classical/quantum states.}
    \label{fig_TBFlowChart}
\end{figure}

As discussed in Section (\ref{Sec_Introduction}), TBQ preparation requires the creation of a coherent superposition of a photon, i.e., an optical mode, in two orthogonal time-bin basis states. We envision the TBQ preparation as a two-step process: (\textit{i}) In the first step, single pulses or pulse pairs are generated at a chosen clock rate from a laser source; (\textit{ii}) Subsequently, these optical modes are either attenuated or undergo a non-linear interaction with a suitable system to produce different types of TBQs. The flowchart in Fig.~(\ref{fig_TBFlowChart}) outlines different methods to generate various types of TBQs, all starting with either a pulsed or a CW laser source.

\textbf{Step (\textit{i})}: If a pulsed laser is employed, the pulse train can either be passed to the next step, thus completing step (\textit{i}), or the emitted pulses can be split into pairs of pulses with a well-defined and stable phase difference. The splitting is achieved by directing the pulsed laser output through a DLI, which introduces a time delay, allowing the creation of superposition states between the early and late time bins, as shown in Fig.~(\ref{fig_TBP}) of Section~(\ref{Sec_Introduction}). If a CW laser is used, step (\textit{i}) employs an intensity modulator (IM) to carve out a train of either single pulses or pairs of pulses from the continuous wave. In the former case, the single pulses can be directed to the next step or to a DLI, just as for the pulsed-laser output. For the latter case, the IM directly generates pulse pairs, which conveniently omits the need for a DLI in the preparation.

\textbf{Step (\textit{ii})} involves a broader range of choices. As a general theme, one may use either single-pulses (right-hand side of Fig.~\ref{fig_TBFlowChart}) or pulse-pairs (left-hand side of Fig.~\ref{fig_TBFlowChart}) from step (i) to generate single-photon level signals. However, if single pulses are used, a DLI will be part of step (ii) to encode the TBQ. The simplest option is to attenuate the train of pulse pairs from step (i) to the single-photon level to generate a stream of weak-coherent TBQs. Following the common theme, the same can be achieved by first attenuating a train of single pulses and subsequently directing these weak-coherent pulses through a DLI to create weak-coherent TBQs.

Rather than deriving the time bins directly from the weak-coherent signals, they can also be employed for another approach, which relies on a strongly coupled two-level system, such as quantum dots or atoms in cavities. In this case, the attenuated weak-coherent pulses drive a strongly coupled two-level system that emits a single photon that is subsequently passed through a DLI to form a single-photon TBQ. If, instead, attenuated pulse-pairs are employed to drive the strongly coupled two-level system, entangled TBQ pairs will be generated.

Instead of attenuating the pulses from step (i), they can also be used to drive single emitters that output single photons, which are then passed through a DLI to create a single-photon TBQ. Bright pulse pairs can also drive single emitters to produce either a single-photon TBQ or an entangled TBQ, depending on the single-emitter type and pumping protocol. 

Finally, the pulses from step (i) can be exploited to drive multi-photon processes, such as spontaneous parametric down-conversion (SPDC) or spontaneous four-wave mixing (SFWM), in a nonlinear medium. A train of pulses from step (i), will generate correlated pairs of photons, of which one can be used to herald the existence of the other. Hence, such sources are coined a heralded single photon source. Passing the heralded photon through a DLI, thus, generates heralded TBQs. If a train of pulse-pairs drives the non-linear medium, the process results in the creation of correlated photons in time-bin states, i.e., entangled TBQs.

In the following Sections (\ref{Sec_CLS}-\ref{Sec_SPS}), we provide a more comprehensive discussion of the various methods of TBQ generation from different light sources, outlined in Fig.~(\ref{fig_TBFlowChart}), along with their key properties.

To determine which laser source is more suitable, one can consider the pulse properties achievable directly from the source and subsequent parts of the preparation. For example, pulsed lasers typically achieve higher peak pulse intensities, which can be advantageous for driving single-emitter systems and non-linear processes. However, if peak power is not of importance, e.g., if pulses are attenuated or drive a strongly coupled two-level system, then the flexibility, cost-effectiveness, and broader choice of CW lasers may make it the preferred choice.

Note that when using CW laser sources with coherence length shorter than the time-bin separation but longer than the desired time-bin width, it is advantageous to carve out a single pulse first and then create a time-bin superposition using a DLI. This approach preserves coherence between the two output modes, which is only limited by the phase stability of the DLI and frequency/wavelength stability of the laser source over the measurement timescale as given by Eq.(\ref{Eq_FreqtoPhaseUncertain}). On the other hand, if the coherence length of the CW laser significantly exceeds the time-bin separation, time-bin superpositions can be either directly generated using intensity modulators or the former method can be employed. In this way, the time-bin parameter choice may be intimately linked to the coherence properties of the CW laser source (see Sec.~\ref{Sec_ParameterSelection}).

\subsection{Coherent Source}
\label{Sec_CLS}

Laser is a coherent source and the emitted coherent state ($|\boldsymbol{\alpha}\rangle$) \cite{Book_fox, Book_Gerry-Knight} can be expressed in the Fock basis (i.e., a basis with a specific number of photons in a well-defined mode \cite{FockState_Mandel1966}), as follows:
\begin{equation}
|\boldsymbol{\alpha}\rangle =e^{-\frac{|\boldsymbol{\alpha}|^2}{2}}\sum_{n=0}^{\infty}\frac{\boldsymbol{\alpha}^n}{\sqrt{n!}}|n\rangle,
\end{equation}
where $n$ is the number of photons, and $|\boldsymbol{\alpha}|^2=\mu$ is the mean photon number. The probability of measuring $n$ photons for a given $\mu$ in a small time window follows the Poissonian statistics \cite{Book_fox, Book_Gerry-Knight} as given below
\begin{equation}
p(n|\mu) = \frac{\mu^n}{n!} e^{-\mu}.
\label{eq_poissondist}
\end{equation}

\begin{figure}[htb!]
    \centering
    \includegraphics[width=0.6\columnwidth]{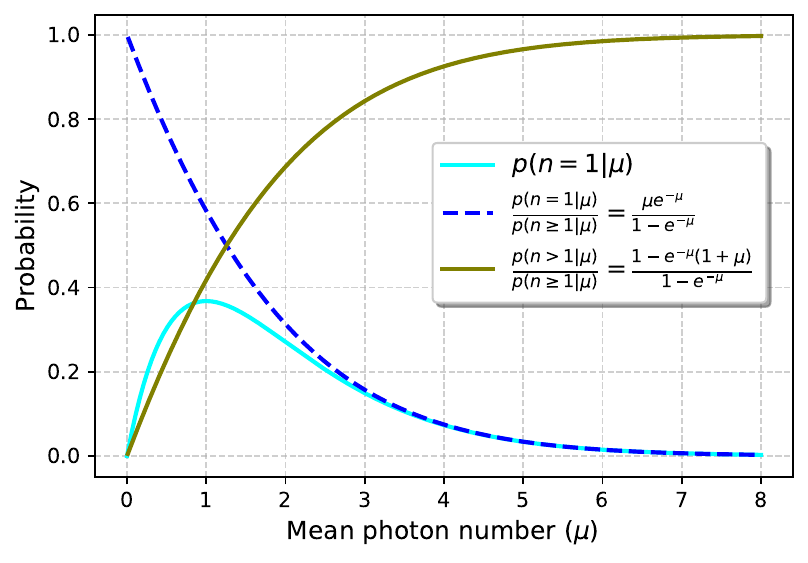} 
    \caption[Poissonian probability distribution for a weak-coherent pulse.]{Poissonian probability of weak-coherent pulses with mean-photon $\mu$ containing a single-photon $p(n = 1|\mu)$, and the fraction of non-vacuum pulses ($p(n\geq 1|\mu)$) containing single photons ${p(n = 1|\mu)}/{p(n \geq 1|\mu)}$, and multi-photons ${p(n > 1|\mu)}/{p(n \geq 1|\mu)}$ as a function of $\mu$.}
    \label{fig_poisson}
\end{figure}

Poissonian statistics is uniquely characterized by its mean value ($\mu$). From Eq.~(\ref{eq_poissondist}), it follows that the standard deviation ($\Delta n$) equals the square root of the mean; i.e.,  $\Delta n = \sqrt{\mu}$. Thus, a coherent light source with steady intensity is the most stable form of classical light \cite{Book_fox}. A plot of Poissonian distribution is shown in Fig.~(\ref{fig_poisson}). It is evident that as the mean-photon number increases, the probability of multi-photon events increases, and the probability of a non-vacuum pulse containing a single photon exponentially decreases. Thus, the weak-coherent pulse is limited by the multi-photon contributions from the Poissonian statistics of the laser source. A TBQ prepared using weak-coherent pulse can only approximately mimic a true qubit state for a low mean photon number ($\mu << 1$). 
Alternatively, a coherent source can interact strongly with a two-photon transition in a suitable medium such that all multiphoton terms in the photon-number distribution can be suppressed, allowing maximum single-photon probability $p(1|\mu)$ to exceed 80\% \cite{SPS_Huang2012}.

\textit{Setting the mean-photon number ($\mu$):} In the pulsed mode operation, a weak-coherent state ($|\boldsymbol{\alpha}\rangle$) is prepared \cite{QKD-WCP_Huttner1995, QKD-WCP_Bennett92} by attenuating a train of coherent pulses to the extent that the mean photon number per pulse ($\mu$) becomes less than one. The attenuation of coherent pulses is achieved using a set of variable optical attenuators (VOAs), with the attenuation level settings as discussed below. Suppose that coherent pulses are generated at a repetition rate of $f_{rep}$~[Hz] and have an average input power of $P_{in}$ [W] as measured on an optical power meter. Let us assume that the intended mean photon number per pulse is $\mu$. The average power contained in such weak-coherent pulses is given by
\begin{equation}
P_{f_{rep}}=  \frac{\mu f_{rep}\, h c}{\lambda},
\end{equation}
where $h$ is Planck's constant, $c$ is the speed of light in a vacuum, and $\lambda$ is the wavelength of laser. 
Thus, the attenuation ($\eta$) required for reaching the mean photon per pulse $\mu$ is given by
\begin{equation}
\eta (dB) = 10~\log_{10}\left[\frac{P_{f_{rep}}}{P_{in}}\right] = 10~\log_{10}\left[\frac{\mu f_{rep}\, h c}{\lambda P_{in}}\right].
\end{equation}

As an example, a coherent pulse generated at $f_{rep} = 100~\text{MHz}$, $\lambda=1550$ nm,  and $P_{in}=1$ nW requires an attenuation of $-25.91$ dB to reach the mean photon number $\mu=0.2$ per pulse. This weak-coherent pulse is an approximate substitute for an ideal SPS input to the DLI in Fig.~(\ref{fig_TBP}). Note that when preparing a TBQ superposition via direct attenuation of a pair of coherent pulses, a factor of two must be taken into account when setting $\mu$. In this section, we will discuss the preparation of the TBQ using a CW laser and a directly modulated laser, and in Section~(\ref{Sec_SPS}) using a single photon source.

\subsubsection {Continuous-Wave laser}
\label{Sec_CWLaser}

The simplest and most popular method to prepare a weak-coherent TBQ uses a CW laser followed by an IM \cite{EOM_Lo2019} driven by an RF source to create a phase-coherent `early' and `late' optical pulse sequence \cite{QKD_TB_generator_Morales}, which the VOA attenuates to create TBQs in the $|+\rangle$ state: $(|e\rangle+|l\rangle)/\sqrt{2}$. A phase modulator is used to provide a phase shift to the late-time bin to generate the state: $(|e\rangle + e^{i\phi} |l\rangle)/ \sqrt{2}$. When $\phi=\pi$, we obtain the $|-\rangle$ state: ($|e\rangle-|l\rangle)/\sqrt{2}$. For generating only early or late or a sequence of these qubits, the RF modulation is required to be ON only for one of these time bins. Generation of the most general state: $|\psi\rangle=\alpha |e\rangle + \beta e^{(i\phi)}|l\rangle$ requires controlling the relative intensity of early and late time-bins. 

\begin{figure} [ht!]
    \centering
\includegraphics[width=\columnwidth]{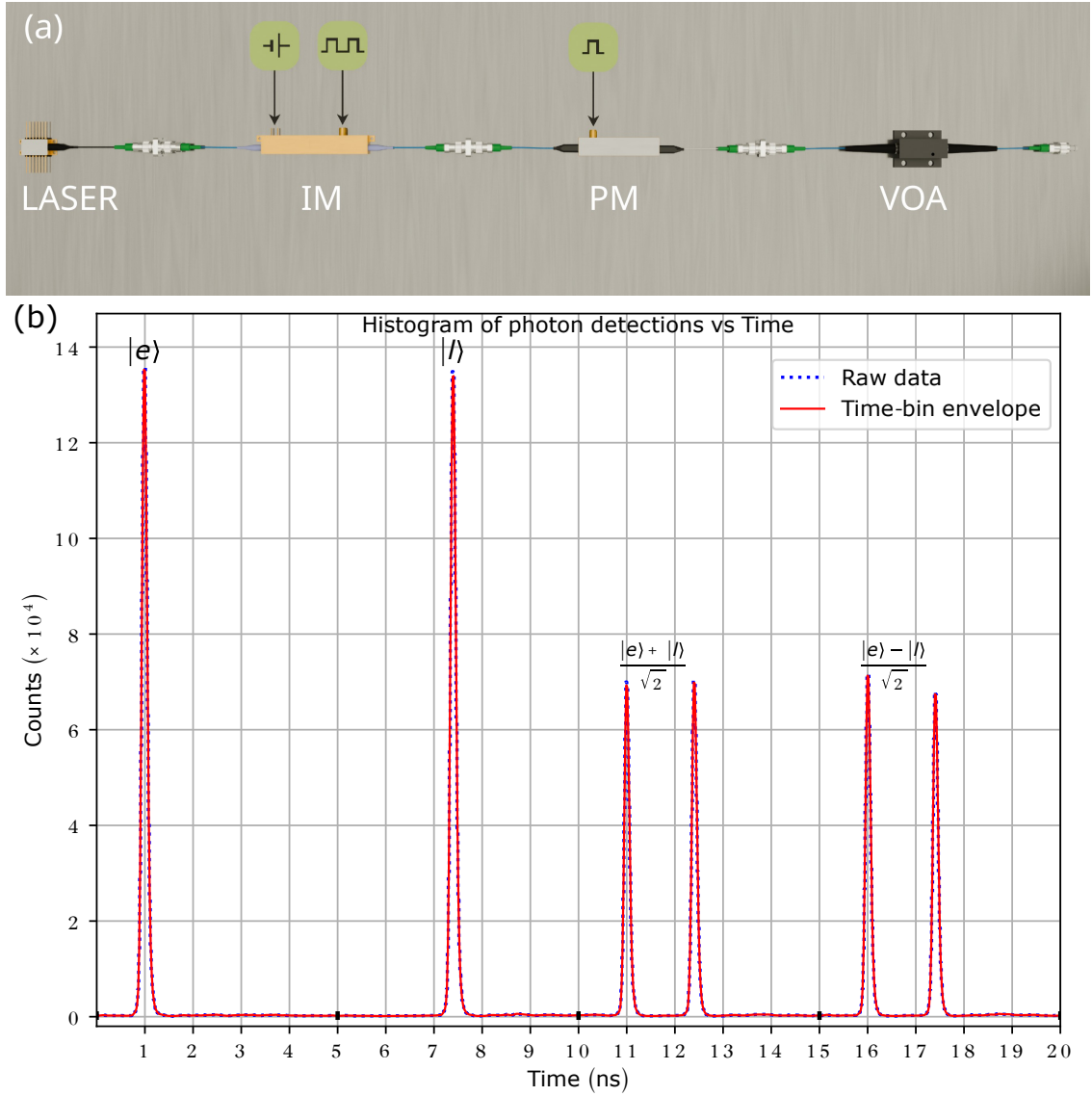}
\caption[Time bin qubit preparation using a CW laser source.]{(a) Schematic for time bin qubit preparation. An IM is used for carving out pulses from a CW laser source, which is then passed through a PM, producing arbitrary time-bin states. (b) Accumulated photon detections for TBQs generated in the states $|e\rangle$, $|l\rangle$, $|+\rangle=(|e\rangle+|l\rangle)/\sqrt{2}$, and $|-\rangle=(|e\rangle-|l\rangle)/\sqrt{2}$, respectively, using a CW laser and IM. The peak height of $|\pm\rangle$ states is half of the $|e\rangle$ and $|l\rangle$ states to maintain the same mean photon number per qubit. In the Z-basis, the $|+\rangle$ and $|-\rangle$ states are indistinguishable. To differentiate them, measurements must be performed on a superposition basis using delay-line interferometers (DLIs)}. 
\label{fig_TB_prep}
\end{figure}

The schematic of the experimental setup for the TBQ preparation and the prepared qubit states are shown in Figs. (\ref{fig_TB_prep}a) and (\ref{fig_TB_prep}b), respectively. The coherence time ($t_p$) of the pump laser source is typically much larger than the time-bin separation ($\tau_{el}$), such that the weak-coherent TBQ prepared in this manner shares a common phase coherence across multiple contiguous qubits, which is a prerequisite in distributed phase reference QKD protocols \cite{DPR-QKD_Kumar2024}. Hence, this method is predominantly used in distributed phase reference QKD, such as COW-QKD (see Section~\ref{Sec_COW-QKD}) and DPS-QKD \cite{DPS-QKD_Inoue2002, DPS-QKD_Inoue2003} systems.

The extinction ratio of time bins is one of the key parameters for characterizing the qubit. It is defined in decibels (dB) as follows:
\begin{equation}
    \text{Extinction Ratio (dB)} = 10 \log_{10}\left(\frac{\text{Max-counts}}{\text{Min-counts}}\right),
\end{equation}
where Max- and Min-counts denote the maximum number of counts in the time-bin peak and the average (background) counts between time bins, respectively, as can be determined from Fig.~(\ref{fig_TB_prep}b). The extinction ratio of TBQs is directly related to the quantum bit error rate (QBER) in the QKD protocols \cite{TB-QKD_Boaron2018}. Therefore, it is recommended to generate qubits with a high extinction ratio ($ > 30~dB$) to minimize the QBER. It is worth noting that the extinction ratio is a characteristic feature of the IM used to carve the time bins. 

Alternatively, one can use a CW laser and intensity modulator to create a single weak-coherent pulse and then use a DLI to create the phase-coherent pulse pair (time-bin superposition), followed by attenuation to prepare the TBQs. However, this method has the disadvantage that both IM and DLI require active stabilization. Hence, the generation of time bins using only a high-bandwidth IM is preferred as it is easier to implement.  In terms of preparation, time bins with sharp rise and fall times that are carved from a CW laser using a modulator also have requirements. The modulator requires a high bandwidth RF source and amplifier that can generate the narrow RF pulses having $V_{\pi}$ voltage to drive the electro-optic modulators (IM and PM).  The bandwidth of electro-optic modulators \cite{EOM_Lo2019} should be much larger than the pulse width and rise/fall time  (see Sec.~\ref{Sec_TempWidthTB}a).

\paragraph{Laser linewidth and frequency drift consideration:} 
We will distinguish between two contributions to the frequency variations of the laser source. The first is the fundamental uncertainty and short-term fluctuations of the laser frequency, which we refer to as the linewidth $\Delta\nu_l$ and which is associated with the laser's coherence time and length. The second is the long-term drift of the laser's mean frequency, $\Delta\nu_d$.
The frequency stability of the laser plays a crucial role in governing the phase stability of TBQs.

For example, in the case of weak-coherent TBQ, the phase error $\Delta \phi_s$ arising due to source laser is directly dependent on the time-bin separation $\tau_{el}$, the laser linewidth ($\Delta \nu_l$), and long-term frequency drift ($\Delta \nu_d$) as follows \cite{QCom_Shields2021, MDI_Woodward2021} :
\begin{equation}
\Delta \phi_s = 2 \pi \tau_{el}\sqrt{\Delta \nu_d^2+\Delta \nu_l^2}  =  2 \pi \tau_{el} \Delta \nu_s
\label{Eq_FreqtoPhaseUncertain}
\end{equation}
where $\Delta\nu_s$ is the source laser frequency error.  Therefore, to minimize the phase error of the qubit, the time-bin separation and the laser frequency error of the CW laser should be as small as possible. In Sec.~(\ref{Sec_Phase}), we elaborate on how this relates to and affects the choice of other parameters of the TBQs.

\subsubsection{Pulsed laser source}
\label{Sec_WCPS}
 
Pulse trains can also be generated directly by lasers through techniques such as mode-locking or direct laser modulation. These generated pulses can either be attenuated and passed through a DLI to produce a weak-coherent TBQ, or used to drive a non-linear medium, resulting in various types of TBQs, as illustrated in Fig.~(\ref{fig_TBFlowChart}). A mode-locked laser generates ultra-short pulses of light by locking the phases of different longitudinal modes of the laser cavity, resulting in constructive interference at regular intervals. This results in a train of coherent pulses with high peak powers and precise temporal separation, ideal for pumping nonlinear sources. However, the repetition rate is typically fixed and not easily scalable to higher rates, making mode-locking less suitable for scalable TBQ preparation.

Alternatively, optical pulses can be generated through direct laser modulation by varying the laser's drive current, a technique commonly known as gain switching. This method of encoding is often used in classical optical communication \cite{Review_Cheng2018}. In this method, the current is modulated periodically to drive the laser above and below its lasing threshold \cite{Laser_Siegman}, generating pulses as short as a few picoseconds. However, gain switching introduces frequency chirp and timing jitter due to inherent laser dynamics. Optical injection locking addresses these limitations by employing a primary-secondary laser configuration \cite{QCom_Shields2021, DirectLaserMod_Zhou2026}. Here, periodic emissions from the primary laser seed the secondary laser cavity, controlling the wavelength and phase of the stimulated emission. This results in reduced temporal jitter, relative intensity noise, and frequency chirp. Optical injection locking is particularly useful for improving the photon indistinguishability between different sources, a critical requirement for quantum information processing. Interestingly, gain-switched pulsed laser sources inherently produce phase-randomized pulses, making them valuable for applications such as quantum random number generation (QRNG) \cite{QRNG_Abellan2014}, and QKD \cite{TB-QKD_Alberto2018}. An alternative method has been demonstrated for generating phase-randomized states using steady-state emission of gain-switched laser diodes \cite{GainSwitching_Yuan2014} in the scalable implementation of temporal and phase-encoding QKD \cite{QKD-BB84TB_Francesconi2023}. Recently, a low-error encoder for time-bin and decoy states generation using nested Sagnac and Mach-Zehnder interferometers and a phase modulator has been demonstrated for QKD \cite{TBEncod_Scalcon2025}. By removing the need for active stabilization, the encoder offers a simplified optical architecture capable of producing quantum states in any dimension.

\subsection{Single Photon Source (SPS)}
\label{Sec_SPS}

\begin{figure} [ht!]
    \centering
    \includegraphics[width=0.9\linewidth]{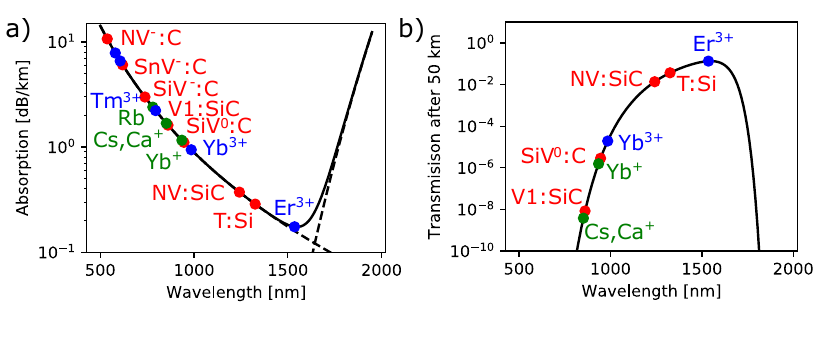}
    \caption[Optical emitters based on wavelength and fiber loss.]{(a) Optical emitters and absorption coefficient (black solid line) of ultrapure ``dry'' silica fiber caused by Rayleigh scattering and infrared absorption (black dashed lines). (b) Transmission after 50 km of optical fiber.
 \textbf{Figure reproduced with permission from \cite{QNet_Colloquium_Reiserer}}.}
    \label{Fig_Emitters}
\end{figure}

Single-photon sources (SPSs) can be classified into two main categories based on their emission characteristics: (i) probabilistic/heralded SPS, and (ii) deterministic or on-demand SPS. In probabilistic SPS, pairs of correlated photons are simultaneously generated in two distinct modes (polarization, frequency, spatial, etc.), and one photon is used to herald the presence of the other photon. These sources are known as probabilistic SPSs due to the fundamentally unpredictable nature of photon pair generation. SPDC \cite{SPDC-review_Zhang2021} and SFWM \cite{SFWM-Fiber_Palmett2023} are two methods for implementing probabilistic SPSs. In deterministic SPS, an external control, usually optical pumping as shown in Fig.~(\ref{fig_TBFlowChart}), is used to excite a single emitter system to emit a single photon through relaxation to its ground state. Examples of these sources \cite{SPS-Rev_Lounis2005, SPS_Buckley2012,  SPS-SS_Reimer2019, SPS-Rev_Sinha2019, SPS-Rev_Sinha2023, SPS-Rev_Stefan2009, SPS_Achar2024} include single atom \cite{SPS-Atom_Hijlkema2007, SPS-Atom_Higginbottom2016}, single ions \cite{SPS-Ion_Maurer2004, SPS-Ion_Barros2009}, quantum dots \cite{SPS-QD_Somaschi2016, SPS-QDot_Senellart2017, SPS-Qdot_Review2020, QDotReview_Loredo2026}, and colour centers \cite{SPS-CC_Khramtsov2018, SPS-CC_Andrini2024, SPS-CC_Patel2025}. Figure~(\ref{Fig_Emitters}) showcases various deterministic quantum light emitters along with attenuation in silica fibers at the corresponding wavelength. Transmission loss at visible wavelengths makes it prohibitive for fiber-optic communications. Telecom C-band turns out to be the best candidate with 0.2 dB/km fiber transmission loss. We discuss the methods for preparing TBQs using an SPS in sections  (\ref{Sec_QDotTB}) and (\ref{Sec_HeraldedTB}).

An ideal SPS \cite{SPS-Rev_Stefan2009, SPS-SS_Reimer2019} emits exactly one photon in a well-defined optical mode, ensuring efficient coupling and delivery for an experiment. Thus, an ideal source has two essential characteristics: (i) it emits at most one photon per emission, and (ii) it achieves perfect single-photon emission efficiency ($\eta_{\text{source}}=1$) \cite{SPS-Book_Migdall2013}. Such sources exhibit sub-Poissonian photon statistics \cite{Book_fox} as characterized by the condition: $\Delta n < \sqrt{\mu}$, where $\Delta n$ represents the standard deviation of the photon number distribution, and $\mu$ is the mean photon number. This inequality indicates reduced fluctuations compared to Poissonian statistics, a signature of non-classical light. Therefore, for a given mean photon number, the sub-Poissonian distribution is narrower than the Poissonian distribution. 

A convenient method to characterize photon number statistics is to measure the second-order intensity-intensity correlation function, $g^{(2)}(\mathcal{T})$ \cite{Book_fox, Book_Gerry-Knight}. It is measured in the Hanbury Brown and Twiss experiment \cite{HBT_Expt} and defined as
\begin{equation}
    g^{(2)}(\mathcal{T}) = \frac{\langle I(t)I(t+\mathcal{T})\rangle}{\langle I(t)\rangle \langle I(t+\mathcal{T})\rangle},
    \label{eq:g2define}
\end{equation}
where $I(t)$ is the intensity of the light field at time $t$. The Symbol $\langle I(t) \rangle$ indicates the time average of $I(t)$ computed over a long period. 

All SPSs exhibit anti-bunching in the second-order intensity-intensity auto-correlation function, i.e., $g^{(2)}(0)<1$. The $g^{(2)}(0)$ value is used to specify the purity of the SPS, and $g^{(2)}(0)=0$ is a defining characteristic of an ideal (pure) SPS. A coherent source, on the other hand, will exhibit a $g^{(2)}(\mathcal{T})=1$ \cite{SPS-Book_Migdall2013}.
Additionally, the quality and performance of SPSs are assessed by spectral brightness (photons emitted/s/nm), and indistinguishability of the generated photons \cite{SPS-Rev_Sinha2019, SPSD_Bienfang2023}. 
All practical SPSs have certain limitations/imperfections arising either from the underlying physical principle used for the generation of the `single-photon state' or experimental implementations thereof \cite{SPS-Book_Migdall2013, SPS-Rev_Eisaman2011, SPS-Rev_Scott2020, SPS-Qdot_Review2020, SPDC-review_Zhang2021}. 
Typical imperfections are the probabilistic nature of the source, including multi-photon emission, as stated, photon generation and collection efficiency, and spectral and temporal stability. For most SPSs, as the source brightness increases, the purity tends to decrease, indicating a trade-off between these two figures of merit. For single emitter sources, this trade-off is not inherent to the system, but caused by experimental imperfections, such as multiple excitations by the pump laser, which can be inhibited by shortening the excitation pulse. More details about the advantages and limitations of probabilistic single photon sources can be found in, e.g., references \cite{Ent-SPDC_Osung2013, SPDC_Burnham1970, SPS-Rev_Eisaman2011, SPS-Rev_Scott2020, SPS-Book_Migdall2013, SPS_Oxborrow2005}. However, owing to the ease of preparation of  weak-coherent pulses, they are widely used in QRNG, QKD, and other QIP protocol implementations \cite{QRNG_Ma2016, QKD-Decoy_HKL2004, QKD-RMP_Scarani2009}. 

\subsubsection{Deterministic or on-demand SPS}
\label{Sec_QDotTB}

We utilize ``single-emitter'' quantum systems \cite{SPS-SS_Aharonovich2016, SPS-SS_Esmann2024} as on-demand SPSs. Though these approaches use distinct material systems (see Fig.~\ref{Fig_Emitters}), they all operate on roughly similar principles. To emit single photons, an external control process drives the system into an excited state that emits a photon upon return to a lower energy state. Optical cavities are commonly used to enhance emission efficiency by channelling photons into a well-defined spatial mode. 
In this section, we will discuss quantum dots as a representative example of single-emitter systems for TBQ generation. The following discussion applies to any two-level system with a lifetime short enough, possibly through Purcell enhancement in a cavity, to be feasible for time-bin encoding. 

\begin{figure}[htb]
\begin{center}
\includegraphics[clip, trim= 1cm  1cm  2cm  0.5cm, width=0.5\columnwidth]{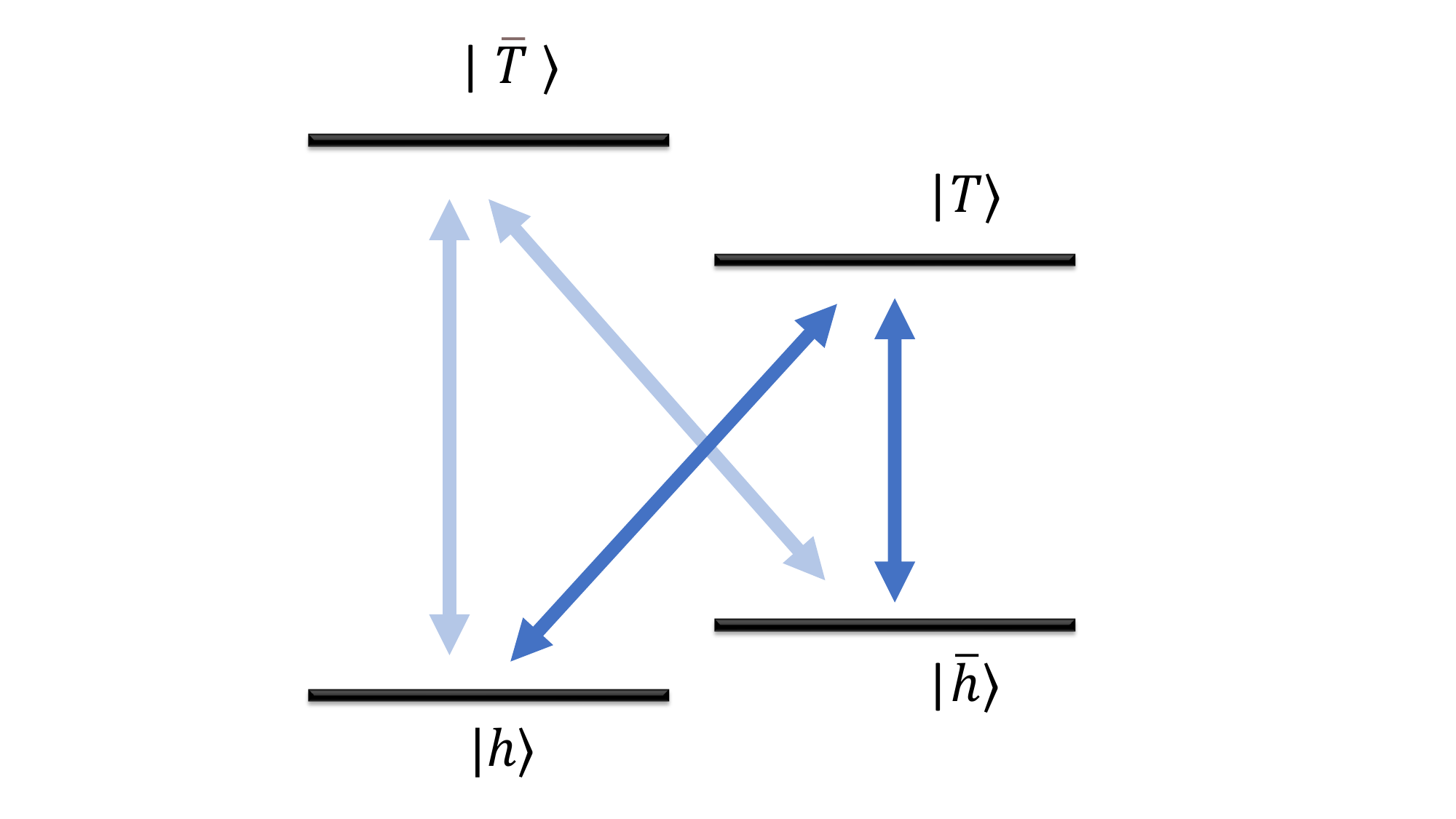}
\caption[Two-Raman process for TBQ generation in a quantum dot.]{Energy level diagram of quantum dot and fields to implement the two-Raman process for TBQ generation. Here, $|h\rangle$ and $|\bar{h}\rangle$ are orthogonal hole states and $|T\rangle$ and $|\bar{T}\rangle$ are the positive trion states}.
 \label{qbit-qdot}
\end{center}
\vspace{-2 em}
\end{figure}

Quantum dots \cite{QDotReview_Heindel23, QDotReview_Remesh2026} are nanoscale fabricated semiconductor systems with atom-like discrete energy levels, often approximated by a quantum-well potential formed in the semiconductor band structure. Electrons in the valence band of a quantum dot can be excited by an optical pulse to the conduction band, which leads to the formation of electron-hole pairs. When the electron-hole pair recombines, a single photon will be emitted. This single photon can be passed through a DLI to generate a TBQ as sketched in Fig.~(\ref{fig_TBP}). To avoid relying on phase-stabilized DLIs, quantum dots can also be used to generate TBQs in a more direct way using cavity-enhanced Raman scattering \cite{TB-Qdot_Lee2018}. This reference uses a single hole-charged InAs quantum dot in a Voigt geometry magnetic field to generate a double-lambda system of hole-positive trions, as shown in Fig.~(\ref{qbit-qdot}). A narrow bandwidth laser source drives the transitions between the hole and positive trion. 

After generating the holes, two resonant pulses drive the diagonal transition, generating a photon in a superposition of two time bins. The pulse sequence that is used to generate the arbitrary TBQs can be described as follows: The system is initialized in the $|h\rangle$ state by a non-resonant optical pumping pulse. The Raman transition occurs with a probability $p_1$ by a resonant pulse at time $t_1$ and results in a state $\sqrt{1-p_1}|h\rangle|0_{t_1}\rangle+\sqrt{p_1}|\bar{h}\rangle|1_{t_1}\rangle$, where $|1_{t_i}\rangle$ ($|0_{t_i}\rangle$) denotes the presence (absence) of a photon at time $t_i$. Another Raman transition occurs with a second resonance pulse at time $t_2$ with probability $p_2$, which results in the state $\sqrt{1-p_1}\sqrt{1-p_2}|h\rangle|0_{t_1}\rangle|0_{t_2}\rangle+\sqrt{p_1}|\bar{h}\rangle|1_{t_1}\rangle|0_{t_2}\rangle+\sqrt{1-p_1}\sqrt{p_2}|\bar{h}\rangle|0_{t_1}\rangle|1_{t_2}\rangle$.
If a photon is detected, the $| h \rangle$ term can be ignored and the state $|\bar{h}\rangle(\sqrt{p_1}|1_{t_1}\rangle|0_{t_2}\rangle+\sqrt{1-p_1}\sqrt{p_2}|\rangle|0_{t_1}\rangle|1_{t_2}\rangle)$ is obtained through post-selection. For the probability of creating a photon in each time bin to be equal, the second pulse should be brighter than the first pulse. If $p_1=2p_2=1$, the probability of being in each time-bin is equal, and we have $\frac{1}{\sqrt{2}} |\bar{h}\rangle (|1_{t_1}\rangle|0_{t_2}\rangle + |0_{t_1}\rangle|1_{t_2}\rangle)$.  

The benefit of using quantum dots is that they are on-demand sources for TBQ generation, making it easier to create a controlled, repeatable system for quantum information tasks. 
Moreover, since quantum dots can be integrated into existing semiconductor technology \cite{QDot-PIC_Norman2018}, they are potentially easier to scale up than other types of qubits. However, quantum dot-based TBQs also face challenges. For example, they require cryogenic temperatures for ideal operation. In addition, it is important to consider the decay time of the quantum dots for TBQ implementation since it sets a limit on the emission rate of the quantum dots. It should be noted that quantum dots are subject to various sources of dephasing and decoherence, which can decrease the coherence properties of TBQs. For example, acoustic phonons and charge noise can lead to dephasing and reduce the qubit coherence time. Decoherence limits the time duration for which the qubit information remains intact, affecting the overall qubit fidelity and hence transmission distance \cite{DecohControl_Roszak2015}.

\subsubsection{Strongly coupled two-level systems} 
\label{Sec_SCTLS}

The deterministic SPSs considered in the previous section require strong optical pump fields. However, if single-emitters are strongly coupled to an optical mode, e.g., using cavities or plasmonic enhancement, even a weak optical pump field can realize such SPS \cite{SCTLS_NLO_Review_Lukin_2014}.
Such systems, which generally consist of a neutral atom, defect center, quantum dot, or trapped ion, with an optical transition coupled to a cavity mode, provide unique capabilities for manipulating and controlling quantum states.

The strong coupling regime is realized when the coupling strength between the emitter and the cavity surpasses the decay rates of both the emitter and the cavity. This regime facilitates coherent energy exchange between the emitter and the cavity, resulting in the formation of hybrid light-matter states known as polaritons. In these cavity quantum electrodynamics systems, optical nonlinearity arises from the discrete energy-level structure of the atom. However, the electronic states of atoms typically exhibit short coherence times \cite{SCTLS_NLO_Review_Lukin_2014}. To address this limitation, metastable states are employed, allowing quantum coherence to be stored for extended durations.

The strongly coupled two-level systems exhibit the property of reflecting an incoming photon if the coupled atomic level is in the excited state. By leveraging this phenomenon, spin-photon entanglement can be generated as described here, e.g., \cite{ SCTLS_BellState_Rempe_2017}. The procedure begins with the preparation of the strongly coupled two-level system in a superposition state. This is followed by a pair of weak-coherent pulses, representing a TBQ, sent to the cavity. The initial product state of the spin and the TBQ; $\ket{\psi_i} = (\alpha \ket{e} + \beta \ket{l})(\ket{\uparrow} + \ket{\downarrow})$, is ``carved out'' into the Bell state by inverting the spin with a $\pi$ pulse between time-bins. Thus, the reflected photon is entangled with the spin, $\ket{\psi_f} = \alpha \ket{e\downarrow} + \beta \ket{l\uparrow}$. Such a system is highly valuable for quantum networks, enabling the distribution of entanglement over large distances via the photonic qubit, as well as long-term storage of information in the long-lived spin qubit \cite{SCTLS_QNet_Protocol_1997, SCTLS_QNet_Rempe_2015}. This approach has been successfully demonstrated with, e.g., neutral atoms \cite{SPS-AtomCavity_Reiserer2015}, silicon-vacancy (SiV) centers \cite{SCTLS_QNet_SiV_Lukin_2019, SCTLS-QNet-SiV_Lukin2024}, and nitrogen-vacancy (NV) centers \cite{SCTLS_QNet_NV_Hanson_2013} in diamond photonics.

\subsubsection{Probabilistic sources based on spontaneous parametric processes}
\label{Sec_HeraldedTB}

TBQ can be prepared using heralded SPSs such as SPDC or SFWM. Under these processes, a laser source pumps a nonlinear material, generating a pair of correlated photons. One of the photons from the pair, say the idler, is sent to a detector, where its detection heralds the presence of a signal photon. The signal photon is passed through a DLI (as shown in Fig. (\ref{fig_TBP})) to generate a time-bin superposition of `early' and `late' states at the output port of the interferometer, thus creating a TBQ \cite{TB-SPDC_Erhan2012}. However, the probabilistic nature of the pair-generation process sets a trade-off between the qubit generation rate and the purity of the state (defined by the auto-correlation function $g^2(0)$) due to the multiple photon-pair generation in the non-linear processes \cite{SPDC_Takeoka2015, SPDC_Broome2011}. A continuous wave pump-based source can also generate TBQs using post-selection. However, the limitation is that one cannot temporally filter the photons, thus limiting the qubit generation rate. 

\begin{figure}[ht!]
\centering
\includegraphics[width=0.80\columnwidth]{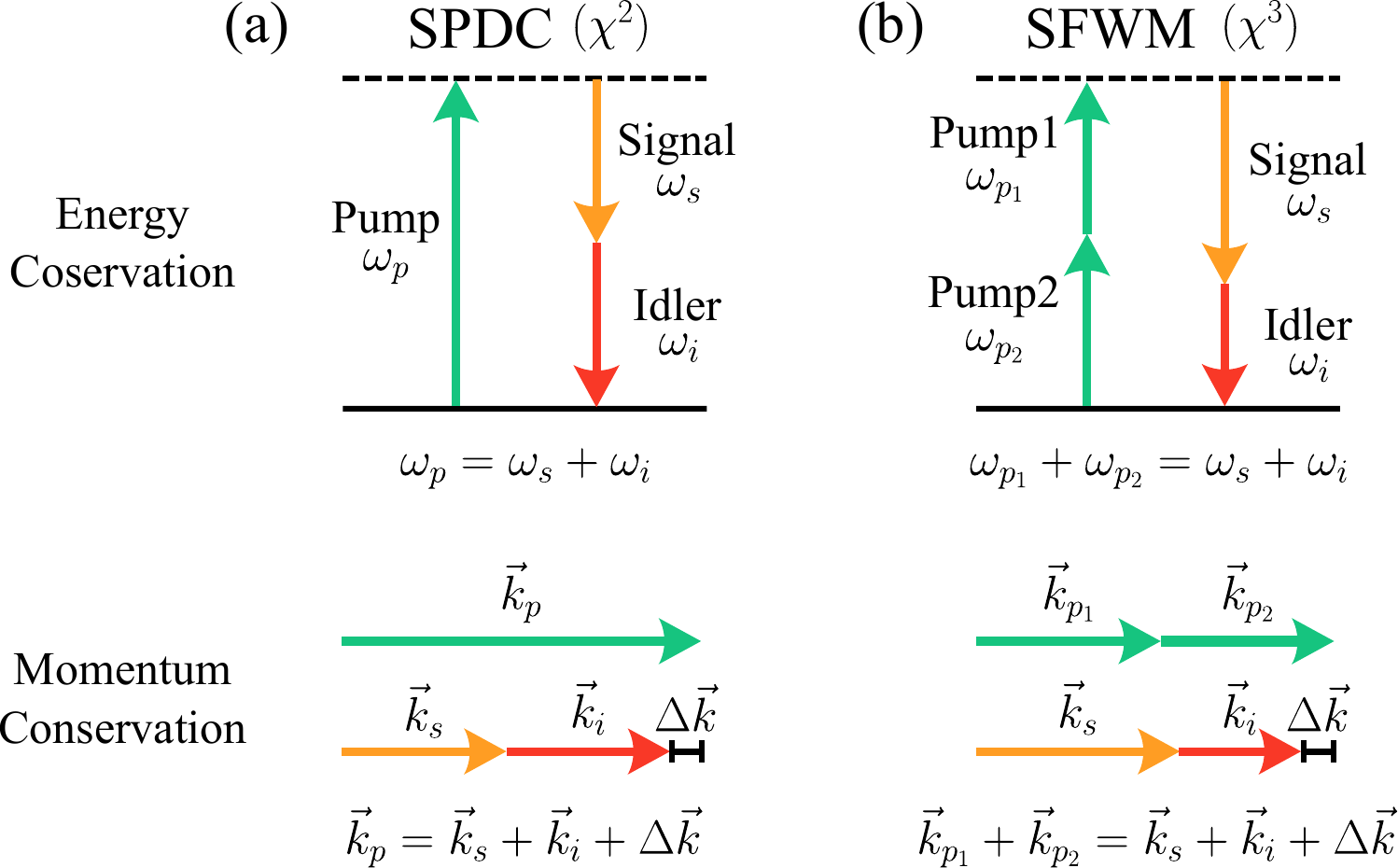}
\caption[Phase matching condition in SPDC and SFWM processes.]{Phase matching condition for (a) SPDC and (b) SFWM process. The $\omega_p (\Vec{k_p}), \omega_s (\Vec{k_s}), \omega_i (\Vec{k_i})$ are the energy (momentum) of the pump, signal, and idler photons, respectively, and $\Delta k $ is the phase mismatch term, which should be zero for perfect phase matching.}
\label{fig_PhaseMatch}
\end{figure}

\textit{SPDC} is a second-order nonlinear ($\chi^{(2)}$) optical process \cite{NLO_Boyd2008}, wherein a strong pump field interacting with a nonlinear medium induces the spontaneous generation of correlated photon pairs (signal and idler) from the vacuum fluctuations \cite{SPDC_Louisell1961, SPDC_Harris1967, SPDC_Burnham1970, SPDC_Couteau2018}.
It satisfies the conservation of energy and momentum, collectively referred to as the \textit{phase-matching condition}, as depicted in Fig.~(\ref{fig_PhaseMatch}\,a). However, typically, quasi-phase matching is used to compensate for phase mismatches and prevent destructive interference in the long waveguides. This is achieved by periodically modulating the nonlinear crystal's properties (e.g., through periodic poling), enabling constructive interference and efficient photon pair generation, which allows for tunable, high-efficiency entangled photon sources in quantum technologies. Critical phase matching can also be employed by utilizing birefringent material and dispersion. An in-depth discussion on various implementations of SPDC for time-bin entanglement generation can be found in Section (\ref{Ent_SPDC}).

\textit{SFWM} is a third-order ($\chi^{(3)}$) non-linear process \cite{SFWM_Yariv1977, NLO_Boyd2008, SFWM-Fiber_Palmett2023} in which two pump photons interact to spontaneously generate a signal-idler photon pair.  As the pump light propagates through a $\chi^{(3)}$ material, photon pairs are generated from the vacuum in sidebands. 
Hence, the generated entangled pair is close to the pump laser wavelength.
Notably, all optical materials exhibit some degree of $\chi^{(3)}$ non-linearity, with commonly used materials including Si$_3$N$_4$, GaAs, Si and SiO$_2$. 
However, typically $\chi^{(3)}$ coefficient weaker compared to $\chi^{(2)}$ coefficient of non-linear materials.
SFWM is also commonly used in integrated photonic circuits, leveraging the inherent non-linearity and cavity enhancement provided by micro-ring resonators \cite{SFWM-Review_Wang2023}. 
This is advantageous because photon-pair generation, filtering, and encoding can all be done on commonly used optical platforms such as Si$_3$N$_4$ and Si.

The SFWM is characterized by two distinct regimes: degenerate and non-degenerate. Degenerate SFWM occurs when the pump photons of the same frequency ($\omega_{p_1} = \omega_{p_2}$) result in a pair of photons. 
However, the non-degenerate case requires pump photons from different frequencies ($\omega_{p_1} \neq \omega_{p_2}$). 
The SFWM process also follows the phase matching conditions as shown in Fig~(\ref{fig_PhaseMatch}b). 
Various techniques can be employed to achieve efficient phase matching, such as dispersion engineering of the material or waveguide, utilizing birefringent material, quasi-phase matching, and setting the pump frequency near anomalous dispersion \cite{NLO_Boyd2008, SFWM-Review_Wang2023}. 
In the phase-matching condition, the SFWM process becomes more efficient, leading to enhanced photon pair generation and improved conversion efficiency. Similar to SPDC, the phase-matching condition results in the generation of a broad spectrum. Consequently, spectral filtering of the signal and idler photons is necessary.

SFWM requires the annihilation of two pump photons, resulting in the single-photon and the coincidence count rate scaling quadratically with the pump power ($\mu\sim P^2$). In contrast, in the SPDC, they scale linearly with the pump power ($\mu\sim P$). 
However, the coincidence-to-accidental ratio (CAR) decreases with increasing mean photon-pair number $\mu$ (CAR $\sim 1/\mu$) and, hence, scales with pump power as $\sim 1/P^2$ ($\sim 1/P$) for SFWM (SPDC) in the high power regime. This results in an increase in the second-order autocorrelation function $g^{2}(0)$ at higher powers, mainly due to multi-photon events.
Hence, the pump power cannot be increased arbitrarily, which limits the photon generation efficiency and, consequently, the count rate and the heralding efficiency of heralded single-photon sources. An in-depth discussion of various implementations of SFWM for time-bin entanglement generation can be found in Section (\ref{Ent_SFWM}). A recent review \cite{SPS-IntPhot_Wang2021} outlines key concepts of parametric photon pair sources, current advances, and future prospects in integration and multiplexing.

\section{Time-bin qubit transmission}
\label{Transmission}
\FloatBarrier

In quantum communication protocols, TBQs are typically transmitted over free-space or fiber-optic channels. The suitability of TBQs for a practical application in long-distance quantum communication is greatly influenced by the characteristics of the transmission channel. Channel-induced noise, such as dispersion, scattering, loss, and other environmental noise, could impact the fidelity, coherence, and indistinguishability of the time-bin. In many cases, it is desirable for TBQs to coexist with classical optical signals; however, engineering challenges do exist in terms of minimizing crosstalk for reliable quantum information transfer. Hence, understanding these channel-dependent limitations is essential when putting TBQs into practice. This section addresses such challenges involved in transmitting TBQs. Many of the concepts discussed here are broadly relevant to other encoding schemes as well.

\subsection{Fiber-optic transmission}\label{Fiber-optic transmission}

Signal attenuation, chromatic dispersion, polarization mode dispersion, and noise-inducing spontaneous Raman scattering (SRS) due to potentially coexisting classical signals are the detriments to the fiber-optic transmission of TBQs. As we will see in Section~(\ref{Sec_PMD}), the TBQs are particularly insensitive to polarization mode dispersion as the information is encoded in the arrival time of photons. 

\subsubsection{Signal attenuation}
\label{Sec_FiberAtt}

A (quantum) signal propagating through an optical fiber undergoes attenuation due to scattering and absorption, which causes an exponential decay with transmission distance. If a signal with mean photon number $\mu_{\text{in}}$ is launched into a fiber of length $L$ (km), then the mean photon number at the output port \cite{Book_agrawal1} is given by:
\begin{equation}
\mu_{\text{out}} = \mu_{\text{in}} e^{-\upalpha L},
\label{Eq_ChanelAtt}
\end{equation}
where $\upalpha$ is the fiber attenuation coefficient that captures the fiber loss per km from all sources. Conventionally, it is expressed in the units of dB/km as follows:
\begin{equation}
\upalpha_\mathrm{dB}=-\left(\frac{10}{L}\right)\log_{10}\left[\frac{\mu_{out}}{\mu_{in}}\right] = 4.343~\upalpha.
\label{Eq_ChanelAttdB}
\end{equation}
This attenuation of optical signal power is naturally equivalent to a non-unity qubit transmission probability, resulting in a reduced photon detection rate given by Eq.~(\ref{Eq_ChanelAttdB}).
The attenuation constant depends on the wavelength of light and reaches a minimum value of about $0.2$~dB/km near 1550 nm for standard single-mode fibers (SMF-28). For other wavelengths, the attenuation increases considerably as shown in figure \ref{Fig_Emitters} and reaches a level of a few dB/km in the visible range. For extended transmission distances, the significance of dark counts originating from the SPDs at the receiver becomes evident, leading to a reduction in the signal-to-noise ratio (SNR). Therefore, repeater-less fiber-based communication links are limited to a few hundred kilometers. For longer distances, one must adopt quantum repeaters (see Sec.~\ref{Sec_QNetworks}) or free-space quantum communication \cite{QNwtwork_Simon2017} as discussed in Section~(\ref{Sec_FSTransmission}).

\subsubsection{Chromatic dispersion}
\label{Sec_ChromDisp}

In a dispersive medium, the phase and group velocities of light depend on the signal frequency. Consider an optical pulse with spectral width $\Delta\omega$ (where $\Delta\omega= 2\pi\Delta\nu_\tau$, and $\Delta\nu_\tau$ is the spectral bandwidth of the pulse defined in Eq.~(\ref{Eq_TBP})), launched into an SMF of length L. Due to frequency-dependent group velocity, spectral components travel at different velocities and reach the fiber output at different times, causing temporal mode broadening or chromatic dispersion \cite{Book_agrawal1, Book_agrawal2}. For an optical pulse of spectral width $\Delta \omega$, the temporal broadening due to chromatic dispersion in a fiber of length L is given by 
\begin{equation}
\Delta T = \frac{dT}{d\omega}\Delta\omega = L\frac{d^2\beta}{d\omega^2}\Delta\omega = L\beta_2\Delta\omega,
\end{equation}
where $\beta$ is the fiber propagation constant, and $\beta_2 = d^2\beta/d\omega^2$ is known as the group velocity dispersion. Alternatively, $\Delta T = L D \Delta\lambda$ in terms of the wavelength spread, where $D = -\frac{2\pi c}{\lambda^2}\beta_2$ is known as the dispersion parameter (unit: ps/nm-km). Furthermore, the dispersion parameter can be decomposed into the material dispersion ($D_M$) and the waveguide dispersion ($D_W$). The material dispersion depends primarily on the optical properties of the propagation medium, whereas the waveguide dispersion arises from the spatial distribution of the mode along the waveguide geometry. A waveguide with a smaller mode area exhibits stronger dispersion effects, and it can be engineered by modifying the waveguide geometry.
For standard silica-based single-mode fibers (SMF28), the dispersion parameter is the smallest ($D \sim 1$ ~ps/nm-km) at 1.3 $\mu m$ range and reaches $\sim$ 17~ps/nm-km at 1.5 $\mu m$ wavelength range. Dispersion limits the maximum bit rate ($B$) permissible since the product $B \Delta T$ should be less than unity. 

\textit{Dispersion compensation methods:} The temporal broadening of pulses, a consequence of dispersion \cite{Disp_Potasek1986, Disp_Anritsu2004, Disp_wandel2005, Disp_Gruuner2007, Disp_Malekiha2015, Disp_Ruan2021, Disp_Czerwinski2024}, has far-reaching implications for quantum communication, especially in QKD. An increase in pulse width can reduce the orthogonality of time-bin modes (see Fig.~\ref{fig_TBplot}), resulting in increased quantum bit error rates and a drop in overall fidelity of quantum information processing tasks. Consequently, addressing dispersion becomes pivotal for maintaining the integrity and efficiency of quantum communication protocols.

\begin{figure}[ht!]
    \centering
    \includegraphics[width=\columnwidth]{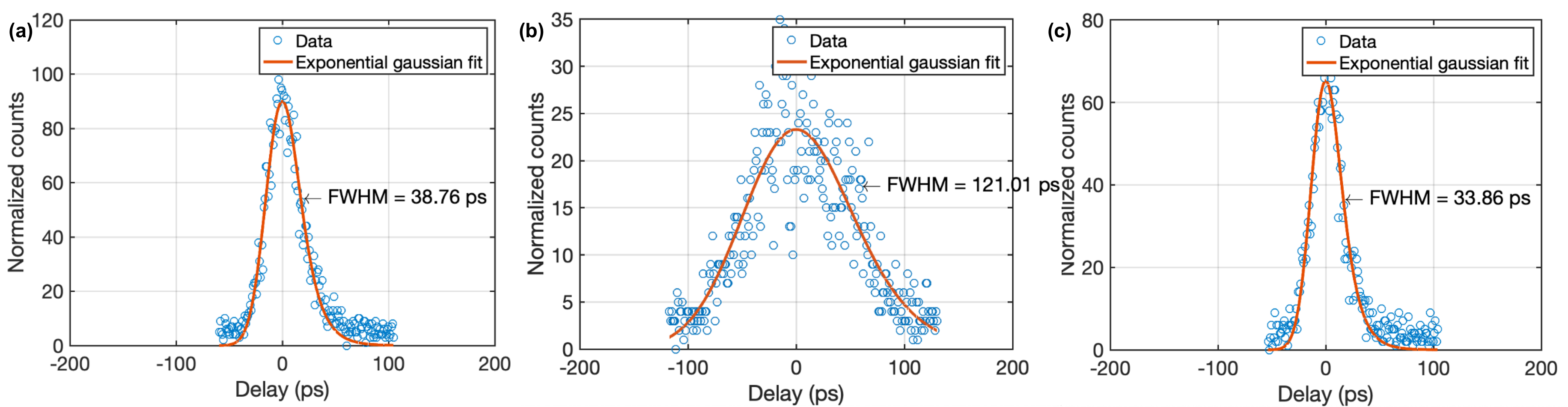}
    \caption[Pulse broadening and compensation in optical fiber transmission.]{Histograms of single-photon detection times demonstrating pulse broadening and compensation in optical fiber transmission. (a) Initial pulse width before transmission through the fiber. (b) Pulse broadening as observed after transmission through an 11 km fiber-spool due to chromatic dispersion. (c) Pulsewidth restoration after applying a Dispersion Compensation Module (DCM) to counteract the dispersion accumulated in the fiber and recover the temporal shape of the original pulse.}
\end{figure}

Within the domain of QKD, existing telecommunication devices offer a feasible avenue for dispersion compensation. Dispersion compensation has been explored using dispersion compensating fiber  \cite{TB-QKD_Alberto2018} and fiber-Bragg gratings \cite{Disp-FBG_Hill1994, Disp-FBG_Sumetsky2007, Disp-QKD_Dynes2012}. 
Choosing a fiber with the opposite sign of group velocity dispersion and a suitable length can compensate for the whole of the dispersion experienced by the photons. 
However, the solution can only work for a fixed-length quantum channel, and changes need to be made to the length of the compensation fiber for varying channel lengths, thereby making this method less flexible. 
On the other hand, chirped fiber Bragg gratings, where different wavelengths reflect from different sections of the gratings, thereby allowing us to overcome the broadening due to dispersion. This method also offers flexibility to compensate for different channel lengths by tuning the device's temperature. However, associated losses typically range from 6 to 10 dB, depending on the required amount of dispersion correction, thus posing a formidable challenge. Another common approach is the integration of dispersion compensation directly into the sender's (Alice's) setup before attenuation to weak-coherent pulses \cite{Disp-QKD_Pathak2023}. This allows for the transmission of pre-compensated pulses, effectively circumventing the issue of compensation device losses and preventing their contribution to overall channel losses. 

Interestingly, non-local dispersion compensation techniques have also been developed, particularly for time-energy entanglement distribution (see Sec.~\ref{Sec_TECharMethods}). By applying a negative dispersion to one photon \cite{Disp_Neumann2021}, we can effectively compensate for the dispersion experienced by the entangled photon pair as shown in Fig.~(\ref{fig_time-energy3}) . This approach, demonstrated by various research groups, holds promise in maintaining high communication rates over extended fiber distances while minimizing losses. An alternative strategy involves generating photons around the zero dispersion wavelength. Through careful preparation, where one photon carries positive dispersion and the other negative, a natural compensation of the two-photon wave packet occurs as they traverse the fiber. One can also employ dispersion-shifted fiber (DSF), which is designed to have zero dispersion at the signal wavelength \cite{Disp-DSF_Zhang2009}, effectively minimizing the pulse broadening induced by chromatic dispersion.

\subsubsection{Polarization mode dispersion}
\label{Sec_PMD}

Optical fibers are typically birefringent due to asymmetrical cross-sections caused by manufacturing imperfections (e.g., SMF) or by design (e.g., PMF). Consequently, orthogonal polarization components experience distinct effective group velocities. 
Therefore, an optical pulse containing both polarization components experiences broadening during transmission. This phenomenon is known as polarization mode dispersion \cite{PMD_Nelson2004}. For a PMF of length $L$, the broadening due to polarization mode dispersion is given as 
\begin{equation}
    \Delta T = \left| \frac{L}{v_{gx}} - \frac{L}{v_{gy}} \right| = L (\beta_{1x}-\beta_{1y})=L\Delta \beta_1,
\end{equation}
where x and y denote two orthogonal polarization modes and $\Delta \beta_1$ is related to the group velocity difference in the two principal states of polarization \cite{PMD_Nelson2004}. In conventional SMF, birefringence varies randomly along the length of the fiber. Thus, the polarization state of the light changes in a random fashion, which induces pulse broadening. 

The broadening induced by polarization mode dispersion in SMF is described by the root mean squared value of $\Delta T$, which captures the average behaviour of the random changes in birefringence. The variance $\sigma^2_T=\langle \Delta T ^2\rangle$ \cite{Book_agrawal1} is given by
\begin{equation}
    \sigma^2_T(z)=2(\Delta\beta_1)^2l^2_c \left[\exp(-z/l_c) +z/l_c-1 \right],
\end{equation}
where the two polarization components remain correlated over the length $l_c$ known as the correlation length ($\sim 10~m$). For short distances ($z<<l_c$), $\sigma_T= (\Delta \beta_1) z$ and longer distances ($z = L >>l_c$), $\sigma_T \approx \Delta\beta_1 \sqrt{(2l_c L)} \equiv D_p \sqrt{L}$, where $D_P$ is known as the polarization mode dispersion parameter. For modern spooled fiber, $D_p\approx 0.1~ps/\sqrt{km}$ and $\sigma_T\approx 1$ ps/(100 km fiber), which is much smaller than the broadening due to chromatic dispersion. Hence, for pulses longer than 10 ps, polarization mode dispersion effects can be ignored. 
Hence, only if chromatic dispersion is compensated (see Sec.~\ref{Sec_ChromDisp}), the remaining temporal broadening due to polarization mode dispersion will limit the rate of polarization-encoded data transmission at sufficiently high bit rates ($> 10$ GHz) and over long fibers ($> 60$ km). Hence, it is only in this limit that time-bin encoding provides an advantage, in that it is inherently insensitive to polarization mode dispersion.

\subsubsection{Spontaneous Raman Scattering}
\label{Sec_SRS}

\begin{figure}[ht!]
    \centering
   \includegraphics[width=\columnwidth]{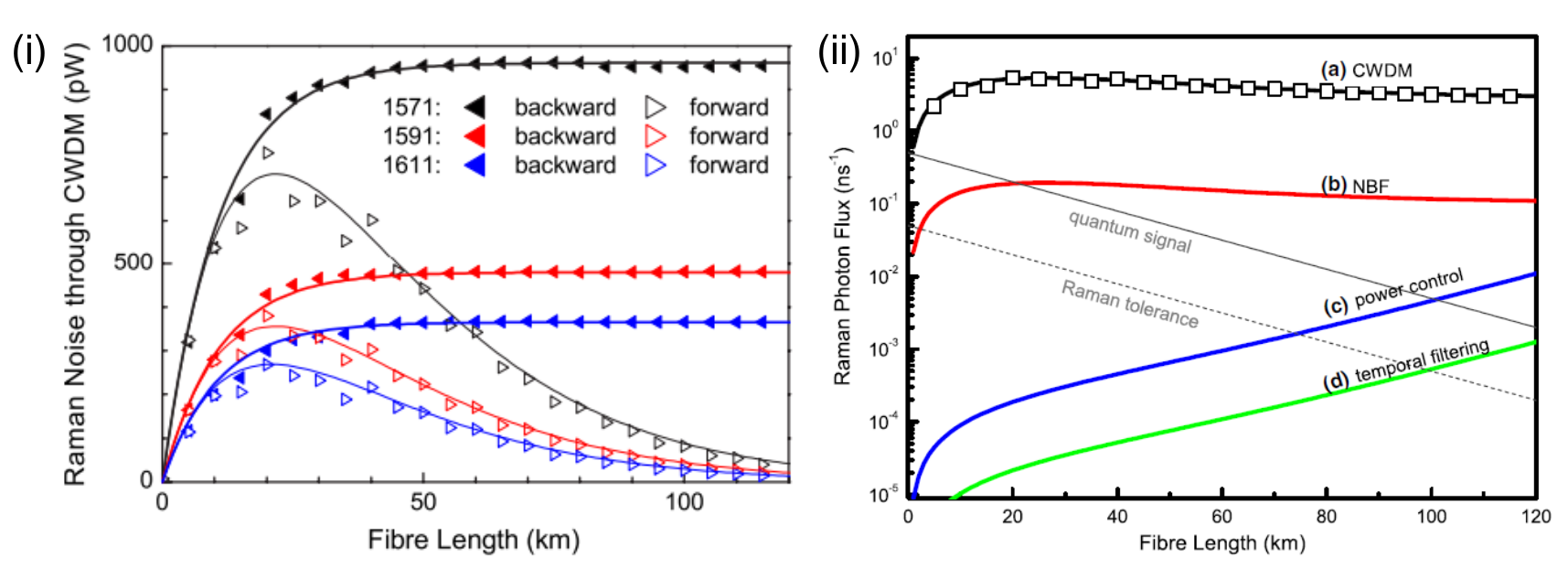}
    \caption[Spontaneous Raman scattering noise in optical fibers and its mitigation.]{(i) Forward and backward Raman scattering from the classical channel at different wavelengths and fixed 0~dBm launch power to the quantum channel at 1550 nm filtered through coarse wavelength division multiplexing (CWDM $\sim 20$~nm pass-band). \textbf{Figure reproduced under Creative Commons license from \cite{QKD-WDM_Patel2012}}. (ii) Raman noise at the receiver's single-photon detector. (a) Experimentally measured (symbols) and theoretically estimated (solid line) Raman noise after Bob's CWDM. (b) Raman noise after the narrow-band filter (NBF). In both cases, 0-dBm lasers at 1591 and 1611 nm are launched from Alice and Bob. (c) Raman noise decreases with reduced launch power. (d) Raman noise detected within the gated detector's active window. Solid lines in (b)-(c) show calculated values. \textbf{ Figure reproduced under Creative Commons license from \cite{QKD-WDM_Patel2012}}.}
    \label{fig:SRS-TempFilter}
    \end{figure}

It is advantageous to use fiber that also carries classical signals rather than use fibers specifically devoted only to quantum communications, which is referred to as ``dark'' fiber.
This allows the existing network of deployed optical fibers to be utilized for quantum communication using wavelength division multiplexing, thereby reducing the cost of deploying a global quantum network. However, SRS noise arising from the interaction of the classical signal with the phonon modes of the optical fiber degrades the signal-to-noise ratio of the quantum channel, which poses a critical challenge in building quantum networks \cite{QKD-WDM_Patel2012, QKD-WDM_Silva2014, QKD-WDM_Wang2017, QCom-WDM_Kim2022, QKD-WDM_Wang2023}. Although one can readily filter out the classical signal wavelength, however, scattered photons generated by SRS or nonlinear processes (e.g SFWM) in the fiber are impossible to suppress entirely due to their spectral overlap with the quantum signal. Here, we briefly describe the standard experimental impact and techniques for SRS-noise suppression in quantum communication, which are summarized by Fig.~(\ref{fig:SRS-TempFilter}).

\textit{(a) Minimizing launch power for classical signals:} Minimizing the launch power of the classical signal while using high-gain/sensitivity detectors at the receiver reduces the SRS noise, which scales linearly with the power, cf. Eq.~(\ref{eq:SRScocounter}). This SRS mitigation approach relies on achieving a balance between lowering the classical signal launch power and maintaining the signal strength at the end of the channel above the detection threshold set by the classical receiver. Although this approach can effectively reduce SRS noise, excessive reduction in launch power may conflict with the requirements of communication standards, leading to signal degradation at the receiver.

\textit{(b) Consideration of co-propagating vs. counter-propagating classical signals:} For a laser with power $P_0$ incident into an optical fiber of length $L$, the SRS noise per nanometer scattered on the quantum channel at the fiber's output can be described as:
\begin{equation}\label{eq:SRScocounter}
\begin{split}
    & P_{\text{SRS, co}} = P_0 \beta_f L \exp(-\upalpha L),\\
    & P_{\text{SRS, counter}} = \frac{P_0 \beta_b L}{2\upalpha}\left[1-\exp(-2\upalpha L)\right],\\
\end{split}
\end{equation}
where $\beta_f$ and $\beta_b$ are the effective SRS coefficients for forward and backward scattering. Backward scattering is stronger than forward scattering in long fibers as the classical signal travels back to the SPD at the receiver without being subjected to attenuation, whereas forward scattering diminishes due to fiber attenuation. Counter-propagating signals can thus result in higher Raman noise at the receiver, especially in longer fiber spans. 

\begin{figure}[ht!]
\centering
\includegraphics[width=1\columnwidth]{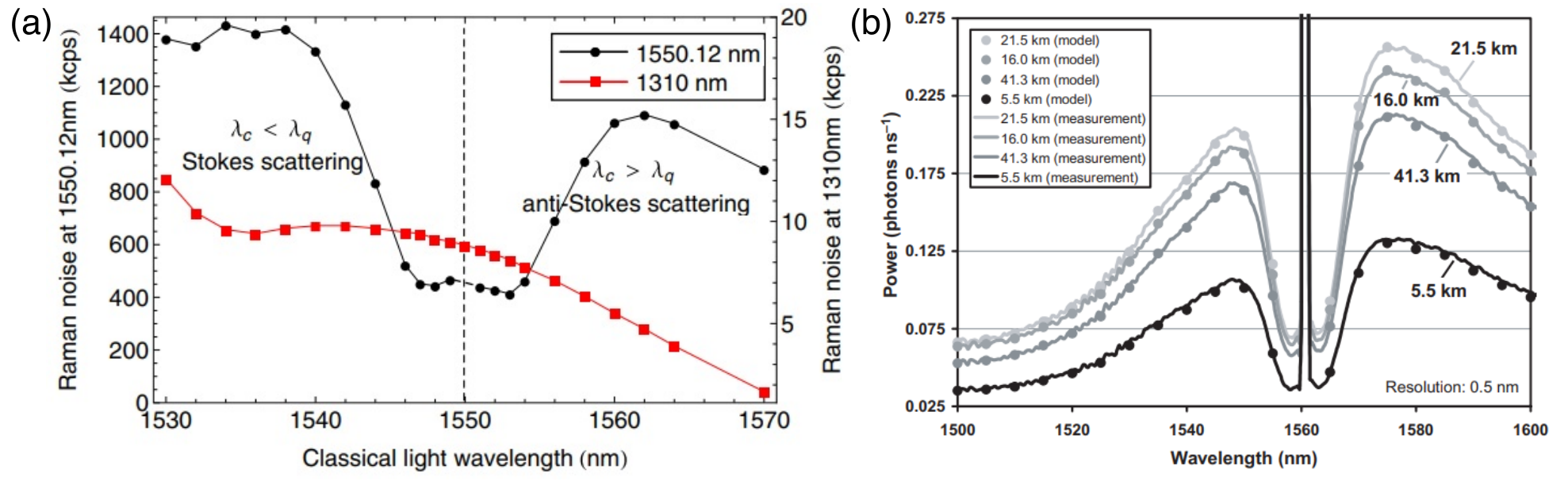}
\caption[Raman noise at different wavelengths.]{(a) Raman scattering from classical light in the C-band into quantum channels at 1550.12 nm and 1310 nm. \textbf{Figure reproduced with permission from \cite{QKD-WDM_Wang2017}}. (b) Experimentally measured and simulated Raman spectra produced by a continuous-wave pump propagating through varying lengths of SSMF. \textbf{Figure reproduced under Creative Commons license from \cite{SRS_Peters2009}}.}
\label{fig:RamanHump}
\end{figure}

\textit{(c) Wavelength selection:} SRS noise in the quantum channel can be minimized by proper selection of the wavelengths of the classical and quantum channels. The SRS spectrum, depicted for a 1560~nm classical channel in Fig.~(\ref{fig:RamanHump}), has distinct maxima on each side. The anti-Stokes side of the SRS spectrum (wavelength lower than the wavelength of the classical light) is generally favourable for quantum channels, as shown in Fig.~(\ref{fig:SRS-TempFilter}b) for single-mode fiber. In general, a larger wavelength separation between the classical and quantum channels is also favourable, with an arrangement of a quantum channel in the telecom O-band and classical channels in the C-band commonly suggested \cite{QKD-WDM_Patel2012}. Although the SRS spectrum drops near the classical channel wavelength, suggesting a suitable wavelength range for the quantum channel, this may be negated by the presence of FWM or side-band modulation if multiple classical channels are simultaneously present. In general, the SRS spectrum can contain complex features, such that displacement of the quantum channel by only a few tens of nanometers can reduce the SRS noise by an order of magnitude \cite{SRS_Thomas2023}. Specifically, when the classical channel is set at 1550 nm, the quantum channel at 1290 nm experiences significantly lower noise than the one at 1310 nm \cite{SRS_Thomas2023}. Recently, Zischler et al. \cite{CoexistingQKD_Zischler2025} developed a semi-analytical propagation model incorporating Raman scattering, four-wave mixing, spatial crosstalk, and Rayleigh back-scattering in single-mode and space-division-multiplexed (SDM) fibers, and showed that interference on coexisting quantum channels is minimized at higher transmission bands, with four-wave mixing being negligible for counter-propagating but potentially relevant for co-propagating classical signals.

\textit{(d) Spectral filtering:} Narrow-band filtering using devices such as Fiber-Bragg gratings (FBG) can isolate the quantum signal from a broad SRS noise spectrum. Although effective, this technique requires high precision and fine-tuning to ensure the overlap of the narrow filter pass-band with the quantum signal.

\textit{(e) Temporal filtering:} As shown in Fig.~(\ref{fig:SRS-TempFilter}), using gated detectors, the system can temporally filter out SRS noise~\cite{QKD-WDM_Patel2012}. This method relies on synchronizing the detection time window with the arrival of quantum signals, reducing the chances of detecting SRS noise that falls outside this window. This technique improves signal integrity by minimizing unwanted noise (including SRS) detection during off periods.

\textit{(f) Coherent filtering:} Spectral and temporal filters discriminate modes based on their intensities, which contain only partial information. Coherent filtering, that is, mode selection based on the entire properties of the mode, which includes phase, can allow improved noise selection \cite{Filtering_Raymer2020}.
This may be achieved using $\chi^{(2)}$ or $\chi^{(3)}$ processes in nonlinear crystals for instance.

\textit{(g) Cooling the fiber:} Lowering the fiber's temperature can reduce the phonon population in the fiber, which in turn suppresses SRS. Thus, cooling the fiber can lead to a significant reduction in SRS noise. However, the requirement to reduce the temperature of an entire deployed fibre would add significant complexity and operational cost, making it unlikely to find practical use in quantum communication \cite{Ent-SRS_Takesue2006}. Furthermore, fibers are constantly upgraded through doping and other techniques to reduce the SRS noise.

\textit{(h) Polarization:} The SRS noise photons are expected to be predominantly generated in the polarization mode, the same as the classical signal. Therefore, launching the classical and quantum light in orthogonal polarizations will minimize SRS noise. This technique requires monitoring fluctuations of the source laser polarization and will be particularly beneficial for forward-scattered SRS photons. 

\subsection{Free-space transmission}
\label{Sec_FSTransmission}

For fiber-based quantum communication, the achievable range is generally limited to a few hundred kilometers due to photon transmission in the optical fiber decreasing exponentially with length \cite{Fiber-QKD_Hiskett2006}, as discussed in Section~(\ref{Fiber-optic transmission}). For a free-space link through vacuum, the signal attenuation scales as $\sim 1/r^2$ due to beam divergence, thus promising a more favourable loss scaling as compared to optical fibers over long distances. Over the past few years, several free-space quantum communication experiments, including links to satellites, involving weak coherent laser pulses and entangled photons, have been performed, extending to larger distances and achieving higher bit rates \cite{Ent-HD_Steinlechner2017, Teleportation_Yin2012, BBM92Pol_Mishra2022, QKD_Padgett2023, VacuumBeam_Jiang2024}.  

In practice, the atmosphere introduces additional losses and complications, which pose additional challenges to free-space quantum communication.
Amongst these are obstructions in the line of sight, including those caused by the Earth's curvature, signal attenuation caused by diffraction, ambient light pollution \cite{LightPollution_Yastremski2025}, and atmospheric turbulence \cite{Turbulence_Jaouni2025}. In an atmospheric channel, variations in temperature, wind, and pressure create turbulence, causing the light beam to spread, wander, or become distorted, leading to attenuation. In general, the spatial modes of photons are distorted during free-space transmission due to these atmospheric effects, causing the signal to deviate from its original Gaussian profile. 
This makes it difficult and highly lossy to couple photons into single-mode fibers for detection, necessitating the use of multimode fibers.  Additionally, weather conditions such as fog, rain, or dust can absorb or scatter photons, further contributing to signal loss and limiting the communication distance in the atmosphere. The farthest distance achieved in free-space quantum communication through the atmosphere so far is over 144 km \cite{QKD-Ent_Ursin2007, Ent-FreeSpace_Fedrizzi2009}. 

Specifically, for time-bin encoding, such spatial distortion degrades the interference contrast in superpositions of early and late time-bins within a delayed-line interferometer (DLI).
Recently, there have been some attempts to make the interferometers at the receiver robust against multimodal beam profiles \cite{TBM-MM_Jin2018, QKD_Padgett2023}. Using the multi-mode time-bin analyzer, Jin et. al. demonstrated a free-space BB84-QKD over a turbulent depolarizing free-space channel of 1.2 km, yielding 138 bits/second \cite{BB84-FreeSpace_Jin2019}. 
Chen et al. demonstrated a proof-of-principle reference-frame-independent QKD protocol using time-bin encoding over a 2 km intracity free-space link, achieving a low QBER of just 1\% \cite{TB-RFIQKD_Chen2020}.
Recently, Cocchi et. al. performed QKD over free-space horizontal links during nighttime and daytime, achieving 40 kbps secret key rate over 500 m long links \cite{TB-FreeSpaceQKD_Cocchi2025}.
Single-photon interference of TBQs reflected from a rapidly moving satellite up to five thousand kilometers away has been demonstrated at a ground station, achieving a visibility of $67\%$ \cite{TB-SatGroundInt_Vallone2016}. In other experiments, the TBQs are transmitted over a long distance in free space by converting them into polarization encoding or using hyper-entanglement between time-bin and polarization bases \cite{Ent-Hyper_Hong2018, Ent-Hyper_Joseph2022}. More details on free-space signal attenuation in general, which similarly affects TBQ transmission, can be found in Refs. \cite{FreeSpaceCom_Amr2016, QCom_Ghalaii2022, FreeSpace_Kim2001, FreeSpaceCom_Kaushal2017}.

Using satellites as nodes for free-space quantum communication allows the link to approach near-vacuum conditions and additionally overcomes line-of-sight limitations due to the Earth's curvature. Up and downlinks to ground-station transceivers only effectively correspond to a few tens of kilometers of transmission through the atmosphere \cite{QCom_Pfennigbauer2005}. Satellite-based quantum communication \cite{QCom_Juan2017, Teleportation_Ren2017} has been achieved to record distances up to 12900 kilometres \cite{Satellite-QKD_Li2025} using polarization qubits.
However, the normal challenges of free-space transmission near the ground station (atmospheric absorption, turbulence, and light pollution) remain, albeit less significant. In addition, specific challenges related to satellites, including phase shifts due to the satellite's inherent motion and the Doppler shift effect, need to be addressed \cite{SatComReview_Lu2022, SatCom_Goswami2025}. These issues are particularly important for time-bin encoded qubits, which are sensitive to temporal jitter and phase variations during free-space propagation. Furthermore, the need for quantum memory integration and the complexity of establishing satellite-to-satellite quantum links further complicate satellite-based quantum transmission \cite{DaytimeQCom_Gruneisen2021, Thesis_Vedovato2015, DopplerShift_Schlake2025}. For a detailed review of the challenges involved in satellite quantum communication, please see reference \cite{SatComReview_Lu2022}.


\section{Time-bin measurement systems and characterization methods}
\label{Sec_Measurement}
\FloatBarrier

The measurement of TBQs is performed by projecting them onto different mutually unbiased bases such as $Z = \{|e\rangle, |l\rangle \},~X= \{ |+\rangle, |-\rangle \}$, and $Y=\{ |+i\rangle, |-i\rangle\}$, where $|\pm \rangle = (|e\rangle \pm |l\rangle)/\sqrt{2}$, and $|\pm i\rangle = (|e\rangle \pm i |l\rangle)/ \sqrt{2}$. A projection in $Z$-bases is performed by recording the arrival time of the photons on a SPD with respect to a reference clock using a time-to-digital converter (TDC). Accumulating the arrival time of the photons results in a histogram with two peaks corresponding to the `early' and `late' time bins, as shown in Fig.~(\ref{fig_TB_prep}b).
For qubits prepared in the $X$- or $Y$-basis, the early and late states are detected with equal probability in the $Z$-basis measurement, as shown in Fig.~(\ref{fig_TB_prep}b) 
A projection in a superposition basis is performed by inducing a time delay of the early bin equal to the time-bin separation $\tau_{el}$, such that it interferes with the late bin and, thus, reveals information on their relative phase. This delay can be achieved with a DLI (Sec.~\ref{Sec_MeasInt}) or using light matter interfaces (Sec.~\ref{Sec_AltApproach}).

We begin by reviewing the characteristics of single-photon detection systems, followed by various methods for performing TBQ measurements in superposition bases using measurement interferometers, light-matter interfaces,  and time-bin to polarization conversion. Finally, we discuss characterization techniques for TBQ states, including QST, state fidelity analysis, and visibility measurements.

\subsection {Single photon detection system} 
\label{Sec_SPDS}

A single-photon detection system mainly consists of an SPD and a TDC, which is used for detecting and time-tagging the arrival time of single photons.  The choice of SPDs and TDC directly affects the performance of the underlying protocol.
The performance of SPDs \cite{SPD_Hadfield2009, SPSD_Bienfang2023} are characterized by key parameters such as detection efficiency, dead-time, jitter, dark count rate, after-pulsing probability, etc. Quantum information applications mainly utilize the following two types of SPDs: (i) Single-photon avalanche diodes (SPADs) and (ii) Superconducting Nanowire Single-Photon Detectors (SNSPDs). 

SPADs are based on the avalanche multiplication principle \cite{SPAD_Ceccarelli2020}, where an incident photon creates an electron-hole pair that triggers an avalanche breakdown, leading to a detectable electrical signal.
Silicon SPADs \cite{SPAD_Acconcia2023} are suitable for the visible to near-infrared (NIR) wavelength range (400 - 1100 nm), whereas InGaAs SPADs are used for the NIR range (900 - 1700 nm). 
High-end Si-SPADs have high detection efficiency ($\sim 90\%$), low dead time ($\sim 40$ ns), low jitter ($\sim 40$ ps), and low afterpulse probability ($<0.1\%$). 
InGaAs SPADs \cite{SPAD-InGaAs_Ribordy98} have lower detection efficiencies ($\sim 30\%$), higher dead time (on the order of microseconds), and higher jitter ($\sim 100$ ps) as compared to Si-SPADs. SPADs tend to have a compact size and operate at room temperature, and they are thus highly convenient. 
The low detection efficiency and high dead-time of InGaAs SPADs limit the performance of quantum information protocols in the near-infrared wavelength range.
On the other hand, SNSPDs consist of nanowires that are cooled to approximately $\sim 0.8-2$ K to bring them into a superconducting state \cite{SNSPD-SOTA_Zadeh2021}. 
When a photon is absorbed by the nanowire, it becomes resistive, producing a detectable voltage signal. 
These detectors can be designed to operate at visible or NIR wavelengths and exhibit high efficiency ($> 95\%$), short dead-time ($\sim 10$ ns), low jitter ($< 10$ ps), and negligible afterpulsing. Although SNSPDs have a relatively broadband detection efficiency compared to SPADs, their efficiency still depends on the wavelength. For instance, an SNSPD optimized for telecom wavelengths typically shows reduced efficiency at visible or NIR wavelengths and vice versa. SNSPDs inherently exhibit polarization-dependent single-photon absorption, leading to polarization-dependent detection efficiency and necessitating polarization control before the detectors. Strategies for minimizing this effect \cite{SNSPD-PolDep_Dorenbos2008} and accurate characterization of polarization-dependent efficiency \cite{SNSPD-PolDep_Fei2022} have been reported.

The jitter of SPDs and TDC broadens the observed width of each time-bin, thus limiting the minimum allowable time-bin separation below which temporal overlap occurs, as discussed in Section~(\ref{Sec_TBSep}). Therefore, high-resolution (low-jitter) detectors enhance time-bin encoding by enabling smaller time-bin separations, increasing information density, and relaxing interferometric stabilization requirements. The dead time of SPDs further constrains the minimum time-bin separation, as photons arriving in consecutive bins may not be registered, thereby limiting the detection of $|\psi^+\rangle$ patterns in the BSM, as discussed in Section~(\ref{Sec_BSM}). Dark counts occur randomly and are uniformly distributed across the time-bin histogram, introducing a background floor that limits the measured extinction ratio and interference visibility. In this context, the superior detection efficiency, timing resolution, and low-noise characteristics of SNSPDs make them the preferred platform for high-rate, high-fidelity single-photon detection in modern quantum information experiments.

 \subsection{Measurement using Delay Line Interferometers} 
 \label{Sec_MeasInt}
 
\begin{figure}[ht!]
\begin{center}
 \includegraphics[clip, trim= 6cm  4cm  2cm  6cm, width=0.85\columnwidth]{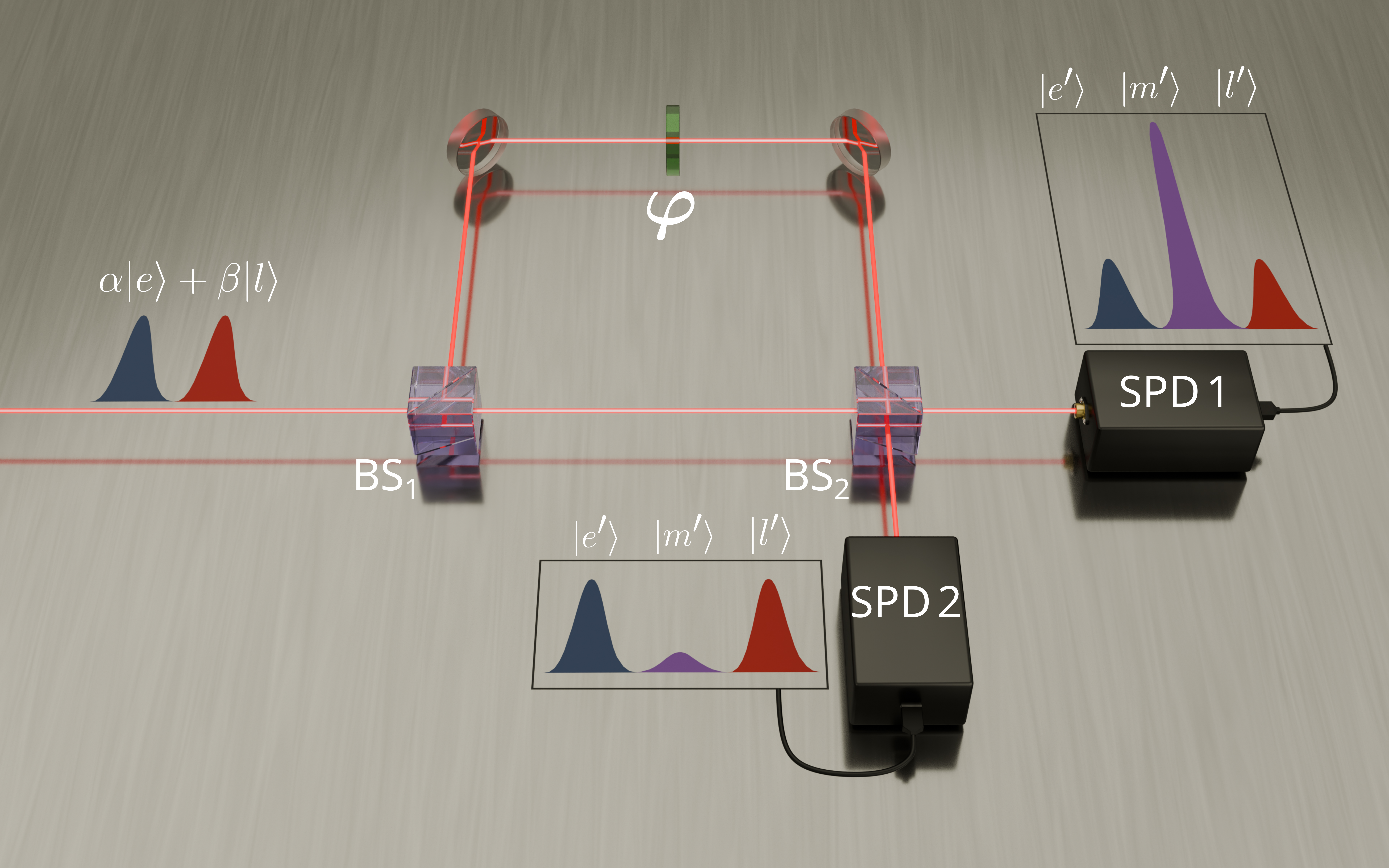}
\caption[Schematic of a TBQ measurement setup using a free-space unbalanced MZI]{Schematic of a TBQ measurement setup using a DLI in free-space unbalanced MZI configuration. Both beam splitters ($\text{BS}_1$ and $\text{BS}_2$) have a 50:50 splitting ratio. BS: Beam splitter, SPD: Single photon detector.}
\label{fig_TBM-UMZI}
\end{center}
\end{figure}

When a TBQ in a superposition state is injected into a DLI, with a path difference the same as time-bin separation ($\tau_{el}$), followed by an SPD at each of the two output ports, the resulting histogram of the photon's arrival time produces three equispaced peaks (separated by $\tau_{el}$), as illustrated in Fig.~(\ref{fig_TBM-UMZI}).
The blue (red) side peak corresponds to photons in the early (late) time-bin travelling along the short (long) path of DLI. The central interference peak (magenta) arises from the first-order interference between photons from the early time-bin travelling along the long path and photons from the late time-bin taking the short path. The blue and red peaks represent projections on the Z-basis, whereas the interference peak corresponds to a projection on a superposition basis. 

\color{black}
We may systematically assess the probabilities for clicks in the different output temporal and spatial modes of the DLI. 
Consider a general TBQ: $|\psi\rangle = \alpha |e\rangle + \beta |l\rangle$ being measured by a DLI as shown in Fig.~(\ref{fig_TBM-UMZI}). As we know, DLI has short ($S$) and long ($L$) paths, with the long path introducing a relative-phase shift $\varphi$. At the first beam splitter ($BS_1$), each time bin transforms as: $|e\rangle \rightarrow \frac{1}{\sqrt{2}}( |e_S\rangle + |e_L\rangle ),~ |l\rangle \rightarrow \frac{1}{\sqrt{2}}( |l_S\rangle + |l_L\rangle )$. Accounting for the phase shift in the long arm, the state after the first beam-splitter is: $|\psi_{BS_1}\rangle = \frac{\alpha}{\sqrt{2}}( |e_S\rangle + e^{i\varphi}|e_L\rangle ) + \frac{\beta}{\sqrt{2}}( |l_S\rangle + e^{i\varphi}|l_L\rangle )$.
At the second beam splitter ($BS_2$), each time-bin mode evolves as $|x_S\rangle \rightarrow \frac{1}{\sqrt{2}}(|x_S\rangle_1 + |x_S\rangle_2)$ and $|x_L\rangle \rightarrow \frac{1}{\sqrt{2}}(|x_L\rangle_1 - |x_L\rangle_2)$; giving each amplitude an overall factor of $1/2$, where subscripts $1$ and $2$ label the output spatial modes.
In each spatial mode, the output is divided into three time bins: early $(|e'\rangle)$ from $|e_S\rangle$, middle $(|m'\rangle)$ from the interference of $|e_L\rangle$ and $|l_S\rangle$, and late $(|l'\rangle)$ from $|l_L\rangle$.  The states at the two-output ports of  the beam splitter are given as
\begin{equation}
\begin{aligned}
    \ket{\psi_{D1}} &= \frac{\alpha}{2}\ket{e'} + \frac{1}{2}(\alpha e^{i\varphi} + \beta)\ket{m'} + \frac{\beta e^{i\varphi}}{2}\ket{l'},\\
    \ket{\psi_{D2}} &= \frac{\alpha}{2}\ket{e'} + \frac{1}{2}(\beta - \alpha e^{i\varphi})\ket{m'} -\frac{\beta e^{i\varphi}}{2}\ket{l'}.
\end{aligned}
\end{equation}
The corresponding detection probabilities for the middle (interference) bins are given as: $P_{D_1}(|m'\rangle) = \frac{1}{4}\big|\alpha e^{i\varphi} + \beta\big|^2$ and $P_{D_2}(|m'\rangle) = \frac{1}{4}\big|\beta - \alpha e^{i\varphi}\big|^2$. For a qubit with equal superposition: $\alpha = \beta = 1/\sqrt{2}$, constructive interference in port-1 ($\varphi = 0$), the probabilities of detections in different temporal modes are: $P_{D_1}(|e'\rangle) = \frac{1}{8},~ P_{D_1}(|m'\rangle) = \frac{1}{2},~ P_{D_1}(|l'\rangle) = \frac{1}{8}$. Thus, the ratio of early to middle bin detection probability is $1:4$. The DLI, thus, maps the TBQ onto three temporally resolved outputs per port, with interference in the middle bin enhancing its probability relative to the non-interfering early and late bins.

\color{black}
For a phase setting $\varphi$ of the phase shifter in the measurement interferometer, the interference peaks at the two output ports of DLI correspond to projections along $|\varphi^\pm\rangle=(|e\rangle \pm \exp(i\varphi)|l\rangle)/\sqrt{2}$.
Therefore, for X-basis (Y-basis) projections, the relative phase of the DLI (relative phase acquired by the photons taking either of the two paths) should be set to $\varphi=\pi$ ($\pi/2$) as detailed in Refs. \cite{TBM-UTBA_Bussieres2010, TBM-UTBA_Bussieres1_2006, TBM-UTBA_bussieres2_2006, TB_Pittman2013}.  
Using two detectors simultaneously at the two output ports of the interferometer enables access to projections onto both orthogonal superposition basis states at the same time. 
Alternatively, a single detector can be used, though this results in a 50\% signal loss and allows access to projection onto a single superposition basis state at a time. 
The DLI facilitates random projections in either the Z-basis or the superposition basis, which is often useful for QKD protocols such as BB84 \cite{QKD-BB84TB_Tang2023}. 
However, a deterministic Z-basis projection can be achieved by bypassing the DLI, thus avoiding its insertion loss. Likewise, a deterministic projection on the X-basis can be achieved at the output of DLI if the first beamsplitter in Fig.~(\ref{fig_TBM-UMZI}) is replaced by a fast optical switch, such that the early bin is always directed through the long arm and the late bin is always through the short arm of the DLI. 

\begin{figure}[ht!]
\centering
\includegraphics[clip, trim=2.25cm 7cm 2.5cm 7cm, width=0.60 \columnwidth]{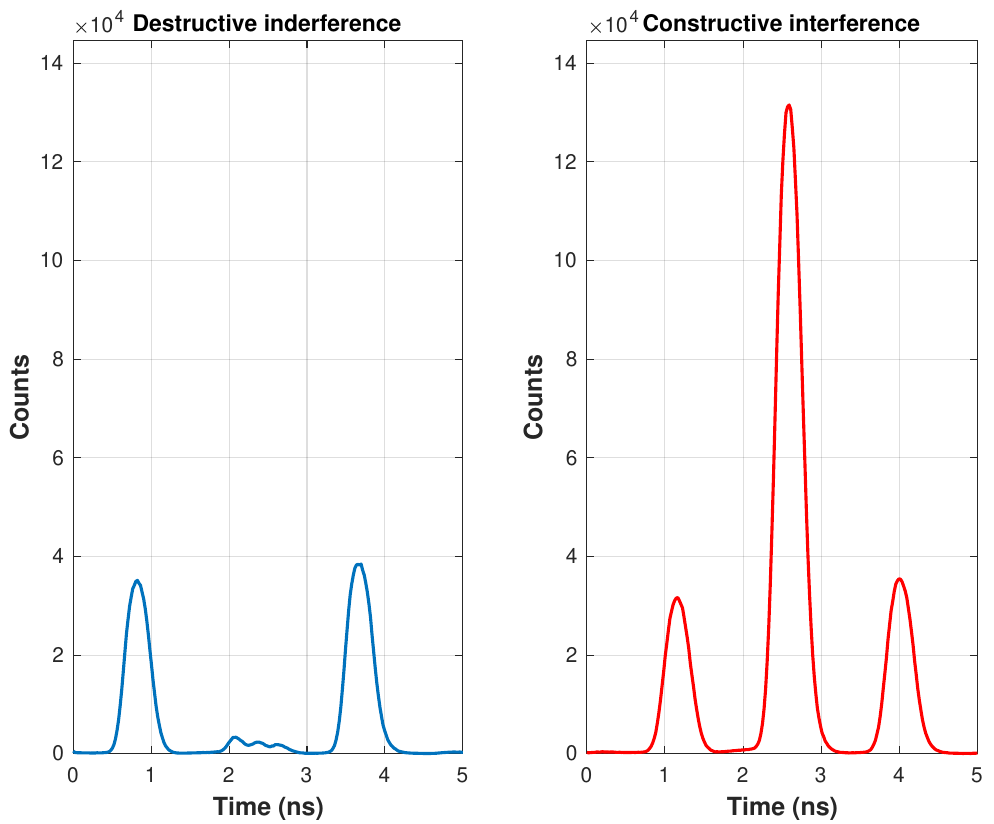}
\caption[TBQ measurement results in X-basis.]{Histogram of photon detections in the measurement of TBQ prepared in $|+\rangle$ state. The central bin in the plot corresponds to X-basis projections on $|+\rangle$ and $|-\rangle$ for the right-hand and left-hand plots, respectively. Side-bins correspond to $|e\rangle$ and $|l\rangle$ projections in the Z-basis. For an ideal setup, all X-basis detections should correspond to $|+\rangle$; however, imperfect matching of the delay in the measurement DLI and mismatch between the temporal amplitude/width of the early and late time-bins causes a small residual signal in the $|-\rangle$ output.)}
    \label{fig_TBMX}
\end{figure}

As an example, a histogram of TBQs prepared in a superposition state $|+\rangle=(|e\rangle+|l\rangle)/\sqrt{2}$ and projected on the X-basis is shown in Fig.~(\ref{fig_TBMX}).  
In the case of constructive interference, the interference peak is four times the intensity of the side peaks, whereas it is close to zero for destructive interference, as evident in Fig.~(\ref{fig_TBMX}). 
The imperfect interference in the superposition basis, visible as a small leakage in the middle bin of the left plot of Fig.~(\ref{fig_TBMX}), is due to slight imperfections in the time-bin preparation and measurement interferometers. In the following, we will explore the implementations of some of the representative measurement interferometers in both free-space and fiber-based configurations, offering insights into the advantages and limitations of different phase stabilization methods \cite{IntStabilization_Toliver2014}.

 \subsubsection{Free-space interferometers}
 \label{FreeSpaceInt}

\begin{figure}[ht!]
    \centering
 \includegraphics[clip, trim= 6cm  0cm  6cm  0cm, width= 0.9\columnwidth]{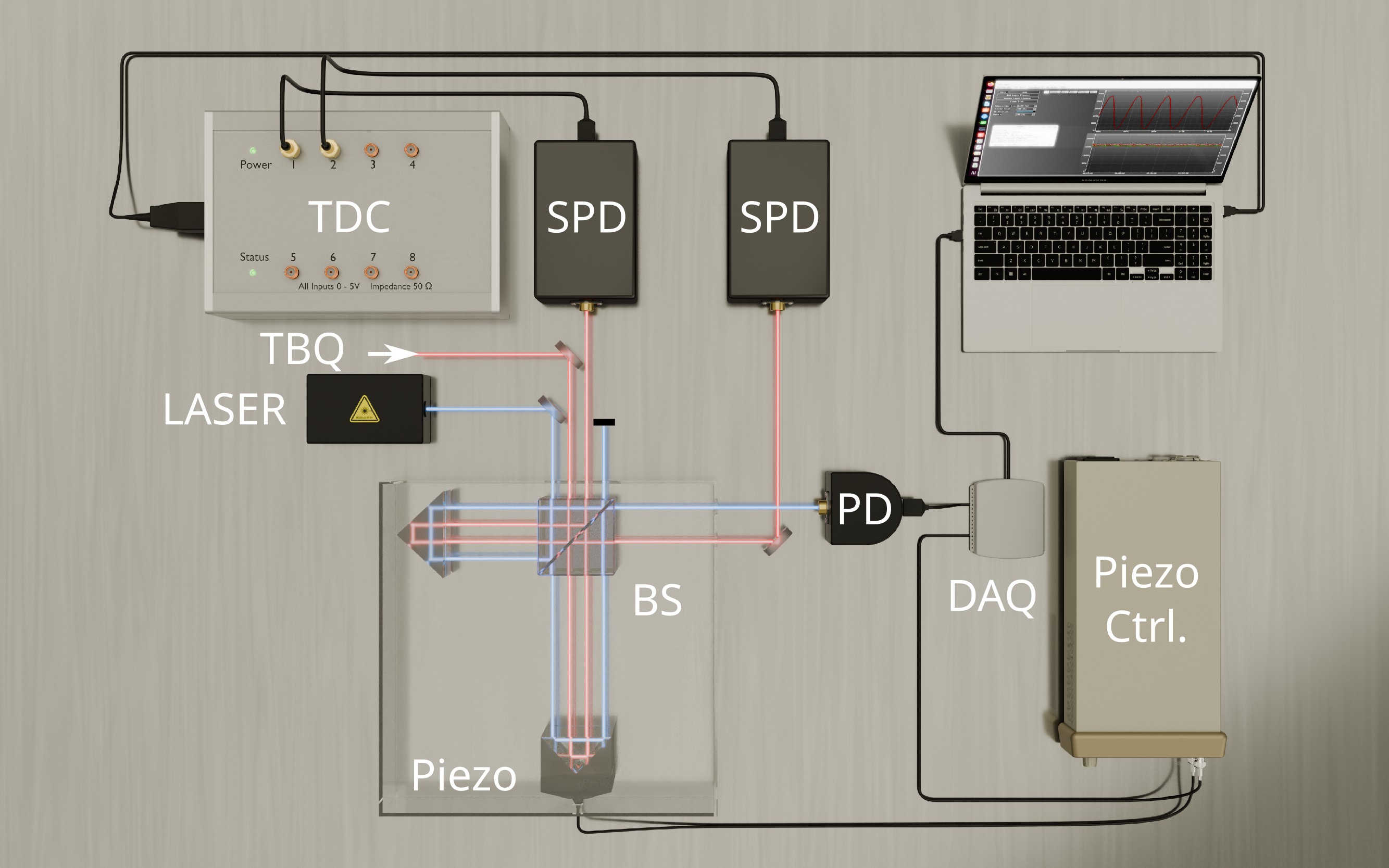}
\caption[Free-space unbalanced Michelson interferometer.]{Free-space unbalanced Michelson interferometer for time-bin measurement. A PID-based active phase-locking mechanism is used to stabilize the phase of the interferometer using a reference laser (blue) input. The TBQ is launched from a spatially segregated input (red) to the interferometer. BS: Beam splitter, DAQ: Data acquisition card, Piezo Ctrl: Piezo controller, PD: Photo detector, SPD: Single photon detector, TBQ: Time-bin qubit. TDC: Time to digital converter.}
\label{fig_FSUMI}
\end{figure}

Among free-space DLIs, the most commonly used configurations are the unbalanced Mach-Zehnder interferometer (UMZI) and the unbalanced Michelson interferometer (UMI) \cite{FreeSpaceDLI_Li2011}. Here, we will briefly discuss the phase stabilization of such unbalanced interferometers, using UMI as an example, and associated challenges. 
A UMI consists of a 50:50 beam splitter and two retro-reflectors positioned at its output ports with a relative path difference of $\tau_{el}/2$ from the BS, which redirect the beams parallel to the input path and reflect them back to the input beam splitter, where interference takes place, as illustrated in Fig.~(\ref{fig_FSUMI}). For enhanced thermal stability, all optical components are mounted on a Zerodur glass plate, having a near-zero thermal expansion coefficient, and the entire assembly is housed within a temperature-controlled enclosure \cite{TBM-UTBA_Bussieres2010}.
The UMI can be actively stabilized using the interference pattern of a reference or locking laser (see Sec.~\ref{Sec_PhaseStableMethods}). For free-space interferometers, one may conveniently use another spatial mode for the locking laser, which propagates parallel to the qubit mode.  This is easily achieved using a retroreflector. 
 
One important practical consideration is that any back reflections of the surfaces, e.g., of the central beam splitter or the retro-reflector front surfaces, can cause the formation of undesired optical resonances, which induce additional frequency-dependent modulation of the signal.
The heating of the interferometer due to ambient temperature changes will affect the locking laser signal through both interference path-length variations and resonator modulation.
This essentially causes distortion to the interference pattern and prevents reliable locking of the interferometer phase.  To mitigate the cavity effect, the optical elements (that have surfaces normal to the beam) should be placed at an angle such that back-reflections do not form a resonator. Angles of incidence of 5-10\% are typically sufficient and do not significantly change the optical properties of the elements. Indeed, these types of reflections also cause additional reflections of the qubits, leading to photons arriving at unexpected times, which can be treated as loss events if they do not arrive within the expected time bins for the qubit. For an in-depth discussion on phase stabilization methods, please refer to Section~(\ref{Sec_PhaseStableMethods}).

\subsubsection{Fiber-based interferometers}
\label{Sec_FiberInt}

The MZI is not very practical in fibers because the polarization evolution of the two input fields needs to be matched at the output beamsplitter for the best interference visibility. 
Alternatively, one can use a fiber-based UMI with Faraday mirrors at the two ends, which inherently ensures that polarization evolution due to fiber birefringence is compensated \cite{MI_Ferreira1995, TB-QKD_Muller1997, MI_Mo2005}. Any polarization changes accumulated in the forward path are automatically compensated in the backward path due to the $90^\circ$ polarization rotation offered by Faraday mirrors, thus producing stable interference with high visibility \cite{TE-Ent_Tittel1997}. 
Moreover, in fiber-based interferometers, spectral or temporal mode separation is the only way to isolate the reference locking laser (see Sec.~\ref{Sec_PhaseStableMethods}) from the qubit signal. Therefore, DWDM filters or optical switches are used at each input/output port. However, if suitable components, such as Faraday or Bragg mirrors, circulators, and other optical components, are not easily available, e.g. commercially, for the wavelength of the locking laser and qubits, the MZI may be the only viable option.

\begin{figure}[htbp!]
    \begin{center}
 \includegraphics[clip, trim = 0cm 6cm 0cm 10cm, width= 0.9 \columnwidth]{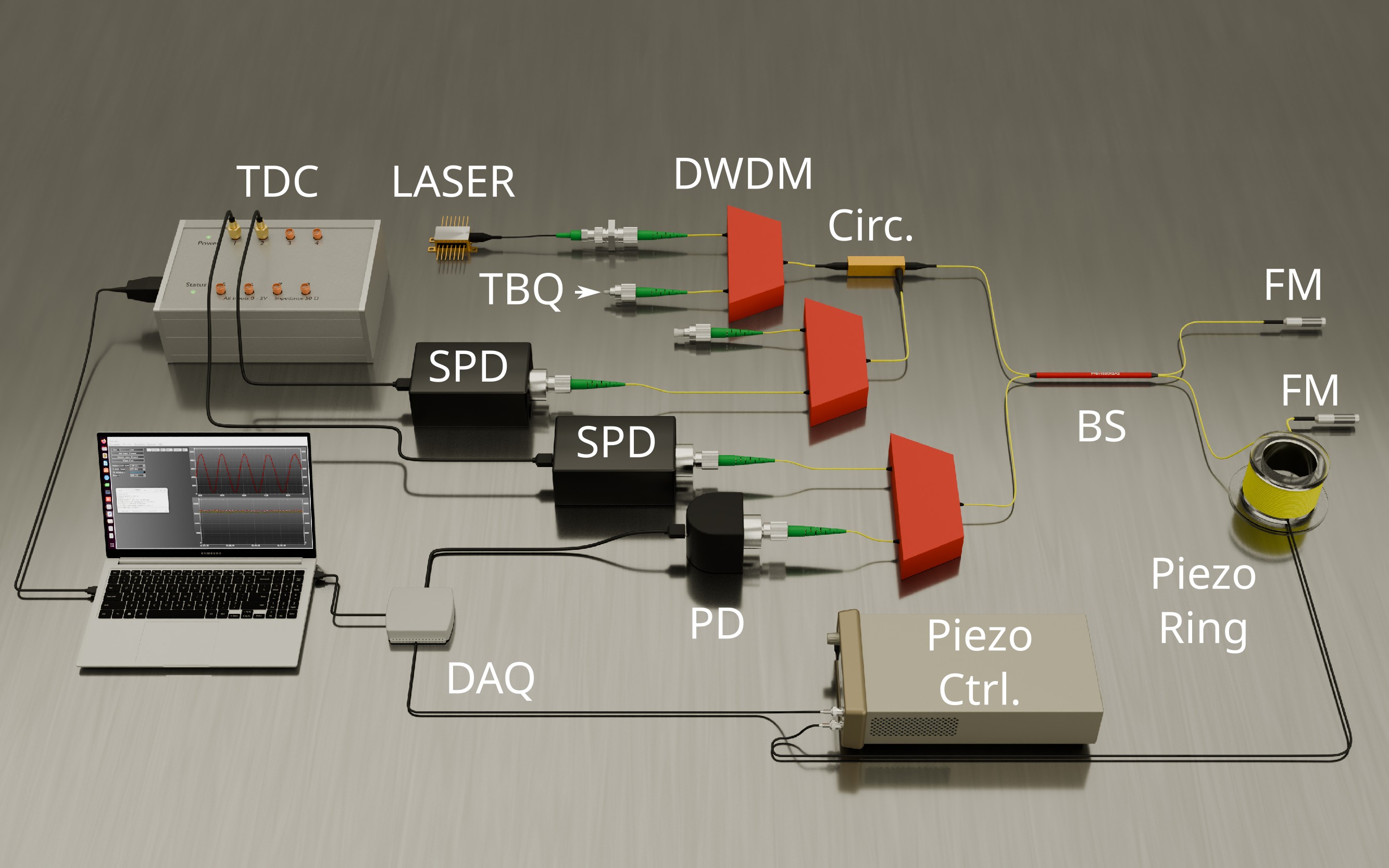}
 \caption[Fiber-based unbalanced Michelson interferometer.]{Fiber-based UMI for TBQ preparation or measurement. The reference laser beam and TBQs are multiplexed using a DWDM filter and launched into the UMI. At the output ports, these signals are demultiplexed, with the reference signal directed to a photodetector. The DAQ card processes the resulting signal to generate an error signal, which is amplified by the Piezo amplifier and applied to the Piezo for phase-locking the interferometer via PID control. SPDs detect the demultiplexed TBQs, and time-bin analysis is performed using a time-to-digital correlator. BS: Beam splitter, DAQ: Data acquisition card, FM: Faraday mirror, Piezo Ctrl: Piezo controller, PD: Photo detector, SPD: Single photon detector, TBQ: Time-bin qubit, TDC: Time to digital converter.}
    \label{fig_FiberUMI}
    \end{center}
\end{figure}

The schematic of a fiber-based UMI for time-bin analysis is shown in Fig.~(\ref{fig_FiberUMI}). The UMI consists of DWDM filters at each input (and output) port, followed by a circulator at the input port, a 50:50 beam splitter, and Faraday mirrors at the two ends \cite{Ent-dist_Kim2021}. 
A frequency-stabilized laser with a wavelength offset from the qubit signal is attenuated to a low power level (a few $\mu W$) and launched at the input port of the UMI. 
At the output port, the locking laser is demultiplexed, and a PID-controlled phase-locking algorithm is implemented  \cite{PhaseLocking_Svarc23}. The PID algorithm provides an error signal to the piezo controller that stretches the fiber looped around the piezo-ring to lock the interferometer \cite{Ent-dist_Kim2021} at a given set point. For an in-depth discussion on phase stabilization methods, please refer to Section~(\ref{Sec_PhaseStableMethods}).

\subsubsection{Integrated photonics-based interferometers}

Integrated photonic-based DLIs offer several advantages due to their compact footprint, making them inherently more stable against environmental perturbations like vibrations and temperature fluctuations. The DLIs are typically in MZI configuration, as the Michelson interferometer inevitably requires a mirror. 
The small footprint also limits the achievable time-bin separations up to a few ns, posing challenges for certain quantum communication protocols. 
Typically, to compensate for the loss difference between the two arms, the short arm is equipped with a variable optical attenuator, realized using another MZI \cite{PIC_TBQ_DLI_Grange2024}. 
Additionally, the choice of material platform is crucial, as it dictates factors like the usable wavelength range, bend radii and compatibility with other integrated components. 
The phase of the interferometer can be tuned using the thermo-optics effect \cite{Ent-FWM_Zhang2018, Ent-FWM_Xiong2015}, the electro-optic effect \cite{PIC_TBQ_DLI_Grange2024}, or the carrier depletion effect \cite{BB84_Sibson2017, BB84_Sax2023}.
Platforms such as Lithium niobate, lithium tantalate, and barium titanate are particularly promising due to their excellent electro-optic properties for ultra-fast phase tunability, along with thermal stability. 
Nonetheless, challenges such as propagation losses and efficient in-out coupling must be carefully managed.
The integrated photonic platform has been used for the generation of two-qubit \cite{Ent-FWM_Zhang2018, Ent-FWM_Xiong2015, TBE_Chen2018} as well as high-dimensional \cite{HD-QKD_Yu2025} time-bin entanglement, along with QKD protocols \cite{BB84_Sibson2017, BB84_Sax2023, QKD-PIC_Sibson2017}.

\subsubsection{Interferometer phase stabilization methods and implementations} 
\label{Sec_PhaseStableMethods}

In this section, we discuss different noise sources that contribute to the phase fluctuations in a DLI and illustrate some representative examples of phase stabilization methods.
The relative phase acquired through a DLI of optical path delay $L$ is given as
\begin{equation}
    \phi = 2\pi \frac{L}{\lambda_Q} = 2\pi \frac{\mathrm{n} \tilde{L}}{\lambda_Q},
\end{equation}
where $L=\mathrm{n}\tilde{L}$, $\mathrm{n}$ is the refractive index of the DLI medium, $\tilde{L}$ is the DLI path delay, and $\lambda_Q$ is the wavelength of the TBQ laser source.
Considering small changes in the refractive index $\delta \mathrm{n}$, DLI path delay $\delta \tilde{L}$, and TBQ source laser wavelength $\delta \lambda_Q$, the resulting phase displacement is given as:
\begin{equation}
\delta \phi = 2\pi \left[ \frac{\tilde{L}}{\lambda_Q} \delta \mathrm{n} + \frac{\mathrm{n}}{\lambda_Q} \delta \tilde{L} + \frac{\mathrm{n} \tilde{L}}{\lambda_Q^2} \delta \lambda_Q \right]
. \label{Eq_DLIPhaseDisplace}
\end{equation}
Note that the last term of Eq.~(\ref{Eq_DLIPhaseDisplace}) corresponds to the phase uncertainty given in Eq.~(\ref{Eq_FreqtoPhaseUncertain}) using $\mathrm{n}\tilde{L} = c \tau_{el}$.

In Section~(\ref{Sec_MeasInt}), we have learnt that the DLIs are designed using bulk free-space optics, optical fibers, or integrated photonic structures. These media exhibit a refractive index change ($\delta \mathrm{n}$), and path delay variation ($\delta \tilde{L}$) with air convection or temperature change. Furthermore, any vibrations and mechanical stress can also cause small changes in the optical path lengths.  Practically, it is not possible to distinguish the phase fluctuations arsing from refractive index change and path delay variations independently from the shift in the interference pattern. Therefore, for all practical purposes, the variation in the optical path delay $\delta L$ is what matters. This leads to a more practical expression for the phase fluctuations as
\begin{equation}
\Delta \phi = \frac{2\pi}{\lambda_Q} \sqrt{(\Delta L)^2  + \left(\frac{L}{\lambda_Q} \Delta \lambda_Q\right)^2} .\label{Eq_DLIPhaseInstability}
\end{equation}

 The phase fluctuations in DLI hamper the speed and security of time-bin quantum cryptography and quantum communication protocols by increasing the bit error rate. The contribution of the phase mismatch $\Delta\phi$ to the quantum bit-error rate (QBER) follows a $\sin^2(\Delta\phi)$ dependence \cite{IntPhaseTrack_Makarov2004}, exhibiting a quadratic scaling for small $\Delta\phi$.
Since the DLIs are inherently phase-unstable, they require either an active closed-loop phase stabilization using the interference pattern of a reference laser or the use of materials with near-zero thermal expansion coefficients at room temperature. In addition, they should be housed in a temperature-controlled, vibration-free environment to avoid phase fluctuations and for enhanced phase stability.  
High phase stability, long-term reliability, continuous operation, and little cross-talk between the quantum signal and the reference signals are all desirable characteristics of an ideal phase locking. 
This section provides an overview of the various phase-locking schemes for DLIs using a reference laser and also discusses purely passive or temperature-control-based phase-stabilization methods adopted by commercial DLI manufacturers.

\textit{(a) Single-colour PID locking:} The simplest approach to phase stabilization of a DLI is to use a frequency-stabilized reference laser source and lock the interferometer phase using its interference pattern with a PID control algorithm as schematically shown in Fig.~(\ref{fig_FiberUMI}). In such a single-colour PID locking scheme, the intensity of the reference signal ($I(\phi_R)$) varies with the phase of the DLI as seen by the reference signal ($\phi_R$) as follows:
\begin{equation}
 I(\phi_R)=I_{mid}[1 + \mathcal{V} \cos(\phi_R) ],
\end{equation}
where $I_{mid}=(I_{max}+I_{min})/2$, and $\mathcal{V} =(I_{max}-I_{min})/(I_{max}+I_{min})$ is the visibility of the DLI - a characteristic feature which is typically independent of the wavelength in its operation range. 
The wavelength of the reference laser ($\lambda_R$) is slightly offset from that of the quantum signal ($\lambda_Q$) for proper segregation at the output; the phase relationship of the quantum ($\phi_Q$) and reference ($\phi_R$) signals is given as $\phi_Q/\lambda_Q=\phi_R/\lambda_R$. Therefore, the intensity of the quantum signal (counts/second in case of TBQs) varies as
\begin{equation}
    I(\phi_Q) = I_{mid} \left[1 + \mathcal{V}\cos\left(\frac{\phi_R\lambda_Q}{\lambda_R}\right) \right].
\end{equation}

Typically, reference and quantum signals at two different wavelengths are derived from different laser sources. Thus, their relative wavelength drift also causes an error in the preparation or measurement of the TBQs. Since the interferometer phase is actively stabilized using the reference laser at $\lambda_R$, when the wavelength $\lambda_R(t)$ changes, the PID adjusts the optical path delay $L(t)$ such that $\phi_R$ is constant (or nearly so). However, this adjusted path delay $L(t)$ appears as a different phase $\phi_Q=2\pi L(t)/\lambda_Q$ at quantum wavelength. Assuming only wavelength fluctuation in the two laser sources as the primary sources of error, the phase error in the quantum signal can be expressed as
\begin{equation}
    \delta \phi_{Q} = \frac{2\pi L}{\lambda_Q} \left( \frac{\delta\lambda_{R} }{\lambda_R} +\frac{\delta\lambda_Q}{\lambda_Q} \right) \Rightarrow \Delta \phi_{Q} = \frac{2\pi L}{\lambda_Q} \sqrt{\left( \frac{\Delta\lambda_{R}}{\lambda_R} \right)^2 + \left(\frac{\Delta\lambda_Q}{\lambda_Q}\right)^2},
\end{equation}
where the first term accounts for the phase error arising from fluctuations in the reference signal's wavelength, while the second term reflects the impact of wavelength fluctuations in the quantum signal.
As an example, for $L = 30$ cm (or $\tau_{el}=1$ ns), a wavelength drift of $\delta\lambda_Q=0.1$ pm (12.7 MHz) at $\lambda_Q=1535.04$ nm, and  $\delta\lambda_R=0.1$ pm (12.5 MHz) at $\lambda_R=1535.04$ nm at $\lambda = 1550$ nm causes a phase error in the TBQ  $\Delta\phi_Q = 6.45^\circ$ in the superposition-basis projection of time-bins. Thus, one requires an ultra-stable locking laser. Alternatively, one can use the same locking laser for stabilizing the DLI used for preparing the TBQ as well as the measurement DLI, thus cancelling the phase error due to the locking laser's wavelength drift.

Although the implementation of the single colour PID locking scheme is straightforward, it faces a few critical problems. For example, the sensitivity of this single-color feedback system \cite{Int_cho2009} is given as $S\propto dI(\phi_R)/d\phi_R=V I_{mid}\sin(\phi_R)$. Thus, the sensitivity of the locking is phase-dependent, and locking the interferometer to the maximum and minimum of the fringe ($0$ and $\pi$ phases) is unstable as the sensitivity at these points vanishes. Moreover, multiple phases of the interferometer ($\pi/2, 3\pi/2$) map to the same intensity of the reference laser, creating a phase ambiguity.

\textit{(b) Two-colour locking Scheme:} A solution to the phase-dependent locking sensitivity problem is provided by a two-colour feedback scheme \cite{Int_Piotr} consisting of in-phase and quadrature components, making the locking sensitivity constant for all phases. 
Moreover, it also provides access to all phases in the range $ [0, 2\pi]$. Please see Ref. \cite{Int_Piotr} for more details on the two-colour locking scheme.  Importantly, whenever a reference laser is used for locking a DLI, the phase stability is dependent on the absolute wavelength stability of the reference laser, similar to that given in Eq.~(\ref{Eq_FreqtoPhaseUncertain}). 
Another method uses frequency-shifting of the locking laser for arbitrary-phase locking of the fiber DLIs \cite{PhaseLocking-Int_Chen2024}.  Recently, a novel phase-locking technique has been reported for fiber-based MZIs that uses discrete single-photon detections \cite{PhaseLockInt_Hacker2023}.
Moreover, the improved phase stability of the DLI can also be achieved by decreasing the path difference of the DLI, thereby reducing the separation of time bins. However, the minimum time-bin separation is dictated by the measurement devices' jitter and channel dispersion as given by Eq.(\ref{eq_jitter}). 

\textit{(c) Temperature stabilized DLIs:} Alternatively, one can use passively temperature-stabilized commercially available DLIs such as Optoplex DPSK demodulators~\cite{Int_Optoplex-DLI} and Kylia MINT \cite{Int_Kylia-DLI}.
These DLIs mainly utilize a Michelson interferometer configuration as shown in Fig.~(\ref{fig_FSUMI}). 
 The key requirement in such devices is to minimize the thermal expansion of the optical paths due to ambient temperature fluctuations. 
 Often, manufacturers use materials with a low thermal expansion coefficient to make them robust against temperature changes and optimize the glass and air pathways, which intrinsically compensate for the thermal expansion and chromatic dispersion effects. Additionally, the interferometers are placed in a hermetically sealed box.  In such cases, phase stability depends on the temperature stability of the DLI, which can be improved by passively isolating the DLIs from the environmental temperature fluctuations or actively stabilizing the DLI temperature using the PID algorithm.  Towards an integrated photonics platform, PLC Mach-Zehnder interferometers \cite{PLC_Himeno1998, PLC-MZI_Honjo2004, PLC_Ren2018, PLC-DLI_Kawashima2012} are other alternatives that have been used for TBQ measurements.
The relative phase of the interferometer is controlled by applying a voltage across the resistive heater located near one of the retroreflectors. 
 These DLIs are well-suited for TBQs in single spatial mode (Gaussian), as needed in most single-mode fiber-based experiments. 
 For multi-mode TBQ analyzer implementation, please see references \cite{TBM-MM_Jin2018, TB-QKD_Tannous2025}.

 \textit{(d) Thermally insensitive DLI:} Temperature-insensitive fiber-based DLI \cite{TempInsensitiveDLI_Shi2022} has been designed using a hollow-core fiber with low thermal sensitivity in one arm of DLI and a single-mode fiber with a relatively high thermal sensitivity in the other arm. By properly matching the length ratio with the thermal sensitivity, a thermally insensitive DLI has been designed.

\subsection{Measurement using light-matter interfaces} \label{Sec_AltApproach}

The task of inducing a time delay to project TBQs onto one of the superposition bases can also be achieved by other means through unbalanced interferometers.  One method to create a time delay is to store the time-bin encoded photon in a quantum memory and retrieve it with only 50\% probability at one time and then retrieve it with 50\% probability after time $\tau_{el}$.
The memory-assisted measurement approach has been employed for several experiments \cite{QMemory_Saglamyurek2011, QMemory_Saglamyurek2015, QMemory_Riedmatten2015}, mostly using the atomic frequency comb quantum memory, which inherently has a large multimode storage capacity. Since 100\% storage efficiency is not achievable, in practice, the time-bin projection is achieved by setting two storage times, which differ by the time-bin delay and feature equal (but less than 50\%) probability of recall, each. One advantage of the memory-assisted approach is that the projection basis can be set precisely by the memory preparation and recall procedure, removing the need for path-length stabilization of an interferometer. Although quantum memories have a higher loss than interferometers, if a specific application already requires quantum storage, the memory-assisted approach avoids additional elements for performing the measurement.

\subsection{Measurement by conversion to polarisation qubits}
\label{Sec_TB-POl}

As quantum communication systems are pushed to run at higher rates, the time-bin separation ($\tau_{el}$) must be shortened and along with this, the chosen detectors must also exhibit lower jitter ($\tau_{JD}$) in order to correctly detect the time-bin state \cite{BSM_Raju2014}. For example, a 100 GHz clocked system would necessitate the use of state-of-the-art detectors \cite{SPD_Hadfield2009}, which are not broadly available and complex to operate. One method to bypass this requirement, is to transform TBQs into polarization qubits using linear or non-linear optical components and then perform measurements in the polarization bases. Importantly, photonic entanglement is preserved when qubits are transformed from one encoding to another \cite{Ent-Conversion_Zukowski1991, TBM-UTBA_Bussieres2010}. An added advantage of such conversion is that the polarization degree of freedom facilitates straightforward measurements along any other direction on the Bloch sphere, including the computational and superposition bases.



\begin{figure}[htbp!]
    \centering
   \includegraphics[width= \columnwidth]{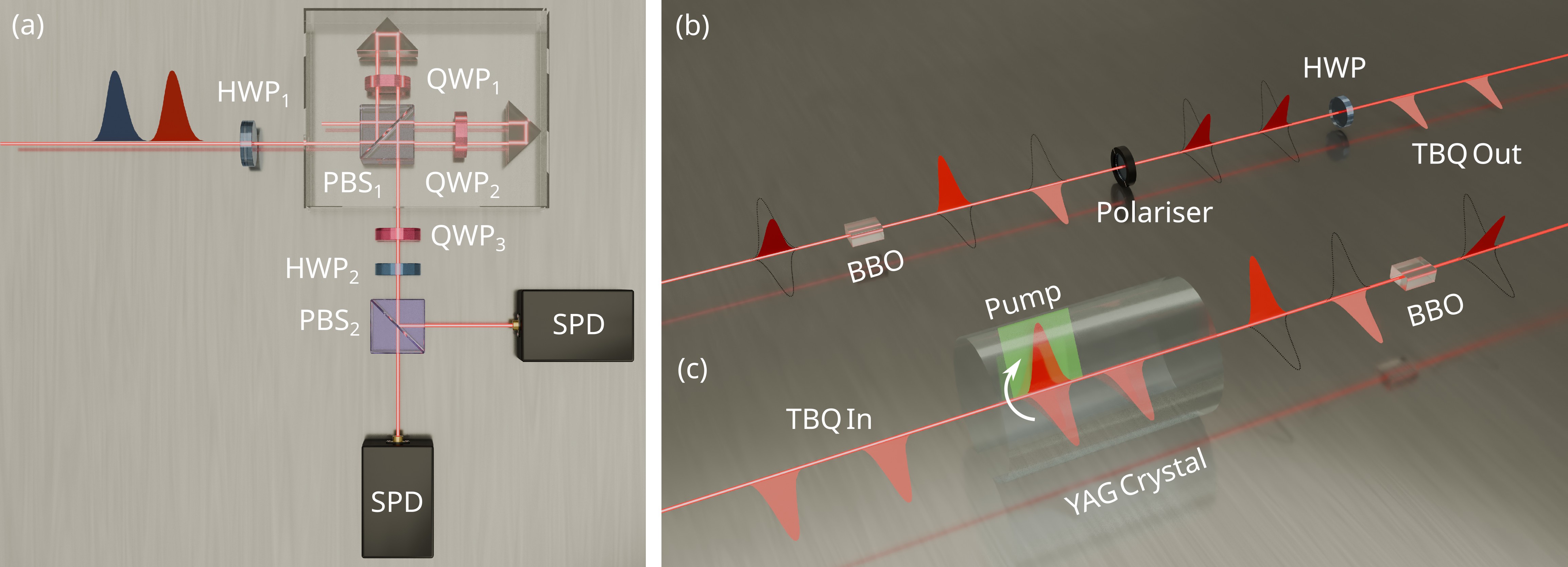}
    \caption[Qubit conversion and measurement.]{Qubit conversion and measurement. (a) Free-space time-bin to polarization conversion and polarization basis analysis. \textbf{Redrawn figure from \cite{TBM-UTBA_Bussieres2010}}. The solid box shows the components enclosed within a temperature-controlled box. (b) Polarization to time-bin, and (c) time-bin to polarization conversion. \textbf{Redrawn figure from \cite{TBM-PolConv_Kupchak2017}}. BBO: Beta-barium borate, HWP: Half-wave plate, PBS: polarizing beam splitter, QWP: Quarter-wave plate, SPD: Single-photon detector.}
    \label{fig_TB-Pol}
\end{figure}

One method for converting time-bin to polarization encoding utilizes passive optics combined with post-selection \cite{TBM_Sanaka2002, TBM_Takesue2005, TBM-UTBA_Bussieres2010, Ent-Hyper_Li2015}. While straightforward, this approach introduces additional losses and compromises the fidelity of the quantum state.
The schematic of a time-bin to polarization converter (also called a universal TBQ analyzer) \cite{TBM-UTBA_Bussieres2010} is shown in the Fig.~(\ref{fig_TB-Pol}a).  The polarization state of the input TBQ is rotated to $45^\circ$ (with respect to the horizontal polarization state transmitted by $\text{PBS}_1$) using a half-waveplate ($\text{HWP}_1$). The two polarization components of each TBQ are separated in orthogonal directions by $\text{PBS}_1$ and travel along the short and long arms of the UMI, having a round-trip delay of $\tau_{el}$. After passing through the quarter-waveplates ($\text{QWP}_1$ and $\text{QWP}_2$) in the respective paths, they exit through $\text{PBS}_1$. At the output port, photons emerge in three time-slots: namely, `early', `middle', and `late', separated by $\tau_{el}$. 
In the middle bin, the input TBQ: $|\psi\rangle=\alpha|e\rangle+\exp(i\phi) \beta|l\rangle$ gets converted to a polarization qubit: $|\psi\rangle=\alpha|V\rangle+\exp(i[\phi+\phi_a])\beta|H\rangle$, where the phase $\phi_a$ is an additional phase accrued due to the analysis interferometer \cite{TBM-UTBA_Bussieres2010}. This polarization qubit is then analyzed using $\text{QWP}_3$, $\text{HWP}_2$, and $\text{PBS}_2$ combination for projection and detected using two SPDs.

A method for interconverting polarization and ultrafast TBQs is illustrated in Fig.~\ref{fig_TB-Pol}(b,c).
For polarization to TBQ conversion, an arbitrary polarization qubit passes through a birefringent medium e.g. $\alpha$-BBO (Barium Borate), which temporally separates the horizontal and vertical polarization components, creating the early and late time-bin modes \cite{TBM-PolConv_Kupchak2017} as shown in Fig.~(\ref{fig_TB-Pol}b). The polarizer oriented along the diagonal direction makes the two time-bins co-polarized at $50\%$ loss, and then HWP prepares horizontally polarized time-bins. 
For TBQ to polarization conversion, an arbitrary TBQ is an input to the optical kerr shutter \cite{OKS_Yu2003} as shown in Fig.~(\ref{fig_TB-Pol}c). A pump field is applied to the late bin and focused both in the Kerr medium (YAG crystal) to rotate the polarization from horizontal to vertical. Subsequently, the orthogonal time bins are passed through an $\alpha$-BBO (Barium Borate) with an optic axis rotated by $90^\circ$ to overlap the two time bins into one temporal mode, thereby creating a polarization qubit \cite{TBM-PolConv_Kupchak2017}. Using this method, picosecond TBQs were generated and used for the BB84 QKD protocol \cite{TBM_qudit_Bouchard2023}. Qubit interconversion has been demonstrated to be useful for cross-encoded BB84 QKD protocol \cite{CrossEncQKD_Scalcon2022}.

An alternative strategy employs active optical switches, such as electro-optic modulators or Pockels cells, to achieve time-bin to polarization conversion by rotating the polarization of a selected time-bin \cite{TBM-PolConv_Kupchak2017}.
TBQs can be converted into polarization qubits using a 1x2 fast optical switch [as discussed in Section (\ref{Sec_Ent-switch})] and PBS. For each TBQ input to the switch, it is biased in such a way that the early state is transmitted to port `c' and the late state is reflected to port `d'.  These time-bins are superposed on a PBS after providing a delay to the early, and polarization rotation by $\pi/2$, creating a polarization qubit. Such a setup effectively uses an interferometer, and therefore, its phase needs to be stabilized. Notably, the optical switches are better suited for $\sim$ nanoseconds or longer time-bin separations, whereas the optical Kerr shutter-based approach is best for a few tens of picoseconds separations.
More recently, nonlinear techniques have been explored to detect temporal-mode-encoded states through novel mechanisms \cite{TBM-Ufast_Donohue2013, TBM_Ansari2017, TBM-fs_Kochi2022, TBM-fs_Kochi2022CLEO, QComp_Bouchard2024}; however, these methods are currently constrained by low operational efficiencies.

\color{black}
\subsection{Measurement using interference with reference states}

An interferometer-free approach for measuring the TBQ or qudit state was recently demonstrated \cite{TB-Robust_Simon2025}. In this scheme, an unknown target qubit ($\ket{\psi_{in}}$) is interfered with a set of known reference states ($\ket{\psi_{ref}}$) on a 50:50 beam splitter. To enable HOM interference (see Sec.~\ref{Sec_HOM}), the reference states must be indistinguishable from the target state in all degrees of freedom except for time and path. The wavefunction overlap between the two states directly determines the photon anti-bunching probability ($P_{ab}$) given by:
\begin{equation}
    P_{ab} = \frac{1 - |\langle\psi_{in}|\psi_{ref}\rangle |^2}{2}.
\end{equation}

By estimating the anti-bunching probabilities with reference states from mutually unbiased bases, the unknown state can be reconstructed using standard maximum-likelihood estimation techniques. For a $d$-dimensional state,  $d+1$ reference states are required when using a complete set of mutually unbiased bases. Consequently, this scheme is particularly advantageous for higher-dimensional time-bin states, as it requires only two detectors. While this method eliminates the need for a phase-stabilized interferometer, it effectively shifts the experimental complexity from implementing arbitrary projective measurements in high-dimensional Hilbert spaces to the preparation of high-dimensional, well-controlled reference states. In this work \cite{TB-Robust_Simon2025}, arbitrary TBQs and their corresponding reference states are prepared by coherently mapping a single photon into programmable temporal modes using polarization-controlled time delays, enabling independent control over the relative amplitudes and phases of the time bins without relying on interferometric stability.


\color{black}
\subsection{Time-bin qubit characterization methods}

TBQs can be characterized by estimating their fidelities with respect to the intended states or by performing QST. This is achieved by measuring the qubits along various bases. 


\subsubsection{Quantum state tomography (QST)} 
\label{Sec_QST}

QST is a method to characterize the state of an unknown quantum state by performing measurements on a tomographically complete basis set \cite{QST_Review2019}.  
The concept of QST was first introduced and implemented in Ref. \cite{QST_James2001} for polarization qubits \cite{QST_Altepeter2004}, and later QST was experimentally demonstrated for TBQs \cite{QST_Takesue2009, QST_Wang2012, QST_Ansari2017, QST_Pilnyak2019, QST_Sedziak2020} and qudits \cite{QST_qudit_Samantha2016, QST_qudit_Ikuta2017}. 
A single-qubit has three independent parameters and requires a minimum of four measurements (one extra for normalization) for the complete characterization of the state (density matrix). However, using a tomographically overcomplete measurement set can enhance the accuracy of tomographic reconstructions \cite{QST-MeasSets_Burgh2008}. 
The density matrix of a single-qubit can be expressed as
\begin{equation}
\hat{\rho} = \frac{1}{2}\sum_{j=0}^{3}s_j \hat{\sigma_j}, 
\label{Eq_QSTDM}
\end{equation}
where $\{\hat{\sigma}_0, \hat{\sigma}_1, \hat{\sigma}_2, \hat{\sigma}_3\}$ denote the Pauli operators (with $\hat{\sigma}_0 \equiv \mathbb{I}$), and the real-valued $s$- parameters $\{s_0, s_1, s_2, s_3\}$ are given as $s_j = \mathrm{Tr}[\hat{\sigma}_j \hat{\rho}]$.
These $s$-parameters satisfy the normalization condition: $\sum_{j=1}^3s_j^2=1$ ($-1 \leq s_1, s_2, s_3 \leq 1$), and $s_0=1$. The parameters \{$s_i$\} are directly related to the projection probability of the qubit in the three mutually unbiased bases: $\{|e \rangle, |l\rangle \},~\{|+\rangle, |-\rangle\}$, and $\{ |+i\rangle, |-i\rangle \}$ \cite{QST_James2001, QST_Takesue2009, QST_Review2019} as follows: $s_0 = P (|e\rangle) + P(|l\rangle), ~ s_1 = P (|+\rangle) - P (|-\rangle), ~ s_2 = P(|+i\rangle) - P(|-i\rangle),~ \text{and}~s_3 = P(|e\rangle) - P(|l\rangle)$, where $P(|x\rangle)$ is the projection probability of the qubit in the basis-state $|x\rangle,~x=e,l,+,-,+i,-i$.  In the case of polarization-qubits, parameters \{$s_j$\} are commonly known as Stokes parameters \cite{QST_James2001}.
Due to experimental errors and imperfections, the experimentally reconstructed density matrix (\ref{Eq_QSTDM}) may not be physical. 
Therefore, the Maximum Likelihood Estimation method \cite{MLE_Banaszek1999, QST_James2001, QST_Takesue2009} can be adapted to generate a physical density matrix from the set of noisy or imperfect measurements. 
For a detailed discussion on accessing different projection bases using one-bit DLI and implementation of complete QST for a time-bin entangled photon pair, please see Reference \cite{QST_Takesue2009}.

\subsubsection{State fidelity measurement}

The state fidelity of a TBQ quantifies how close its actual state ($|\psi\rangle$) is to the intended or target state ($|\tilde{\psi}\rangle$), and is an important metric to assess the performance of TBQs in QIP tasks. A pure but imperfectly prepared TBQ can be modelled as \cite{MDIQKD_Chan2014}
\begin{equation}
    \ket{\psi} ~ = ~ \frac{1}{\sqrt{(1 + 2 b)}} \left( \sqrt{\alpha+b} ~\ket{e} + e^{ i \phi}\sqrt{1 - \alpha + b}~\ket{l} \right), 
\end{equation}
where parameter $\alpha$ determines the probability amplitude of either state, parameter $b$ represents the background photons, and $\phi$ is the relative phase difference between early and late state. In an ideal scenario, $\alpha$ takes the values 0 or 1 for Z-basis states, 0.5 for superposition-basis states, and $b=0$ for all states, indicating no background noise. However, due to imperfections in the preparation process, such as a qubit prepared using an interferometer with reduced visibility, $b$ is typically nonzero, introducing background photons into the system. The parameters $m$ and $b$ can be estimated by measurement in the Z-basis, while $\phi$ can be estimated in a superposition basis. The fidelity of the qubit state with respect to the target qubit ($|\tilde{\psi}\rangle$) can be calculated as, $\mathcal{F} = |\langle\tilde{\psi}|\psi\rangle|^2$. For instance, the fidelity of a target  $|e \rangle$ qubit $\mathcal{F} = (\alpha+b)/(1+2b)$. The measured fidelity can be used to estimate the performance of protocols; the quantum bit error rate can be estimated in QKD protocols.



\section{Parameter selection for time-bin qubits}
\label{Sec_ParameterSelection}
\FloatBarrier
 
From an experimental point of view, TBQs have two key parameters: the shape of the time-bins (which consists of packet profile and width corresponding to each time bin) and the temporal delay between bins. Each of these parameters influences the performance of the qubit, and the correct choice depends on the specific application. In the following subsections, we provide a brief guide to selecting these key parameters for TBQ encoding, considering the combined insight from Sections~(\ref{Sec_Preparation}) and (\ref{Sec_Measurement}). A summary of these considerations is provided in Table~(\ref{Tab-parameter_selection}).
Additional factors, which affect the parameter choice for time-bin entangled states, are outlined in Sec.~(\ref{Sec_CoinMeasCons}b).
Note that, although the quantum information is encoded in the temporal mode, it is also essential to ensure that the other photonic modes, e.g., polarization and spatial modes, remain well-defined.

\begin{table}[htbp]
\centering
\renewcommand{\arraystretch}{1.5}
\begin{tabular}{|p{2cm}|p{6cm}|p{6cm}|}
\hline
\textbf{Parameter} & \textbf{Lower Bound Factors} & \textbf{Upper Bound Factors} \\
\hline
\multirow{4}{=}{\textbf{Time-bin width} \hspace{10pt} ($\Delta \tau$) \vfill} 
  & \textbullet{} Bandwidth of preparation setup & \textbullet{} Time-bin separation ($\tau_{el}$)\\
  & \textbullet{} Laser pulse-width or emitter lifetime & \textbullet{} Spectral filter bandwidth  \\ 
  & \textbullet{} Quantum memory bandwidth & \\
\hline
\multirow{3}{=}{\textbf{Time-bin separation} ($\tau_{el}$)} 
  & \textbullet{} Measured pulse-width ($\Delta \tau_M$) & \textbullet{} Source laser coherence time \\
  & \textbullet{}$\,\,\Pi$-pulse duration in spin-photon entanglement ($\tau_\pi$) & \textbullet{} TBQ phase stability \\
\hline
\end{tabular}
\caption{A summary of factors that bound TBQ parameter selection: time-bin width and separation. Lower bounds are governed by preparation and timing constraints, while upper bounds are limited by coherence and bandwidth.}
\label{Tab-parameter_selection}
\end{table}

\subsection{Wavepacket Shape}

The shape of each time-bin wavepacket consists of two main components: the wavepacket profile (e.g., Gaussian, rectangular, etc.) and the temporal width of the time bins. Here we will discuss the considerations regarding these two components.

\subsubsection{Wavepacket Profile}

The wave packet profile of each time bin determines the qubit's spectral properties.
Gaussian wavepackets are usually used for time-bin encoding; however, other profiles can also be considered depending on their source or the application.
For example, sinc-shaped temporal pulses may be used in cases where spectral flatness is needed \cite{DEVASAHAYAM2004110, soto2013optical}. On the other hand, exponentially increasing or decaying pulses may be preferred due to their compatibility with certain quantum memory protocols \cite{PhysRevLett.98.243602}

Different wavepacket profiles also affect the ease of distinguishing time bins and the sensitivity of the qubits to temporal and spectral noise. For example, Gaussian pulses, with their smoother temporal boundaries, are less prone to sideband distortion when passing through dispersive media (see Sec. \ref{Sec_ChromDisp}) compared to sharper-edged waveforms such as rectangular pulses. On the other hand, rectangular pulses have well-defined start and end times, making them easier to use for precise time synchronization in communication systems.
In this paper, we mostly discuss the Gaussian wavepacket. 

\subsubsection{Temporal Width of Time Bins}
\label{Sec_TempWidthTB}

The prepared temporal width of each time bin ($\Delta\tau$) determines the qubit's resilience to effects such as dispersion and needs to balance multiple practical factors. 
The temporal width of the time bins is inversely related to their spectral bandwidth, as governed by the time-bandwidth product. The general time-bandwidth product relation is given by
    \begin{equation}
    \Delta\tau \Delta \nu_\tau = c_0,
    \label{Eq_TBP}
    \end{equation}
where $\Delta \nu_\tau$ is the pulse's spectral bandwidth. 
$c_0$ is a shape-dependent constant, and its values vary depending on the pulse shape--for example,
$c_0=0.441$ for a Gaussian pulse and 
$c_0=0.886$ for a rectangular pulse. This interplay between temporal and spectral widths influences the preparation and application of the time bins.

\textit{(a) Preparation schemes:} 
Depending on the TBQ source, the temporal width may be limited or adjustable.
For example, using a pulsed laser for time-bin generation results in a fixed pulse width determined by the laser's characteristics. 
On the other hand, when carving pulses from a continuous-wave (CW) laser, the pulse width is adjustable. For very short pulses, a sufficiently high bandwidth of the RF-signal source and electro-optic modulator is required. 
As an example, the preparation of 100 ps narrow optical pulses requires at least Nyquist-limited bandwidths of the total corresponding RF source, amplifier, and modulator circuit ($>20$ GHz). 
However, these limitations are not just determined by the source laser or modulators. In strongly coupled two-level systems or single emitters, such as quantum dots, the temporal width is constrained by the inherent lifetime of the system.

\textit{(b) Applications:}
The pulse-width through the time-bandwidth relationship becomes specifically important in the interference of two TBQs. The ability to synchronize the time-bin generation across potentially large distances will limit how small $\Delta \tau$ can be while ensuring the simultaneous arrival of the time-bins. Conversely, in the spectral domain, when two independently generated time bins are interfered, for example, in HOM measurement, spectral indistinguishability is more easily maintained if $\Delta \tau$ is small.
This relationship is also important in entanglement generation through nonlinear processes like SPDC. 
Because of the wide range of energy- and momentum-conserving processes that can occur within SPDC, the resulting photon pairs (signal and idler) exhibit a broad spectrum. 
To generate high-fidelity time-bin entanglement, spectral filtering is often employed (see Sec. \ref{Ent_SPDC}). The highest spectral purity is achieved when the bandwidth of the filtered photons is narrower than the pump's bandwidth, ensuring that only a limited subset of the broad spectrum contributes to the time-bin entanglement.

 When storing TBQs in quantum memory, the bandwidth of the quantum memory adds a limitation to the time-bin width choice. The memory's bandwidth must be equal to or larger than the spectral bandwidth of the TBQs. If the qubit bandwidth exceeds the memory bandwidth, the storage efficiency will be reduced. Therefore, the qubit's temporal width and the quantum memory's bandwidth should be compatible.

In summary, choosing the time-bin width involves accounting for multiple factors, including the preparation scheme, spectral bandwidth, time-bin orthogonality, and quantum memory limitations. Balancing these constraints is essential to achieving optimal performance in quantum communication and computation systems, particularly in long-distance or high-fidelity scenarios.  

\subsection{Time bin separation} 
\label{Sec_TBSep}

In general, the temporal width of and delay between time bins must be chosen to balance the phase stability and orthogonality of the time-bin modes. 

\subsubsection{Orthogonality of time-bin modes} 
\label{Sec_TBOrthogonality}

As discussed in Section~(\ref{Sec_TBEncodingDefs}), to maintain the orthogonality of the temporal modes, the time separation between the bins should be greater than the prepared temporal width of each time bin, $\tau_{el} > \Delta \tau$. As noted above, $\Delta \tau$ may be more or less adjustable depending on the preparation, e.g., creating pulses from a CW laser (Section \ref{Sec_CWLaser}) or employing single photon emitters with intrinsic life-times (Section \ref{Sec_QDotTB}). In addition, when the time-bins propagate through fiber the pulses broaden due to chromatic dispersion by an amount $\Delta \tau_{CD}$ proportional to the fiber length and inversely proportional to the time-bin width (see Section~(\ref{Sec_ChromDisp})).  

Additionally, the total measured time-bin width is affected by the jitter of the detectors $(\Delta \tau_{JD})$ \cite{Jitter_Allmaras} and time-tagging electronics ($\Delta \tau_{JT}$). Here, timing jitter refers to the uncertainty or statistical variations in the timing of the device's output signal following a detection event, which sets the timing resolution. 
Therefore, SPDs and corresponding electronics must allow high temporal resolution.
For instance, state-of-the-art SNSPD can offer a small possible jitter (a few ps) \cite{Jitter_Esmaeil2020, Jitter_Korzh2020, Jitter_Taylor2022}. Accounting for all the sources of broadening, the measured pulse width $\Delta\tau_M$ can be expressed as:
\begin{equation}
\Delta\tau_M=\sqrt{\Delta\tau^2+\Delta\tau_{CD}^2+\Delta\tau_{JD}^2+\Delta\tau_{JT}^2}.
\label{eq_jitter}
\end{equation}
Therefore, to distinguish the time bins, one must ensure that the delay $\tau_{el}$ be larger than the measured bin width $\Delta\tau_M$. 

As an example, consider a 1550~nm wavelength Gaussian pulse having a 100~ps pulse width (FWHM) propagating through a 100~km long single-mode fiber with a chromatic dispersion coefficient of 17~ps/(nm-km).  The pulse width at the output end of the fiber due to chromatic dispersion becomes 160~ps.  When it is detected using a SPD ($\Delta\tau_{JD}=50$ ps) and time-tagger ($\Delta\tau_{JT}=30$ ps), its measured pulse width becomes $\Delta\tau_M=170$~ps. Therefore, as discussed in Section~(\ref{Sec_TBEncodingDefs}), to ensure that the mode overlap between the early and late time bins remains as low as $0.01\%$, a temporal separation of $\tau_{el} = 6\,\sigma=2.55\,\Delta\tau_M = 433$~ps is used.

\subsubsection{Relative phase stability}
\label{Sec_Phase}

Generally, the phase stability of the time-bin will determine the upper limit to $\tau_{el}$. As discussed in Section~(\ref{Sec_PhaseStableMethods}), source laser frequency inaccuracy (linewidth, $\Delta\nu_l$, and frequency drift, $\Delta\nu_d$) generates TBQ phase error. 
The phase error due to the source laser frequency inaccuracy $(\Delta\phi_s)$ can be calculated as in Eq.~(\ref{Eq_FreqtoPhaseUncertain}), which scales linearly with $\tau_{el}$. 
Therefore, to minimize the phase error of the qubit, the time-bin separation, the linewidth and the frequency drift of the source laser should be as small as possible.
For a concrete example, consider a weak coherent TBQ with $\tau_{el}=1$~ns carved from laser light having a frequency drift of $\Delta \nu_d = 10$~MHz. This produces a phase error of at least $3.6{\degree}$. On the other hand, since the linewidth of most of the lasers is below the MHz range, the phase error due to the source laser linewidth would be negligible. 

The relative phase between the early and late time bins, and its stability, are affected by both the preparation and measurement apparatus.
In the preparation, a phase error, $\Delta\phi_{\textit{g}}$, may be introduced through fluctuations of the electronic signal amplitude into the phase modulator (Section~(\ref{Sec_CWLaser})) or through pathlength instability of a DLI (Section~(\ref{Sec_PhaseStableMethods})).  For the former $\Delta\phi_{\textit{g}}$ is independent of $\tau_{el}$, whereas for the latter Eq.~(\ref{Eq_DLIPhaseInstability}) illustrates that the phase-error will proportional to the time-bin delay, $\tau_{el}$. Hence, the phase stability of interferometers constrains the maximum feasible time-bin separation. 
Finally, errors in the projection basis of the measurement apparatus may contribute to the observed phase-error, $\Delta\phi_{\textit{m}}$ of the state (Section~(\ref{Sec_PhaseStableMethods})). Since a DLI is most commonly employed for this purpose, a smaller time-bin separation will be advantageous.~~~~~

Additionally, in the context of spin-photon entanglement generation as described in Section~(\ref{Sec_SCTLS}), a $\pi$-pulse is usually required in between early and late pulses to flip the spin qubit \cite{SCTLS_BellState_Rempe_2017}. This sets a lower limit on the time-bin separation of the generated entangled pair, such that the $\pi$-pulse duration $(\tau_{\pi})$ should be smaller than the time-bin separation, $\tau_{\pi}<\tau_{el}$. 


\section{Time-bin entanglement} 
\label{Sec_TBE}
\FloatBarrier

Quantum entanglement represents the defining characteristic of quantum mechanics, as suggested by Schr\"{o}dinger in 1935 \cite{Ent_Schrodinger1935}. Historically, it has been at the forefront of the discussion about the nature of reality and the completeness of quantum mechanics \cite{EPR_Einstein1935, EPR_Bohr1935, EPR_Bell1964}. 
Entanglement has also been proven to be a quintessential resource for several quantum information applications \cite{Book-QIP_chuang, Book-QIP_wilde} such as quantum communication \cite{Ent-QTech_AOPReview2024}, metrology \cite{QMetrology_Giovannetti2004, QMetrology_Giovannetti2011, QMetrology_Huang2024}, sensing \cite{QSensing_Pirandola2018, QSensing_Liu2024}, and distributed quantum computation \cite{QComp-Dist_Main2025, QComp-Dist_Barral2025}. 
Accordingly, there is an increased interest in developing methods for preparing and characterizing entanglement in bipartite \cite{Ent-Rev_Horodecki2009, Ent-Rev_RMP2009, Ent-Char_Ali2021}, multipartite \cite{Ent-MP_Walter2016}, and higher-dimensional \cite{HD_Erhard2020}  systems. 
In this section, we will review different experimental methods for generating bipartite and multipartite time-bin entanglement and the characterization techniques.


\subsection{Bipartite qubit entanglement preparation}

The bipartite time-bin entangled qubit states --- the simplest representation of time-bin entanglement --- can be generated using SPDC, SFWM, a quantum dot, and an optical switch. We explained how SPCD, SFWM, and quantum dots can be used for TBQ generation in Sections (\ref{Sec_HeraldedTB}) and  (\ref{Sec_QDotTB}). Here, we focus on entanglement generation.

\subsubsection{SPDC}  
\label{Ent_SPDC}

One of the most common methods for preparing time-bin entanglement utilizes the SPDC process \cite{Ent_Brendel1999, Ent_Marcikic2002, Ent_Ma2009} in a second-order nonlinear ($\chi^{(2)}$) crystal. 
Conceptually speaking, when a pump TBQ in the superposition state $(|e\rangle_p+\exp(i\phi)|l\rangle_p)/\sqrt{2}$ undergoes the SPDC process (see Fig.~\ref{fig_PhaseMatch}), it results in the following probabilistic transformation:
\begin{equation}
\frac{|e\rangle_p+e^{i\phi}|l\rangle_p}{\sqrt{2}}\ \xrightarrow{SPDC}\ \frac{|e\rangle_s|e\rangle_i+ e^{i\phi}|l\rangle_s|l\rangle_i}{\sqrt{2}}.   
\label{eq_SPDC}
\end{equation}
where subscript $s\,(i)$ denotes a photon at the signal (idler) wavelength for the given time-bin. However, the efficiency of the SPDC process is low ($\leq$ 1 in a million); therefore, a pair of classical pump pulses occupying early and late time bins is typically sent through the nonlinear crystal to increase the pair generation rate. 
 The down-converted photons follow thermal statistics \cite{SPDC-Thermal_Paleari2004}, hence the pump power is tuned such that most down-conversion events produce a vacuum, some produce a pair, and only a small amount produce more than one pair.


Owing to the broad range of signal and idler frequency pairs satisfying the phase-matching condition for a given pump photon, the SPDC process typically results in a broadband spectrum.
Therefore, spectral filtering of signal and idler photons is often performed to generate high-fidelity time-bin entanglement.
The two key considerations for spectral filtering are the central wavelength and the bandwidth of the down-converted photons. 
The central wavelength of the filters for the signal and idler wavelengths should match the energy conservation for a given pump wavelength. 
For a given pump pulse duration $\Delta t_p$, the pump bandwidth is given by the time-bandwidth product relation in Eq.~(\ref{Eq_TBP}). 
Note that in cavity-enhanced SPDC, the cavity naturally suppresses the production of pairs that are non-resonant with the cavity, yielding a more efficient filtering step but at the expense of the complexity of designing and aligning a cavity \cite{SPDC-Cavity_Slattery2019, SPDC-Cavity_Moqanaki2019, SPDC-Cavity_Kojouri2023}.

\begin{figure}[htbp!]
    \centering
    \includegraphics[width=\linewidth]{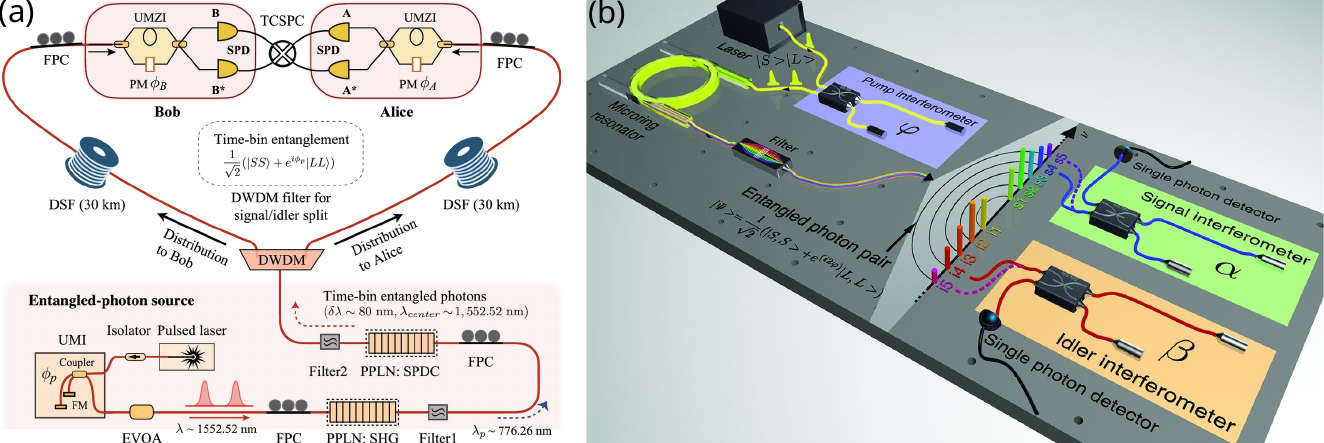}
    \caption[Time-bin entanglement generation using spontaneous parametric processes.]{Time-bin entangled photon generation using spontaneous parametric processes. Schematic of the experimental setup for  (a) 
   SPDC-based time-bin entangled photon generation and distribution. \textbf{Figure reproduced with permission from Ref. \cite{Ent-dist_Kim2021}}. (b) SFWM-based time-bin entangled photon generation and characterization. \textbf{Figure reproduced with permission from Ref. \cite{Frequency_Reimer2016}}.}
    \label{fig_Ent-SPDC-SFWM}
\end{figure}

Recently, Xie et al. demonstrated time-bin entangled photon pair generation and detection using the Quantum Instrumentation and Control Kit on RFSoC-FPGA, achieving CAR $>150$ and visibility $>95\%$, matching conventional setups \cite{TBEnt-SPDC_Xie2023}.
Time-bin entangled states have also been generated via SPDC on integrated photonic platforms, including a periodically poled lithium niobate (PPLN) crystal on lithium niobate on insulator (LNOI) \cite{PIC_TBQ_DLI_Grange2024} and a 3C-silicon carbide-on-insulator platform \cite{TBEnt-SPDC_Li2024}.
A cascaded second-order non-linear process of SHG and SPDC \cite{TBEnt_Hunault2010, Ent-Cascade_Zhang2021} has been used to generate time-bin entanglement in the same wavelength band as the pump, as shown in Fig.~(\ref{fig_Ent-SPDC-SFWM}a). 
Such entangled states have been successfully distributed over long-distance fibers \cite{Ent-dist_Thew2002, Ent-dist_Marcikic2004, Ent-dist_Honjo2007, QCom-WDM_Kim2022} for quantum communication applications. 
Extending these efforts to realistic network environments, Seo \textit{et al.} demonstrated noise-resilient distribution of time-bin entanglement over a 40 km deployed metropolitan fiber network using time-bin-frequency hyperentanglement and frequency-resolved detection, achieving significant noise suppression and recovery of Bell nonlocality under high-loss conditions \cite{TB-HyperEnt_Seo2026}. 
Earlier, a time-bin entanglement purification protocol based on sum-frequency generation was proposed to mitigate imperfections inherent to SPDC-based sources \cite{TB-EntPurification_Yan2021}, enabling purification through post-selected suppression of multi-photon events and effective reduction of bit- and phase-flip errors, thereby improving the fidelity of the resulting entangled state.

\subsubsection{SFWM}
\label{Ent_SFWM}

Time-bin entanglement generated via SFWM is similar to SPDC, with the main differences lying in energy and phase matching, as discussed in Section (\ref{Sec_HeraldedTB}). As for SPDC, the spectral purity of the generated entangled photons in SFWM primarily depends on the effectiveness of filtering the signal and idler photons relative to the pump bandwidth. Additionally, spontaneous Raman scattering from the strong pump degrades the purity of the generated pair \cite{SFWM-Purity_Wang2025}. Hence, greater care is needed when filtering the photon pair from the pump, as the photons are spectrally closer to the pump in contrast to SPDC. Interestingly, near-unity spectral purity has been achieved using cavity-enhanced SFWM \cite{SFWM-Purity_Liu2019} in silicon micro-ring resonators. 

The initial demonstration of time-bin entanglement generation using SFWM dates back more than two decades, employing dispersion-shifted fibers \cite{Ent-FWM_Takesue2005}. 
Since then, advancements in integrated photonics enabled SFWM demonstrations on Si nano-wires, long waveguides \cite{TBEnt-SFWM_Yan2023}, and micro-ring resonators \cite{Ent-FWM_Fang2018, Ent-FWM_Xiong2015, Ent-FWM_Takesue2007, Ent-FWM_Takesue2014, Ent-FWM_Wakabayashi2015, Ent-FWM_Afzal2024}. 
Micro-ring resonators offer compact size and high-quality factor, enabling efficient light confinement and enhancing photon interaction. 
Interestingly, Si$_3$N$_4$ emerges as a prominent material for SFWM due to its higher $\chi^{(3)}$ non-linearity compared to Si. 
Initially, long double-strip Si$_3$N$_4$ waveguides were utilized as the nonlinear medium \cite{Ent-FWM_Zhang2018}, but recent developments \cite{Ent-FWM_Fujiwara2017, Ent-FWM_Samara2019, Ent-FWM_Lu2019, Ent-FWM_Ren2023} involve Si$_3$N$_4$ micro-ring resonators for SFWM-based time-bin entanglement generation, as shown in Fig. (\ref{fig_Ent-SPDC-SFWM}b). 
Notably, a recent breakthrough demonstrates heterogeneous laser integration for SFWM in the InP and Si$_3$N$_4$, showcasing its potential for quantum frequency conversion \cite{Ent-FWM_Mahmudlu2023}. The SFWM process has also been employed to generate multiplexed polarization- and time-bin-entangled photon-pairs \cite{Ent-Multiplexed_Li2016} and hyperentangled photon pairs in various degrees of freedom, including time-polarization \cite{HyperEnt-Pol-ET_Suo2015}, frequency-time-bin \cite{TB-HD-QIP_Reimer2018}, and time-bin-frequency-bin \cite{TB-HyperEnt_Congia2025}.

\subsubsection{Quantum dot}
\label{Sec_Qdot}

Although quantum dots are predominantly employed as single-photon emitters, they have long shown promise for realizing deterministic photon-pair sources.
One method for creating time-bin entanglement from a QD \cite{Ent-Qdot_Philip2022} uses a pulsed laser, which is passed through a DLI to generate a phase-coherent pair of early and late pulses.
These pulse pairs resonantly excite the biexciton state in the InAsP quantum dot embedded in a nanowire, which results in two photons emitted in succession through a biexciton-exciton cascade process. 
Addressing the emitter with a pump pulse that has a very low probability of exciting the quantum dot ensures that a maximum of one photon pair is generated. 
Post-selecting the biexciton-exciton photon-pair detection events \cite{Ent-Qdot_Weihs2017, Ent-Qdot_Philip2022, TB-Qdot_Bracht2024} produces the time-bin entangled state
\begin{equation}
\begin{aligned}
    |\phi\rangle = \frac{|e\rangle_b|e\rangle_x+\exp(i\Delta\phi)|l\rangle_b|l\rangle_x}{\sqrt{2}}
     \equiv \frac{|ee\rangle+\exp(i\Delta\phi)|ll\rangle}{\sqrt{2}},
    \end{aligned}
\end{equation}
where subscripts $x$ and $b$ denote the exciton and biexciton, respectively, and the relative phase $\Delta \phi$ is controlled by a phase plate in the pump interferometer. 
A theoretical framework for the measurement of time-bin entangled photons from quantum dots can be found in Ref. \cite{TB-Qdot_Bracht2024}.

\begin{figure}[ht!]
    \centering
\includegraphics[width=0.70\columnwidth]{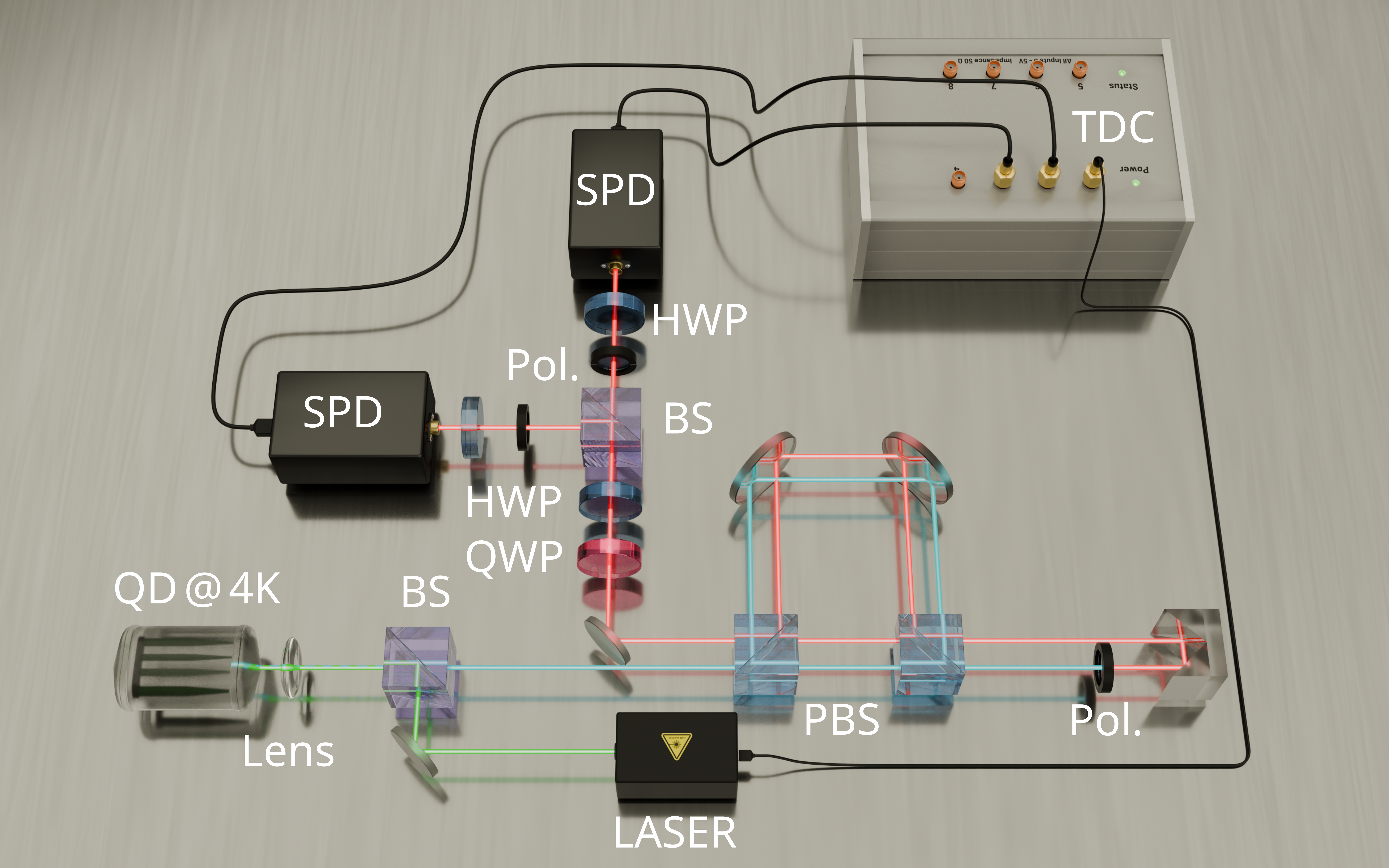}
\caption[Time-bin entanglement generation from quantum dot source.]{Schematic of the quantum dot-based polarization entanglement generation and conversion to time-bin entanglement using UMZI and polarizer. BS: Beam-splitter, HWP: Half-wave plate, PBS: Polarizing beam splitter, Pol.: Polarizer, QD: Quantum Dot, QWP: Quarter-wave plate, SPD: Single-photon detector, TDC: Time to digital converter. \textbf{Redrawn figure from \cite{Ent-Qdot_Versteegh2015}}.}
\label{fig_ent_qdot}
\end{figure}

Another method \cite{Ent-Qdot_Versteegh2015} uses a quantum dot to generate a pair of polarization-entangled photons and subsequently converts it into time-bin entanglement using a UMZI followed by a $45^\circ$ polarizer (see Fig.~\ref{fig_ent_qdot}). 
When a pair of entangled photons in polarization $(|HH\rangle+ \exp(i\phi)|VV\rangle)/\sqrt{2}$ passes through UMZI consisting of a pair of PBSs (inside the dotted box in Fig.~(\ref{fig_ent_qdot}), the H-polarization component gets transmitted from the PBS and traverses through the short path, and the V-polarization component gets reflected and traverses through the long path. 
Thus, the H-component of the photon pair reaches as the early state ($|ee\rangle$) and the V-component as the late state ($|ll\rangle$). 
Both of these time bins are projected onto a polarizer at $45^\circ$, resulting in the time-bin entangled state $(|ee\rangle + \exp(i\phi') |ll\rangle)/\sqrt{2}$ at the loss of 50\% of the input signal intensity. 
We refer the reader to the following references for additional implementations of quantum dot-based time-bin entanglement \cite{Ent-Qdot_Simon2005, TBEntQdot_Jayakumar2013, Ent-Qdot_Jayakumar2014,  Ent-Qdot_Versteegh2015, Ent-Qdot_Huber2016, Ent-Qdot_Gines2021, Ent-Qdot_Lee2019, Ent-Qdot_Appel2021} and hyperentanglement \cite{HyperEnt-Qdot_Maximilian2018} generation.  Furthermore, a recent study has experimentally demonstrated that resonance fluorescence from a weakly coupled two-level emitter can be manipulated using beam splitters, delay lines, and post-selection to produce maximally entangled time-bin entanglement \cite{TBEnt-ResFlu_Hu2025}.

\subsubsection{Entanglement generation using a fast optical switch and post-selection}  \label{Sec_Ent-switch}

One way to generate a time-bin entangled state is by combining two TBQs from independent sources by using an optical switch \cite{Ent-Switch_Takesue2014}. This method is similar to polarization entanglement generation using a polarization beam splitter.
In photonic quantum information experiments involving polarization qubits, a polarizing beam splitter plays a crucial role in interference \cite{Ent-MP-Rev_Pan2012}. 
It is a two-input two-output device; for any polarization state input at each port, it transmits one polarization component and reflects the orthogonal component, usually horizontal and vertical, respectively. 
For the TBQ, an optical switch is similar to the action of PBS for polarization qubits, specifically on early and late states. 
 There are different types of optical switches, such as electro-optic, acousto-optic, and semiconductor optical amplifier switches, that can be used for TBQ switching due to their high speed and precision. However, slower switches such as Micro-Electro-Mechanical Systems and thermo-optic types are generally unsuitable for such applications due to their slower response times.

 \begin{figure}[ht!]
    \centering
   \includegraphics[width=0.7\columnwidth]{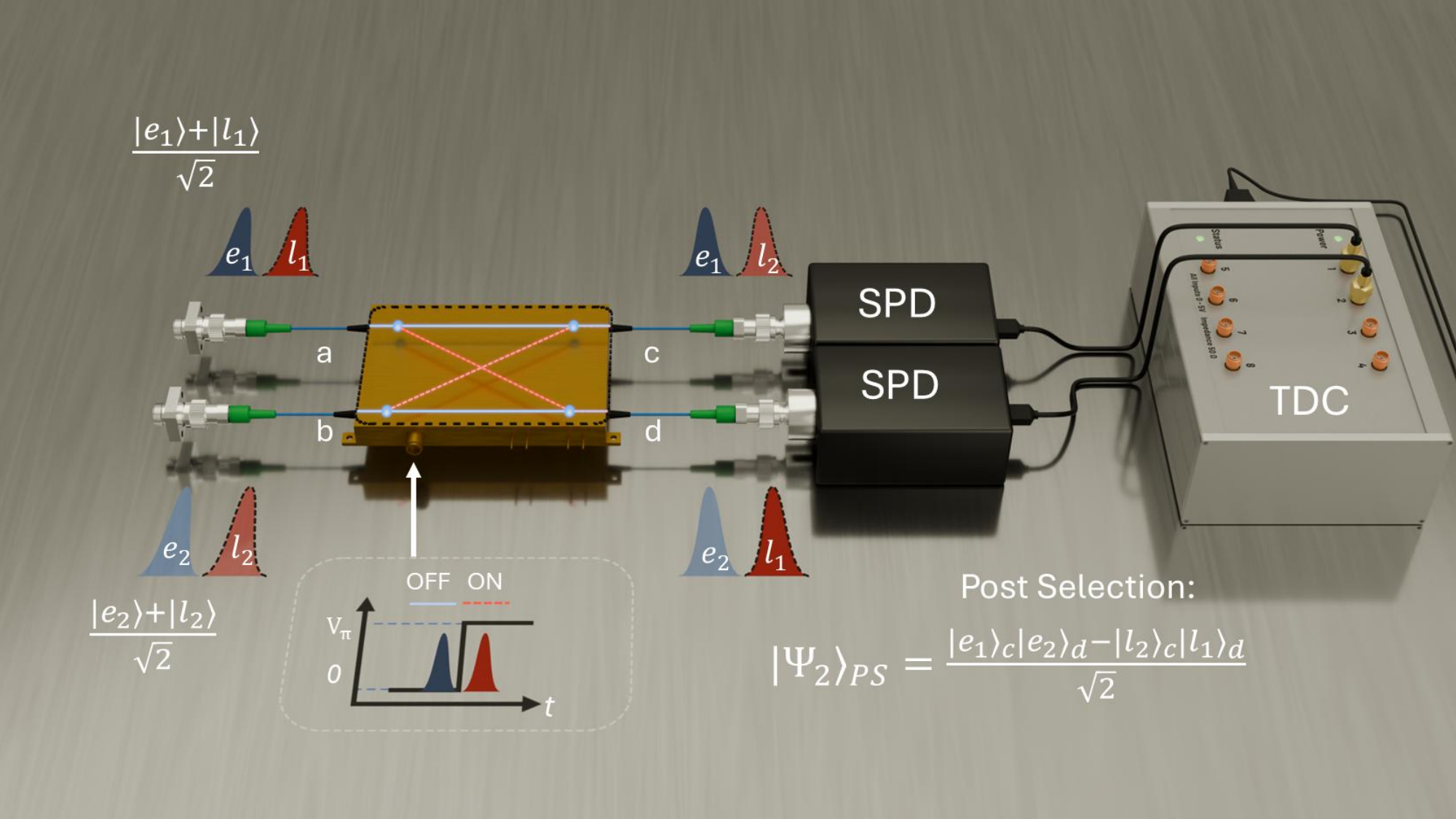}
    \caption[Time-bin entangled state generation on an optical switch.]{Schematic for time-bin entangled state generation on a $2\times 2$ optical switch. A TBQ is incident on each input port of the switch. The switch operates in a way that it transmits any photon in `early' time-bins, i.e., from mode $a \rightarrow c$ and $b \rightarrow d$, whereas it reflects those in `late' time-bins, i.e., from mode $a \rightarrow d$ and $b \rightarrow c$. By post-selecting coincidence, at both the output ports of the switch, a time-bin entangled state is heralded. SPD: Single-photon detector, TDC: Time to digital converter. (\textbf{Redrawn figure from \cite{GHZ_Leili2024})}.} 
    \label{fig_switch}
\end{figure}
 
To understand how the switch operates on a TBQ, consider the following example of switching using a 2x2 electro-optic switch: Labeling the input ports as `a' and `b' and the output ports as `c' and `d', assume the phase is initially set (via a DC bias) so that light passes from `a' to `c' and `b' to `d'. When an RF pulse with a peak-to-peak voltage equal to the modulator's $\pi$ voltage ($V_\pi$) is applied, it briefly alters the paths, switching the light from `a' to `d' and `b' to `c'. If the RF pulse lasts longer than the bin duration and is applied between the arrival of the early and late bins, the transformations in Eq. (\ref{eq_switch}) are achieved, as shown in Fig.~(\ref{fig_switch}).
\begin{equation}
\begin{aligned}
        &|e\rangle_{a} \rightarrow|e\rangle_{c}, 
        &|l\rangle_{a} \rightarrow e^{i\phi}|l\rangle_{d},~~~~ 
        &|e\rangle_{b} \rightarrow|e\rangle_{d}, 
        &|l\rangle_{b} \rightarrow -e^{-i\phi}|l\rangle_{c}.
\end{aligned}
\label{eq_switch}
\end{equation}
Time-bin entanglement can be generated by injecting two TBQs into the switch \cite{Ent-Switch_Takesue2014} as follows. Consider input to the switch is a separable state of two qubits
\begin{equation}
|\Psi_1\rangle=\left(\dfrac{|e_1 \rangle_a + |l_1 \rangle_a}{\sqrt{2}}\right) \otimes \left(\dfrac{|e_2\rangle_b + |l_2\rangle_b}{\sqrt{2}}\right),   
\label{eq_BS1}
\end{equation}
After the switching action, the output state becomes
\begin{equation}
|\Psi_2\rangle=\dfrac{|e_1\rangle_c |e_2 \rangle_d -e^{-i \phi}|e_1\rangle_c|l_2 \rangle_c+ e^{i\phi}| e_2\rangle_d l_1\rangle_d -|l_2\rangle_c| l_1\rangle_d}{2}.
\label{eq_BS2}
\end{equation}
Post-selecting the events that have photons at both output ports of the switch produces a Bell state: 
\begin{equation}
|\Psi_2\rangle_{PS}=\dfrac{|e_1\rangle_c |e_2 \rangle_d  -|l_2\rangle_c| l_1\rangle_d}{\sqrt{2}}.
\label{eq_BS3}
\end{equation}
To generate a maximally entangled state, the input time-bin states must be indistinguishable; therefore, the temporal or spectral bandwidth of the input time-bin states must be matched for entanglement generation through interference on the switch. It should be noted that the extinction ratio of the switch will limit the entanglement visibility.

\subsection{Multipartite entanglement}

The size of entangled quantum systems can be increased in two ways --- either by increasing  (i) the dimensions of individual quantum systems, i.e,  going from qubits to qudits, or (ii) the number of entangled particles --- resulting in high-dimensional and multipartite quantum systems, respectively. 

Photonic multipartite entanglement is a useful resource in multiparty quantum privacy protocols such as conference key agreement \cite{QCKA_Proietti2021, QCKA_Pickston2023}, quantum secret sharing \cite{QSS_Hillery1999, QSS_Anders1999, QSS-HDE_Tittel2001, QSS_Gaertner2007, QSS_Hwang2011, QSS_Basak2023},  quantum voting \cite{PhysRevApplied.18.014005}, quantum secure direct communication \cite{QSDC_Jin2006, QSDC_Man2006, QSDC_Hassanpour2014}, and efficient QKD protocols \cite{MultipartyEnt_Kempe1999, Ent-MP_Epping2017}. Moreover, it has also been employed in foundational studies, such as tests of quantum nonlocality without relying on Bell-type inequalities \cite{Ent-MP3_Pan2000}. 


Generating arbitrarily large multipartite time-bin entangled states of light is practically challenging, as it requires the implementation of quantum gates between individual qubits/qudits. This is probabilistic, and implementations are lossy and can be noisy. Thus, the generation rates of such states decrease exponentially with the increasing number of particles in the entangled state, a step that is more challenging for qudits compared to qubits. A significant amount of work has been devoted to studying the simplest multipartite entangled systems beyond two qubits, specifically three-qubit states. We will discuss the methods for generating and characterizing multipartite and high-dimensional photonic time-bin entangled states in Section  (\ref{Sec_GHZ-State}) and (\ref{Sec_HD-Ent}), respectively.

There are two different classes of genuine tripartite entangled states --  GHZ (Greenberger-Horne-Zeilinger) states and W-states -- which cannot be transformed into each other by even stochastic local operations and classical communication \cite{Ent-MP_Dur2000}. 
The three-qubit GHZ state exhibits strong three-party correlations but lacks any pairwise (two-party) correlations, meaning the entanglement is collectively shared among all three qubits. In contrast, the $W$ state also demonstrates three-party correlations but differs in that it retains pairwise entanglement; even if one qubit is lost, the remaining two qubits remain entangled. This makes the $W$ state more robust against particle loss \cite{GHZ-W-Conversion_Zheng2020}.  
Moreover, if one qubit of a GHZ state is measured in the computational basis, the states of the remaining qubits become fully determined, but this is not the case for the $W$ state. These properties make the GHZ state a strong candidate for multipartite quantum communication. 
Hence, in the following sections, we focus on GHZ and more general multipartite states, which have more applications and for which experimental work with time bins is available. 

\subsubsection{GHZ-state}
\label{Sec_GHZ-State}

A $M$-qubit GHZ-state is given as $|\psi^{(M)}\rangle = 1/\sqrt{2}\left(|0\rangle^{\otimes M} + |1 \rangle^{\otimes M}\right),~M \geq 3$. 
Multipartite GHZ-states can be generated either by concatenating sources of bipartite entangled photons \cite{TE_Agne2017} or by combining bipartite entangled photons from independent sources \cite{GHZ_Lo2023, GHZ_Davis2023CLEO, GHZ_Davis2023Optica, GHZ_Leili2024}. We will outline recent experiments based on the latter approach, which have demonstrated progress towards three- and four-partite time-bin GHZ states.

\begin{figure}[ht!]
    \centering
   \includegraphics[width= \columnwidth]{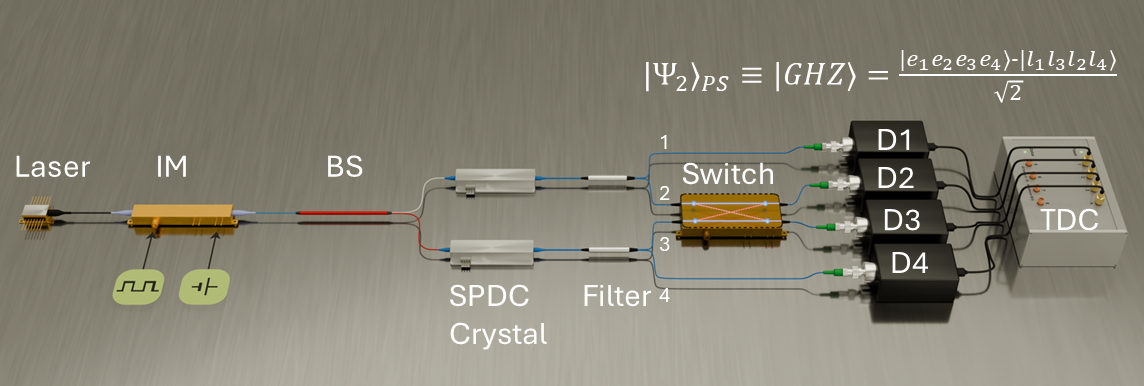}
    \caption[Time-bin GHZ-state generation by SPDC, interference, and post-selection.]{Schematic of the time-bin GHZ-state generation via SPDC, interference on an optical switch, and post-selection. A pair of time-bin entangled sources is generated by sending a classical superposition of early and late pulses to a pair of SPDC sources and spectral filtering of signal and idler photons. One photon from each source is entangled on an optical switch, and post-selection across all four detectors results in a time-bin GHZ state. BS: Beam-splitter, D1, D2, D3, D4: Single-photon detectors, IM: Intensity Modulator, TDC: Time-to-digital converter.}
  \label{fig_GHZ}
\end{figure}

Consider the schematic in Fig.~(\ref{fig_GHZ}) for four-partite GHZ-state generation. At first, two entangled pairs are generated, e.g. by SPDC.
The joint state of such a four-qubit system can be written as
\begin{equation}
|\Psi_1\rangle=\left(\dfrac{|e_1 e_2\rangle + |l_1 l_2\rangle}{\sqrt{2}}\right) \otimes \left(\dfrac{|e_3 e_4\rangle + |l_3 l_4\rangle}{\sqrt{2}}\right).   
\label{eq_GHZ1}
\end{equation}
Assume one TBQ from the pair source is directed to one of the inputs of a fast optical switch, and another qubit from another pair source is directed to the other input of the switch.
If the switch is biased and changes its state using a synchronized RF pulse akin to that discussed in Sec.~(\ref{Sec_Ent-switch}), that is, if the transformation in Eq.~(\ref{eq_switch}) is implemented, then the following state is produced:
\begin{equation}
|\Psi_2\rangle=\dfrac{|e_1 e_2 e_3 e_4\rangle -e^{-i\phi}|e_1 e_2 l_2 l_4\rangle+ e^{i\phi}|l_1 l_3 e_3 e_4\rangle -|l_1 l_3 l_2 l_4\rangle}{2}.
\label{eq_GHZ2}
\end{equation}
Post-selecting on the outcome when all four detectors click \cite{Pascal_Thesis, GHZ_Leili2024} results in a four-qubit GHZ-state given as
\begin{equation}
|\Psi_2\rangle_{PS}\equiv|GHZ\rangle=\dfrac{|e_1 e_2 e_3 e_4\rangle - |l_1 l_3 l_2 l_4\rangle}{\sqrt{2}}.  
\end{equation}

High-fidelity GHZ-state generation requires that the spectral bandwidth of the photons from two different SPDC sources interfering with the switch be smaller than the bandwidth of the pump pulse undergoing SPDC. 
Nevertheless, this protocol produces the GHZ states only after detection, i.e., it post-selectively generates the state.
However, this is suitable for a variety of quantum information protocols, such as secret sharing.

Similar implementations to generate three-partite GHZ-states \cite{GHZ_Lo2023, GHZ_Davis2023Optica, GHZ_Davis2023CLEO} combine a time-bin entangled photon pair with a weak coherent TBQ instead, again using a $2\times 2$ intensity modulator. To ensure wavelength indistinguishability, the weak coherent TBQ is generated from a laser that is also used to generate the bipartite entangled photons through a nested SHG and SPDC process. In this experiment, the GHZ state had a fidelity of $71.3\%$, which is well above the classical limit of $50\%$ \cite{GHZ_Lo2023}. 
Earlier, Zhang et al. \cite{Frequency_Reimer2016} demonstrated the generation of genuine four-photon time-bin entangled states using an integrated Kerr frequency comb. Pulsed pumping of a microring resonator produces entangled photon pairs via SFWM, and by post-selecting coincidences across different comb lines, the resulting state exhibits genuine multiphoton correlations, providing a scalable platform for complex quantum networks. 

\subsection{Entanglement characterization}
\begin{figure}[ht!]
\centering
\includegraphics[clip, trim= 1cm 1cm  6cm  0cm, width=0.8\columnwidth]{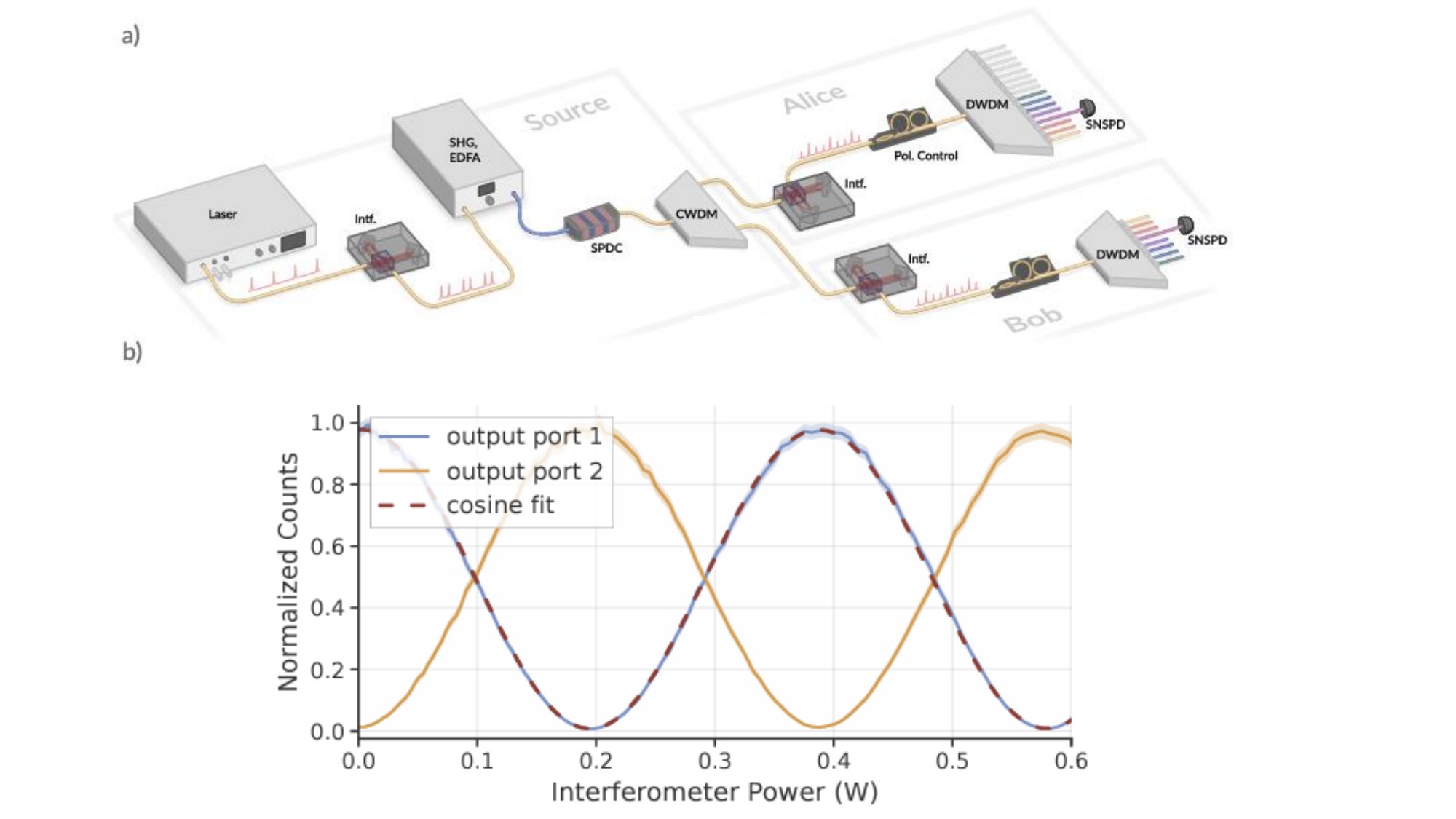}
\caption[Time-bin entanglement characterization.]{(a) An example of time-bin entanglement characterization. Signal and idler pairs are generated using SPDC. (b) The interference patterns observed at the center time bin are solely due to the rate of coincidental events. Given the strong correspondence between the observed fringe patterns and a cosine curve, it is assumed that the phase linearly correlates with the electrical power supplied to the interferometer's phase shifter. \textbf{Figure adapted from \cite{HighRateEnt_Mueller2024} \textcopyright~ Optical Society of America}. CWDM: Coarse wavelength division multiplexing, DWDM: Dense wavelength division multiplexing, EDFA: Erbium-doped fiber amplifier, Intf.: Interferometer, SHG: Second-harmonic generation, SPDC: Spontaneous parametric down-conversion.}
\label{fig_TBEChar}
\end{figure}

Entanglement characterization requires the TBQs to be sent through measurement interferometers (see Section \ref{Sec_MeasInt}) and to perform certain correlation measurements. The schematic of a two-qubit entanglement characterization setup is shown in Fig.~(\ref{fig_TBEChar}).

\subsubsection{Measurement considerations}
\label{Sec_CoinMeasCons}

Before discussing the characterization methods, we will first consider key aspects of coincidence measurements, including post-selection, the conditions required for high-fidelity entanglement generation and detection, and the impact of DLIs on coincidence rate scaling.

\begin{figure}[ht!]
    \centering
    \includegraphics[width=0.70\linewidth]{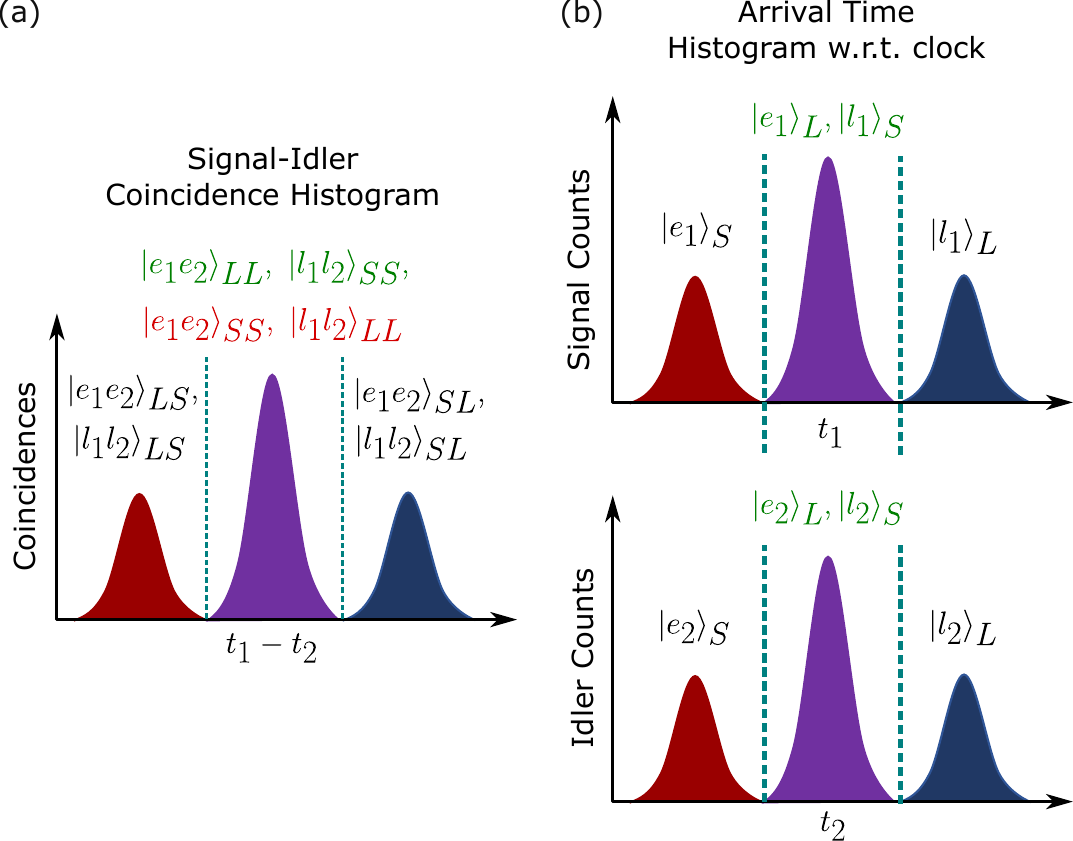}
    \caption[Coincidence measurement in entanglement characterization.]{Coincidence measurement in entanglement characterization. (a) Coincidence measurement between signal and idler photon pairs. (b) Arrival time of Signal and Idler photons recorded with respect to a reference clock synchronized with the qubit generation.}
    \label{fig_TBECoincMeas}  
\end{figure}

\textit{(a) Coincidence measurement and post-selection:} When a two-qubit time-bin entangled state $|\psi_{12}\rangle=(|e_1e_2\rangle+\exp{(i\phi_0)}|l_1l_2\rangle)/\sqrt{2}$ is passed through a pair of DLIs, the following transformation occurs: 
\begin{equation}
\begin{aligned}
   \ket{e_1e_2} &\xrightarrow{\text{DLIs}} \frac{1}{2}\left[\textcolor{red}{\ket{e_1e_2}_{SS}} + e^{(\phi_a+\phi_b)}\textcolor{green}{\ket{e_1e_2}_{LL}} + e^{(\phi_b)}\ket{e_1e_2}_{SL} + e^{(\phi_a)}\ket{e_1e_2}_{LS}\right], \\
   \ket{l_1l_2} &\xrightarrow{\text{DLIs}} \frac{1}{2}\left[ \textcolor{green}{\ket{l_1l_2}_{SS}} + e^{(\phi_a+\phi_b)}\textcolor{red}{\ket{l_1l_2}_{LL}} + e^{(\phi_b)} \ket{l_1l_2}_{SL} + e^{(\phi_a)} \ket{l_1l_2}_{LS}\right], 
   \end{aligned}
   \label{Eq_TBDLI}
\end{equation}
where subscript S(L) denotes the short (long) path of DLI taken by the photon in a given temporal mode, and $t_1$($t_2$) represents the arrival time of the signal (idler) photon. 

It is evident from Eq.~(\ref{Eq_TBDLI}) that a direct coincidence measurement between signal-idler photon results in three coincidence peaks, where the middle peak has contributions from two early and two late photons taking short or long paths, as shown in Fig.~(\ref{fig_TBECoincMeas}a). Among these, only events corresponding to both early photons taking a long path ($\ket{e_1e_2}_{LL}$) and late photons taking a short path ($\ket{l_1l_2}_{SS}$) are indistinguishable and give rise to interference. Therefore, when direct coincidence measurement is performed, the entanglement visibility, defined as
\begin{equation}
 \mathcal{V} \coloneqq \frac{C_{\text{max}} - C_{\text{min}}}{C_{\text{max}} + C_{\text{min}}},
 \label{Eq_EntVisibility}
\end{equation}
is limited to $50\%$. Here, $C_{\text{max}}$ and $C_{\text{min}}$
 are the maximum and minimum coincidence counts, respectively.
This problem can be solved by performing a triple coincidence measurement with a reference clock synchronized with the pump pulse used for qubit generation. Alternatively, first recording the arrival time of signal and idler photons with respect to the reference clock and then performing a coincidence measurement between middle bins as shown in Fig.~(\ref{fig_TBECoincMeas}b). The middle bin has contributions only from two early photons taking long paths and late photons taking the short paths, resulting in $100\%$ entanglement visibility in the ideal case.

\textit{(b) Conditions for generation and detection of high-fidelity entanglement:} The following conditions must be met: 
(i) The path imbalance of Alice's and Bob's DLIs must be equal to the time-bin separation and small compared to the coherence time of the pump laser ($L_A-S_A=L_B-S_B=c\tau_{el} < c\,t_p$). Here,
$ L_A\ (L_B) \quad \text{and} \quad S_A\ (S_B)$
denote the lengths of the long and short arms of Alice's (Bob's) DLI, respectively. Additional considerations on the role of pump coherence in the two-photon interference can be found in Ref.~\cite{TB-TPI_Liang2011}. 
(ii) The path imbalance of the DLIs must be large compared to the coherence length of the down-converted photons ($L_A-S_A=L_B-S_B > c\tau_{cd}$) to avoid single-photon interference. 
(iii) The width of measured coincidence peaks ($\Delta\tau_{mc}$) or the coincidence window ($\Delta T_{coin}$) must be smaller than the time-bin separation ($\Delta\tau_{mc}, \Delta T_{coin} < \tau_{el}$) to distinguish the neighbouring peaks and post-select the interference terms. 
(iv) The inverse of the repetition rate of the qubit generation ($f_{rep}$)  should be larger than twice the time-bin separation ($1/f_{rep} > 2 \tau_{el}$) to avoid overlap/interference of photons from successive qubit cycles. 
Lastly, the wavelength of the pump laser and the phase of the DLI must be stable over the measurement timescale (see Section ~\ref{Sec_ParameterSelection}).

\textit{(c) Coincidence rate scaling due to measurement DLIs:} Notably, the TBQ measurement through DLI involves signal splitting into different spatial and temporal modes, and thus a reduction in the detected coincidence count rate. As an example, consider the maximally entangled state $|\psi_{12}\rangle = 1/\sqrt{2}\left(|e_1e_2\rangle + \exp(i\phi_0) |l_1l_2\rangle \right)$. As evident from Eq.~(\ref{Eq_TBDLI}), the coincidence count rate between the two middle (superposition) bins of a pair of equivalent outputs varies as
\begin{equation}
P^\text{TB}_\text{Coin}=\frac{1}{16} \left[1 + \cos(\phi_a+\phi_b-\phi_0)\right], 
\end{equation}
where $\phi_a$ and $\phi_b$ represent the phase settings of two interferometers. When $\phi_a+\phi_b=\phi_0$, the coincidence probability becomes $1/8$. The same reasoning applies to the second equivalent pair, with both being complementary to the coincidence rate for inequivalent pairs. Consequently, only a quarter of all emitted photon pairs result in both photons being detected in the superposition basis, while another quarter is detected in the Z-basis \cite{Ent-Qdot_Weihs2017}. In the remaining half of the cases, one photon is detected in the Z-basis and its pair in the superposition basis.

\subsubsection{Bipartite entanglement characterization}
\label{Sec_BipartiteEntChar}

Entanglement inherent in quantum states can be quantified or witnessed by measuring the quantum state in different bases and evaluating a suitable function based on the measurement outcomes \cite{Ent-Rev_Horodecki2009}. We cover some of the standard experimental methods for entanglement verification \cite{EntVerification_Van2007}, including QST, entanglement fidelity, entanglement visibility, Bell-violation, and entanglement witness operator measurement as discussed below. 

\textit{(a) QST:} 
A most general n-qubit system has $2^{2n}-1$ independent parameters. Therefore, QST of a two-qubit entangled state is carried out by sending each qubit through a DLI and performing joint measurements in a complete set of 16 basis states \cite{QST_Takesue2009}. The density matrix $\rho^{\text{exp}}$ is then reconstructed using a maximum likelihood estimation procedure. All relevant properties of the state --- such as entanglement measures \cite{EntMeas_Plenio2007}, fidelity, purity, etc. --- can be computed from its density matrix. 

\textit{(b) Entanglement fidelity:} 
Fidelity of an experimentally generated state $\rho^\text{exp}$ determines the closeness of the state to a target state $\rho^T$, and it is defined as 
\begin{equation}
\mathcal{F}(\rho^T, \rho^\text{exp}) = \left(\text{Tr}\sqrt{\sqrt{\rho^T}\rho^\text{exp}\sqrt{\rho^T}}\right)^2, \label{Eq_Fidelity}
\end{equation}
where $\mathcal{F}$ takes values between zero and one. For an ideal case, $\rho^T=\rho^\text{exp}$ and $\mathcal{F}=1$. According to the fidelity definition (\ref{Eq_Fidelity}), knowledge of the density matrix is required, which requires full QST. In many cases, an experimenter aims to prepare one of the Bell states, and in such cases, fidelity is called Bell-state fidelity or entanglement fidelity.

Entanglement fidelity of the TBQ can be estimated by visibility measurement in two complementary bases (e.g., along X- and Y-basis) using DLIs. One of the qubits is projected along a given basis, say $|+\rangle$  or $|+i\rangle$, using the corresponding interferometer, and the phase of the other interferometer ($\phi$) is scanned in the range $[0, 2\pi]$. The coincidence counts at the output ports of the interferometers are recorded as a function of the phase. The plot of coincidence vs. phase gives rise to a sinusoidal pattern, and visibility is calculated. Then, a lower bound on the entangled fidelity of the generated state \cite{Ent_fink2017, Ent-Char_Ali2021} is given as the average visibility of the entanglement in two complementary bases.

\textit{(c) Entanglement witness:}
A hermitian operator $\hat{W}$ is said to be an entanglement witness operator \cite{EntWitness_Lewenstein2000, EntWitness_Brub2002, Ent-Rev_Horodecki2009} if it has a non-negative expectation value for separable states ($\text{Tr}[\hat{W}\rho_{s}] \geq 0~ \forall ~\rho_{s}$) and a negative expectation value for at least one entangled state ($\text{Tr}[\hat{W}\rho_{e}] < 0~\exists ~\rho_{e}$). Thus, observation of $\text{Tr}[\hat{W}\rho_{e}] < 0$ implies that state $\rho_e$ is entangled. Notably, a positive expectation value does not guarantee that the state is separable, as a given witness operator can detect only some entangled states. 

For any bipartite system, an entanglement witness operator can be decomposed as $\hat{W}=\sum_{ij}c_{ij}\hat{A}_i\otimes\hat{B}_j$, where $c_i$ are real numbers, and operators $\{\hat{A}_i\}$ and $\{\hat{B}_j\}$ are local positive operator valued measures (POVMs) on the Hilbert space of subsystems A and B, respectively. Thus, entanglement witnesses can be measured by local measurements only. This makes entanglement witness experimentally measurable observables, making them a practical and effective tool for detecting entanglement. Moreover, entanglement witnesses can cost-effectively detect entanglement as they require fewer measurements than a tomographically complete basis set.
Recently, entanglement witness and QST have been used to detect two-qubit time-bin entanglement \cite{TBE-Witness_Hwang2023}, demonstrating a resource advantage with entanglement witness.

\textit{(d) Bell-inequality violation:}
One of the strongest ways to characterize the entanglement of a state is by performing a Bell test and quantifying the associated degree of violation of a Bell inequality \cite{Ent-Review_Guhne2009, TBE-PSL_Francesco2018}. 
One commonly used inequality is the Clauser-Horne-Shimony-Holt (CHSH) \cite{CHSH1969} version of the Bell inequality. The Bell parameter, commonly denoted as the $S$-parameter, is defined as
\begin{equation}
    S=\left|E(\phi_a, \phi_b)-E\left(\phi_a, \phi_b^{\prime}\right)+E\left(\phi^{\prime}_a, \phi_b\right)+E\left(\phi^{\prime}_a, \phi^{\prime}_b\right)\right|,
    \label{Eq_CHSH}
\end{equation}
where classically $S \leq 2$. The correlation coefficient $E(\phi_a,\phi_b)$ \cite{BellAngles_Garuccio1981} is given as 
\begin{equation} 
E(\phi_a,\phi_b) = \frac{C(\phi_a,\phi_b)-C(\phi_a,\phi_b+\pi) - C(\phi_a+\pi,\phi_b) + C(\phi_a+\pi,\phi_b+\pi)} {C(\phi_a,\phi_b) +C(\phi_a+\pi,\phi_b) + C(\phi_a,\phi_b+\pi) + C(\phi_a+\pi,\phi_b+\pi)},
 \label{Eq_CorrCoeff}
\end{equation}
where $C(\phi_a,\phi_b)$ is the measured coincidence when qubit A(B) is projected along $\phi_a$ ($\phi_b$). Note that the angles $\phi_a$ and $\phi_b$ corresponding to optimal Bell-violation are dependent on the entangled state. 
  The $S$-parameter is calculated, and if it exceeds 2 by a statistically significant margin, a violation of the Bell inequality is confirmed. While such a violation implies the presence of entanglement, the degree of violation does not generally quantify the amount of entanglement \cite{Bell-Ineq_Munro2001, Bell-Ineq_Acin2004}, except in specific cases such as pure two-qubit states \cite{EM_Singh2020}.

The two TBQs are sent to two separate DLIs as discussed in Sec.~(\ref{Sec_MeasInt}). 
The coincidence detection rate of the middle bins at the output of each interferometer, referred to as A and B, is measured for varying interferometer phases. This coincidence rate is given by 
\begin{equation}
C(\phi_a,\phi_b)=C_0\left[1+\mathcal{V} \cos(\phi_a+\phi_b)\right],
\label{Eq_CoinVis}
\end{equation}
where $C_0$ is the average coincidence rate, $\mathcal{V}$ is the measured interference visibility (entanglement visibility) and $\phi_a$ and $\phi_b$ are the relative phases of the corresponding interferometers \cite{QCom_Houwelingen2008}.
Here, the visibility is found by fitting coincidence from multiple measurements of $\phi_a$ and $\phi_b$ using Eq.~(\ref{Eq_CoinVis}). 

An ideal Bell test is performed under a strict set of conditions (i.e., ideal Bell tests are loophole-free), which are not always met in experimental tests \cite{Bell-RMP_Brunner2014, Bell-Loophole_Wiseman2022, Bell-Loophole_Kaiser2022}. Notably, even in the ideal implementation of Franson-type Bell-violation using time-bin entanglement, Alice and Bob post-select indistinguishable events in the middle bin as shown in Fig.~(\ref{fig_TBECoincMeas}b), giving rise to a post-selection loophole \cite{TE-LocalRealism_Aerts1999}. 
This loophole can be closed by modifying the measurement interferometers either by replacing the first beam splitter with a fast optical switch that enables early (late) time bins to take long (short) paths deterministically \cite{TBE-PSL_Francesco2018} or by using an interferometer in a hug-configuration \cite{TB-Ent-PSFree_Bacchi2025}. Nevertheless, it is the consensus in the community that performing a Bell test without these conditions met is still sufficient for entanglement characterization.


\textit{(e) Entanglement visibility:}  
Since the maximum visibility is limited to one, Eqs.~(\ref{Eq_CorrCoeff}) and (\ref{Eq_CHSH}) yield
\begin{equation}
    S=2\sqrt{2} \mathcal{V} \leq 2.
\end{equation}
For $\mathcal{V} > \frac{1}{\sqrt{2}}$ the CHSH inequality is violated.  Therefore, entanglement visibility also quantifies the magnitude of entanglement in a state. The measured visibility of time-bin entanglement gets impacted by multiphoton pair emissions, DLI path delay mismatch, and group velocity dispersion as discussed in Ref.~\cite{TBEntVisMod_Zhong2024}.  

\subsubsection{Multipartite entanglement characterization}

General methods to quantify entanglement for arbitrary multiqubit states are still an active area of research. 
 Some methods have been proposed to characterize multipartite entangled states \cite{Ent-MP_McCutcheon2016, Ent-Cert_Friis2018}. Losses associated with generating and detecting photonic TBQs render performing these measurements challenging due to the low coincidence detection rates. 
 Furthermore, one needs to ensure a high degree of interferometer stability over the measurement timescale.
 
\textit{(a) Mermin Inequality:}  
One way to assess multipartite entanglement is to use Bell inequalities that have been generalized to multiple qubits.
Some of these include the set of Mermin \cite{Ent_Mermin1990, Ent_Alsina2016, Ent-qutrit_Alsina2016} and Svetlichny \cite{SvetlichnyInequality_1987, Ent-Sep_Collins2002, SvetlichnyInequality_Lavoie2009}  inequalities. The Mermin inequality detects any multipartite nonlocality under local realism, while the Svetlichny inequality tests for genuine multipartite nonlocality by excluding hybrid local-nonlocal explanations \cite{BellNonlocality_Scarani2019}.
 GHZ-class states produce the largest violation of both the Mermin and Svetlichny inequalities \cite{MerminInequality_Chen2004}. Notably, the W-state does not violate the Svetlichny inequality and only weakly violates Mermin's inequality \cite{Bell-RMP_Brunner2014}.
 

For instance, for three TBQs, suppose each qubit is measured in either the $X$ or $Y$ basis. 
The three-qubit Mermin inequality can be expressed as
\begin{equation}
 M_3  =| \langle X_1Y_2Y_3 \rangle+ \langle Y_1 X_2Y_3 \rangle+\langle Y_1Y_1X_3 \rangle-\langle X_1X_2X_3 \rangle| \leq 2. 
\end{equation}
where the subscripts $1,2,3$ refer to the corresponding qubits. 
The violation of this inequality, $\langle M_3 \rangle \geq 2$, witnesses the entanglement. 
Note that quantum mechanics predicts $\langle M_3 \rangle \leq 4$. 
There are mappings between different multipartite Bell inequalities, for instance, between Mermin and Svetlichny \cite{Ent-Tripartite_Ghose2009}, and other inequalities have been proposed for qudits, such as those in References \cite{Ent-HDBell_Polozova2016, CHSH-3State_Kaszlikowski2002, Ent-CustomBell_Barizien2023}. 

\textit{(b) Entanglement Visibility:} 
Entanglement in a $N$-qubit system can be detected when entanglement visibility $\mathcal{V}>\frac{1}{\sqrt{N}}$ \cite{MP-Ent_Pezze2016}.  Notice that, with increasing qubits $N$, the required minimum visibility to detect entanglement decreases. Genuine multipartite entanglement is detected when $\mathcal{V}>\sqrt{(1-\frac{1}{N})^2+\frac{1}{N^2}}$ for N-particle \cite{MP-Ent_Pezze2016}. Thus, the minimum visibility required for genuine $N$-partite entanglement detection is larger than the entanglement detection itself and increases with $N$. 

\textit{(c) Entanglement Fidelity:} Fidelity between a target state $\rho^T$ and the experimentally generated state $\rho$
is a similarity metric defined as: $F(\rho,\rho^T)=Tr(\rho,\rho^T)$. 
When the target state is a GHZ state, a fidelity value greater than 0.5 implies genuine multipartite entanglement \cite{Ent_sackett, Parity_Leibfried2005}. 
The GHZ fidelity $F_{GHZ}$ is obtained by measuring two observables, namely, Population $P$ and Coherence $C$, as given below\cite{GHZ_Mooney2021}:
\begin{equation}
    F_{GHZ} := \frac{P+C}{2},
\end{equation}
where $P=\rho_{00...0, 00...0}+\rho_{11...1, 11...1}$ is directly measured by projecting each qubit in Z-basis, 
and $C=|\rho_{00...0, 11...1}|+|\rho_{11...1, 00...0}|$ is obtained from parity oscillations \cite{Parity_Leibfried2005, Parity_Thomas2011} or multiple quantum coherence \cite{MQC_Ken2020}. 
In the case of TBQs, coherence is measured by projecting each qubit through DLI and calculating the visibility of the interference peak in coincidence counts.


\textit{(d) Entanglement Witness:} 
Instead of fully characterizing the multi-partite state, it is often sufficient to establish if entanglement is present. To that end, an entanglement witness $\hat{W}$ is an operator that remains non-negative for all separable (or bi-separable) states. Hence, if $\text{Tr}[\hat{W}\rho] < 0$, it indicates the presence of entanglement, specifically genuine multipartite entanglement \cite{GHZ-Witness_Acin2001, EntWitness_Guhne2003, EntWitness_Bourennane2004, EntWitness_Sperling2013}. Entanglement witness offers significant advantages over QST for verifying entanglement in multipartite and high-dimensional quantum systems. While the experimental cost of quantum tomography scales exponentially with system size, entanglement witnesses can detect genuine multipartite entanglement using only a few measurements. 


\section{Interference of time-bins from independent sources}
\label{Sec_Interference}
\FloatBarrier

We examine the HOM interference effect and BSM techniques, which exploit the interference of qubits \cite{Interf_Rainer2006}, typically produced by independent sources. These techniques are fundamental to a wide range of modern quantum information protocols, including quantum teleportation, entanglement swapping, quantum repeaters, and the development of the quantum internet \cite{Quinternet_Yu2015, Quinternet_Pirandola2016, Quinternet_Dur2017, QNwtwork_Simon2017, Quinternet_Cacciapuoti2020_1, Quinternet_Cacciapuoti2020_2, Quinternet_Marcello2020, Quinternet_Huberman2020, Quinternet_Singh2021}.

\subsection{Hong-Ou-Mandel (HOM) Effect}
\label{Sec_HOM}

\begin{figure} [ht!]
    \centering
    \includegraphics[width=\linewidth]{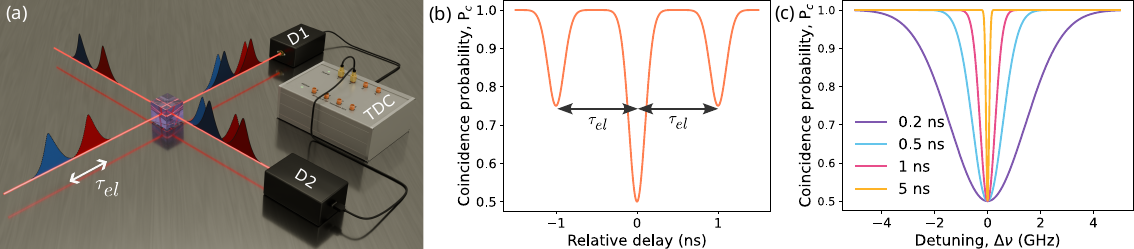}
    \caption[Schematic of the HOM-dip measurement Setup.]{(a) Schematic of HOM-dip measurement setup \cite{TB-Robust_Simon2025}. (b) HOM dip for timing scan of weak-coherent TBQ in a superposition basis, where the middle dip corresponds to the perfect overlap of both early and late modes, while the side dips correspond to the overlap of early-late and late-early temporal modes. (c) HOM  dip for frequency scan of weak-coherent TBQ with different Gaussian pulse widths. Note that maximum HOM-visibility for weak-coherent TBQ is limited to 50\%, resulting in a minimum coincidence probability of 0.5. However, for a single-photon source, the maximum visibility reaches 100\%, corresponding to a minimum coincidence probability of zero. D1, D2: Single-photon detectors, TDC: Time-to-digital converter.} 
    \label{Fig_HOMSetup} 
\end{figure}

The HOM effect \cite{HOM_Mandel1987, HOM_review_Bouchard} is a second-order interference effect which is observed when two independent photons arrive simultaneously on a $50:50$ beam splitter as shown in figure \ref{Fig_HOMSetup} (a). If the photons are identical in polarization, timing, frequency, and spatial mode (except for directions into the beam splitter), they bunch together at either output port. Thus, it is also referred to as two-photon interference. In practice, the HOM effect is observed by varying one degree of freedom (usually arrival time or frequency) to render the photons distinguishable and monitoring the change in coincidence detection events at the two detectors at the beam splitter outputs, as illustrated in Fig.~(\ref{Fig_HOMSetup}b). The relative reduction in the coincidence detection rates is known as the HOM visibility, calculated as $100\times(C_{max} - C_{min})/C_{max}$, and is often used to quantify the distinguishability of incoming photons. 

The coincidence probability for input TBQ states $\ket{\psi_{in}}$ can be estimated with respect to the electric field operators at the two output ports ($\hat{E}_{o_1}$ and $\hat{E}_{o_2}$) of BS as:
\begin{equation}
    P^{1,1}_{D_1,D_2} = \bra{\psi_{in}}\hat{E}^{\dagger}_{o_1} \hat{E}^{\dagger}_{o_2} \hat{E}_{o_2} \hat{E}_{o_1}\ket{\psi_{in}},
\end{equation}
where $\hat{E}_{o1} = (\hat{E}_{i_2}-i \hat{E}_{i_1} )/\sqrt{2}$ and $\hat{E}_{o_2} = (\hat{E}_{i_1} -i \hat{E}_{i_2})/\sqrt{2}$, for the input field operators $\hat{E}_{i_1}$ and $\hat{E}_{i_2}$. For TBQ, the electric field operator can be written as a superposition of early ($\hat{a}_e$) and late ($\hat{a}_l$) temporal modes as:

\begin{equation}
    \begin{aligned}
        \hat{E}_k = \xi_{k}(\omega,t) \hat{a}_{e,k}+ \xi_{k}(\omega,t-\tau_{el}) \hat{a}_{l,k},
    \end{aligned}
\end{equation}
 where
\begin{equation}
    \xi_{k}(\omega,t) = \frac{1}{\Delta\tau\sqrt{2\pi}} e^{-\frac{t^2}{2\Delta\tau^2}} e^{-i (\omega_k t  + \phi_k)},
\end{equation}
for input port index $k = i_1, i_2$, with $\Delta\tau$ representing the time-bin width, $\omega_k$ the central frequency, and $\phi_k$ the phase of each input photon. The above expression is valid for single photons; however, it can be modified for other photon sources by taking into account the associated photon statistics.

Finally, the above formalism can be used to compute the coincidence probability $P_c$ of two inputs modes in superposition state $\ket{\psi_{in}} = (\ket{e} + e^{i\phi}\ket{l})/\sqrt{2}$ with relative temporal delay ($t$) and spectral detuning ($\omega_{i_1} - \omega_{i_2}$) as follows: 

\begin{equation}
    P_{c} = 1 - V_{\text{max}} \left( 
    \frac{1}{2}e^{-\frac{(t - \tau_{\text{el}})^2}{2\Delta\tau^2}} +
    e^{-\frac{t^2}{2\Delta\tau^2}} +
    \frac{1}{2}e^{-\frac{(t + \tau_{\text{el}})^2}{2\Delta\tau^2}} 
\right) e^{-\Delta \tau^2 (\omega_{i_1} - \omega_{i_2})^2},
\label{equ:hom_conci}
\end{equation}

where $V_{\text{max}}$ denotes the maximum HOM visibility in the perfectly indistinguishable case. The $V_{\text{max}}$ depends on the photon statistics; for instance, it is unity for true single photons, resulting in $100\%$ visibility. 
However, the maximum HOM interference visibility with either coherent or thermal light is limited to $0.5$ or $0.33$, respectively, because of the multi-photon events associated with the photon statistics \cite{HOM_review_Bouchard, HOM_Khodadad2023}. 
The three exponential terms in the parentheses of \,\eqref{equ:hom_conci} represent the interference contributions arising from the overlap of different combinations of early and late temporal modes, also represented in Fig.~(\ref{Fig_HOMSetup}b). 
The exponential term in frequency mismatch highlights the sensitivity of TBQ interference to spectral distinguishability. 
For instance, a narrower temporal shape has a wider frequency bandwidth according to the time-bandwidth product as given in Eq.~(\ref{Eq_TBP}), thus resulting in a wider spectral HOM dip as shown in Fig.~(\ref{Fig_HOM-BSM}c). 

The HOM interference also depends on the qubit state of the incoming time bins. For instance, $\ket{e}$ and $\ket{l}$ qubits arriving on the beam splitter would not result in HOM dip as they are temporally separated; hence, the two wave-packets don't interfere. The width of the HOM-dip is given by $\sqrt{2}$ times the width of the individual input pulses, reflecting the convolution of the two photon wavepackets in the interference process. Importantly, the measured HOM-dip width is independent of the detector jitter, as the dip is obtained through cross-correlation of the detection events rather than relying on single-event timing alone. This makes the HOM measurement a robust indicator of temporal indistinguishability between photons, even in setups with detectors exhibiting finite recovery (dead) times.


\subsection{Bell-State Measurement (BSM)} 
\label{Sec_BSM}

Bell-states \cite{Book-QIP_chuang, Book-QIP_wilde} are a set of four maximally entangled two-qubit states, given by $\ket{\phi^{\pm}} = (\ket{ee} \pm \ket{ll})/\sqrt{2}$ and $\ket{\psi^{\pm}} = (\ket{el} \pm \ket{le})/\sqrt{2}$.
They also form an orthonormal basis set on the Hilbert space of two qubits.
A BSM (also known as a Bell measurement or a Bell-basis measurement) \cite{BSM_Markus1996, BSM_Kwiat1998, BSM_Lutkenhaus1999, BSM-Hyper_Walborn2003, BSM-Hyper_Schuck2006} is a joint projection of two qubits onto one of the four Bell states; in other words, a measurement projects the state of the two qubits onto one of the Bell states. 
BSMs are commonly employed in gate-based quantum computing, in which the BSM is performed by a combination of CNOT and Hadamard gates followed by measurement in the Z-basis \cite{Book-QIP_chuang}. 
BSM is also essential for quantum communication protocols such as teleportation (see Sec. \ref{Sec_Teleport}), quantum repeaters, entanglement swapping (see Sec. \ref{Sec_EntSwap}), and MDI-QKD (see Sec. \ref{Sec_MDI-QKD}). Achieving a complete BSM of photonic qubits \cite{BSM-Full_Elena2024} requires mapping onto matter qubits to perform the required one and two-qubit gates, use of a number of ancilla qubits \cite{BSM-Ancilla_Bayerbach2023, Boosted-BSM_Hauser2025}, or hyperentanglement \cite{BSM-Hyper_Schuck2006, BSM-HyperEnt_Li2016, HyperEnt_Deng2017}.

\begin{figure}[ht!]
    \centering
    \includegraphics[width= 0.8\columnwidth]{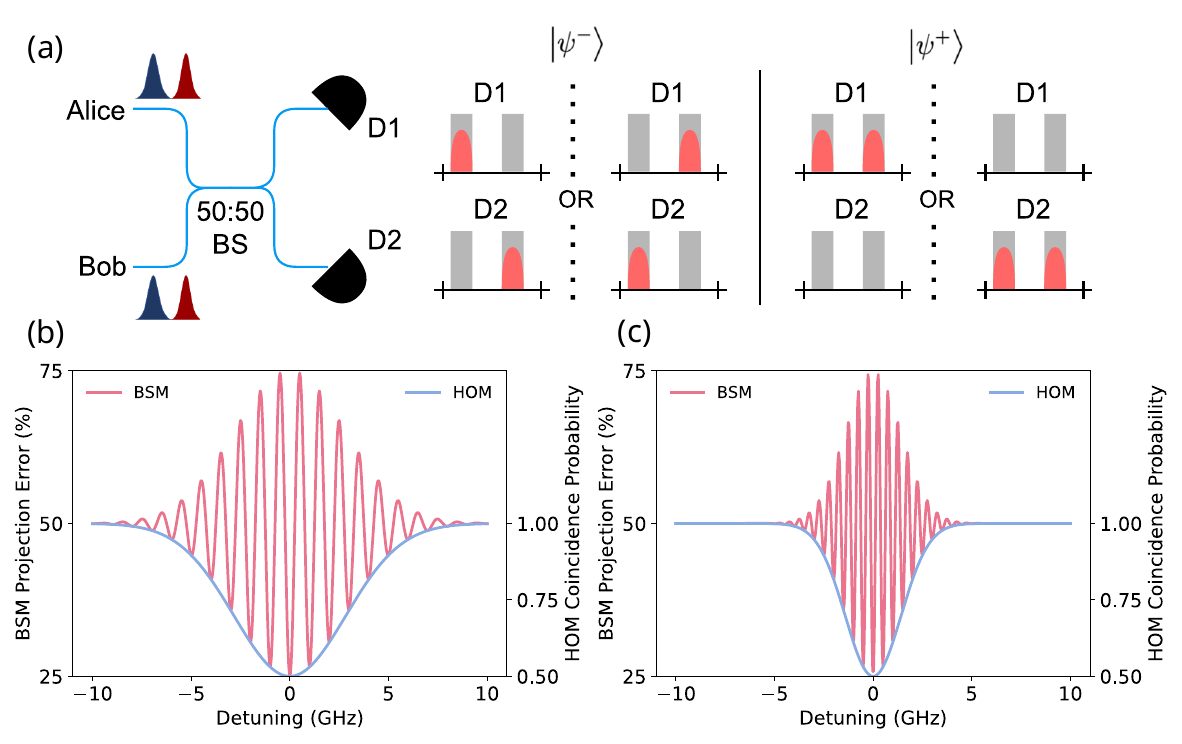}
    \caption[A partial BSM scheme for TBQs.]{(a) The partial BSM scheme for TBQ where two independent qubits interfere on a 50:50 BS and are detected by single photon detectors with the detection pattern for $\ket{\psi^-}$ and $\ket{\psi^+}$ projection. (b) and (c) illustrate the BSM projection error in the X-basis resulting from frequency mismatch between incoming weak-coherent TBQs, with a 1 ns time-bin separation and FWHM widths of 100 ps and 200 ps, respectively.}
    \label{Fig_HOM-BSM}
\end{figure}

Quantum communication protocols are often limited to a partial BSM due to the inability to distinguish the output modes using linear optics alone \cite{BSM_Lutkenhaus1999, BSM_Calsamiglia2001}. 
A partial BSM can be easily performed by interfering two photonic qubits on a $50:50$ beam splitter similar to HOM interference, as illustrated in Fig.~(\ref{Fig_HOM-BSM}a). 
Photon indistinguishability in degrees of freedom other than time is crucial for BSM, and thus, HOM typically serves as a precursor to BSM.
 
To understand the BSM, we first examine the representation of TBQ states in the Bell basis, as summarized in Table~(\ref{tab_BSM}). This table lists the transformation of various two-qubit combinations, originally defined in the Z and X bases, into linear combinations of the four Bell states.

\begin{table}[ht!]
    \centering
    \renewcommand{\arraystretch}{1.5}
  \begin{tabular}{|>{\centering\arraybackslash}p{0.08\columnwidth}|>{\centering\arraybackslash}p{0.10\columnwidth}|>{\centering\arraybackslash}p{0.10\columnwidth}|>{\centering\arraybackslash}p{0.4\columnwidth}|}
        \hline
   \textbf{S.N.}  &  \textbf{Qubit 1} & \textbf{Qubit 2} & \textbf{Bell basis representation} \\ 
        \hline
      1 &  $\ket{e}$ & $\ket{e}$ & $( \ket{\phi^+} + \ket{\phi^-})/\sqrt{2}$ \\ 
         \hline
      2 &   $\ket{e}$ & $\ket{l}$ & $( \ket{\psi^+} + \ket{\psi^-})/\sqrt{2}$ \\  
         \hline
     3 &   $\ket{e}$ & $\ket{+}$ & $( \ket{\phi^+} + \ket{\phi^-} + \ket{\psi^+} + \ket{\psi^-})/2$ \\
         \hline
    4 &    $\ket{e}$ & $\ket{-}$ & $( \ket{\phi^+} + \ket{\phi^-} - \ket{\psi^+} - \ket{\psi^-})/2$ \\
         \hline
     5 &   $\ket{l}$ & $\ket{e}$ & $( \ket{\psi^+} - \ket{\psi^-})/\sqrt{2}$ \\
         \hline
     6 &   $\ket{1}$ & $\ket{l}$ & $( \ket{\phi^+} - \ket{\phi^-})/\sqrt{2}$ \\
         \hline
      7 &   $\ket{1}$ & $\ket{+}$ & $( \ket{\phi^+} - \ket{\phi^-} + \ket{\psi^+} - \ket{\psi^-})/2$ \\
         \hline
     8 &   $\ket{1}$ & $\ket{-}$ & $( -\ket{\phi^+} + \ket{\phi^-} + \ket{\psi^+} - \ket{\psi^-})/2$ \\
         \hline
      9 &  $\ket{+}$ & $\ket{e}$ & $( \ket{\phi^+} + \ket{\phi^-} + \ket{\psi^+} - \ket{\psi^-})/2$ \\
         \hline
     10 &   $\ket{+}$ & $\ket{l}$ & $( \ket{\phi^+} - \ket{\phi^-} + \ket{\psi^+} + \ket{\psi^-})/2$ \\
         \hline
      11 &  $\ket{+}$ & $\ket{+}$ & $( \ket{\phi^+} + \ket{\psi^+})/\sqrt{2}$ \\
         \hline
     12 &   $\ket{+}$ & $\ket{-}$ & $( \ket{\phi^-} - \ket{\psi^-})/\sqrt{2}$ \\
         \hline
      13 &   $\ket{-}$ & $\ket{e}$ & $( \ket{\phi^+} + \ket{\phi^-} - \ket{\psi^+} + \ket{\psi^-})/2$ \\
         \hline
      14 &  $\ket{-}$ & $\ket{l}$ & $( -\ket{\phi^+} + \ket{\phi^-} + \ket{\psi^+} + \ket{\psi^-})/2$ \\
         \hline
     15 &   $\ket{-}$ & $\ket{+}$ & $( \ket{\phi^-} + \ket{\psi^-})/\sqrt{2}$ \\
         \hline
     16 &   $\ket{-}$ & $\ket{-}$ & $( \ket{\phi^+} - \ket{\psi^+})/\sqrt{2}$ \\
         \hline
    \end{tabular}
    \caption{Two-qubit transformation from Z/X basis to Bell basis.}
    \label{tab_BSM}
\end{table}

Furthermore, the projection of two-qubit states onto specific Bell states can be understood through the action of a beam splitter, as described in Eq.~(\ref{Eq_BSM}). 
\begin{equation}
    \begin{aligned}
    \label{Eq_BSM}
        \ket{\phi^+} & \xrightarrow{\text{BS}} \frac{i}{2} \left(\ket{ee}_{D1,D1} + \ket{ee}_{D2,D2} + \ket{ll}_{D1,D1} + \ket{ll}_{D2,D2}\right), \\
        \ket{\phi^-} & \xrightarrow{BS} \frac{i}{2} \left(\ket{ee}_{D1,D1} + \ket{ee}_{D2,D2} - \ket{ll}_{D1,D1} - \ket{ll}_{D2,D2}\right), \\
        \ket{\psi^+} &  \xrightarrow{BS} \frac{i}{2} \left(\ket{el}_{D1,D1} + \ket{le}_{D2,D2}\right), \\
        \ket{\psi^-} &  \xrightarrow{BS} \frac{i}{2} \left(\ket{el}_{D1,D2} - \ket{le}_{D1,D2}\right).
    \end{aligned}
\end{equation}

Under the beam splitter transformation, the Bell states are mapped onto distinct two-photon detection signatures at the two output ports (denoted as $D_1$ and $D_2$) of the beam splitter, as shown in Fig.(\ref{Fig_HOM-BSM}a). Thus, by analyzing the detection patterns at the output ports, the Bell state of the input two-qubits can be effectively identified, enabling the successful implementation of the BSM.

The partial BSM measurement scheme can unambiguously project the qubits onto $\ket{\psi^{-}}$ and $\ket{\psi^{+}}$ only based on the detector pattern from Eq.~(\ref{Eq_BSM}) and represented in Fig.~(\ref{Fig_HOM-BSM}a). 
However, projections onto $\ket{\phi^+}$ and $\ket{\phi^-}$ cannot be distinguished due to similar detection patterns as given in Eq.~(\ref{Eq_BSM}). 

Since HOM interference underpins BSM, photon indistinguishability is crucial for error-free projections. 
Additionally, the Bell-state projection error for TBQs encoded into weak coherent states is lower bounded at $25\%$ due to multi-photon events governed by Poissonian statistics and is captured as follows:

\begin{equation}
    e_x = \frac{1}{2} - \frac{1}{4} e^{-2\pi \Delta \tau^2(\omega_{i_1} - \omega_{i_2})^2} \cos[2 \pi \tau_{el} (\omega_{i_1} - \omega_{i_2})],
\end{equation}

where, $\Delta \tau$ is the time-bin width, $\omega_{i_1} - \omega_{i_2}$ is the frequency detuning between two input TBQ and $\tau_{el}$ is the time-bin separation. The projection error in the BSM for weak-coherent TBQ is further illustrated in Fig.~\ref{Fig_HOM-BSM}(b) and (c), corresponding to different time-bin widths. 
To minimize the error, the time-bin separation should be kept as small as possible, and the frequency difference between the two sources must be minimized as discussed in Section~(\ref{Sec_ParameterSelection}). 
This constrains the time-bin separation as discussed previously in Sec~(\ref{Sec_TBSep}).

The detection of $\ket{\psi^+}$ from TBQs is also possible if the SPD has a response time faster than the time-bin separation.
Alternatively, a fast optical switch can be used to separate early and late (as shown later in Fig. \ref{fig_switch}) after the interference of qubits at the 50:50 beam splitter \cite{TB-PathConverter_Ren2025}. 
However, two additional SPDs are required for this approach, where $\ket{\psi^+}$ corresponds to detector clicks from the same switch while $\ket{\psi^-}$ is from a different switch. 
The additional loss in the measurement due to the optical switches is a limiting factor in this scheme; fortunately, integrated photonics platforms can overcome this problem to some extent. Furthermore, optical nonlinearity can be employed to efficiently execute full BSM using integrated photonic components, as recently demonstrated via mirroring resonators \cite{BSM-NLO_Kwiat2025}.
 

\section{Time-bin qudits} 
\label{TBQudits}

Although the conventional unit of quantum information is the qubit, i.e. information is encoded in a two-level quantum system, more levels can also be used.
To use a classical analogy, nature utilizes a four-letter alphabet for DNA \cite{DNA_Szathmary2003}, arguably the most vital information storage system \cite{HD_Erhard2020}. 
A natural extension of a qubit to higher dimensions is the $d$-dimensional ($d>2$) system called a qudit. 
The most general state of a pure $d$-dimensional qudit can be written as 
\begin{equation}
| \psi_d\rangle = \sum_{j=0}^{d-1} \alpha_j e^{i\phi_j}|t_j\rangle,
\end{equation}
where $\alpha_j$ is the probability amplitude of mode $|t_j\rangle$ satisfying $\sum_{j=0}^{d-1}|\alpha_j|^2=1$, and $\phi_j$ is the phase of $j$th mode. 
Here, each mode corresponds to a distinct time-bin that is separated from the nearest time-bin by $\tau_b$. An example of a time-bin ququart ($d=4$) state is shown in Fig.~(\ref{quditint}).

Higher-dimensional quantum systems \citep{HD_Erhard2020, HD-QC_Daniele2019} provide advantages in several quantum information applications. 
For example, a qudit can encode $\log_2(d)$ bits of classical information, implying each qudit has a larger information capacity as compared to a qubit \cite{HD-TB_Thomas2016, HD_Wang2020}. 
Moreover, high-dimensional states possess enhanced robustness to quantum cloning and offer increased resilience to noise \cite{HD-QKD-Sec_Cerf2002} and eavesdropping attempts \cite{HD-QKD-ErorTol_Nikolopoulos2006}, permitting a larger error tolerance in QKD systems than that afforded by qubits \cite{HD-QKD_Sheridan2010, HD-QKD_Bechmann2000, HD-QKD_Vagniluca2020}. 
The enhanced noise resilience of qudits also offers benefits when they are entangled. 
It has been shown that increasing the dimension of each quantum system without changing the number of entangled particles produces more robust entanglement, whereas increasing the number of particles for a fixed dimension renders entanglement fragile \cite{Ent-multiqudit_Liu2009}. 
Utilizing qudits in quantum information tasks has advantages in terms of security, computational power, and efficiency, overcomes some challenges in qubit-based systems, and has a role in studying the foundations of quantum mechanics \cite{HD-QKD-ErorTol_Nikolopoulos2006, TE_Ecker2019, HD-QC_Daniele2019}. 
In this section, we discuss methods for preparing and measuring time-bin qudits.

\subsection{Time-bin qudit preparation}
\label{Qditprep}

A coherent superposition of a single photon in more than two ($d \geq 3$) orthogonal time-bin basis states is required to form a qudit. 
The methods to generate time-bin qudits naturally extend from those for TBQs (as discussed in Sec. \ref{Sec_Preparation}). For example, an attenuated laser pulse traversing a multi-path interferometer \cite{RobustDLI_Islam2017, Duality_Englert2008} constitutes a weak-coherent time-bin qudit at the output. Alternatively, photons heralded from an SPDC or SFWM source or a single-emitter source can be input to the multi-path interferometer. 
Notably, an on-demand source of arbitrary time-bin encoded qubits, qutrits, and ququads has been demonstrated \cite{TB-AtomCavity_Jones2013} using a strongly coupled atom-cavity system \cite{SPS-Cavity_Kuhn2010, SPS-Cavity_Jones2011}.
Similarly, carving a train of pulses from a strongly attenuated CW laser beam using an amplitude modulator provides the qudit equivalent to weak-coherent TBQ. 
Finally, a qudit can be generated by pumping a two-level system, e.g. a quantum dot, with $d$ phase-coherent pulses, akin to the method described in Section~(\ref{Sec_QDotTB}) for qubits. 

\begin{figure}[ht!]
    \centering
    \includegraphics[clip, trim= 0.6cm  2cm  0.4cm  3cm, width=0.60\columnwidth]{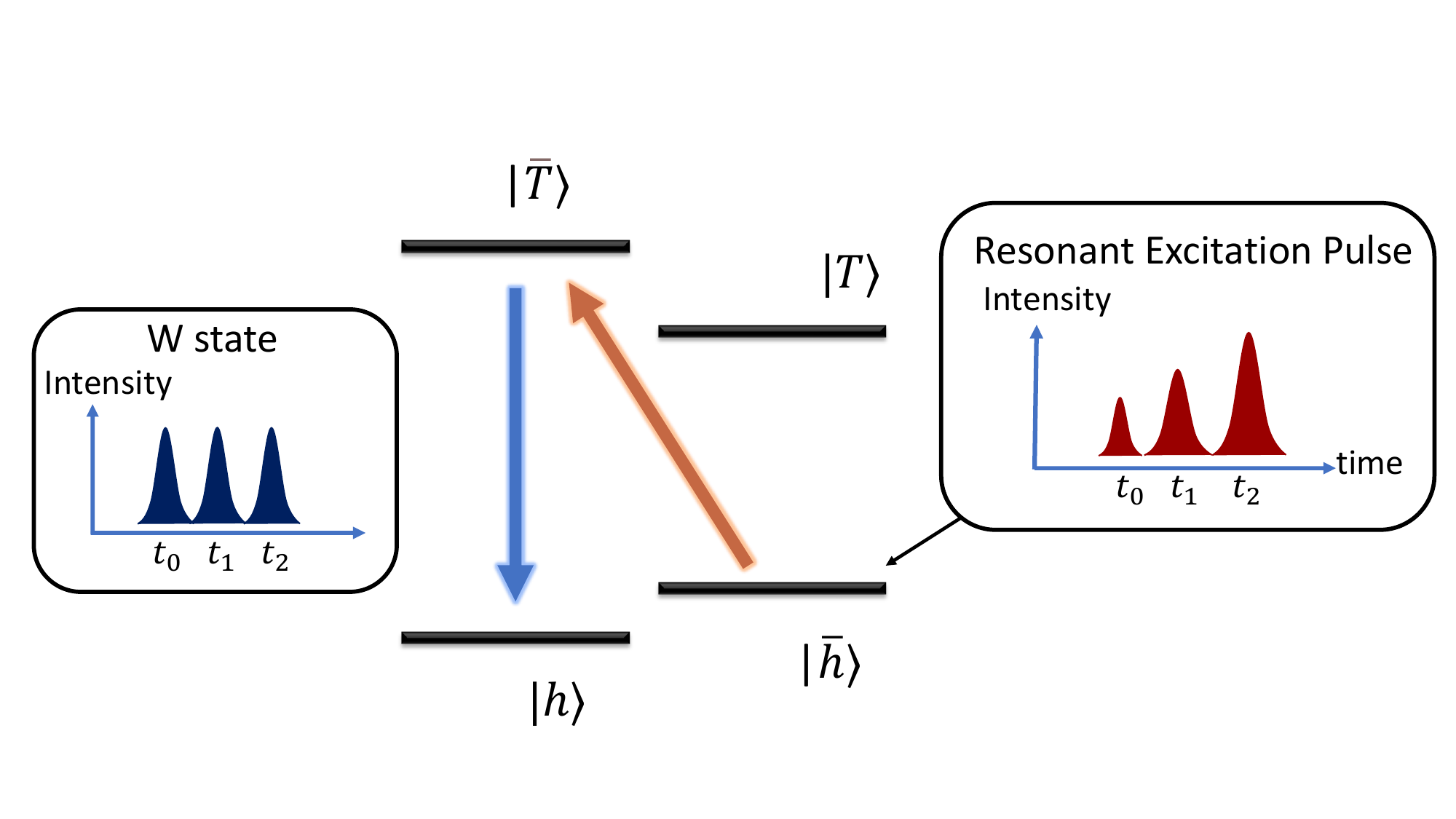}
    \caption[Time-bin qutrit generation using a quantum dot.]{Time-bin qutrit or single photon W-state generation using a quantum dot. A sequence of three pulses is used to excite a quantum dot. The intensity of these pulses is chosen so that the probability of decay in each time-bin is equal \cite{W-State_Lee}.}
    \label{fig_Wstate}
\end{figure}

In Ref.~\cite{W-State_Lee}, a time-bin qutrit (\( d = 3 \))  has been generated using quantum dots. 
As shown in Fig. (\ref{fig_Wstate}), a series of three weak resonant pulses probabilistically excite the quantum dot, driving the \(\ket{\bar{h}} \rightarrow \ket{\bar{T}}\) transition. 
Consequently, the system can decay \(\ket{\bar{T}} \rightarrow \ket{h}\) into each of these three time-bins with equal probability, resulting in a time-bin qutrit.
Recently, a probabilistic scheme for generating arbitrarily high-dimensional time-bin states based on quantum walk dynamics has been demonstrated~\cite{TB-Robust_Simon2025}. In this approach, the quantum walk is implemented in the time domain, while state control is achieved through the polarization degree of freedom. This method enables the generation of arbitrary time-bin qudit states, including both pure and mixed states, by optimizing the set of polarization operations.

\subsection{Time-bin qudit measurement}
\label{Qditmeas}

A $d$-dimensional time-bin qudit is measured in the \( Z \)-basis by measuring the arrival time of a photon in a specific temporal bin \( \{|t_n\rangle\in\{|t_0\rangle,|t_1\rangle,...,|t_{d-1}\rangle\}\) (shown for ququarts in Fig.~\ref{fig_time-energy2}c) Additionally, time-bin qudits can be prepared and measured in mutually unbiased basis states:
\begin{equation}
    |\psi_n\rangle= \frac{1}{\sqrt{d}}\sum_{m=0}^{d-1}
    e^{\left(\frac{i 2\pi m n}{d}\right)}\,
    |t_m\rangle, \quad\quad n=0,...,d-1.
\end{equation}

\begin{figure}[ht!]
  \centering
  \includegraphics[width=0.95\textwidth]{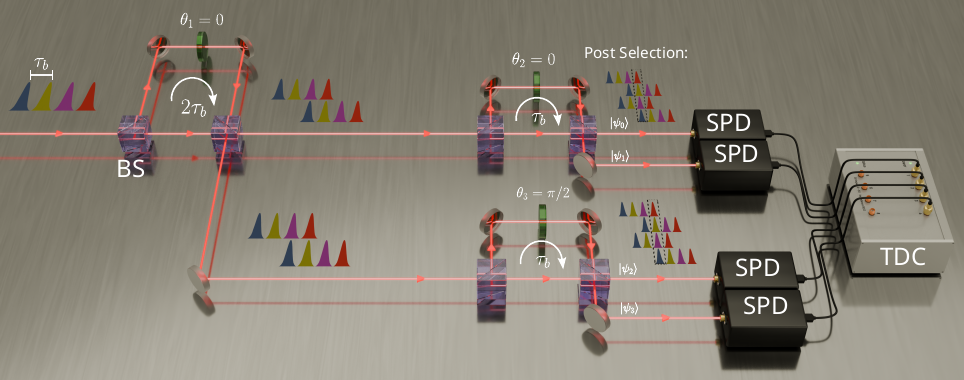}
  \caption[An optical network of DLIs for measuring time-bin qudits in mutually unbiased bases.]{An optical network of delay line interferometers for measuring time-bin qudits in mutually unbiased bases \cite{TBHD-MUB_Ikuta2022}. The network is depicted with 4 time-bins ($d=4$). The detection of a photon at any output corresponds to projecting onto one of the basis states $(|\psi_n\rangle)$. Note that different colours used for time bins are just for illustrative purposes. BS: Beam-splitter, SPD: Single-photon detector, TDC: Time-to-digital converter. } 
  \label{quditint}
\end{figure}

To measure time-bin qudits in mutually unbiased bases, multi-path DLIs with $d$ paths are required \cite{Duality_Englert2008, BellTestQutrit_Thew2004}. 
Alternatively, for a $d = 2^k$ dimensional system, where $k$ is an integer, a linear network of DLIs with different delays may be used \cite{QKD_Brougham2013, TBM_qudit_Bouchard2023}. 
In this configuration, a cascade of DLIs is employed, where the output of one interferometer feeds into the next.
To simplify the setup, only one branch of the full tree can be used, projecting the qudit onto one of the states $|\psi_n\rangle$ at a time. 
For example, to measure a  $d=4$ qudit, either a 4-path DLI or a cascade of 1-bit DLI (DLI with a path difference of $\tau_b$)  with phase setting of $\theta_1=0$ and two 2-bit DLIs (DLI with a path difference of $2\tau_b$) with $\theta_2=0$, $\theta_3=\pi/2$ can be used, as depicted in Fig.~(\ref{quditint}).   The general method for constructing an $N = 2^d$ qudit can be achieved by recursively applying this technique: the two outputs of an interferometer, with a delay of $\frac{\tau_b d}{m}$ and phase $\theta_1 =\phi$, are fed into subsequent interferometers. These subsequent interferometers have a delay of $\frac{\tau_b d}{2m}$, with phases set as $\theta_2 = \frac{\phi}{2}$ and $\theta_3=\frac{\phi+\pi}{2}$ \cite{QKD_Brougham2013}. 
By post-selection, the state can be projected into all required states, allowing for QST  \cite{QST_qudit_Ikuta2017} to characterize the state. 
Furthermore, an experimental realization of a scalable scheme for measurement in $(d+1)$ mutually unbiased bases for d-dimensional QKD using $\log_p(d)$ interferometers in prime power dimensions $d = p^N$ has been reported \cite{TBHD-MUB_Ikuta2022}.

One major disadvantage of using cascaded DLIs is that the probability of measuring a $d$-dimensional superposition scales as $1/d$ due to the increasing number of possible paths in the nested DLI configuration. 
To overcome this problem, the temporal Talbot effect \cite{Talbot_Jannson1981} has been employed to detect time-bin qudit superpositions efficiently \cite{TBQudit_Widomski2024}.  
The measurement of time-bin qudits has also been discussed in Ref. \cite{TBM_qudit_Lukens2018}. 
There are also ways to convert time-bin qudits to hybrid encodings, which simplifies the distinction between different states. In \cite{TBM_qudit_Bouchard2023}, the authors achieve this using a method akin to the time-bin to polarization conversion discussed in Section~(\ref{Sec_TB-POl}). In a recent work, Daynes et al. \cite{TBM_Danese2026} demonstrated programmable generalized measurements on high-dimensional time-bin photonic states by exploiting spatio-temporal mode coupling in a multimode fiber. By mapping engineered spatial superpositions onto distinct dispersive delays, the fiber functions as a stable, common-path interferometric processor, enabling a wide class of programmable temporal projections without cascaded interferometers.

\subsection{Scaling overheads from qubits to qudits}

We have discussed the key considerations for parameter selection of time-bin encoding in Section~(\ref{Sec_ParameterSelection}). Building on the insights from that section, here we develop an intuitive framework for the practical limits of time-bin encoding by treating a $d$-dimensional qudit as a temporal frame of duration $T_{\mathrm{frame}} = d\,\tau_b$, containing $d$ mutually coherent temporal modes separated by a fixed bin spacing $\tau_b$. 
Assuming ultra-narrow pulse ($\Delta\tau$) preparation capability, the achievable dimension $d$ is jointly constrained by the source repetition rate $f_{\mathrm{rep}}$, detection system timing resolution [predominantly limited by the SPD jitter $\Delta\tau_{JD}$, see Eq.~(\ref{eq_jitter})], the coherence time ($t_p$) of the laser source, and interferometric phase stability. Assuming independent Gaussian broadening mechanisms, the measured temporal width $\Delta\tau_M$ follows Eq.~(\ref{eq_jitter}). Temporal resolvability of adjacent bins  requires $\tau_b \gtrsim \kappa\,\Delta\tau_M$, with $\kappa \approx 3$--$5$ to suppress inter-bin cross-talk [see Sec. \ref{Sec_TBEncodingDefs}]. For state-of-the-art SNSPDs with $\Delta\tau_{JD} \sim 10$--$30\,\mathrm{ps}$, this implies practical bin spacings $\tau_b \sim 50$--$200\,\mathrm{ps}$.

To avoid inter-frame overlap in a pulse train operating at repetition rate $f_{\mathrm{rep}}$, the total frame duration must satisfy $d\,\tau_b \lesssim 1/f_{\mathrm{rep}}$, leading to the bound $d \lesssim 1/(f_{\mathrm{rep}}\,\tau_b)$. In addition, detector dead time $\tau_{\mathrm{dead}}$ imposes $\tau_b > \tau_{\mathrm{dead}}$ in high-rate regimes, although for typical SNSPDs this constraint is weaker than the jitter-limited condition in closely-spaced bin implementations. Together, these constraints define the repetition-rate-limited dimensional ceiling for a given temporal spacing.

Coherent manipulation and projective measurement rely on cascaded DLIs whose maximum path delay scales as $\tau_{\max} = (d-1)\tau_b$. Maintaining high interference visibility requires this delay to remain within the source coherence time, i.e., $(d-1)\tau_b < t_p$. Interferometric phase noise (instability) accumulated over the largest imbalance must satisfy $\delta\phi_{\mathrm{rms}} \ll 1$~rad, and since total phase excursion scales with optical path difference, the stability requirement tightens approximately inversely with dimension. For $d \sim 16$--$32$ and $\tau_b \sim 100$--$500\,\mathrm{ps}$, the largest separation reaches $1.6-16$~ns, corresponding to several meters of optical path imbalance, demanding sub-milliradian stability and therefore active phase stabilization with ultra-stable frequency, narrow-linewidth quantum laser sources $1.74\,\mathrm{MHz} - 1.74\,\mathrm{kHz}$ [see Eq.~(\ref{Eq_FreqtoPhaseUncertain}), Sec.~(\ref{Sec_CWLaser})].

As a concrete example, consider a GHz-rate platform with $f_{\mathrm{rep}} = 1\,\mathrm{GHz}$ and $\Delta\tau_{JD} \approx 20\,\mathrm{ps}$. Choosing $\tau_b \approx 50\,\mathrm{ps}$ satisfies temporal distinguishability and yields $d \lesssim 1/[(1\,\mathrm{GHz})(50\,\mathrm{ps})] \approx 20$; allowing practical margins typically restricts operation to $d \sim 8$--$16$. The longest interferometric delay is then $(d-1)\tau_b \sim 0.4$--$0.8\,\mathrm{ns}$, corresponding to optical path differences of roughly $8$--$16\,\mathrm{cm}$. The symbol rate becomes $R_{\mathrm{sym}} = f_{\mathrm{rep}}/d$, i.e., $60$--$125\,\mathrm{MHz}$ for $d = 8$--$16$, yielding raw photon information rates scaling as $R_{\mathrm{sym}}\log_2 d$, on the order of several hundred $\mathrm{Mbps}$ to a few $\mathrm{Gbps}$ under ideal single-photon detection. These coupled constraints produce a fundamental dimensionality--rate trade-off: although the information per photon increases as $\log_2 d$, the symbol rate decreases as $f_{\mathrm{rep}}/d$, and interferometric sensitivity increases with $(d-1)\tau_b$, causing practical performance to saturate beyond $d \sim 16$--$32$ in current implementations.

\subsection{Time-bin qudit entanglement}
\label{Sec_HD-Ent}

Entangled qudit states carry many of the advantages of qudit encoding discussed earlier, such as high information capacity per photon \cite{HD-QC_Daniele2019}, robustness against noise \cite{Ent-HD_Zhu2021}, and higher error-rate tolerance in QKD \cite{HD-QKD_Sheridan2010, HD-QKD_Groblacher2006, HD-QKD_Mafu2013, TE-HD-QKD_Zhong2015}. Moreover, entangled qudits offer relaxed conditions on the detector efficiency for closing the detection loophole in tests of quantum non-locality  \cite{Ent-HD_Ikuta2016, BellTest_Tamas2010}, and in the implementation of the device-independent QKD protocols \cite{DI-HDQKD_Huber2013}. A practical advantage of high-dimensional entanglement as compared to the multipartite equivalent (i.e., with the same total dimension) is the higher coincidence rate, as it requires the generation and detection of only two photons. 

\begin{figure}[ht!]
    \centering
       \includegraphics[width=0.9\linewidth]{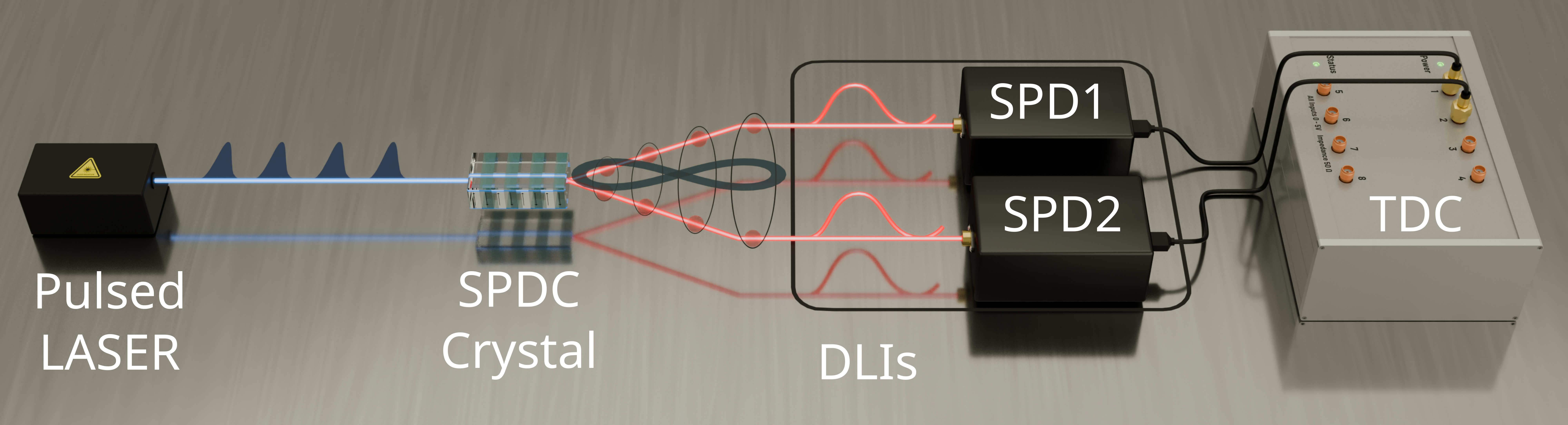}
   \caption[High-dimensional time-bin entanglement.]{Schematic diagram of the experimental setup for creating and characterizing high-dimensional time-bin entanglement. A sequence of $n$-pulse with uniform intensity and time-bin separation is used to pump a SPDC crystal. When exactly one photon pair is produced from these pulses, it results in an $n$-dimensional qudit-qudit entangled state. \textbf{Redrawn figure from \cite{TB-HD-EntChar_Martin2017}}. DLIs: Delay-line interferometers, SPD: Single-photon detector, SPDC: Spontaneous parametric down-conversion, TDC: Time-to-digital converter.}
    \label{fig_HD-TBE1}
\end{figure}

Time-bin entangled qudits can be generated by pumping a nonlinear medium with a sequence of $n$-pulses with uniform intensity and time-bin separation $\tau_b$. The pulses can be generated using either a CW laser and intensity modulator or direct laser modulation, as discussed in section \ref{Sec_CLS}. For example, the SPDC process can be used for non-linearity as shown in Figure \ref{fig_HD-TBE1}. Suppose exactly one photon pair is produced at the output of the crystal during excitation using these $n$-pulses. In that case, the resulting entangled state can be expressed as
\begin{equation}
    |\Psi\rangle = \frac{1}{\sqrt{d}}\sum_{j=0}^{d-1} \alpha_j e^{i\phi_j} \ket{t_j,t_j},
\end{equation}
where $\ket{t_j,t_j}$ represents the states in which both photons occupy the $j$th time-bin, characterized by a probability amplitude $\alpha_j$, and $\phi_j$ is the relative phase. 
Similar to the case of TBQs, the coherence time of the pump laser ($t_p$) must be larger than the time-spread of $d$-pulses (i.e., $t_p > \tau_b\, d$) to maintain a constant phase relationship between these pulses. In such a case, $\alpha_j$ and $\phi_j$ can be considered constant within the pulse sequence $\tau_b\,d$. Furthermore, the pump power is adjusted such that the probability of generating more than one photon pair within a $d$-pulse train is negligible. 

In a recent work, Schiffer et al. \cite{Ent-TBHD_Schiffer2026} have used a CW pump laser and SPDC to generate high-dimensional time-bin entanglement from a continuous temporal photon-pair stream, where the effective time-bins and dimensionality are defined a posteriori through temporal discretization and certified using nested Franson interferometry \cite{Ent-TBHD_Kanitschar2024}, enabling a bright, stable, and flexibly scalable HD time-bin entangled source.
High-dimensional time-bin entanglement has been demonstrated using SPDC \cite{HDTB-Ent_Riedmatten2002, Ent-HD_Riedmatten2004, Ent-HD_Stucki2005, Ent-HD_Richart2012, TB-HD-EntChar_Martin2017}, SHG-SPDC cascaded process \cite{Ent-HD_Takuya2018, Ent-HD_Takuya2018}, and the SFWM process \cite{Ent-FWM_Takesue2014, QST-HD_Samantha2015, QST_qudit_Samantha2016, HD-QKD_Yu2025}. Additionally, hyper-entanglement of time-bin and polarization degree of freedom is demonstrated using the SPDC process \cite{Ent-HD_Steinlechner2017}.

\subsection{Time-bin qudit entanglement characterization}

Analogous to entangled qubits, entangled qudits can be analyzed to quantify their degree of entanglement in the system. 
For this purpose, each photon is passed through a DLI. 
The temporal delay between the short and long arm of the DLI is set to be $\tau_b$ or $2\tau_b$ to check coherence between the two-neighbouring ($j~\&~j+1$) or two next-neighbouring ($j~\&~j+2$) temporal modes, respectively. 
These interferometric delays must be much larger than the temporal duration of each pulse or time-bin ($\Delta\tau$), and the coherence time of the down-converted photons ($t_{dc} =c\,\lambda^2/\Delta\lambda_f$, where $c$ is the speed of light, and $\Delta\lambda_f$ is the bandwidth of the filter) to avoid any first-order interference. Under these conditions, second-order interference is observed in the coincidence rates in the interference peak at the output ports of the two interferometers. Each DLI acts as a local projection onto the state $ \langle j,j|+ e^{i(\phi_a+\phi_b)} \langle j+i, j+i|$, where $i=1,2$, and $\phi_a~\&~\phi_b$ are the relative phases between the two arms of the corresponding interferometers \cite{TB-HD-EntChar_Martin2017}. 

The entanglement visibility is determined by scanning the phase of one of the two interferometers to identify the maximum and minimum coincidence rates, corresponding to constructive and destructive interference, respectively. 
Tomographic reconstruction of the full-density matrix is impractical as it would require $n$-DLIs \cite{TB-HD-EntChar_Martin2017}. However, one can provide a lower bound on the entanglement present in the system using methods discussed in Ref. \cite{TE-HD-EntChar_Alexey2017}. Firstly, one measures the coincidence events on a time-of-arrival basis, obtaining the diagonal density matrix elements ($\langle j,k|\rho|j,k\rangle$). Secondly, one measures the interference visibility between two neighbouring ($j~\&~j+1$) temporal modes, as well as two next-neighbouring  ($j~\&~j+2$)  temporal modes. Thus, one can estimate the off-diagonal elements $\langle j, j|\rho|j+i,j+i\rangle$ for $i=1,2$. From these measurements, one can obtain a lower bound on the entanglement of formation \cite{Ent-HD_Huber2013} of $\rho$ as follows:
\begin{equation}
    EoF(\rho) \geq - \log_2\left(1-\frac{B^2}{2}\right),
\end{equation}
where B is given as
\begin{equation}
    B = \frac{2}{\sqrt{|C|}} \left( \sum_{j,k \in C, j<k} |\langle jj|\rho|kk\rangle| - \sqrt{\langle j,k|\rho|j,k\rangle \langle k,j|\rho|k,j\rangle} \right) ,
\end{equation}
where $|C|$ represents the cardinality of the set $C$, meaning the number of index pairs $(j,k)$ included in the sum. Note that $B$ serves as a lower bound for the concurrence of $\rho$.

Another implementation employed a Fabry-Perot-like two-photon interferometer \cite{Ent-HD_Stucki2005} for entanglement analysis. Furthermore, a method has been developed for certifying high-dimensional time-bin entanglement in fiber-loop systems \cite{Ent-TBHD_Euler2025}, in which entanglement creation and detection can utilize the same physical components. Moreover, measurements in only two experimentally accessible bases are sufficient to establish a lower bound on the entanglement dimension for both two- and multiphoton quantum states. Higher-dimensional Bell inequalities, such as the CGLMP inequality \cite{CGLMP_Collins2002}, can be used to measure time-bin or time-energy qudit entanglement, as shown for high-dimensional time-bin states in \cite{Ent-HD_Ikuta2016}.


\section{Time-energy entanglement}
\label{Time-Energy}
\FloatBarrier

Time-energy entanglement is a form of quantum entanglement in which the energy and time of arrival of a pair of photons are individually uncertain, yet exhibit strong correlations in their joint properties that arise from strict conservation laws. Specifically, the sum of the photon energies and the difference in their arrival times are well-defined, reflecting non-classical correlations that cannot be explained by local hidden variable theories.
The concept of time-energy entanglement was proposed by J.D. Franson in 1989 to test a Bell inequality \cite{TE_Franson1989}. Time-energy entanglement is a closely related concept to, and precursor of, time-bin and frequency-bin entanglement. While the former involves entanglement in continuous time and energy variables, it can be discretized into either the time or frequency domain using suitable apparatus, resulting in time-bin and frequency-bin \cite{FB-Ent_Ramelow2009, FB-Ent_Olislager2010, FB-QI_Lu2023} entangled states characterized by well-separated time or frequency bins, respectively.

\subsection{Overview}

\begin{figure}[htbp!]
    \centering
  \includegraphics[width= 0.8\columnwidth]{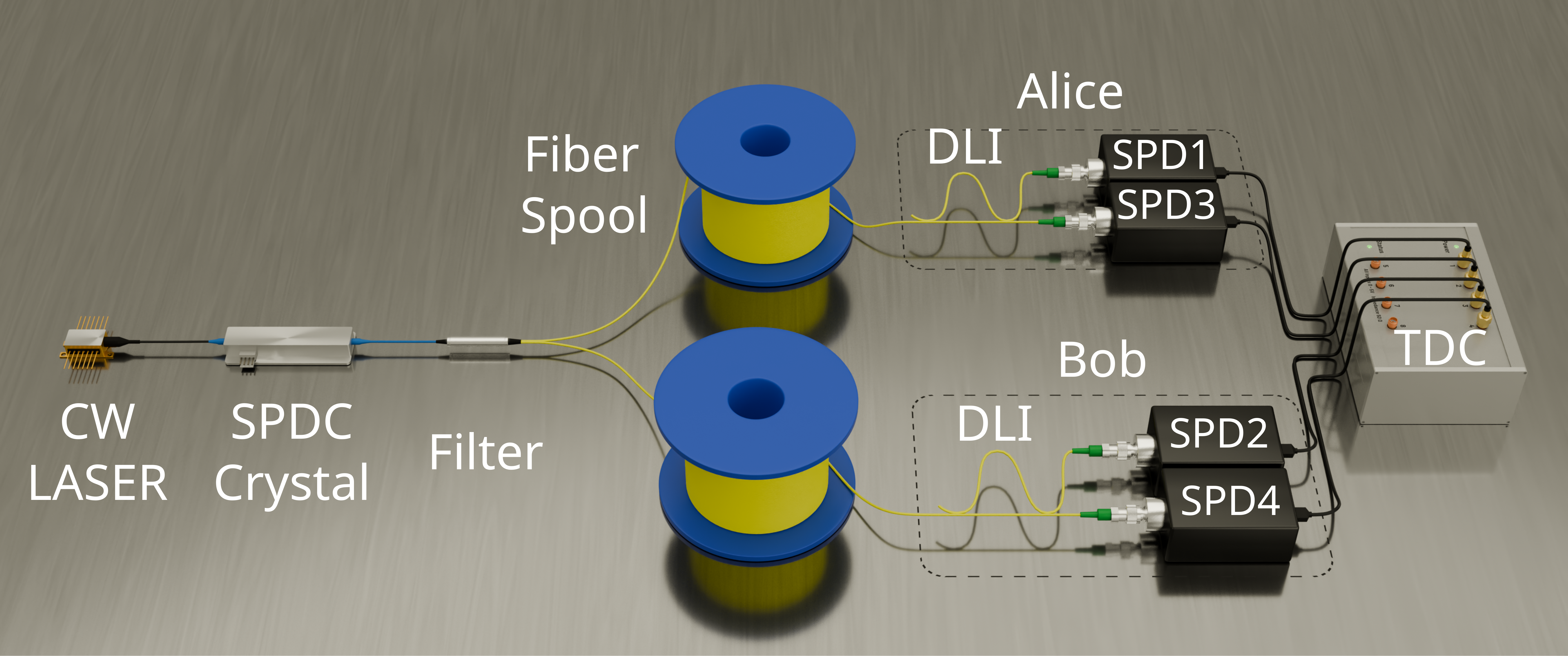}
    \caption[Time-energy entanglement.]{Schematic of the experimental setup for preparing and characterizing time-energy entanglement using Franson interferometers. A CW laser source pumps a SPDC crystal, generating time-energy entangled photon pairs within the coherence length of the pump. When these photons pass through the Franson interferometers, one observes an interference pattern due to the indistinguishability of the photons, whether they take short or long paths. \textbf{Figure concept adapted from \cite{QCom_Gisin2007}}. CW: Continuous-wave, DLI: Delay line interferometer, SPD: Single photon detector, SPDC: Spontaneous parametric down-conversion, TDC: Time-to-digital converter.} 
    \label{fig_time-energy}
\end{figure}

To elucidate the origin of time-energy entanglement, let us consider a continuous-wave pump laser (with frequency $\omega_p$ and coherence time $t_p$) that drives a second-order nonlinear crystal. With a small probability within the pump coherence time, a pair of photons --- denoted as signal (frequency $\omega_s$) and idler (frequency $\omega_i$) --- is generated through the SPDC process, as shown in Fig.~(\ref{fig_time-energy}).
The down-converted photons individually possess finite energy uncertainties ($\Delta \omega_s$ and $\Delta \omega_i$), yet their total energy is strictly determined by energy conservation and is equal to the energy of the pump photon; i.e., $\omega_s + \omega_i = \omega_p$.
Furthermore, both down-converted photons are generated simultaneously, in accordance with the principle of energy conservation.
Yet, the quantum uncertainty regarding their production time ($\Delta t_s$ and $\Delta t_i$) is confined within the long coherence time of the pump laser ($t_p$).
The uncertainty lies in the ``energy'' and ``age'' of the individual photons, yet the total energy sum and the age (time) difference between them are well-defined. Gaining full information about the time of a photon renders its frequency uncertain, and vice versa.
Since energy and time are continuous variables, the correlation in time-energy entanglement is analogous to the idea originally proposed by Einstein, Podolsky, and Rosen \cite{EPR_Einstein1935} for position and momentum correlations of a particle pair.
Importantly, the term ``time-energy entanglement'' should not be interpreted as a correlation between, for example, the energy of one photon and the creation time of the other \cite{TE-Physics_Michael2018}.

A time-energy entangled state can be mathematically represented in the frequency and time domains as
\begin{equation}
\begin{aligned}
    |\psi_{\scriptscriptstyle \mathrm{TE}}\rangle = \int d\omega_s \, d\omega_i \, F(\omega_s, \omega_i) \, |\omega_s\rangle_s |\omega_i\rangle_i
    \equiv & \int dt_s \, dt_i \, G(t_s, t_i) \, |t_s\rangle_s |t_i\rangle_i,
    \end{aligned}
\end{equation}
where \( F(\omega_s, \omega_i) \) and \( G(t_s, t_i) \) denote the \emph{joint spectral} and \emph{joint temporal amplitudes}, respectively.

\begin{figure}[ht!]
    \centering
    \includegraphics[width=0.80\columnwidth]{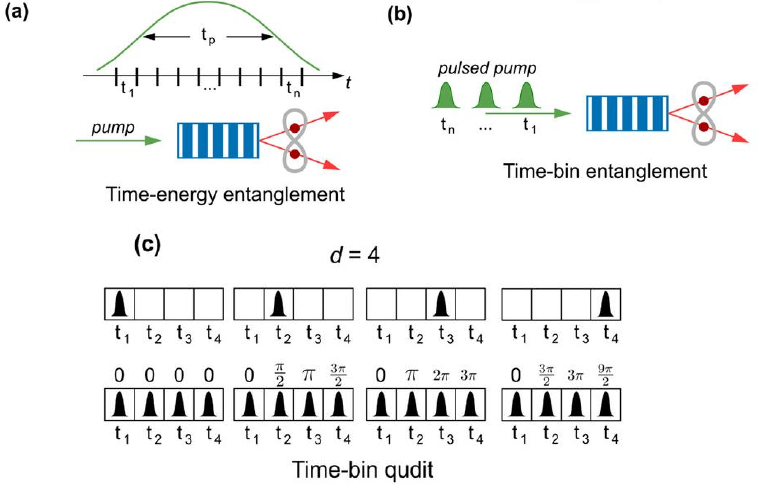}
    \caption[An illustration of time-based encoding schemes.]{An illustration of time-based encoding schemes. (a) Time-energy entanglement:  photon pairs generated within the coherence time of the pump ($t_p$) have a temporal uncertainty ($\Delta t$). Using a long coherence time (narrow-band) pump, one can define time slots $t_1, t_2, t_3... t_n$ during which photon pairs are generated. (b) Time-bin entanglement: This requires a pulsed laser source, and each pulse is considered a different time slot. (c) Time-bin qudit ($d \geq 3$) is a generalization of TBQs to higher dimensions, where a photon can be found in a superposition of $d$ time-bin basis states. \textbf{Figure reproduced under Creative Commons license from \cite{HD-QC_Daniele2019}}.}
    \label{fig_time-energy2}
\end{figure}

Notably, a narrowband pump laser possesses a long coherence time, which permits greater uncertainty in the emission times of the down-converted photon pairs, thereby generating entanglement in the time domain. This state can be written as
\begin{equation}
|\psi_{\scriptscriptstyle \mathrm{TE}}\rangle = \sum_{k=1}^{d} \alpha_k |t_k\rangle_s \, |t_k\rangle_i,
\end{equation}
where $\alpha_k$ is the complex probability amplitude associated with the photon-pairs in the $k$th time slot $|t_k\rangle$ within the coherence time ($t_p$) of the pump \cite{TE_Ecker2019} as shown in the Fig.~(\ref{fig_time-energy2}a).  
Note that in the case of time-energy entanglement, discretization into time states occurs within the pump's coherence time, whereas in time-bin entanglement, the photons are already generated in well-separated time-bin states that share a common phase, making additional discretization unnecessary.
A schematic representation of these concepts is shown in Fig.~(\ref{fig_time-energy2}).

Let us now consider the situation when both down-converted photons having a coherence time $t_{dc}$ are passed through the corresponding DLIs or Franson interferometers \cite{TE_Franson1989}, as shown in Fig.~(\ref{fig_time-energy}). 
Note that SPDC process typically produces broadband photons \cite{SPDC_Couteau2018}; however, when spectral filtering is applied (see Sec.~\ref{Ent_SPDC}), the coherence time of the down-converted photons is determined by the filter bandwidth and is given by $t_{\text{dc}} = \frac{l_c}{c} = \frac{\lambda_{s,i}^2}{c\,\Delta \lambda}$, where $\lambda_{s,i}$ denotes the central wavelength of the signal or idler photon, $\Delta \lambda$ is the filter bandwidth, and $c$ is the speed of light in vacuum.
The state after the DLI is given as
\begin{equation}
|\omega_s\omega_i\rangle \xrightarrow{\text{DLIs}} \frac{1}{2} \left[ |\omega_s\omega_i\rangle_{SS} + e^{i(\phi_a + \phi_b)} |\omega_s\omega_i\rangle_{LL} + e^{i\phi_a} |\omega_s\omega_i\rangle_{LS} + e^{i\phi_b} |\omega_s\omega_i\rangle_{SL} \right],
\label{Eq_TEDLI}
\end{equation}
where $\phi_a$ ($\phi_b$) is the additional phase acquired by the photons propagating through the long path of Alice's (Bob's) DLIs. The subscript $S(L)$ denotes the short (long) path of the corresponding DLI.
Since the two photons were generated simultaneously, there are only two possibilities that they reach the two detectors simultaneously: either both photons travelled through the short arms or the long arms of the corresponding DLIs. 
If the path imbalances (delays) of Alice's and Bob's DLIs are equal (i.e., $L_A-S_A = L_B-S_B$, or more generally $|(L_A -S_A)- (L_B-S_B)| \ll c\,t_{dc}$), and both are much smaller than the coherence length of the pump laser (i.e., $ L_A-S_A, L_B-S_B \ll c\,t_p $), then the two possible paths become indistinguishable. Additionally, to suppress single-photon interference, the path imbalance should be much larger than the coherence length of the down-converted photons (i.e., $ c\,t_{dc} \ll L_A-S_A,~ L_B-S_B$). Under these conditions, quantum mechanics requires that the probability amplitudes corresponding to the indistinguishable processes must be coherently added, leading to two-photon quantum (non-local) interference effects~\cite{QCom_Gisin2007, TE_Kwiat1993}. The coincidence probability of the two photons in the interference peak varies with the phase of the two DLIs as
\begin{equation}
P^\text{TE}_\text{Coin}=\frac{1}{4}\left[1 \pm \cos(\phi_a+\phi_b) \right], 
\label{Eq_TEUncertanity}
\end{equation}
where the (+) sign corresponds to the same set of output ports (SPD1--SPD2, SPD3--SPD4), and the (-) sign corresponds to the crossed set of output ports (SPD1--SPD3, SPD2--SPD4).

\subsection{Characterization methods}
\label{Sec_TECharMethods}

Experimental evidence of time-energy entanglement can be found through correlation measurements of complementary variables and violation of uncertainty relation \cite{TE-CV_Stefano2002}, or nonlocal quantum phenomena such as two-photon quantum interference using Franson interferometry \cite{TE_Kwiat1993}, Bell-inequality violations \cite{TE_Franson1989, TE-Bell_Brendel1992}, and non-local dispersion cancellation wherein temporal distribution of coincidence detections remains unaffected when equal but opposite dispersions are applied to a pair of photons \cite{TE-disp_Franson1992, TE-Disp_Steinberg1992, TE-disp_Li2019}. 

\begin{figure} [ht!]
    \centering
   \includegraphics[width=0.99\columnwidth]{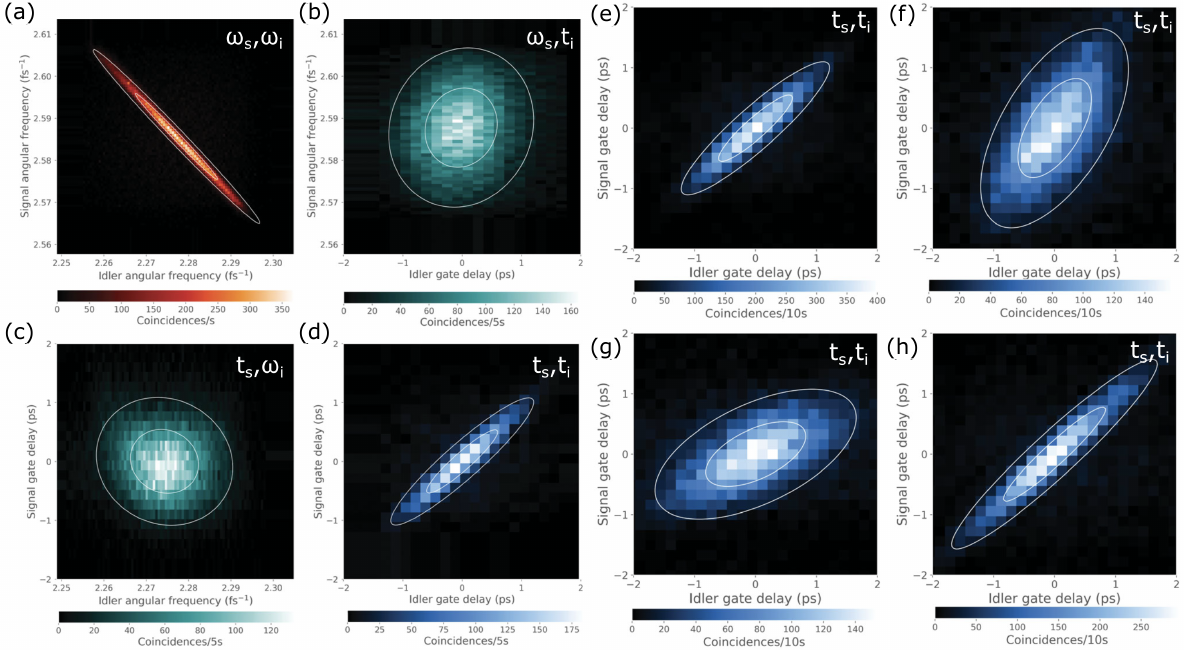}
\caption[Characterization of time-energy entanglement.]{Characterization of time-energy entanglement. (a-d) Spectral and temporal analysis. Coincidence measurements are performed to determine (a) the joint spectral distribution, (d) the joint temporal intensity, and (b),(c) the cross-correlations linking the idler's time (frequency) with the signal's frequency (time). (e-h) Nonlocal dispersion cancellation in the joint temporal distributions. The joint temporal intensity of the signal and idler pair is shown (a) without dispersion, (b) with positive dispersion applied to the signal, (c) with negative dispersion applied to the idler, and (d) with both positive dispersion on the signal and negative dispersion on the idler, illustrating the cancellation effect. \textbf{Figure reproduced with permission from \cite{TE-Char_Jean2018}}.}
    \label{fig_time-energy3}
\end{figure}

\textit{(a) Violation of uncertainty relation:}
Time-energy entanglement has been witnessed through precise spectral and temporal measurements on a pair of photons and the violations of the uncertainty relations \cite{Ent-uncert_Howell2004, Ent-uncert_Edgar2012}. 
For two separable photons, the following inequality \cite{TE-Char_Jean2018, TE-CharUncert_Mei2020} holds:
\begin{equation}
    \Delta (\omega_s+\omega_i) \Delta(t_s-t_i) \geq 1,
    \label{eq_uncert}
\end{equation}
where $\omega_s$ ($\omega_i$) and $t_s$ ($t_i$) are the angular frequency and time of generation of the signal (idler) photon, respectively. Thus, for separable photon pairs, the uncertainty relation (\ref{eq_uncert}) gives a lower bound on the product of the correlations between their frequency sum and arrival time difference. However, the time-energy entangled photon pairs violate this bound, thus providing an entanglement witness. From the spectral and temporal correlations shown in Figs.~(\ref{fig_time-energy3}a) and (\ref{fig_time-energy3}d), the joint uncertainty product $ \Delta (\omega_s+\omega_i) \Delta(t_s-t_i) = 0.290 \pm 0.007$ \cite{TE-Char_Jean2018}, which violates the uncertainty relation (\ref{eq_uncert}) indicating time-energy entanglement.

\textit{(b) Non-local dispersion cancellation:}
The non-local dispersion cancellation \cite{TE-disp_Franson1992} - a phenomenon wherein the temporal distribution of coincidences remains constant when equal and opposite dispersion is applied to each photon - has also been used to witness the time-energy entanglement \cite{TE-disp_Wasak2010, TE-Chip-SiNMRR_Jose2017, TE-Disp_Chae2025}. %


As shown in Figs.~(\ref{fig_time-energy3}e-h), when a positive dispersion of $A_s = (0.0373 \pm 0.0015)~\text{ps}^2$ on the signal and a negative dispersion of $A_i= (0.0359 \pm 0.0014) ~\text{ps}^2$ on the idler is applied, the width of $\Delta(t_s-t_i)$ in Fig.~(\ref{fig_time-energy3}h) remains unchanged as compared to Fig.~(\ref{fig_time-energy3}e) indicating presence of time-energy entanglement \cite{TE-Char_Jean2018}.
Nonlocal dispersion cancellation was also demonstrated by Baek et al. \cite{TE-Disp_Baek2009}, Zhong et al. \cite{TE-disp_Zhong2013} in fiber-based Franson interferometer, and in deployed telecom fibers by Grieve et al. \cite{TE-disp_Grieve2019}.  A recent study \cite{EntTE-Disp_Xiang2022} demonstrates a widely tunable and highly flexible approach to nonlocal dispersion cancellation using wavelength control in time-energy entangled photon pairs.
Nodurft et al. \cite{TE-Disp-MP_Nodurft2020} broadened the study of nonlocal dispersion cancellation to include multi-photon systems, showing that under certain experimental conditions, time-energy entanglement among three or more photons can lead to either full or partial dispersion cancellation.

\textit{(c) Non-local interference, entanglement visibility, and post-selection:}
As evident from Eq.~(\ref{Eq_TEDLI}), in addition to the central interference peak, two non-interfering satellite peaks appear, corresponding to the long-short and short-long paths that a photon pair can take through the DLIs.
Initial measurements of two-photon interference reported visibilities up to $50\%$, limited by the timing resolution (high jitter) of the detectors \cite{TE-Ent_Ou1990, TE-Ent_Kwiat1990, TE-TPI_Franson1991}, which were insufficient to resolve the central interference peak from the satellite peaks. Subsequent two-photon interference measurements using time-resolved detection and short-coincidence window yielded visibilities exceeding the classical limit \cite{TE-TPI_Brendel1991} and approaching the theoretical maximum of 100\%. 
Such high entanglement visibility measurements have been used to verify time-energy entanglement in two \cite{TE_Kwiat1993} and three-qubit systems \cite{TE_Agne2017}. 

Similar to time-bin entanglement, post-selection of middle-bin coincidence events in Franson interferometry results in a post-selection loophole, as discussed in Section~(\ref{Sec_BipartiteEntChar}). 
This post-selection loophole is specific to Franson-type schemes and should be distinguished from the more general detection loophole in Bell tests, which arises from finite detection efficiencies and the fair-sampling assumption, and is known to remain open in most photonic experiments \cite{DetLoophole_Garg1987, DetLoophole_Eberhard1993}. 
This Franson-specific loophole can be eliminated by using the ``hug'' interferometer \cite{TE-Bell-PSLFree_Cabello2009}, employing an optical switch \cite{TBE-PSL_Francesco2018} instead of the first beam splitter in a delayed-line interferometer, or by exploiting hyperentanglement \cite{TE-Ent-PSFree_Strekalov1996} in polarization and time-energy degrees of freedom. 
Notably, Strekalov \textit{et al.} utilized the intrinsic hyperentanglement of photon pairs generated via type-II SPDC and implemented a modified Franson interferometer \cite{TE-Ent-PSFree_Strekalov1996} to obtain a $95\%$ entanglement visibility without temporal post-selection. Rossi \textit{et al.} further demonstrated an interferometric scheme employing a common interferometer for both photons, which likewise removes the need for temporal post-selection \cite{TE-Vis-PSLFree_Rossi2008}.

\textit{(d) Bell-violation:} 
Time-energy entanglement can also be characterized through the violation of Bell's inequality, as outlined in Section~(\ref{Sec_BipartiteEntChar}d). Similar to entanglement visibility measurements, a post-selection scheme is employed to demonstrate the violation of Bell's inequality \cite{TE-Bell_Brendel1992}, including implementations over optical fiber links \cite{TE-Bell_Tapster1994}.
Early long-distance tests of time-energy entanglement demonstrated robust nonlocal correlations over metropolitan-scale fiber links, including Bell inequality violation with photons separated by more than 10 km \cite{TE-Bell_Tittel1998} and stringent lower bounds on the speed of nonlocal influences in tests of ``spooky action at a distance'' over 18 km channel \cite{TE-Bell_Salart2008}.
Recently, a single two-level emitter deterministically coupled to a nanophotonic waveguide \cite{TE-TLS-Bell_Liu2024} has been used for Bell violation with post-selection. 
Notably, Bell-type test has also been performed for high-dimensional states such as time-energy entangled qutrits \cite{BellTestQutrit_Thew2004}, demonstrating a violation of $34\,\sigma$ above the local hidden variable limit. 
The post-selection used even in ideal Franson-type Bell experiments based on time-energy entanglement introduces exploitable loopholes \cite{TE-LocalRealism_Aerts1999, TE-LocalRealism_Jogenfors2014}. A recent attack demonstrated that this vulnerability can be used to fake Bell violations using a classical light source \cite{TE-BellTestHack_Jogenfors2015}. Consequently, Bell violations in such setups cannot be used to certify device-independent security. 


A modified Franson scheme, commonly known as the ``hug'' interferometer scheme, was proposed \cite{TE-Bell-PSLFree_Cabello2009} wherein the short (long) path of one photon and the long (short) path of the other photon end at the same observer, thus eliminating any coincidence due to short-long and long-short events. As the event rejection is local and independent of measurement settings, the scheme avoids the postselection loophole. Different experimental groups \cite{TE-Bell-PSLFree_Lima2010} have utilized the hug-interferometer scheme to address the post-selection loophole in Bell-inequality violation, including entanglement distributed over fiber networks \cite{TE-dist-PSLFree_Cuevas2013, TE-dist-PSLFree_Carvacho2015}.
 Recently, Santagiustina et. al. implemented a post-selection loophole-free certification of entanglement in an integrated photonic chip in the``hug'' scheme \cite{Nonlocality_Santagiustina2024}.
The post-selection loophole in the Franson-interferometry scheme can be overcome, but at a cost of a higher entanglement visibility threshold \cite{TE-LocalRealism_Jogenfors2014}. Such high visibility has been achieved in a non-degenerate SPDC source \cite{TE-Char_Fallon2025}, making it suitable for quantum network applications. 

\subsection{Experimental realizations of time-energy entanglement}

Time-energy entanglement is commonly generated using spontaneous parametric processes such as SPDC \cite{TE-Ent_Tanzilli2002, TE-Ent_Ali2006}, and SFWM \cite{TE-Ent-SFWM_Dong2014, TE-Ent_Oser2020} (see Section~\ref{Sec_HeraldedTB}). Early demonstrations primarily employed bulk nonlinear crystals, while subsequent advances in integrated photonics \cite{SFWM-Review_Wang2023} have enabled improved stability, brightness, and scalability. In particular, SFWM in InGaP photonic crystal cavities has demonstrated ultra-efficient generation of time-energy entangled photon pairs \cite{TE-Ent_Chopin2023}, while ultralow-loss platforms such as silicon \cite{TE-IntSi_Grassani2015} and silicon-nitride microring resonators enable bright, narrowband photon pairs with strong time-energy correlations \cite{PIC-TE-SFWM_Chen2024}. A key advantage with such system is that they support biphoton frequency combs in which time-energy entanglement persists across multiple resonant modes, providing a compact resource for high-dimensional encoding \cite{TE-Chip-SiNMRR_Jose2017}.

Additionally, SPDC has been utilized in periodically poled thin-film lithium niobate waveguides to demonstrated high-quality time-energy entanglement \cite{TE-Ent-LNWG_Zhao2020, TE-Ent-LNOI_Xue2021}. Cascaded nonlinear processes, including sequential second-harmonic generation and SPDC, have also been employed to generate time-energy entangled photons \cite{TE-Ent_Pascal2021, Ent-Cascade_Zhang2021}, enabling operation within the same wavelength band as the pump. Coherent manipulation of frequency-nondegenerate time-energy entangled states further enables direct preparation of frequency-entangled photons, expanding the accessible Hilbert space \cite{TE-Ent_Zhou2014}.

Beyond nonlinear optical platforms, solid-state quantum emitters have emerged as triggered, on-demand sources of time-energy entanglement. Resonantly driven quantum-dot four-level systems produce time-energy entangled photon pairs via biexciton-exciton cascades under continuous-wave excitation \cite{TE-EntQdot_Hohn2023}. Earlier observations of Franson-type interference from driven two-level quantum dots provided direct evidence of nonclassical time-energy correlations \cite{TE-QDot_Peiris2017}. More recently, Bell-nonlocal time-energy entanglement has been realized using two-photon scattering from a two-level emitter coupled to a nanophotonic waveguide \cite{TE-TLS-Bell_Liu2024}.

Time-energy entanglement has also been extended beyond bipartite systems. Three-photon time-energy entanglement was generated using cascaded SPDC and verified via three-photon Franson-type interference \cite{TE-Tripartite_Shalm2012}. High-rate generation of time-energy entangled triphoton W-state has subsequently been demonstrated using spontaneous six-wave mixing in a six-level atomic vapour system with a triple-$\Lambda$ configuration \cite{TE-Ent-WState_Li2024}.

Time-energy entanglement plays a central role in quantum communication and networking. High-fidelity quantum state transfer between photons of different wavelengths while preserving time-energy entanglement has enabled interfacing across heterogeneous quantum systems \cite{QM-Interface_Tanzilli2005}. Applications include high-dimensional entanglement distribution \cite{TE-HD_Bulla2023}, QKD \cite{TE-DispQKD_Liu2019, TE-QKD_Karimi2020, TE-QKD_Liu2023}, quantum random number generation \cite{TE-QRNG_Xu2016, TE-QRNG_Zhang2023}, and quantum clock synchronization \cite{QClockSync_Giovannetti2001, TE-ClockSync_Pelet2025}. Hyperentanglement, which combines polarization/time-energy and discrete-frequency/time-energy degrees of freedom, further expands protocol functionality by enabling simultaneous encoding across multiple photonic Hilbert spaces \cite{TE-Ent-Hyper_Dong2015}. Such hyperentangled states have enabled single-copy entanglement distillation \cite{TE-distt_Ecker2021} and experimental quantum privacy amplification in noisy environments \cite{TE-QPA_Philipp2024}.
Seminal Bell-type experiments using time-energy entangled photons were performed over tens of kilometers of optical fiber \cite{TE-BellTest_Tittel1999}. Subsequent works demonstrated loophole-free or postselection-free Bell violations in deployed fiber networks \cite{TE-dist-PSLFree_Cuevas2013, TE-dist-PSLFree_Carvacho2015, TE_Vergyris2017}, as well as sub-picosecond time-energy correlations probed using nonlinear optical gating \cite{TE-QInterferometry_MacLean2018}. Bell tests on higher-dimensional time-energy entangled states, including qutrits, have also been realized \cite{BellTestQutrit_Thew2004}.

Long-distance distribution of time-energy entanglement has been demonstrated over 100 km of optical fiber \cite{TE-Ent_Zhang2008} and over noisy free-space channels using phase-stable nonlocal Franson interferometry \cite{TE-QKD_Bulla2023}. High-dimensional QKD based on time-energy entanglement has achieved multi-bit encoding per photon pair \cite{TE-HD-QKD_Zhong2012, TE-HDQKD_Irfan2007}, while certification and quantum steering of high-dimensional states have been demonstrated using quantum frequency combs \cite{TE-HD_Chang2024}. Silicon photonic platforms have further enabled high-visibility entanglement distribution over deployed fibers \cite{TE-EntDist_Qin2024, TE-EntDist_Zhao2026} and gigahertz-rate generation of hyperentangled photon pairs \cite{TE-Bright_Gianini2026}. Recent reviews highlight advances in high-dimensional quantum systems, including quantum frequency combs that enable scalable time-energy entanglement across multiple frequency channels \cite{HD-FreqComb_Chang2025}, and on-chip high-dimensional entangled photon sources that enhance scalability, stability, and integration for practical quantum technologies \cite{Ent-HD_Kaur2024}.

Finally, the robustness of time-energy entanglement has been investigated across diverse physical environments, including slow-light media \cite{TE-Ent-Slowlight_Broadbent2008}, warm atomic ensembles \cite{TE-Ent_AtomEnsemble_Park2019}, and Doppler-broadened ladder-type $^{87}\mathrm{Rb}$ systems \cite{TE-Ent-Rb_Lee2019}, and multiuser entanglement swapping architectures based on dense wavelength-division multiplexing and sum-frequency generation \cite{TE-EntSwap_Li2019}. The survival of time-energy entanglement through biological tissue and strongly scattering media further highlights its resilience for real-world deployment \cite{TE-Survival_Lum2021}. Single-photon continuous-variable QKD protocols based on the time-energy uncertainty relation extend their applicability to alternative quantum cryptographic paradigms \cite{TE-QKD_Qi2006}.

\section{Applications of time-bin-encoded quantum states}
\label{Applications}
\FloatBarrier

As previously discussed, TBQs have emerged as a robust and practical approach to quantum information processing due to their resilience to environmental noise, compatibility with existing fiber-optic infrastructure, and ease of implementation.
This section explores key applications of TBQs, including QKD, quantum networking, and quantum computing, highlighting their role in advancing secure communication and scalable quantum systems. We also discuss teleportation, entanglement swapping and quantum repeaters, which are an integral part of the modern quantum information protocols and rely on the interference of qubits generated by independent sources.

\subsection{Quantum Key Distribution (QKD)}
\label{QKD}

QKD is a technique for information-theoretically secure distribution of cryptographic keys by exploiting the principles of quantum mechanics instead of the presumed difficulty of mathematical problems. 
Classical binary key information is communicated using qubits (or qudits); Heisenberg uncertainty and the impossibility of perfect qubit cloning ensure the secrecy and security of the communication between two distant parties, commonly known as Alice and Bob.
Typically, a QKD protocol involves preparation of orthogonal qubits in different bases, transmission over a free-space or optical fiber channel and measurement of qubits. 
Numerous protocols, as well as their modifications, have been proposed over the last few decades \cite{QKD_BB84, QKD_B92, QKD_E91, QKD_BBM92, DPS-QKD_Inoue2002, COW-QKD_Stucki2005, QKD_SSP1998, QKD-MDI_HKL2012, QKD_DI1998}. A detailed review of QKD protocols and security considerations can be found in reference \cite{QKD-RMP_Scarani2009}. 
Furthermore, several experimental demonstrations have been conducted in both laboratory and field environments, as reviewed in \cite{QKD-Advances_Perandola2020, QKD-Review_Xu2020}. 
Some of the challenges involved in building a practical and information-theoretic secure QKD system relate to device imperfections, which lead to side-channel attacks, low rates with long communication distances, and QKD coexisting implementations with classical networks \cite{QKD-Challenges_Diamanti2016, QKD-Security_Sun2022}. 
Although free-space transmission of qubits, including intracity and satellite links, enables long-range communication, it is highly affected by atmospheric turbulence, line-of-sight constraints, light pollution, and weather conditions (see Sec.~\ref{Sec_FSTransmission}). 
While fiber-based transmission overcomes such obstacles, it offers a limited range due to exponential loss scaling (see Sec.~\ref{Sec_FiberAtt}), depolarization, and Raman scattering while coexisting with classical channels (see Sec.~\ref{Sec_SRS}).

\subsubsection{BB84 QKD}

BB84 was the first QKD protocol proposed by Charles Bennett and Gilles Brassard in 1984 \cite{QKD_BB84}. 
The original proposal involves the random preparation of qubits by Alice in horizontal ($\ket{H}$), vertical ($\ket{V}$), diagonal ($\ket{+}$), or anti-diagonal ($\ket{-}$) states of polarization. 
Bob receives and measures the qubits, randomly in either a rectilinear or diagonal basis. 
Alice and Bob further post-select the events when both preparation and measurement were done in the same basis by communicating over a classical channel. 
Successful detection of such an event results in a cryptographic key after post-processing steps such as error correction and privacy amplification to remove any information leakage to an eavesdropper \cite{QKD-PostProcess_Fung2010}. 

\begin{figure}[ht!]
    \centering
    \includegraphics[width = \textwidth]{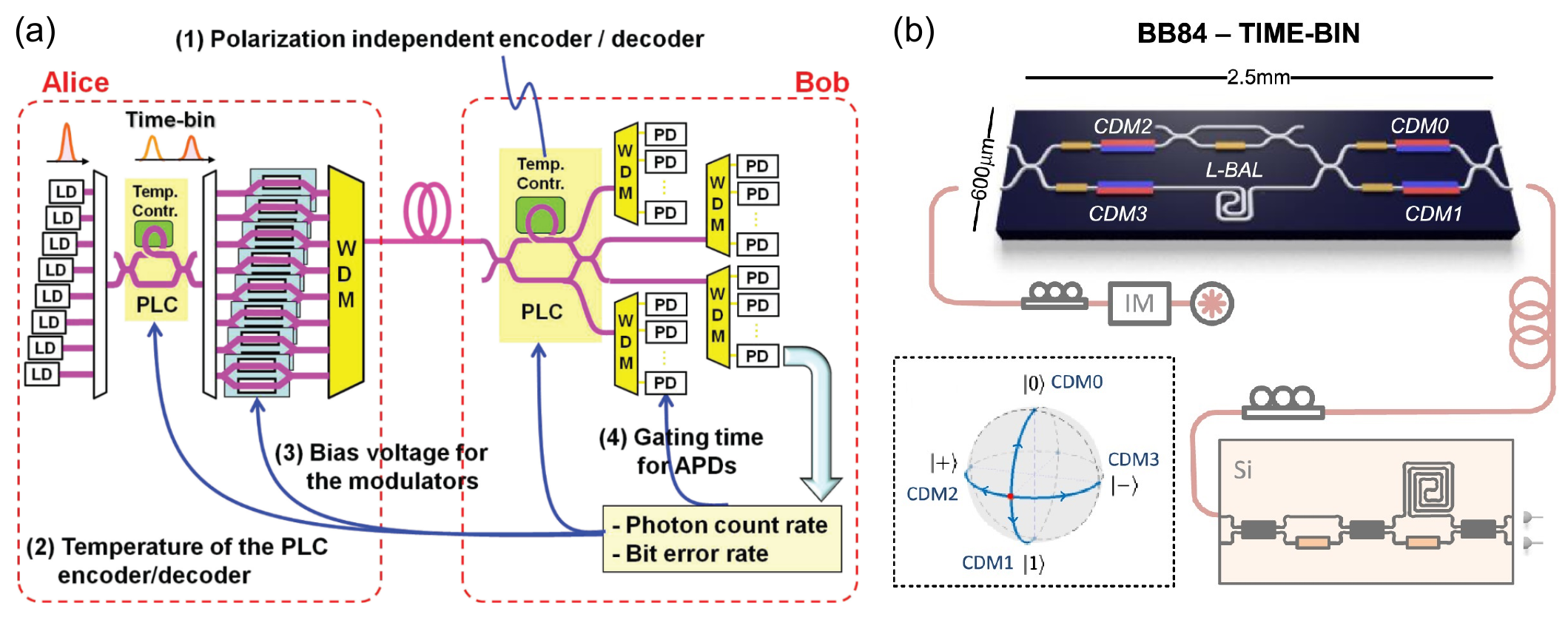}
    \caption[BB84 QKD using time-bin qubit.]{Recent key demonstrations of BB84 QKD using TBQs. (a) Implementation of eight-channel wavelength-division multiplexing for signals in the BB84 protocol, featuring a single time-bin encoder at Alice and a polarization-independent decoder at Bob, achieving higher key rates.\textbf{ Figure reproduced under Creative Commons license from \cite{BB84_Sasaki2015}}. (b) On-chip generation of s using a tunable asymmetric Mach-Zehnder interferometer on silicon, with qubit selection controlled by tuning the path of a second balanced interferometer. \textbf{Figure reproduced under Creative Commons license from \cite{BB84_Sibson2017}}.}
    \label{Fig_BB84-Implementation}
\end{figure}

Originally, polarization encoding over free-space transmission was used to demonstrate the feasibility of the BB84 protocol in 1990 \cite{BB84_experiment}. 
Over the past three decades, numerous implementations have showcased BB84 in various forms \cite{BB84_Zavala2021}. 
Among these, TBQs have emerged as a popular choice for BB84 applications \cite{QKD-BB84TB_Ribezzo2023, QKD-BB84TB_Francesconi2023}, as illustrated in Fig.~(\ref{Fig_BB84-Implementation}). Walton et al. \cite{TB-QKD-DFSS_Walton2003} proposed a BB84-type time-bin QKD protocol implemented within a decoherence-free-subspace framework, combining one-way autocompensation and passive detection to achieve intrinsic immunity to collective phase noise. Three-state BB84 QKD protocol has also been used for field trials over a 21 dB lossy channel in the Florence metropolitan area \cite{ThreeState-QKD_Bacco2019}, generating a secret key rate of 3.4 kbits/s.
Recent advancements include high-speed QKD demonstrated by Sax et al., utilizing TBQs with a 2.5 GHz clock \cite{BB84_Sax2023}. 
Time-bin BB84 has also been successfully integrated into QKD networks for point-to-point communication \cite{BB84_Sasaki2011, BB84_Sasaki2015}. 
Furthermore, the feasibility of TBQs for free-space QKD has been explored by analyzing multi-mode interference visibility \cite{BB84-FreeSpace_Jin2019}. Recently, Qui et al. \cite{TB-QKD_Qiu25} demonstrated time-bin QKD over a 7 km free-space link, showing robust secure key generation under realistic atmospheric conditions and high repetition rates, highlighting the practicality of time-bin encoding for long-distance free-space quantum communication. Additionally, Peter Sibson et al. demonstrated the feasibility of integrated time-bin BB84 QKD on a photonic chip \cite{BB84_Sibson2017}. More recently, on-chip time-bin QKD has been realized over field-deployed fiber using a lithium niobate photonic circuit \cite{BB84-TBChip_Heo2025}, while long-distance operation has been achieved over 120 km employing a telecom quantum dot single-photon source \cite{QKD-TBQDot_Wang2026}. Collectively, these advances highlight the technological maturity and practical viability of time-bin encoded BB84 QKD across integrated, field-deployed, and long-haul fiber platforms.

\subsubsection{Coherent One Way (COW) QKD}
\label{Sec_COW-QKD}

The COW-QKD protocol \cite{COW-QKD_Stucki2005} requires a simple experimental setup, which initially helped tailor it for high-rate state creation and detection. In this protocol, Alice transmits qubits generated using phase-coherent weak-coherent pulses in $\ket{e}$, $\ket{l}$, and $\ket{+}$ states.  Bob randomly measures the qubits in both Z and X bases (same as described in Section \ref{Sec_Measurement}) and stores only the results with identical preparation bases.  The results in the Z-basis are used for key generation, while the X-basis is used as decoy states to monitor eavesdropping by measuring the coherence between successive non-empty pulses. This measure makes the protocol robust against photon-number-splitting attacks. However, the main practical challenge is the requirement of a high-visibility and phase-stable interferometer for X-basis measurements. 

\begin{figure}[ht!]
    \centering
    \includegraphics[width=0.7\linewidth]{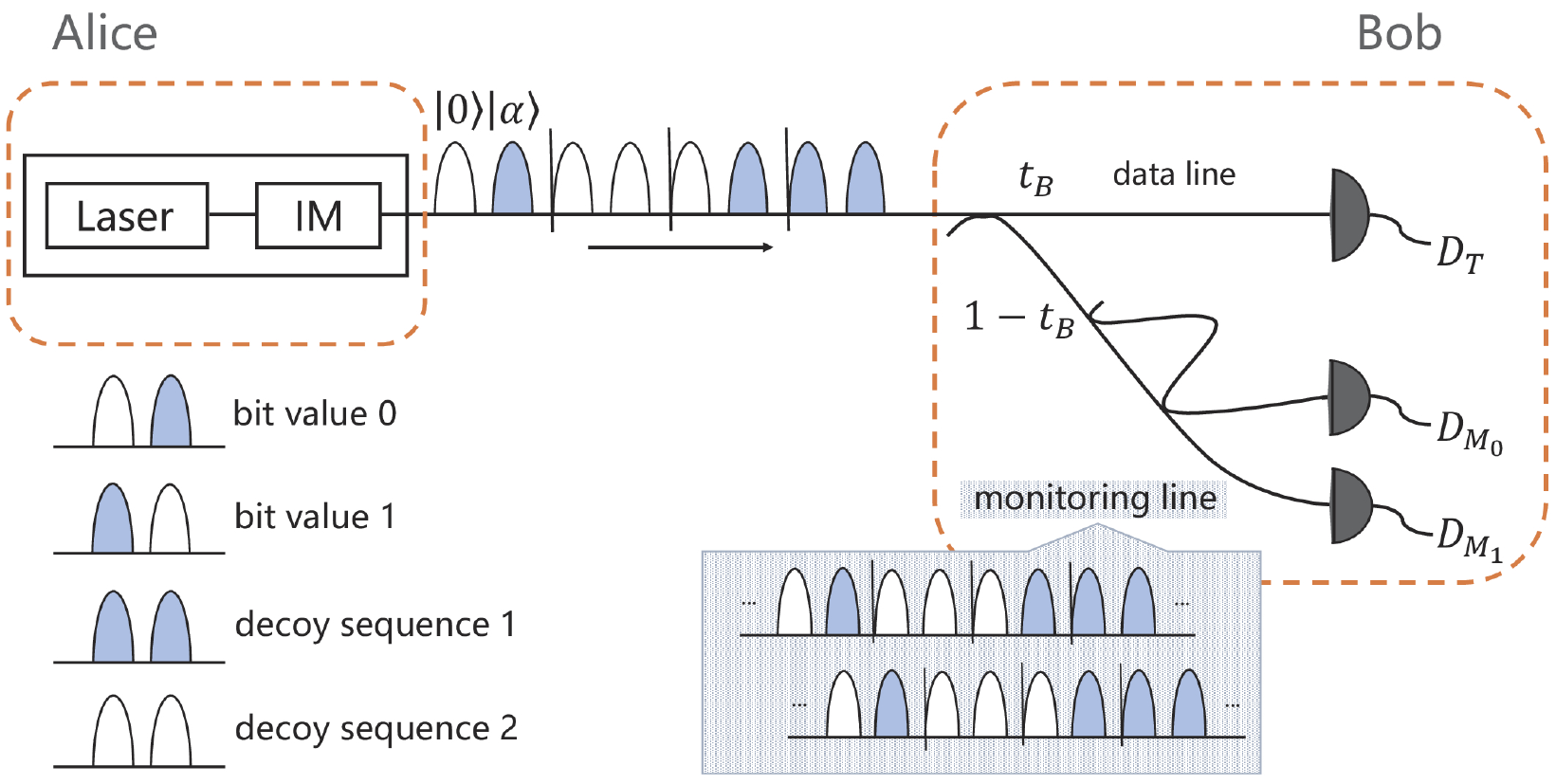}
    \caption[A schematic illustration of the COW QKD protocol.]{A schematic illustration of the COW QKD protocol with decoy and signal weak coherent pulses. \textbf{Figure reproduced under Creative Commons license from \cite{COW-QKD_Gao2022}}}
    \label{fig_COW-QKD}
\end{figure}

Early proof-of-principle demonstrations of COW-QKD were reported by Stucki et al. \cite{COW-QKD_Stucki2005, COW-QKD_Stucki2007}, followed by progressively longer-distance and higher-loss implementations \cite{COW-QKD_Zbinden2009_1, COW-QKD_Zbinden2009_2, COW-QKD_Zbinden2014, COW-QKD_Korzh2015, COW-QKD_Pathak2024, COW-QKD_Abhignan2026}, achieving transmission distances beyond 300 km, channel losses exceeding 50 dB, multi-gigahertz clock rates, and operation with both InGaAs and SNSPDs over fiber links.
A modulator-free implementation was demonstrated, further simplifying the transmitter architecture \cite{COW-ModulatorFree_Roberts2017}.
Furthermore, more integrated systems, wavelength-division multiplexing, and deployment in classical networks were showcased \cite{COW-QKD_Zbinden2014, COW-QKD_Korzh2015, COW-FieldTrial_Wonfor2019, COW-QKD_Spiller2023}. 
For a simpler interferometric implementation, free-space transmission was investigated, and robustness against atmospheric turbulence was characterized using a laboratory-based turbulence simulator \cite{COW-FS_Donalson2022}.
Recent studies have explored enhancements to the COW-QKD protocol, including its optimization for lossy channels \cite{COW-QKD_Lavie2022}, zero-error attack analysis \cite{COW-ZeroErrorAttack_Rey2024}, finite-key security analysis \cite{COW-QKD_Li2024, COW-QKD_Kumar2026, COW-QKD_Pathak2026}, enhanced-security variants \cite{COW-QKD_Dadahhani2025}, simplified implementations with practical security analysis \cite{COW-QKD_Cao2026}, and extensions to high-dimensional QKD \cite{COW-HD_Gokul2024, COW-HDQKD_Sulimany2025}.
Collectively, these advances establish COW-QKD as a practically scalable protocol, combining a simplified preparation architecture with the intrinsic stability of TBQs over fiber, while supporting rigorous security analyses and diverse implementation platforms.

\subsubsection{Entanglement-based QKD} 
\label{Sec_EntQKD}

\begin{figure}[htbp!]
    \centering
    \includegraphics[width=\columnwidth]{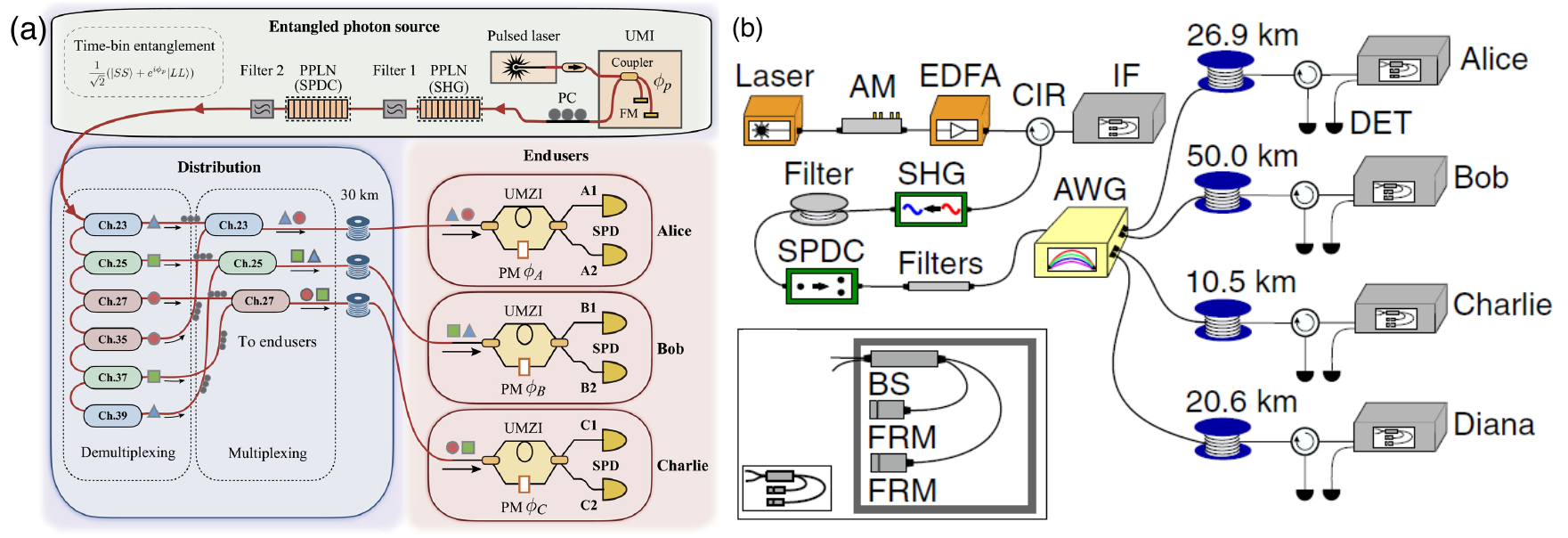}
    \caption[Time-bin entanglement based QKD.]{Experimental setup for time-bin entanglement-based QKD. (a) Demonstration of the E91 protocol utilizing a single SPDC source to generate entangled photon pairs. Spectral multiplexing is employed to distribute entanglement across multiple channels, while unbalanced Mach-Zehnder interferometers (UMZIs) are used for measurement, enabling high-fidelity entanglement-based QKD. \textbf{Figure reproduced under Creative Commons license from \cite{QCom-WDM_Kim2022}}. (b) Implementation of the BBM92 protocol involving four participants, each equipped with an unbalanced Michelson interferometer (UMI) for TBQ measurement. This setup demonstrates multi-user entanglement-based QKD, showcasing its feasibility for scalable quantum communication networks. \textbf{Figure reproduced under Creative Commons license from \cite{QNetwork-BBM92_Fitzke2022}}. }
    \label{fig_E91&BBM92}
\end{figure}

Ekart 91 (E91) is the first QKD protocol utilizing entanglement for secure communication \cite{QKD_E91}. Firstly, an entangled photon pair, e.g., in $\ket{\psi^-} = (\ket{01} - \ket{10})/\sqrt{2}$ Bell state, is distributed to the users, each of whom performs measurements in three different bases separated by $\pi/4$. For instance, Alice chooses basis at an angle of $0$, $\pi/4$ and $\pi/2$ from the rectilinear basis, while Bob chooses $0$, $\pi/4$ and $3\pi/4$. Finally, only the measurements with matching bases are used for key generation. Any eavesdropping can be monitored for CHSH violation, as the error fraction is related to the degree of violation. However, a major challenge for the E91 protocol is the requirement of at least two SPDs per user, in contrast to a single SPD for BB84. Additionally, the key rate is limited by the rate of entanglement generation. A schematic of the BBM92 protocol implemented among four parties is shown in Figure~(\ref{fig_E91&BBM92}b). The E91 protocol demonstrations have primarily focused on polarization qubits over free-space links because of the ease of projections onto multiple bases. Nonetheless, several research groups have demonstrated entanglement distribution and the E91 protocol over fiber channels using TBQs \cite{Ent-dist_Marcikic2004,  QCom-WDM_Kim2022}. 

BBM92 is another entanglement-based protocol \cite{QKD_BBM92} inspired by E91, and it is an entanglement-based variant of the BB84 protocol. 
The only difference from E91 is that users perform measurement randomly and independently on either a rectilinear or diagonal basis only. 
This strategy simplifies the experimental implementation and reduces the overhead due to basis mismatch. 
Interestingly, Fitzke et al. demonstrated the BBM92 protocol among four distant users, utilizing only a single entanglement source as shown in Figure~(\ref{fig_E91&BBM92}a). 
There have been several experimental implementations of BBM92 for polarization \cite{BBM92Pol_Wengerowsky2018, BBM92_Joshi2020Pol, BBM92_Appas2021Pol}, time-bin  \cite{BBM92_Tittel2000, BBM92_Honjo2008_IEEE, BBM92_Honjo2008_Optica, BBM92_Takesue2010, QNetwork-BBM92_Fitzke2022, BBM92_Dolejsky2023, BBM92_Jakob2024}, and time-energy \cite{BBM92_Wen2022, BBM92_Fan2023, BBM92_Bernardi2026} encoding schemes. Overall, the E91 and BBM92 protocols form the foundation of entanglement-based QKD and are inherently compatible with quantum repeater architectures. In this framework, the use of time-bin entangled states makes them particularly well suited for scalable long-distance quantum communication.

\subsubsection{Measurement Device Independent (MDI) QKD} \label{Sec_MDI-QKD}

The practical implementations of various QKD protocols are often vulnerable to side-channel attacks due to device imperfections \cite{QKD-Security_Sun2022, QKD-Sec_Brassard2000}. Side-channel attacks bypass the theoretical security guaranteed by quantum mechanics and leak information to an eavesdropper. Among the side-channel attacks, detection systems are the most vulnerable \cite{QKD-Security_Sun2022}, and MDI-QKD overcomes this vulnerability.
According to the MDI-QKD protocol \cite{QKD-MDI_HKL2012} (and similarly the Side-Channel Free QKD \cite{QKD_MDI_braunstein-2012}), both Alice and Bob prepare their qubits randomly and independently among the four BB84 states, where the probability of the state preparation is optimized for optimal key generation.
The qubits are transmitted to a central node known as Charlie, who performs a partial BSM on the two qubits and post-selects successful measurement on either $\ket{\psi^+}$ or $\ket{\psi^-}$ Bell state. 
Each successful measurement for Z-basis qubits results in a raw key, while X-basis measurements are used for error estimation. 
MDI-QKD provides a higher level of security than traditional QKD methods, as it does not rely on the measurement devices' trustworthiness and thus is immune to detector side-channel attacks.

\begin{figure}[ht!]
\centering
\includegraphics[width = \textwidth]{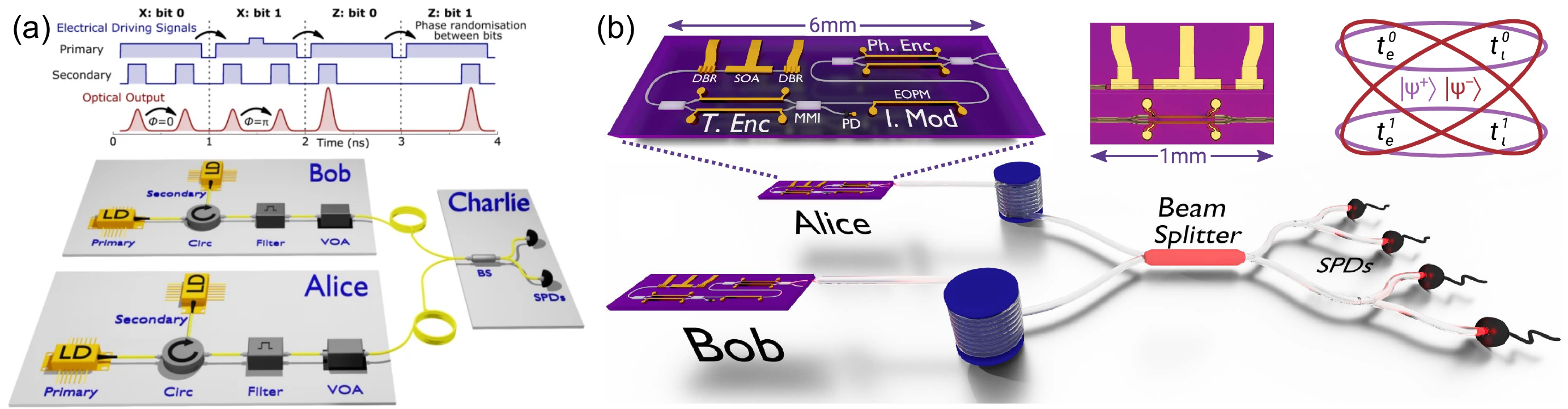}
\caption[Key experimental demonstrations of MDI-QKD.]{Key experimental demonstrations of MDI-QKD. (a) Implementation of preparation using directly modulated lasers in a compact and cost-efficient setup, operating at a 1 GHz clock rate. This approach simplifies the system by eliminating the need for external modulators while maintaining high-speed operation. \textbf{Figure reproduced under Creative Commons license from \cite{MDI_Woodward2021}}. (b) Full integration of source devices utilizing indium phosphide (InP) photonic integrated circuits, enabling scalable and stable MDI-QKD implementations. The use of InP waveguides allows for monolithic integration of laser sources, modulators, and detectors, paving the way for practical and deployable quantum communication networks. \textbf{Figure reproduced under Creative Commons license from \cite{MDI_Semenenko2020}}.}
    \label{Fig_MDI-Implementation}
\end{figure}

Since its proposal by Lo et al. in 2012 \cite{QKD-MDI_HKL2012}, significant efforts have been dedicated to realizing experimental implementations of MDI-QKD. 
The first proof-of-principle experiments were showcased in 2013 using weak coherent TBQs, and decoy states \cite{MDI_Rubenok2013, MDI_Liu2013}.
Fig. \ref{Fig_MDI-Implementation} highlights two notable experiments that utilize TBQs. 
In particular, a record transmission distance of 404 km was achieved using ultra-low-loss fiber \cite{MDI_Yin2016}, while field demonstrations were carried out over a metropolitan network with three users in a star-type topology sharing a central node \cite{MDI_Tang2016}. 
Significant work has also been done towards commercially viable and cost-effective demonstration by simplifying the qubit preparation method and integrating feedback systems to improve photon indistinguishability \cite{MDI_Valivarthi2017, MDI_Anuj2023, MDI_Shao_2025}.  
Further advancements include the development of MDI-QKD systems based on direct diode laser modulation via injection locking \cite{MDI_Woodward2021} and the successful coexistence of MDI-QKD with classical communication networks \cite{MDI_Berrevoets2022}. 
Additionally, reference-frame-independent MDI-QKD was demonstrated by scrambling the polarization of TBQs \cite{MDI_Wang2017}. 
Efforts to develop integrated photonic chips for MDI-QKD using TBQs are also underway, with implementations utilizing silicon for polarization encoding \cite{MDI_Wei2020, MDI_Cao2020} and indium phosphide (InP) waveguides \cite{MDI_Semenenko2020}. Recently, Lu et al. \cite{MDI-QKD_Lu2025} experimentally demonstrated 303 km MDI-QKD incorporating general source-preparation uncertainties via the operator dominance method, achieving enhanced practical security without sacrificing long-distance performance.  In an attempt toward free-space MDI-QKD, Cao et al. \cite{MDI-QKD_Cao2020} demonstrated secure key distribution over a 19.2 km atmospheric channel using adaptive optics and precise source synchronization, paving the way for satellite-based implementations. As evident from these demonstrations, most MDI-QKD implementations employ TBQs rather than polarization qubits to avoid depolarization in single-mode fibers.

\subsubsection{Higher-dimensional QKD}

Most conventional QKD protocols encode information in qubits, which are represented by a two-dimensional Hilbert space. 
Hardware limitations, such as the qubit preparation rate or the long recovery time (dead-time) of single-photon detectors, constrain the secret key rates of the qubit-based protocols. 
These limitations can be overcome by encoding information in higher-dimensional quantum states as discussed previously in Section~(\ref{TBQudits}). 
Additionally, higher-dimensional quantum states are more resistant to noise, meaning that the protocols implemented with these states are tolerant to higher quantum bit error rates \cite{HD-QKD_Sheridan2010}. 
However, using higher-dimensional encoding increases the complexity of the implementation, thus making it more challenging for scalability. 
Similar to other QKD protocols, the high-dimensional time-phase protocol uses a rectilinear basis for the exchange of key information and a superposition basis (phase states) to monitor eavesdropping \cite{HD-QKD_Bechmann2000, HD-QKD-Sec_Cerf2002, HD-QKD_Sheridan2010, HD-QSS_Takesue2006}. 
The preparation and measurement of time-bin qudits follow the discussion described in Sections~(\ref{Qditprep}) and (\ref{Qditmeas}), respectively.

\begin{figure}[ht!]
    \centering
    \includegraphics[width=0.9\linewidth]{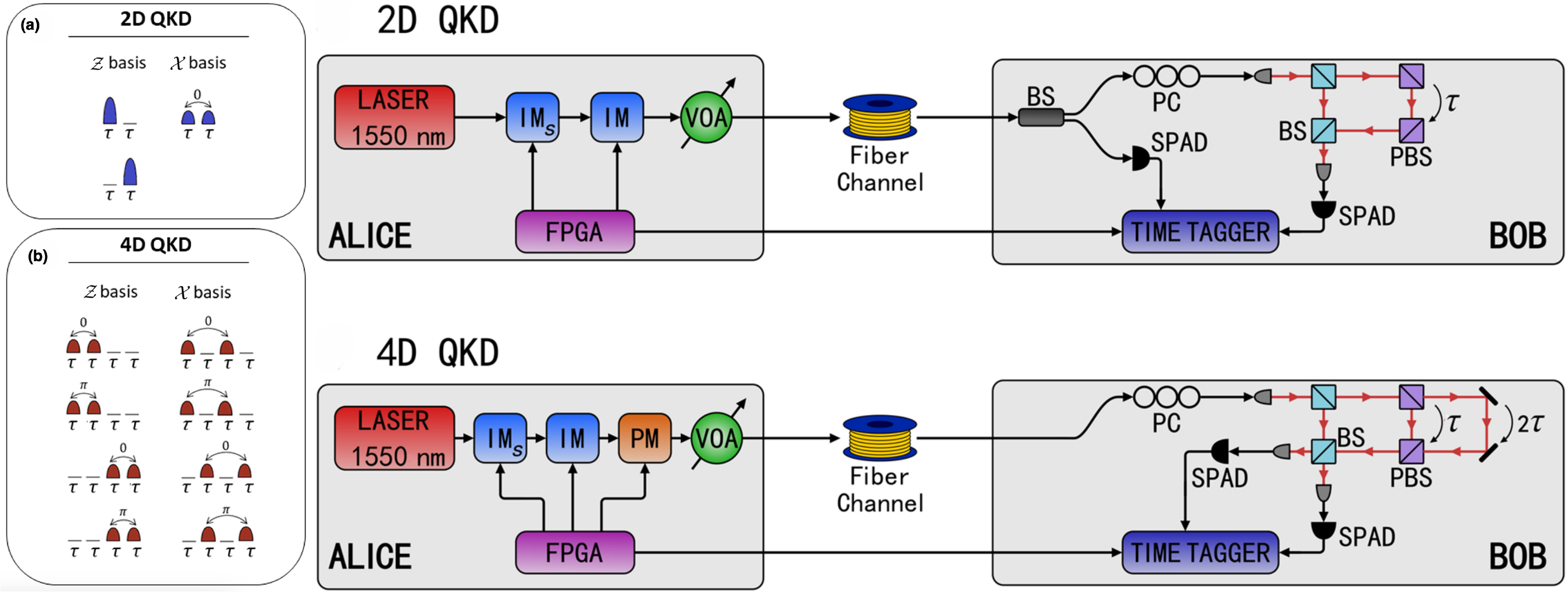}
    \caption[Higher dimensional QKD]{Experimental setups and quantum states for (a) three-state BB84 using qubits and (b) four dimensional (4D) QKD. The two-dimensional implementation requires only three states, while the four-dimensional implementation requires eight states. The time-bin duration and the interferometer delay are denoted by $\tau$. \textbf{Figure reproduced under Creative Commons license from \cite{HD-QKD_Vagniluca2020}}.}
    \label{Fig_HD-QKD}
\end{figure}

A comparison of the required quantum state and practical implementation between two and four-dimensional QKD is shown in Fig.~(\ref{Fig_HD-QKD}) where higher dimensionality requires more sophisticated measurement. 
More demonstrations of QKD systems with high-dimensional time-phase qudits can be found in references \cite{HD-QKD_Lucamarini2013, HD-QKD_Brougham2013, TE-HD-QKD_Zhong2015, HD-QKD_Lee2016, HD-QKD_Islam2017, HD-QKD_Islam2019, HD-QKD_Vagniluca2020, HD-QKD_Sulimany2021, HD-QKD_Chang2023}. 
High-dimensional QKD has also been demonstrated using entangled qudits \cite{HD-QKD_Yu2025}, where the protocol is similar to BBM92, as described in Section~(\ref{Sec_EntQKD}).  Recently, Ogrodnik et al. \cite{HD-QKD_Ogrodnik2025} demonstrated a high-dimensional time-phase BB84 QKD using a single detector per basis, showing 2D and 4D performance in an urban fiber network and highlighting security impacts on key rates.
A significant drawback of time-bin qudit-based QKD schemes is that detecting a $d$-dimensional phase state requires $2d-1$ DLIs, while the state detection efficiency decreases as $1/d$ \cite{HD-QKD_Islam2017}. Consequently, scaling the protocol to larger $d$ becomes challenging.

\subsection{Quantum teleportation} 
\label{Sec_Teleport}

\begin{figure}[ht!]
     \centering
     \includegraphics[width=0.90\columnwidth]{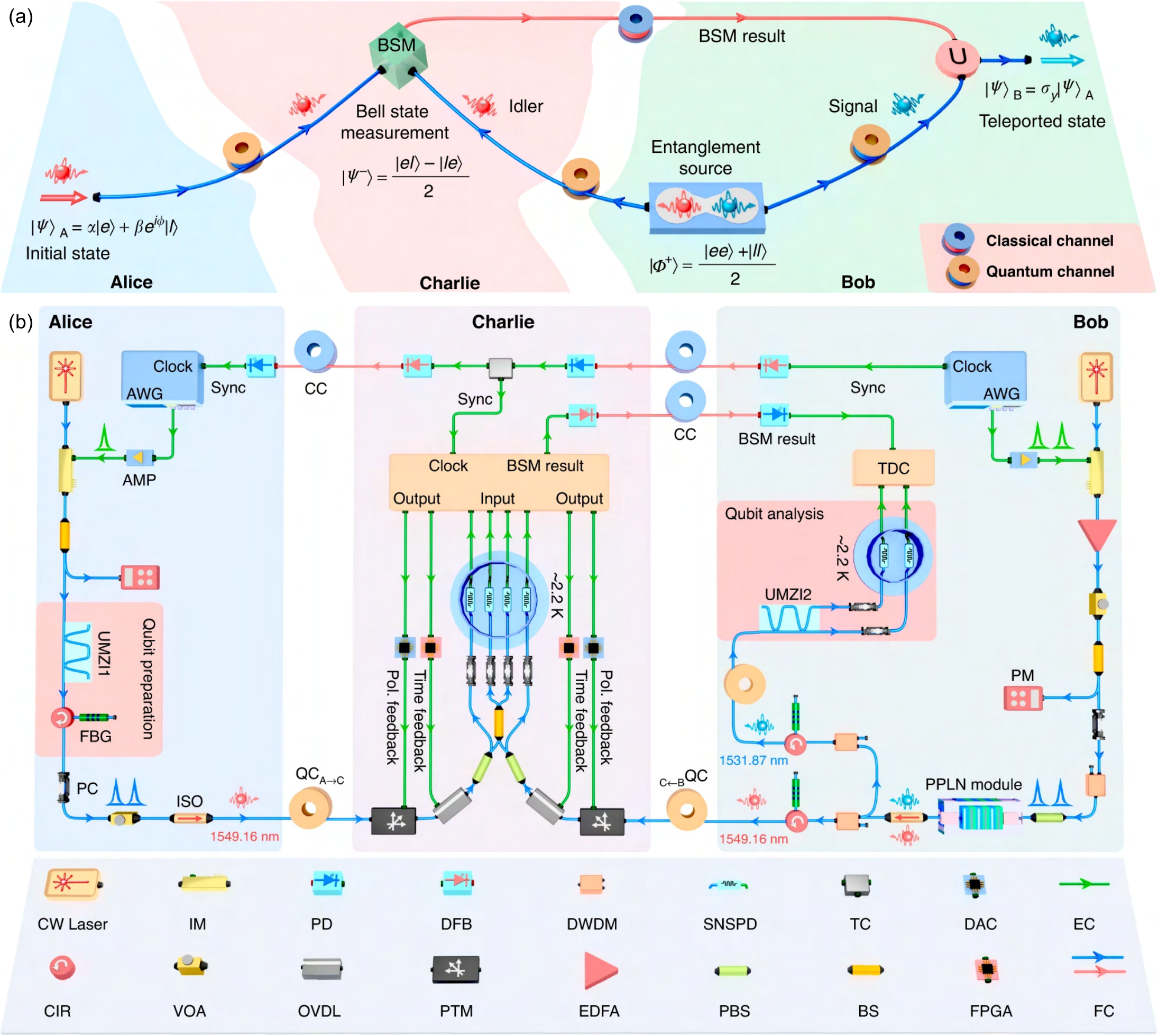}
     \caption[Teleportation over optical fibers using time-bin qubits.]{Implementation of teleportation protocol using TBQ. (a) Scheme of the teleportation protocol where Alice prepares the initial state with a weak coherent source and transmits it to Charlie for BSM with an idler photon from Bob's entangled pair. (b) Detailed experimental setup for teleportation implementation over optical fiber. Alice generates short laser pulses, prepares them using an interferometer, filters and attenuates them, and transmits them to Charlie via a quantum channel. Bob generates pump pulses, creates time-bin entangled photon pairs, filters them, and sends the idler photons to Charlie while storing the signal photons for analysis. Charlie performs Bell-state projection using a beam splitter and single-photon detectors, actively stabilizing arrival times and polarization. Synchronization and security are maintained through feedback systems, optical isolators, and classical communication channels. \textbf{Figure reproduced under Creative Commons license from \cite{Teleportation_Shen2023}}.}
     \label{fig_Teleportation}
 \end{figure}

Quantum teleportation is a method of transmitting an unknown quantum state without physically sending the particle encoding the quantum state. It uses pre-shared entanglement, BSM, and classical communication \cite{Teleportation_Bennet1993}. 
For quantum teleportation between two distant parties, Alice and Bob, first entangled photons are distributed between them. 
Alice performs a BSM on one of the entangled pairs and the qubit to be teleported, $\ket{\psi}_A$ as shown in Fig.~(\ref{fig_Teleportation}). 
The results of the BSM are communicated to Bob for performing local unitary operations on his qubit to recover the teleported state. Since linear-optical quantum teleportation is fundamentally limited by the restricted distinguishability of the four Bell states, allowing at most a 50\% successful BSM, ancilla-assisted schemes have been proposed to boost the teleportation success probability beyond 50\% \cite{BoostedTeleport_Daurelio2025}

The first experimental demonstration of teleportation employed polarization qubits for in-lab free-space transmission and performed the BSM onto $\ket{\psi^-}$ state only \cite{Teleportation_Zeilinger1997, Teleportation_Kimble1998}. A parallel effort also demonstrated teleportation of path-polarization states with distinguishability of all four Bell states \cite{Teleportation_Boschi1998}.
Teleportation has been demonstrated in both free-space \cite{Teleportation_Yin2012, Teleportation_Ursin2004} and fiber optic channels \cite{Teleportation-Review_Iulia2022, Teleportation_Raju2016, Teleportation_Raju2020}; however, free-space involves using expensive, large-aperture optics and complex techniques. 
Thus, optical fiber is a preferred mode of communication for intracity and intercity links, although the losses are relatively higher compared to terrestrial free space links. 
Marcikic et al. showcased teleportation using TBQs at telecom wavelengths, making it compatible with optical fiber networks for long-distance quantum communication  \cite{Teleportation_Marcikic2003, Teleportation-relay_Riedmatten2004}. Technological improvements in experimental devices have fueled the performance of teleportation, enabling network demonstrations \cite{Teleportation_Landry2007, Teleportation_Raju2016, Teleportation_Sun2016}, longer distances \cite{Teleportation_Takesue2015}, and higher clock rates \cite{Teleportation_Anderson2020}. 
Figure~(\ref{fig_Teleportation}) shows an experimental demonstration of the quantum state transfer at a rate of a few Hz over a 64-km-long fiber channel across the University of Electronic Science and Technology of China \cite{Teleportation_Shen2023}
Recently, photonic qubits at telecom wavelengths have been teleported onto matter qubits in solid-state rare-earth-ion-based quantum memories \cite{Teleportation_Rivera2023}. Teleportation has also been demonstrated between a memory-compatible TBQ at 795~nm and a solid-state quantum network node \cite{Teleportation_Iuliano2024}, and the storage of teleported TBQs in erbium-ion ensembles has likewise been realized \cite{Teleportation-TB_An2025}. Beyond qubit systems, teleportation has been extended to high-dimensional photonic states \cite{Teleportation-HD_Luo2019}. Comprehensive reviews of photonic quantum teleportation experiments can be found in Refs.~\cite{Teleportation-Review_Pirandola2015, Teleportation-Review_Xia2017, Teleportation-Review_Iulia2022, Teleportation-Review_Hu2023}.

\subsection{Entanglement swapping} 
\label{Sec_EntSwap}

\begin{figure}[htbp!]
\centering
\includegraphics[width=0.9\columnwidth]{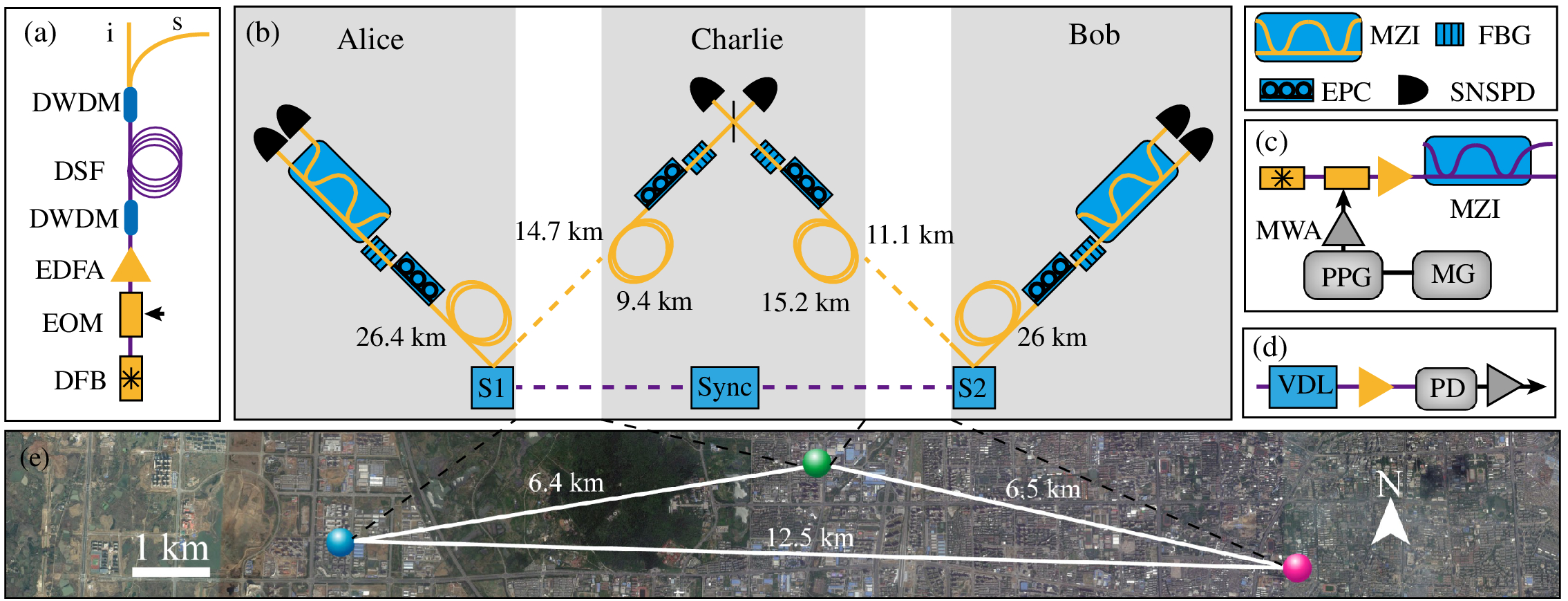}
\caption[Entanglement swapping over optical fibers using time-bin qubits.]{Entanglement swapping with distantly located SFWM-based sources. Qubits are prepared at 1 GHz, with synchronization achieved via clock distribution from Charlie. (a) Time-bin entangled photon-pair source setup utilizing dispersion-shifted fibre (DSF) for SFWM. (b) Experimental realization with coiled idler photons and signal photons transmitted to Charlie. (c) Synchronization signal generation at 500 MHz, doubled via an MZI. (d) EOM driving signal setup. (e) Satellite image of experimental nodes. \textbf{Figure adapted from \cite{Swap_Sun2017} \textcopyright~ Optical Society of America }.}
\label{fig_Ent_swap}
 \end{figure}

Entanglement swapping -- sometimes also denoted as entanglement teleportation -- is a process where two photons from separate entangled pairs become entangled after a BSM is performed on their respective partners \cite{Swap_Zukowski1993}. 
Consider two independent time-bin entangled photon pair sources, denoted as S1 and S2 in Fig.~(\ref{fig_Ent_swap}), each generating signal ($s$) and idler ($i$) photons. 
Charlie performs a joint BSM (as described in Section~\ref{Sec_BSM}) on idler photons, and the entanglement is transferred onto the unrelated signal photons. 
Concatenated entanglement swapping is used to generate entanglement over lossy channels in quantum repeaters \cite{QRepeat_Briegel1998, QRepeat_Simon2007, QRepeat_Sangouard2011, QRepeat_Wu2020, QRepeat_Johannes2020}, enabling long-distance quantum communication \cite{QMemory_Jobez2016, QMemory_Tiranov2015}. 
The first experimental demonstration of entanglement swapping was showcased by Pan et al. in 1998 \cite{Swap_Pan1998} for polarization entangled photons. Subsequently, this technique was also used to showcase the preservation of non-local quantum correlations of entangled photons \cite{Swap_Jennewein2001}.

TBQs have been used in out-of-the-lab entanglement swapping experiments, as shown in Fig.~(\ref{fig_Ent_swap}). 
A characteristic result of entanglement swapping experiments is measuring the visibility of the four-fold coincidence of the two-photon pairs. 
The visibility is proportional to $\sim\left[1+\cos(\phi_a + \phi_b - 2\phi_p)\right]$, where $\phi_a$ and $\phi_b$ are the phases of the two measurement interferometers for signal photons at Alice and Bob as shown in Fig.~(\ref{fig_Ent_swap}) and $\phi_p$ is the time-bin phase of pump pulses \cite{Swap_Sun2017}.
A series of experimental demonstrations has progressively advanced the feasibility of entanglement swapping for quantum networks. 
In 2005, the Geneva group achieved a visibility of 80\%, using TBQs transmitted over a 2.2 km optical fiber link \cite{Swap_Riedmatten2005}. 
They later demonstrated entanglement swapping with photons generated by a continuous-wave pump instead of a pulsed pump, thereby relaxing the stringent synchronization requirements \cite{Swap_Halder2007}. 
A separate experiment utilized SFG to achieve high-fidelity entanglement swapping without the need for post-selection, offering a scalable route for quantum communication \cite{Swap_Sangouard2011}.

Jeongwan et al. demonstrated swapping between time-bin entangled photon pairs at 795 nm and 1533 nm, where the 795 nm photons are compatible with quantum memories--an essential step toward practical quantum repeaters \cite{Swap_Jeongwan2015}. 
Takesue et al. implemented entanglement swapping using entangled pairs generated via SFWM in dispersion-compensated fibers \cite{Swap_Takesue2009}. 
A notable milestone was reached by Sun et al., who reported entanglement swapping over more than 100 km of optical fiber at a 1 GHz clock rate using TBQs \cite{Swap_Sun2017}.
More recently, high-visibility (91\%) entanglement swapping was demonstrated using micro-ring resonators with SFWM and a continuous-wave pump \cite{Swap_Samara2021}, while Davis et al. conducted a demonstration using telecom-wavelength TBQs, marking another step forward for fiber-compatible quantum networks \cite{Swap_Davis2025}. These demonstrations highlight that time-bin encoding not only enables high-fidelity entanglement swapping over long distances but also pave the way for quantum networks.

\subsection{Quantum Networks}
\label{Sec_QNetworks}

Quantum networks enable the coherent generation, transmission, and processing of quantum information across spatially separated nodes. 
At their core, these networks link quantum processors via photonic channels that can distribute entanglement and teleport quantum states, forming the basis for distributed quantum applications. 
As an extension of this concept \cite{Quinternet_Kimble2008}, a ``quantum internet'' has been envisioned--an entanglement-based network capable of performing tasks that are classically impossible \cite{Quinternet_Wehner2018}. 
Even with only modest local quantum processing capabilities, such a network could enable powerful applications, including QKD \cite{QNetwork_Liao2018}, distributed quantum computing \cite{QNetwork_DistriQComp_cuomo_2020}, blind quantum computing \cite{QNetwork_BlindQComp_fitzsimons-2017}, distributed quantum sensing \cite{QNetwork_Qtelescopy_gottesman-2012, QNetwork_QSensing_zhang-2020}, and precision clock distribution \cite{QNetwork_ClockSync_komar-2014}.

The computational advantage of quantum networks stems not only from entanglement distribution but also from their exponentially larger state space. 
Specifically, a classical network of $k$ quantum nodes each with $n$ qubits has a state space of dimension $k2^n$, whereas a fully quantum network achieves a state space of dimension $2^{kn}$, offering dramatically enhanced processing potential for quantum nodes with limited processing capabilities \cite{QNetworks_Luo2024, QNetwork_Bassoli2021, QInt_Kumar2025}. 
A key challenge in realizing such a network is achieving coherent control over the interaction of light and matter at the single-photon level over long distances \cite{Quinternet_Kimble2008}. 
This is essential for maintaining quantum coherence and distributing entanglement in the presence of channel losses and noise. 
To overcome these limitations, quantum repeater protocols have been proposed to enable scalable and fault-tolerant entanglement distribution over lossy channels \cite{QRepeater_DLCZ2001, QRepeat_Briegel1998, QRepeat_Sangouard2011}. 

\begin{figure}[htb!]
    \centering
    \includegraphics[width=0.90\linewidth]{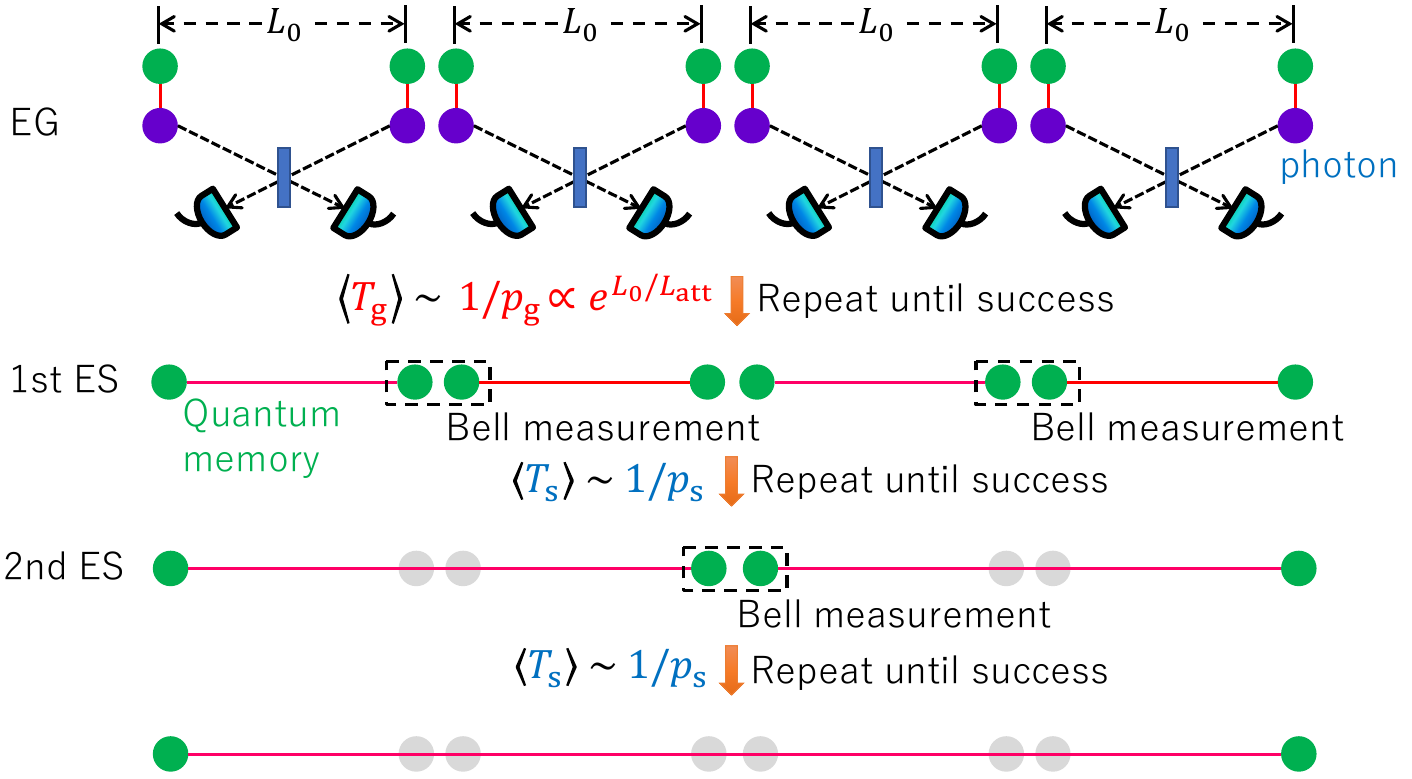}
    \caption[Quantum repeater protocol]{Steps of quantum repeater protocol for three nodes located at regular intervals between Alice and Bob, who are separated by a distance of $4L_o$. The protocol starts by entanglement generation (EG) based on the application of the linear-optical Bell measurement with success probability $p_g$. It is followed by entanglement swapping (ES) with success probability $p_s$. \textbf{Figure reproduced with permission from \cite{Quinternet-RMP_Koji2022}.}}
    \label{fig_QuantumNetworks}
\end{figure}

As illustrated in Fig.~(\ref{fig_QuantumNetworks}), quantum repeater protocols divide the full communication distance into shorter segments by placing repeater nodes at regular intervals between the end users. 
Entanglement is first generated between adjacent nodes using probabilistic BSMs, followed by entanglement swapping operations to extend across the network.
In order to synchronize these operations across all the nodes, quantum memories are required at each node to store entangled states until all neighbouring segments are ready. A detailed review on different types of quantum memories and their implementations can be found here  \cite{QNetwork_Wei2022}. 
Due to the probabilistic nature of both entanglement generation and swapping, these steps must often be repeated many times, with success rates highly dependent on inter-node distances and component efficiencies as highlighted in the Fig.~(\ref{fig_QuantumNetworks}). 
Although resource-intensive, repeater-based architectures significantly improve scalability by transforming the exponential scaling of photon loss in direct transmission into a polynomial scaling, thereby enabling long-distance quantum communication.

The demanding requirements for entanglement distribution, quantum frequency conversion, high-fidelity BSMs, and long coherence times necessary for real-time feed-forward have so far prevented the realization of full-scale quantum repeater architectures. 
However, TBQs provide a feasible way forward through these challenges due to their cross-compatibility with different processes, platforms and existing infrastructure, as discussed previously.
In particular, TBQs facilitate efficient light-matter interactions in devices such as quantum memories, where they offer superior interaction compared to polarization qubits, which may suffer from inefficiencies due to strict polarization alignment. 
Additionally, multi-user frequency conversion-based quantum networking has been demonstrated with time-bin entanglement \cite{QCom-TBEnt_Qi2025}, including fully connected topologies across multiple users \cite{QNetwork-TB16user_Huang2025}, and distribution of high-rate sequential time-bin entanglement in metropolitan fiber networks \cite{Ent-Dist_Achleitner2025}. Next-generation heterogeneous prepare-and-measure networks further demonstrate the feasibility of integrating multiple platforms and protocols in a single network infrastructure \cite{QNetwork_Lu2025}.

Significant progress has been made towards the implementation of elementary quantum repeater schemes employing TBQs \cite{Quinternet-RMP_Koji2022}. 
Notably, demonstrations using nitrogen-vacancy (NV) centers have realized quantum teleportation in early quantum network prototypes with time-bin encoded photons \cite{QNetwork-NV_Pompili2021}, including successful teleportation between non-neighbouring nodes \cite{QNetwork-NV-Teleportation_Hermans2022}. 
Additional platforms utilizing TBQs include silicon-vacancy (SiV) centers \cite{SCTLS-QNet-SiV_Lukin2024} and trapped barium ions serving as quantum memories \cite{QNetwork-BaIons_Saha2025}. 
Recently, Xiang et al. \cite{ET_Ent-QNentwork_Xiang2025} demonstrated a multi-function quantum network using multiplexed time-energy entanglement, enabling simultaneous time synchronization and QKD over 120 km. Their work shows how a single entangled resource can support parallel quantum tasks, advancing practical, scalable quantum networks. A more detailed comparison of various protocols and physical platforms relevant to real-world quantum networks is presented in \cite{QNetwork_Wei2022}. The field has advanced to the point where national and international testbeds now showcase progressively complex elements of quantum networking, as recently reviewed in Refs. \cite{QNetwork-TestbedsReview_Liu2024}. 

Notably, carrier-grade quantum communication infrastructures spanning over 10,000 km have now been demonstrated \cite{QComNetwork_Chen2025}, highlighting the technological maturity required for wide-area deployment. Parallel efforts toward scalable architectures leverage integrated photonics for large-scale quantum communication networks \cite{QNetworks_Zheng2026}, enabling compact, stable, and manufacturable network nodes. At the repeater level, hybrid architectures combining ensemble-based quantum memories with single-spin photon transducers offer a viable pathway toward long-distance entanglement distribution \cite{QRepeater_Gu2025}. Looking further ahead, proposals for a global quantum network integrate ground-based single-atom memories in optical cavities with satellite links to achieve intercontinental connectivity \cite{QNetwork_Ji2026}. 

\subsection{Photonic Quantum Computation}

Time-bin encoding can also be used in linear-optic photonic quantum computing \cite{QComp_Soudagar07, TB_Pittman2013}.
Compared to spatial-mode encoding, time-bin encoding offers a smaller footprint for scalability. 
Additionally, the direct compatibility with quantum communication protocols makes it a strong candidate for distributed quantum computing.
Quantum logic gates for time-bin encoding require phase shifters, beam splitters, interferometers, and switches for non-linear two-qubit operations 
\cite{QComp-TBQ_Humphreys2013, HD-QLogic_Imany2018, QComp-TBQ-Gate_Takesue2020}.  
However, minimizing losses, improving gate fidelity, and maintaining stability remain key hurdles in developing a fault-tolerant photonic quantum computer, independent of the encoding.
Typically, two-qubit gates are challenging to implement in photonic quantum qubits, as photons don't interact and are limited to probabilistic gate operations. 
Recently, McIntyre et. al. proposed a scalable protocol for heralded two-qubit gates using time-bin encoded photons that minimizes errors from photon loss in modular quantum computing systems \cite{QComp-TBQ_McIntyre2025}.
This issue was recently addressed by demonstration of deterministic two-qubit gates using hyper-entanglement of frequency-bin and time-bin encoding \cite{HD-QLogic_Imany2018}.
Moreover, multiple indistinguishable single-photon sources are required for each physical qubit, limiting the scalability of a quantum computer.
Fortunately, the recent demonstration of ultra-fast time-bin encoding \cite{QComp_Bouchard2024}, including high-fidelity quantum operations, can be a potential avenue. 
These systems have demonstrated quantum walks, which represent a specific photonic computational process, with up to 18 steps and over 95\% fidelity, underscoring the promise of time-bin encoding in scalable photonic quantum computing \cite{QWalk_Fenwick2024}.

A recent work demonstrated that $N$-qubit states encoded in a single time-bin qudit could be arbitrarily and deterministically generated, manipulated, and measured using a number of linear optical elements scaling linearly with $N$ \cite{QComp-TBQudit_Delteil2024}, in contrast to earlier approaches requiring $\mathcal{O}(2^N)$ elements \cite{QLogicSim_Cerf1998, UniversalQComp_Garcia2011}.
Higher-dimensional encoding on programmable integrated photonic chips is also being explored for quantum computation \cite{HD-QIP-Review_Chi2023}.
Additionally, Boson sampling, that is, another problem tailored to quantum photonics, has also been demonstrated using time-bin encoding with loop-based interferometers, simplifying system stabilization and enabling applications like complex graph analysis \cite{QComp-TBQ-Boson_Pan2017, QComp-TBQ-Boson_Rohde2014, QComp-TBQ-Boson_Kolthammer2022}. 
This highlights the potential of time-bin encoding in advanced quantum computing tasks. 
For more detailed information on photonic quantum computing, please refer to the following review papers \cite{QIP_Slussarenko2019, QComp-PhReview_Romero2024}. 

\section{Summary and Outlook}
\label{Conclusion}
\FloatBarrier

\subsection{Summary}

In this review, we have provided a comprehensive overview of time-bin encoded quantum states -- namely, TBQs, qudits, and time- and time-energy entanglement -- focusing on the physical principles for their generation and characterization, including experimental implementation, and their applications in modern quantum technologies. Time-bin encoding, with its intrinsic robustness against environmental impacts such as mechanical and thermal perturbations, depolarization induced by refractive index variations, and birefringence in optical fibers is compatible with existing telecommunications optical fiber infrastructure, and it has become one of the most widely used photonic qubit schemes.  

We examined TBQ preparation methods using both coherent sources and true single-photon sources, highlighting the advantages and limitations of continuous-wave and pulsed lasers, as well as deterministic and probabilistic single-photon generation methods.
The transmission of TBQs through fiber and free-space channels was discussed with particular attention to sources of noise and decoherence such as attenuation, chromatic and polarization-mode dispersion, and spontaneous Raman scattering. 
We analyzed different DLI architectures -- free-space, fiber-based, and integrated photonics-based -- for TBQ preparation and measurement. Various design considerations and phase stabilization techniques for DLIs were thoroughly discussed, along with their respective advantages and limitations. An alternative method based on light-matter interfaces for TBQ measurement was also discussed. 
TBQ characterization methods were also reviewed, including QST, fidelity estimation, and techniques for the interconversion of time-bin and polarization basis and measurement. The parameter selection of time bins was discussed with a particular focus on time-bin width and separation, and the constraints imposed by the jitter and bandwidth of the preparation and measurement devices.

We further explored the generation and characterization of bipartite, multipartite, and high-dimensional entanglement using time-bin encoding. Entanglement generation methods include SPDC, SFWM, and single-emitter systems, as well as novel approaches using fast optical switches with post-selection. 
Interference between independently generated TBQs was examined in the context of HOM interference and BSMs. While the former serves as a test of photon indistinguishability, the latter plays a vital role in entanglement swapping and quantum teleportation--key processes for quantum communication and networking applications. The concept of time-energy entanglement was motivated using energy and time correlations in the SPDC process. Unique characterization techniques of such entangled states using time-energy uncertainty, two-photon interference, and non-local dispersion cancellation were discussed. 

Applications of time-bin encoded states span a broad spectrum, including QKD, quantum teleportation, entanglement swapping, and emerging photonic quantum computing and quantum networks architectures. In particular, protocols such as BB84, COW, MDI-QKD, E91, BBM92, and high-dimensional QKD have benefited significantly from the robustness over fiber-optic channels, as well as the scalability of time-bin encoding. 
Moreover, hyperentangled photon pairs produced via the SPDC process in polarization and time-energy degree of freedom enable certain quantum information tasks such as full Bell-state analysis with linear optics, single-copy entanglement distillation, and allow higher key rates over noisy quantum channels.

\subsection{Outlook}

Over the past three decades, TBQs have gone from a neat scientific idea to a leading framework for the implementation of quantum communication protocols. The increasing adoption of TBQs has come on the heels of a set of key enabling technologies, which have alleviated many of the initial practical limitations of TBQs that indirectly favoured polarization encoded qubits. For example, the ability to distinguish TBQ states relies on the temporal resolution of the available single-photon detectors. Conversely, for polarization encoded qubits, readily available polarization beam-splitters direct different qubit states to be discerned on separate single-photon detectors. A naive avenue to overcome the limited temporal resolution of the detectors is simply to increase the time-bin separation; however, as emphasized in Sec.~\ref{Sec_ParameterSelection}, this too comes at a significant expense in terms of the necessary frequency and phase-stability of the lasers and DLIs utilized in the TBQ creation and measurement. 

An examination of the development of single-photon detection capabilities reveals a close link between their performance and photon-wavelength. 
Photo-multiplier tubes (PMTs), which rely on the photoelectric effect, have been used for optical detection for almost a century, and depending on the absorber material, can be optimized for a range of wavelengths to yield quantum efficiencies up to 20-30\% \cite{PMT_Hamamatsu2017}.  
Yet PMTs are bulky -- comprising a vacuum enclosure -- and suffer from relatively large dark-current, causing noise in the photo-detection signal. Fortunately, as semiconductor-based photo-detection was developed in the 1950s, the SPAD emerged as a compact and cost-effective alternative to PMTs. A key constraint of SPADs, however, is the intimate connection between their wavelength sensitivity and their semiconductor material platform, which in turn ties to their performance. Accordingly, SPADs based on silicon yield by far the best performance with $>90\%$ quantum efficiency, $< 10$ dark-counts per second, and down to $\sim 50$~ps timing resolution in a compact package. On the flip-side, silicon-based SAPDs are limited to a wavelength range between 300-1100~nm, which critically does not cover the telecom C-band optimal for fibre-optic channels.

For free-space channels, the photon wavelength of choice for quantum communication has, indeed, been dictated by the efficiency curve of Si-SPADs combined with the atmospheric absorption profile. But, since the polarization state of photons is minimally impacted by free-space transmission, polarization encoding has been the preferred choice, as exemplified by seminal long-distance demonstrations of QKD \cite{QKD-FS_Jacobs1996, QKD-FS_Rarity2001, QKD-global_Kurtsiefer2002, QKD-FS_Schmitt2007, QKD-FS_Hughes2002, QKD-Ent-FS_Marcikic2006, QKD-Ent_Ursin2007} and teleportation \cite{Teleportation_Ma2012, Teleportation_Jin2010}. 
It's important to note that quantum communication relying exclusively on free-space links is highly susceptible to weather conditions and limited to locations with clear line-of-sight, necessitating an alternative based on fibre-optic transmission. However, for transmission over optical fibres, polarization encoding requires complex stabilization, unlike time-bin encoding, which is substantially more robust as detailed in Sec.~\ref{Transmission}. Until recently (10-15 years ago), the bottleneck towards implementing TBQ-based communication over long-distance fibres, was the reliance on InGaAs-SPAD, which are sensitive to wavelengths between 900-1700~nm (covering all telecommunication bands) but face a pronounced tradeoff between efficiency and dark-count noise, while also typically operating in a gated mode. As a consequence, a new detection solution was essential for the widespread adoption of TBQs in fiber-optic-based channels beyond lab-scale demonstrations \cite{Teleportation_Marcikic2003}.

This leads to the introduction of high-performance SNSPDs as the key enabling technology, which has contributed to the fast progress of terrestrial quantum communication through the integration of time-bin encoding at telecommunication wavelengths. Telecommunication wavelength operation in the C-band (around 1550 nm) ensures that fiber attenuation is minimal, at around 0.2 dB/km, and fully compatible with existing optical fibers. At the same time, high-performance SNSPDs have been realized with system detection efficiency above 90\%, timing jitter below 20 ps, dark count rates below 1 cps, and saturation count rates above 100 MHz. 
The combination of low-loss telecom fibers, time-bin encoding, and high-efficiency SNSPDs has made it possible to distribute entanglement over hundreds of kilometers and perform secret key distribution over distances close to the thousand-kilometer scale, making time-bin photonics the leading technology for terrestrial quantum networks. Note that Sec.~\ref{Sec_SPDS} elaborates on the properties and performance of Si-SPADs and SNSPDs in the context of TBQ detection.

Looking ahead, future breakthroughs in integrated photonics are also expected to play a crucial role in the development of time-bin quantum technologies. The integrated photonics platform  \cite{IntQOpt_Tanzilli2011, PIC-Review_Wang2020, PIC-EntReview_Chen2021, PIC-Review_Pelucchi2021, IntPhot-QTech_Giordani2023, IntQPhotonics_Laurent2024, IntQPhotonics_Thyagarajan2025} provides compact, thermally stable, and highly reproducible implementations of unbalanced interferometers, electro-optic phase modulators, time-lens systems, and temporal-mode shaping networks. Time-bin encoding is well matched to these platforms, which tend to have large birefringence and spectral dispersion, making polarization- and frequency encoding schemes more difficult. The recent demonstration of zero-added-loss entanglement multiplexing using time-bin spectral shearing \cite{TB-Ent-Multiplexing_Chapman2025} further illustrates the potential of integrated time-frequency processing for the development of massively parallel quantum networks. Future breakthroughs in interferometric stability, low-loss phase control, and ultrafast modulation will be necessary for the development of scalable time-bin quantum transceivers.

Yet another practical consideration favouring TBQs relates to quantum light-matter interfaces needed for quantum memory and quantum wavelength transduction. Atomic media tend to have a strong polarization-dependent interaction, since it often relies on non-uniform electronic dipole or higher order moment. Hence, only a few platforms enable polarization-independent interaction for interfacing with polarization qubits \cite{QMemory_Jin2015}, and in some cases, separate atomic ensembles have been enlisted to interface with each polarization basis \cite{QMemory_England2012, QStorage_Clausen2012, QStorage_Gundogan2012, QMemory_Zhou2012, QMemory-TB_Meng2026}.  
Although some quantum memory protocols require control pulses temporally matched to the qubit, thus incompatible with time bins, many of the most promising protocols rely on photon echoes with large temporal multimode capacity \cite{QMemory-Echo_Hosseini2009, QMemory_Usmani2010, QMemory_Businger2022, QMemory_Tiranov2016, QMemory_Bonarota2011}. 
Such protocols are also typically compatible with spectrally and spatially multimodal signals, thus allowing frequency-bin or OAM-mode encoding. In that spirit, quantum states multiplexed over multiple photonic degrees of freedom have been interfaced with quantum memories, setting record-breaking total mode counts \cite{QMemory_Saglamyurek2016, QMemory_Yang2018, QMemory_Wei2024, QMemory_Teller2025}.

The extension of time-bin encoding to higher-dimensional Hilbert spaces is a promising approach for enhancing the information capacity per photon, robustness against noise, and security in QKD. These benefits will, of course, also hold for frequency-bin and spatial encoding. Time-bin qudits can be produced in a simple way using multi-path interferometric networks and pulsed optical sources; however, their implementation is still a challenge that requires innovation in ultrafast detection, low-jitter timing electronics, temporal-mode demultiplexing, and loss-resilient decoding architectures. The development of integrated time-to-frequency converters, reconfigurable temporal interferometers, and mode-selective detection systems is anticipated to be a major focus area in this field.

Time-bin encoding is expected to play a central role in future QKD testbeds, particularly in MDI-QKD. The MDI-QKD architecture closely resembles that of quantum repeater-based networks, where an intermediate, potentially untrusted node performs a Bell-state measurement (BSM) on photons transmitted by distant users. Within this topology, time-bin qubits constitute a natural encoding choice due to their compatibility with interferometric BSM modules and their ability to sustain high-visibility two-photon interference over optical fiber, while rigorously eliminating all detector-side channel vulnerabilities.

With the advancement of quantum repeater architectures, time-bin encoding is on the cusp of becoming a natural fit for modular and noise-robust implementations. Its compatibility with optical fiber transmission, quantum memories, and photonic quantum processors makes it the perfect fit for entanglement distribution over heterogeneous platforms. The key areas of research include the development of time-bin-compatible quantum memories using rare-earth doped crystals and atomic ensembles, time-frequency multiplexed repeater nodes, deterministic photon sources using on-chip interferometry, and hybrid matter-photon interfaces using temporal modes. The integration of quantum memories, nonlinear optics, ultrafast photonics, and integrated quantum transceivers is expected to facilitate the development of hybrid architectures that can interface time-bin photons with matter-based qubits, which is a crucial requirement for the development of a fully functional quantum internet.

As the second quantum revolution \cite{SecQRevolution_MacFarlane2003, SecQRevolution_Jaeger2018, SecQRevolution_Ivan2020, QuantumRevolution_Kung2021} gets underway, quantum technologies move from laboratory-scale proof-of-principle experiments to field-deployable systems. Quantum communication is also progressing towards large-scale implementation due to continuous government funding and growing industrial engagement \cite{TQI_Matt2024, QuantumWW_NIA2025, QInitiativesWW_Qureca2025, QIR-MIT_Ruane2025}. Consequently, various metropolitan, national, and global quantum networks are being developed or are already operational  \cite{QNetwork-SECOQC_Peev2009, QNetwork-DARPA_Elliott2018, QNetwork-Cambridge_Dynes2019, QNetwork-UK_Wonfor2021, QNetwork-China_Chen2021, QNetwork-IEQNET_Chung2021, QNetwork-MadQCI_Martin2024, QNetwork-Europe_Brauer2024}. This rapid development is thus ushering in the era of quantum communication as an engineering field, which will increasingly involve a diverse set of professionals other than physicists, such as photonic engineers, network designers, and system integrators  \cite{QWorkforce_Aiello2021, QWorkforce_Kaur2022, QWorkforce_Hughes2022, QWorkforce_Greinert2023, QWorkforce_Greinert2024, QWorkforce_Sander2025, QWorkforce_Li2024, QWorkforce_Adawy2025, QWorkforce_Oliver2025}. In this scenario, the need for standardization of protocols, interfaces, and terminology is thus becoming increasingly pressing. Although initial efforts are being made towards international standards for QKD and quantum networking \cite{QTechStandards_Jenet2020, QTechStandards_Deventer2022, QTechStandards_IEEE2025}, the area remains in its infancy stages \cite{QReady_Purohit2023, QNetwork-BluePrint_Alberto2024}. Until then, technical reviews will continue to be the mainstay for system-level implementation guidance.

In conclusion, time-bin encoded quantum states are on the verge of becoming a central resource for scalable photonic quantum technology. Their natural resilience in optical fibers, strong compatibility with telecommunication infrastructure, smooth integration with photonic circuits, and complementarity with high-performance SNSPDs place them at the very center of future quantum networks. As future QKD protocols, quantum repeater solutions, and distributed photonic quantum processors advance, time-bin encoding is poised to offer a unifying paradigm for modular, high-fidelity, and noise-robust quantum information processing, leading the way towards a globally interconnected quantum internet from the currently fragmented quantum links.



\section*{Acknowledgements}
 We thank Sunil Mahar for his help in redrawing some of the figures. AsS, LE, AnS, and DO acknowledge financial support from the Alberta Major Innovation Fund project on Quantum Technologies, the NSERC CREATE program QUANTA, the NSERC Alliance Quantum Consortia projects ARAQNE and Act2Qrypt, and the NRC Small Teams Initiative QPIC.  

\pagebreak
\phantomsection
\addcontentsline{toc}{section}{Appendix: List of Symbols and Abbreviations}

\section*{Appendix: List of Symbols and Abbreviations}

\begin{center}
\centering
\renewcommand{\arraystretch}{1.1}
\begin{longtable}{| p {0.25\columnwidth} | p{0.65\columnwidth} |}
\caption{\textit{\normalsize{{List of symbols.}}}} \\
\hline
 \textbf{Symbol} & \textbf{Definition} \\ 
\hline
 $\upalpha$ & Attenuation coefficient \\  
 $\alpha$ &  Probability amplitude \\ 
 $|\boldsymbol{\alpha}\rangle$ & Coherent state \\ 
 $\beta_f$ ($\beta_b$) & SRS coefficient for forward (backward) scattering \\
 $\eta$ & Attenuation in $dB$ \\
 $\lambda$ & Wavelength of laser \\
 $\Delta \lambda$ & Wavelength spread (linewidth) of laser \\
 $\mu$ & Mean-photon number \\
 $\Delta \nu_l$ & Laser linewidth\\ 
   $\Delta \nu_d$ & Drift of the laser's mean frequency \\ 
  $\Delta \nu_s$ & Source laser frequency error \\ 
 $\Delta \nu_{\tau}$ &  Pulse's spectral bandwidth   \\ 
 $\omega$ & Angular frequency of  light \\
 $\Delta \omega$ & Frequency uncertainty \\
 $\phi$ & Relative phase between $|e\rangle$  and $|l\rangle$ state or short vs. long paths \\
 $\varphi$ & Phase imparted by the phase shifter in DLI \\
 $\{|\phi^\pm\rangle, |\psi^\pm\rangle \}$  & Bell-states \\
 $\{ \sigma_0, \sigma_1, \sigma_2, \sigma_3$ \} & Pauli Matrices \\
 $\Delta \tau$ & Width of time-bins \\
  $\Delta \tau_M$ & Measured width of time-bins \\ 
 $\tau_b, ~\tau_{el}$ & Time-bin separation \\
 $d$ & Dimension of the quantum system \\
 $D$ & Dispersion parameter \\
 \{ $\{|e\rangle$, $|l\rangle \}$, $|\pm \rangle$, $|\pm i \rangle \}$ & Time-bin basis states for $\{Z, X, Y\}$ basis, respectively.\\
 $f_{rep}$ & Pulse repetition rate \\
 $F$ & Fidelity \\
 $g^{(2)}(\mathcal{T})$ & Intensity-Intensity correlation \\
 $H(V)$ & Horizontal (Vertical) polarization state \\
 $\Vec{k}$ & Momentum vector \\
 $l$ & Orbital angular momentum quantum number \\
 $L$ & Length, distance, or path delay \\
 $\Delta n$ & Standard deviation of the photon number distribution \\
$\{ s_0, s_1, s_2, s_3 \}$ & Stokes parameter \\
 $S$ & Bell-CHSH inequality parameter \\
 t & Time \\
 $t_p (t_{cd})$ & Coherence time of pump laser (down-converted photons) \\ 
 $\Delta t$ & Temporal uncertainty \\
 $u(t)$ & Time-bin envelope \\
 $\mathcal{V}$ & Entanglement Visibility \\
 $v_g$ & Group velocity \\
\hline
\end{longtable}
\label{Symbols}
\end{center}

\vspace{-1cm}
\begin{center}
\centering
\renewcommand{\arraystretch}{1.1}
\begin{longtable}{| p {0.2\columnwidth} | p{0.65\columnwidth} |}
\caption{\textit{\normalsize{List of abbreviations}.}}\\
 \hline
 \textbf{Abbreviation} & \textbf{Definition} \\ 
\hline
BS & Beam Splitter \\
BSM &  Bell-State Measurement  \\
CW & Continuous Wave  \\
dB & Decibel \\
COW & Coherent One-Way \\
CWDM & Coarse Wavelength Division Multiplexing \\
DLI & Delay Line Interferometer \\
DWDM & Dense wavelength division multiplexing \\
HOM & Hong-Ou-Mandel \\
IM & Intensity Modulator \\
MDI & Measurement Device Independent \\
OAM & Orbital angular momentum \\
PBS & Polarizing Beam Splitter \\
PID & Proportional-Integral-Derivative \\
PM & Phase Modulator \\
PMF & Polarization Maintaining Fiber \\
PPLN & Periodically Poled Lithium Niobate \\
QBER & Quantum Bit Error Rate  \\
QIP & Quantum Information Processing \\
QKD & Quantum Key Distribution  \\
QRNG & Quantum Random Number Generator \\
QST & Quantum State Tomography \\
RF & Radio Frequency  \\
RMS & Root Mean Square \\
SFWM & Spontaneous Four-wave Mixing \\
SLM & Spatial Light Modulator \\
SMF & Single-Mode Fiber \\
SNR & Signal-to-Noise Ratio \\
SNSPD &  Superconducting Nanowire Single-Photon Detector \\
SPD & Single Photon Detector \\
SPDC & Spontaneous Parametric Down-Conversion \\
SPS & Single Photon Source    \\
SRS & Spontaneous Raman Scattering \\
TBQ & Time-bin qubit \\
TDC & Time-to-digital converter \\
UMI & Unbalance Michelson Interferometer  \\
UMZI & Unbalance Mach-Zehnder Interferometer \\
VOA & Variable Optical Attenuator \\
\hline
\end{longtable}
\label{Abbreviations}
\end{center}

\setcounter{figure}{0}

\newpage
\sloppy
\addcontentsline{toc}{section}{References}
\sloppy
\bibliography{Reference}


\end{document}